\documentclass[11pt,a4paper]{article}
\pdfoutput=1
\usepackage[nottoc,notlof,notlot]{tocbibind} 
\usepackage[titles,subfigure]{tocloft} 

\usepackage[section]{placeins}
\usepackage{grffile}
\usepackage{pstricks}
\usepackage{multido}
\usepackage{multirow}
\usepackage{verbatim}
\usepackage{amsmath}
\usepackage{amsfonts}
\usepackage{amssymb}

\usepackage{psfrag}
\usepackage{hyperref}
\usepackage{graphicx}
\usepackage{epsf}
\usepackage{epsfig}
\usepackage{epstopdf}
\usepackage{graphicx} 

\usepackage{booktabs} 
\usepackage{array} 
\usepackage{paralist} 
\usepackage{verbatim} 
\usepackage{subfig} 
\usepackage{amsmath}
\usepackage{amssymb}
\usepackage{amsfonts}
\usepackage[section] {placeins}

\usepackage{geometry} 
\begin{document}

 \begin{flushright}
  \begin{tabular}{rl}
     \hspace{-0.5cm}
     \includegraphics[bb= 0in 0in 2in 2in, scale=0.38]{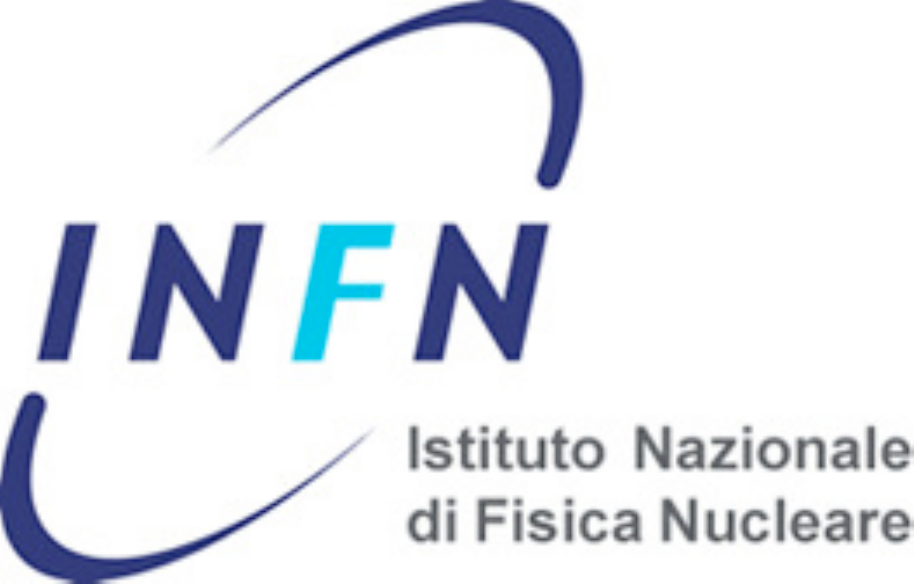} &
       \begin{tabular}{r}
         {\hspace{1cm}~\LARGE\sffamily LABORATORI~NAZIONALI~DI~FRASCATI}\\
         \\
         {\large\sffamily SIDS-Pubblicazioni}\\
       \end{tabular}
       \renewcommand{\arraystretch}{1}
     \end{tabular}
  \vskip 1cm
  
  \renewcommand{\arraystretch}{0.5}
    
      {\underline{\textbf{LNF - 10 / 21(P)}}}\\    
      {\small October 26, 2010} \\      

  \end{flushright}
  \renewcommand{\arraystretch}{1}
  \vskip 2 cm
  \begin{center}
  { \LARGE{\textbf{A 1~mm Scintillating Fibre Tracker Readout by a Multi-anode Photomultiplier}}}
	\vskip 1 cm

   B.D. Leverington, M. Anelli, P. Campana, R. Rosellini
   \vskip 0.25 cm

{\small{\it INFN, Laboratori Nazionali di Frascati, Frascati, Italy}}

\end{center}

\baselineskip=14pt
\vspace{1in}
\begin{abstract}

This technical note describes a prototype particle tracking detector constructed with 1~mm plastic scintillating fibres with a 64 channel Hamamatsu H8500 flat-panel multi-anode photomultiplier readout. Cosmic ray tracks  from an array of 11 gas-filled drift tubes were matched to signals in the scintillating fibres in order to measure the resolution and efficiency of tracks reconstructed in the fibre-based tracker. A GEANT4 detector simulation was also developed to compare cosmic ray data with MC results and is discussed in the note. Using the parameters measured in this experimental setup, modified fibre tracker designs are suggested to improve resolution and efficiency in future prototypes to meet modern detector specifications. 
\end{abstract}

\vspace*{\stretch{2}}
\begin{flushleft}
  \vskip 2cm
{ PACS: 29.40.Mc, 29.40.Gx, 42.79.Pw, 42.81.Pa } 
\end{flushleft}

\thispagestyle{empty}
\newpage
\tableofcontents
\thispagestyle{empty}
\newpage
\setcounter{page}{1}

\section{Introduction}

The rise in occupancy foreseen  in the innermost part of the LHCb tracking stations from a higher LHC luminosity,  suggests  alternative tracking solutions should be explored for the LHCb upgrade~\cite{ref_park}.  A new tracking option will require a more granular and faster detector, capable of withstanding degradation from the higher radiation rates  and must also satisfy the requirement of covering the large area of the innermost region (i.e. $4 \times 4.5$~m$^{2}$) while retaining good spatial resolution with a  low material budget. 

  In this work we present a preliminary study a position tracker made of thick scintillating fibres to replace a large area gas-based tracking detector, made of straw tubes, for the upgrade of the LHCb detector. The use of scintillating fibres in tracking detectors dates to the last couple of decades~\cite{ref_bahr} and  improvements to this technology have and continue to occur. Relatively thick scintillating fibres (in the range of 0.7 to 1~mm) would be coupled with photo-sensor detectors, most likely either Multi-Anode photomultiplier tubes (MA-PMT) or Silicon Photomultipliers (SiPM).  A quite extensive body of literature~\cite{ref_beis}--\cite{ref_bick}, along with the work presented in this report, show  that  single or double cladding scintillating fibres provide  a good light yield, can be resistant to larger radiation doses, while providing ease of operation and construction.  These could then be applied to produce  detector with a spacial resolution  which would meet LHCb tracking requirements with  minimal dead time, so as to avoid pile up effects  in the detector.

The geometry of the current Outer Tracker (OT) detector was chosen to limit the occupancy of the hottest regions in the OT detector to below 10\% at nominal run conditions at $L = 2 \times{} 10^{32}~cm^{-2}s^{-1}$~\cite{ref_terr}. The 5~mm cell diameter of the straws was chosen to limit the drift time within the period of three bunch crossings, while maintaining a manageable number of readout channels. The current OT detector has been designed to withstand irradiation intensities that correspond to an incident rate of 250~kHz per cm of wire length at nominal luminosity (500 kHz/cm$^{2}$)~\cite{ref_bach}. This corresponds to an integrated charge dose of 2.5~C/cm after 10~run-years of $10^{7}$~s. The upgraded detector should be able to withstand irradiation levels in the hottest region corresponding at least up to a luminosity of $L =  2 \times 10^{33}~cm^{-2}s^{-1}$, an increase of one order of magnitude, for 10 years of operation. The present geometry and design of the straw drift tube chambers limit their effectiveness to $L =1 \times 10^{33}~cm^{-2}s^{-1}$ with little or no margin for safety, requiring a device with smaller channel size (or faster response) and better radiation hardness, given the damage caused by the large doses of radiation expected~\cite{ref_tito}.

In terms of geometry, improvements in the occupancy can be gained by using a detector with a smaller profile: a 1~mm scintillating fibre provides a factor of 5 better than the straw drift tube, a 0.7~mm improves by a factor of 7. As well, the decay time of a scintillating fibre is typically 2.5--3.5~ns with a signal delay of 5.33 ns/m~\cite{ref_bicr}. Assuming a fast photodetector, this feature could be used to discriminate particles belonging to different bunch crossings (therefore reducing the pile-up in the detector).

\subsection{Fibre Radiation Hardness}

In gas based devices, high radiation doses result in damage to the anode wire, either creating 
deposits on the wire affecting the electric field around the wire, or causing physical damage to 
the wire surface, resulting in sparking or other effects that increase noise and reduce gain~\cite{ref_bach}. 
These effects are largely difficult to predict and are only discovered after their exposure to 
radiation \textit{in situ}.  The damage to scintillating fibres due to radiation is well understood and 
generally progresses linearly with the dose, but the damaging effects can also be undone via hot air
curing or other processes~\cite{ref_vasi, ref_zorn, ref_wick2, ref_wick, ref_bick}. 

Radiation will typically damage the scintillation light producing fluors which have been added 
to the plastic  core (typically polystyrene) of the fibre. Damage to the fluors results in a 
reduction in the intensity of the light produced by an ionizing particle track. This damage is 
impossible to recover. Different studies by manufacturers and other groups have resulted in a set 
of dyes which are resistant to damage.

Damage to the plastic core of the fibre, typically degrades the optical transmission properties of the fibres below 500~nm, degrading the bulk attenuation length with increased radiation dose. Damage to the polystyrene dominates the light yield loss and typically reduces the optical transmission at wavelengths below 500~nm by 30-60\% after 10 Mrad (6.2$\times10^{14}$~MeV/g)~\cite{ref_zorn}.   However, it is possible to recover large fraction of  optical damage to the core via annealing in hot air~\cite{ref_zorn}.  New types of fast, green, radiation-hard scintillating fibres are available from manufacturers Kuraray(SCF-MJ78(80))  and Saint-Gobain (BCF-20), such that the majority of the spectrum is above 500~nm and the radiation damage effects are mitigated.

Given the current rate of 500~kHz/cm$^{2}$ in the hot regions of the Outer Tracker , the expected rate for an order of magnitude increase in luminosity will be 5~MHz/cm$^{2}$. With a plastic fibre density of $\rho=$~1.05~g/cm$^{3}$~\cite{ref_bicr} and an energy deposit per unit length for minimum ionizing particles of 2~MeV/cm~\cite{ref_pdg}, we can expect 2~MeV$\cdot$g$^{-1}\cdot$cm$^{2}$ in the fibres. At the upgraded luminosity, the hottest spots can expect a dose rate of 10~MHz$\cdot$MeV/g. Over 10 years ($1\times10^{8}~s$) of running, this results in an maximum integrated dose in the fibres of $1\times10^{15}$~MeV/g. After 10 years of running at the upgrade luminosity, the light yield in the plastic fibres will be a little more than 30-60\% reduced without annealing, less so if annealed. Radiation damage to the tracker would result in a lower photoelectron yield, degrading the resolution and efficiency,  as the analog position resolution is expected to be proportional to the inverse of the square root number of photoelectrons, $1/\sqrt{npe}$. Irradiation tests would have to be done to quantify this effect. A digital tracker resolution would be affected by increased false-negative hits due to the decreased light yield in individual channels.

While the rate of thermal neutrons in the LHCb upgrade is currently unknown, the damage to scintillating fibres has been studied previously in References~\cite{ref_asmo}, \cite{ref_asmo2} and \cite{ref_dunn}. The thermal neutron damage results in a reduction in the number of photoelectrons and a degraded attenuation length proportional to the integrated flux of neutrons, similar to ionizing radiation damage.

\subsection{Contributions to the position resolution}

The main contributions to the resolution in a fibre tracker are represented by the
pitch of fibres, the Poisson nature of the photoelectron distribution, and the efficiency
of the photosensor to detect single photons.

\begin{figure}[h!tbp]
\centering
\includegraphics[angle=0,width=0.75\textwidth]{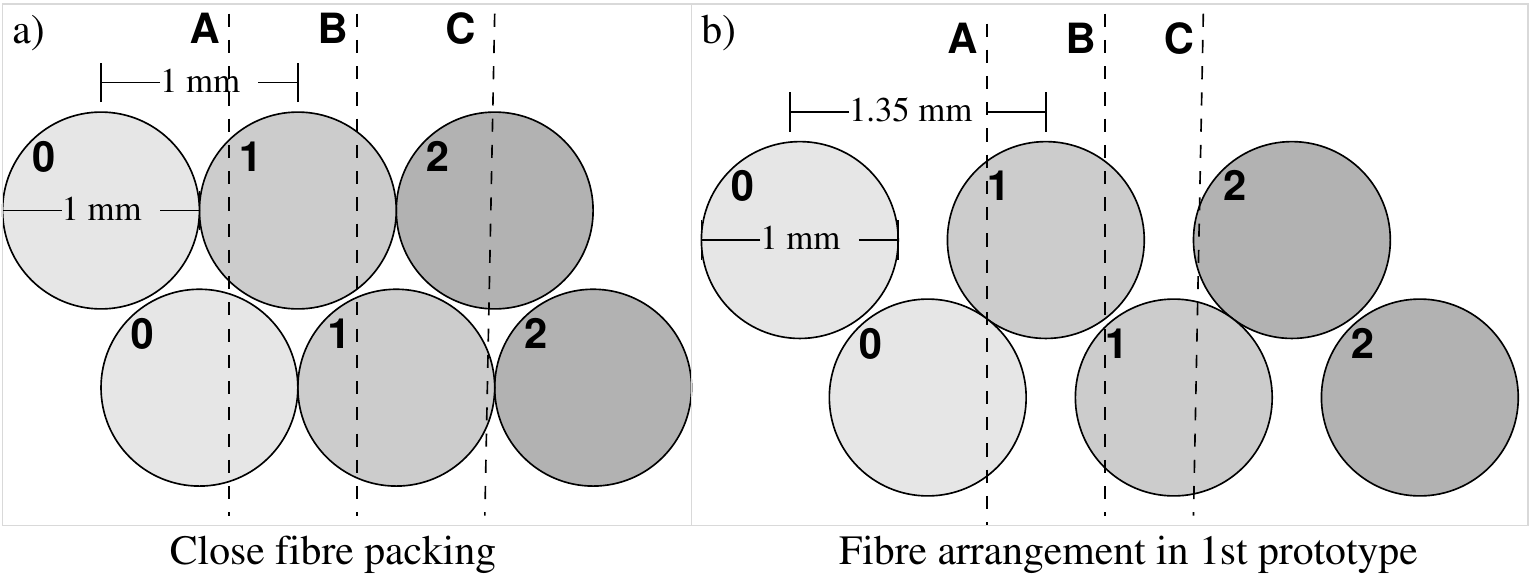} 
\caption{Round fibres overlapped by 50\% provide a resolution near width/$\sqrt{12}$, but the variation in track length in the scintillating material degrades an energy weighted resolution and will not provide more information than a digital hit based resolution. Track A will provide a good signal in both channels, track B is a single channel hit and track C provides a weak signal in one of the shared hit channels.}
\label{fig:fib1}
\end{figure}

In a binary readout mode, where only the channel producing a signal above threshold is recorded, 
the resolution is dominated by the channel width, with a strong contribution from adjacent channel crosstalk. Typically, one can gain from overlapping two 
channels such that the resolution will be proportional to the overlapped regions, as in Fig.~\ref{fig:fib1}. 
Naively, the resolution goes as the width/$\sqrt{12}$ such that two 1~mm fibre channels 
overlapped by 50\% would produce a detector position resolution of 0.15~mm, as on the left of Fig.~\ref{fig:fib1}. 
However, using round fibres, there is very little scintillating material near the edges of 
each channel and therefore, in case of very thin detector layers, each event can undergo
large fluctuation in the amount of traversed active material. This effect, coupled with the  quantum efficiency of the photodetector, light yield attenuation in the fibres and electronics crosstalk, will increase
the number of false-negative and false-positive hits degrading the resolution of the detector~\cite{ref_bahr}.

\begin{figure}[htbp]
\centering
\includegraphics[angle=0,width=0.75\textwidth]{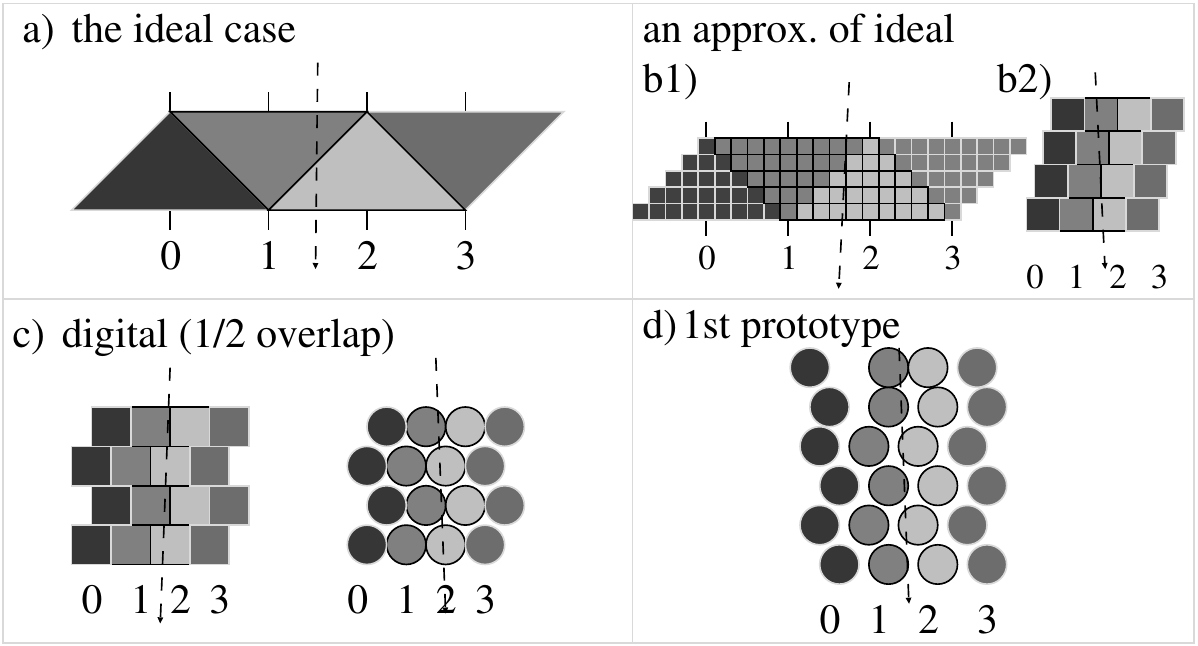} 
\caption{A sketch of 4 different scintillator arrangements. Panel a) is the ideal charge sharing case where the signal amplitude from the total track length in that channel provides position information, with b1) and b2) being good approximations of a) using square fibres. If a digital layout is preferred, c) is the optimal layout with minimal material budget for either square or round fibres, overlapping by 50\%. Panel d) is the effective fibre overlap of the first prototype.}
\label{fig:thickpos}
\end{figure}

An alternative to the binary method, is to use an analog approach. 
If one arranges the fibres  such that the position in the channel is proportional 
to the amplitude of the signal seen in that channel, one can gain additional position information 
with respect to a binary digital readout. The ideal case is in panel a) of Fig.~\ref{fig:thickpos}.  Practically, this is more easily accomplished with a few square fibres per channel as in Fig.~\ref{fig:thickpos}b) where the fibre staggering in each channel is smaller than the contribution to the position resolution from the charge resolution. The analog method will be effective if the  photosensor has good single photoelectron resolution such that one can precisely measure the relative amplitude in each channel and correctly determine the position. An analog readout would require photon counting photosensors along with charge-measuring electronics. This type of analog readout would produce an excellent resolution with larger diameter fibres, reducing the number of channels needed to cover a given region, but at the cost of increased hardware expenses per channel and overall material budget for a good light yield. Square fibres would also be preferred due to ($i$) the lack of gaps between fibres compared to round fibres and ($ii$) the narrower path-length distribution which produces a consistent signal. However, the square fibres cost approximately double compared to round fibres.  Fig.~\ref{fig:thickpos}d is the experimental prototype setup, with only a small amount of fibre overlap. Channels which have shared hits will have very good resolution, but single channel hits will dominate with a degraded resolution.

In both readout cases, the Poisson nature of the photoelectron statistics requires 
the light yield to be large so that  more than a few photoelectrons will be detected. This reduces the probability that no photoelectrons being detected to be negligible. False-negative hits severely degrade the resolution of the digital/binary readout option. 
The binary method would also need a significant light yield at the photodetector, \textgreater{}5 photoelectrons, such that low fluctuations will still exceed the electronics noise of about 1--2 photoelectrons. The analog method would require enough light to precisely 
determine the amplitude of energy deposited in that channel. 
Single clad fibres  typically produce enough light to detect 3-5 photoelectrons/mm of track length (15-25 pe/MeV deposited) with a bialkali readout of 25\% quantum efficiency  near the charged particle track. 
Double clad fibres would improve the trapping efficiency  by at least 50\%. Typical single clad fibres have a trapping efficiency (due to total internal reflection) of 3.5\% of the produced scintillation light while double clad fibres capture 5.6\% (7.3\% for square)~\cite{ref_bicr, ref_ache}. New super- and ultra-bialkali cathode photodetectors have been produced recently by Hamamatsu which have average quantum efficiencies of 30 and 42\%, respectively.

\section{The Test Setup}

\begin{figure}[htbp]
\centering
\includegraphics[width=0.5\textwidth]{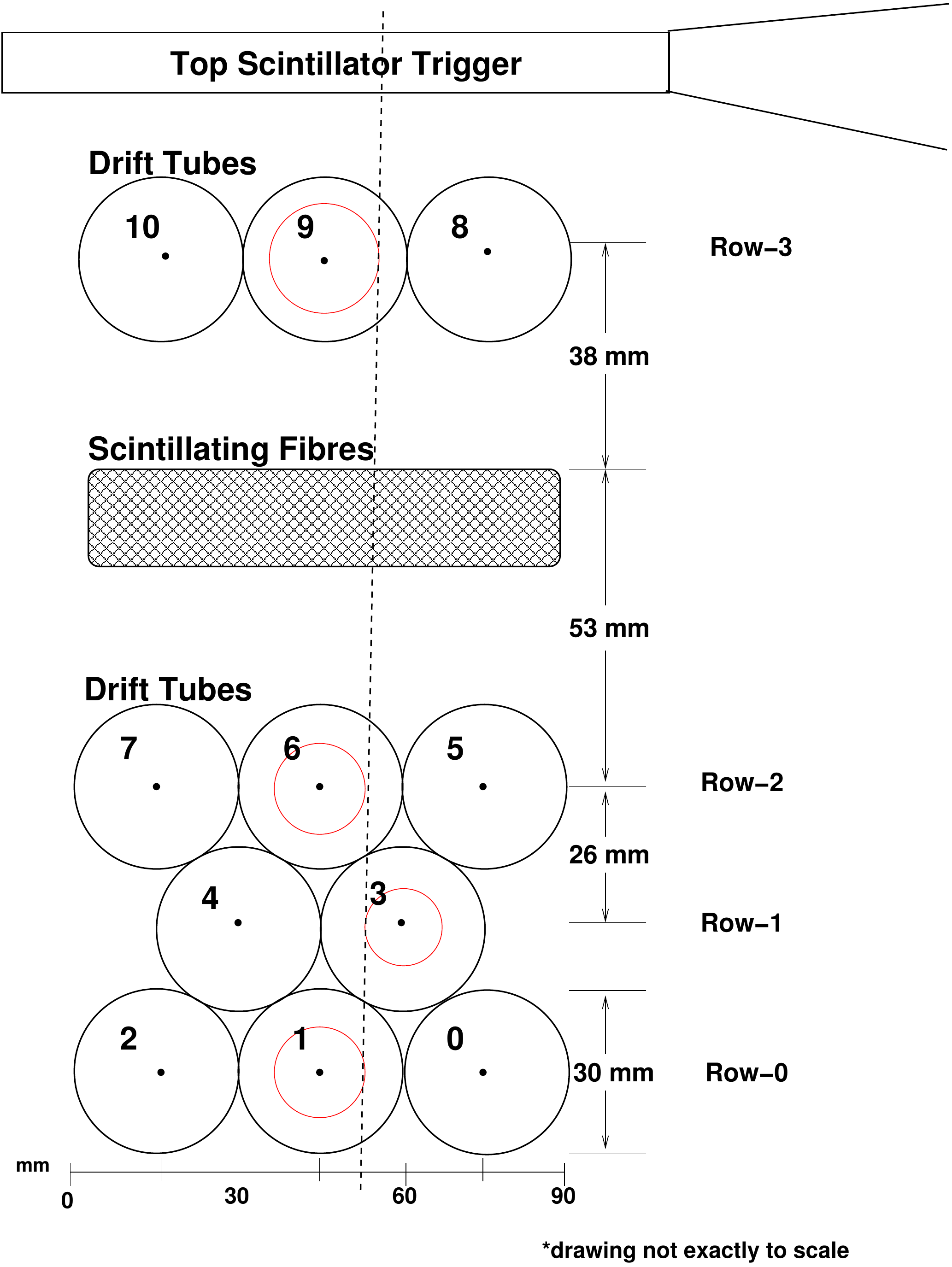} 
\caption{A schematic of the experiment.}
\label{fig:schem}
\end{figure}

A prototype scintillating fibre detector was constructed and tested with cosmic rays at 
Laboratori Nazionali di Frascati dell'INFN (LNF) using 60 parallel columns of six Kuraray~SCSF--81SJ single clad, 1.00$\pm{}$0.05~mm fibres with a  peak emission wavelength of 437~nm (blue). The fibre columns have a 1.35~mm mean horizontal pitch and 0.74~mm vertical fibre pitch. The variation in pitch has not been measured yet, but will have some contribution to the position resolution. The fibres were glued with optical epoxy intermittently along their length, and with the fibres  in each channel ganged together and readout with a 64 channel Hamamatsu 8500 (H8500) Multi-anode  photomultiplier tube (MA-PMT). The test setup, shown in Fig.~\ref{fig:schem}, included four layers of muon drift tubes (MDT) inter-spaced with the fibre detector for tracking purposes.

The 11 drift tubes were 3~cm in diameter, with a 30~$\mu$m diameter Au-W  wire  operated at HV=4600 V. A 40:60 Ar--Isobutane gas mixture in limited streamer regime ensured large amplitude signals and stable operation. Two scintillator paddles in coincidence  formed the trigger/gate for the setup,  with one paddle placed directly on top of the MDTs and one placed 60~cm below the fibre detector, ensuring nearly vertical tracks. The scintillating fibre tracker is placed between the top three tubes and the bottom eight tubes, as shown in Fig.~\ref{fig:schem}, in order to minimize the resolution of the extrapolated track from the drift tubes.

\begin{figure}[htbp]
\centering
\includegraphics[width=0.49\textwidth]{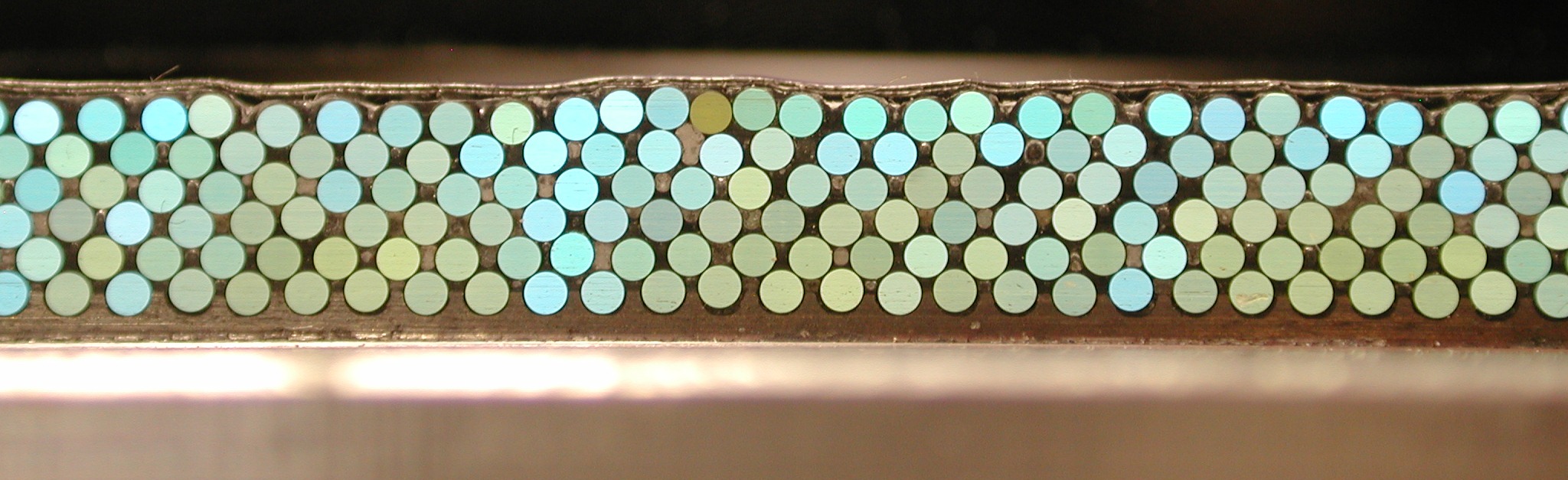} 
\includegraphics[width=0.49\textwidth]{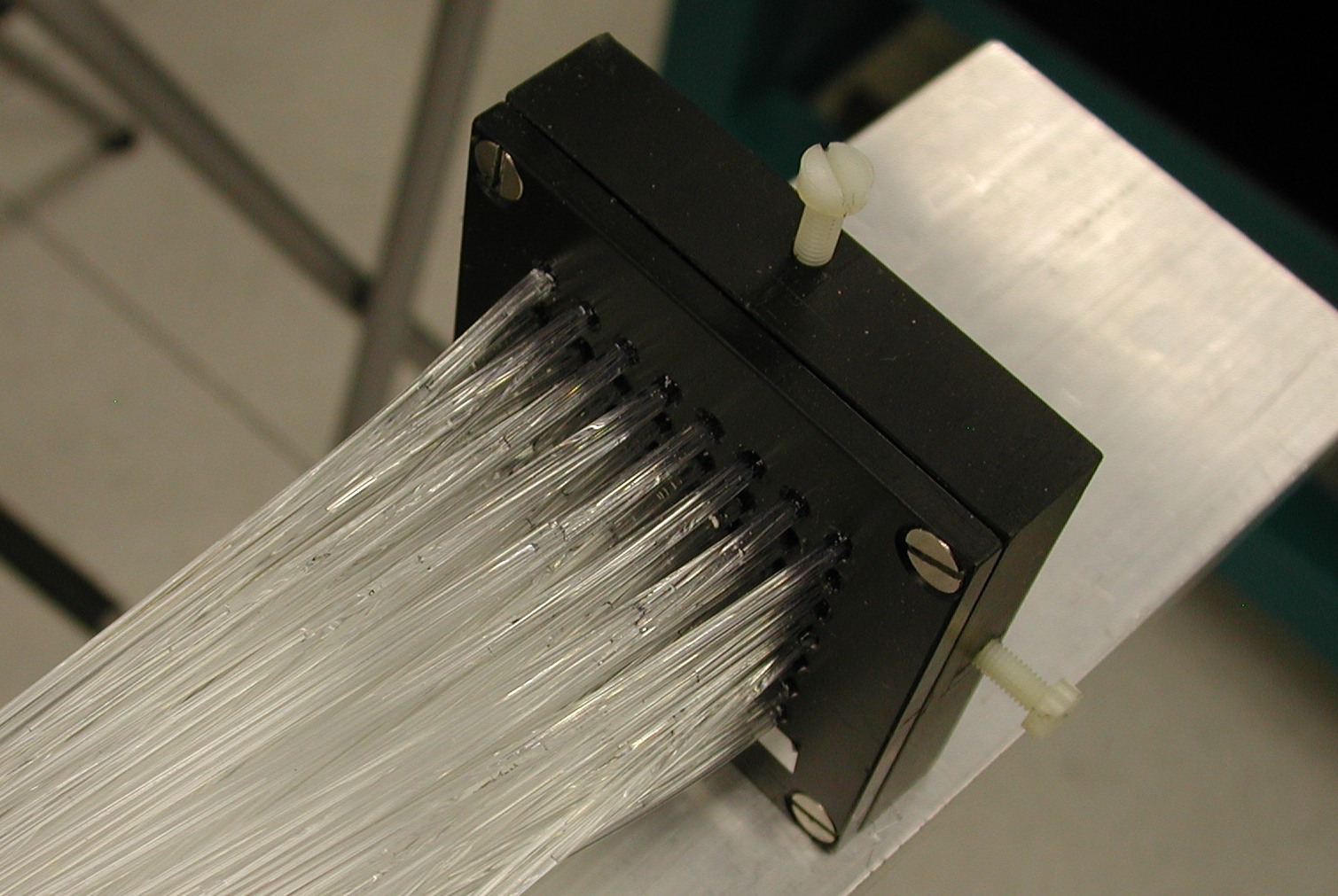}
\caption{(left) A photograph of the fibres packed in the fibre plane. (right) The 60 fibre channels are mounted in an array to be attached to the MA-PMT.}
\label{fig:fibreplanephoto}
\end{figure}

The scintillating fibres are arranged similar to Fig.~\ref{fig:fib1} 
with a horizontal pitch of 1.35~mm and an average vertical pitch of 0.74 mm. 
A photograph of the prototype is shown in Fig.~\ref{fig:fibreplanephoto} (left). The precision of the fibre placement is unmeasured, but from the photograph, it is clear that the precision decreases dramatically from the bottom to top rows of fibres, until it is seen that there is even an extra fibre in a column. The fibres from each channel are bundled and mounted in an array which is glued to the MA-PMT, shown in Fig.~\ref{fig:fibreplanephoto} (right).

\begin{figure}[htbp]
\centering
 \includegraphics[width=0.45\textwidth]{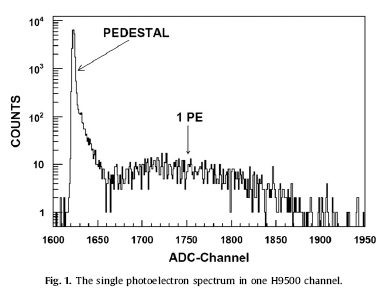}
\includegraphics[angle=90,width=0.54\textwidth]{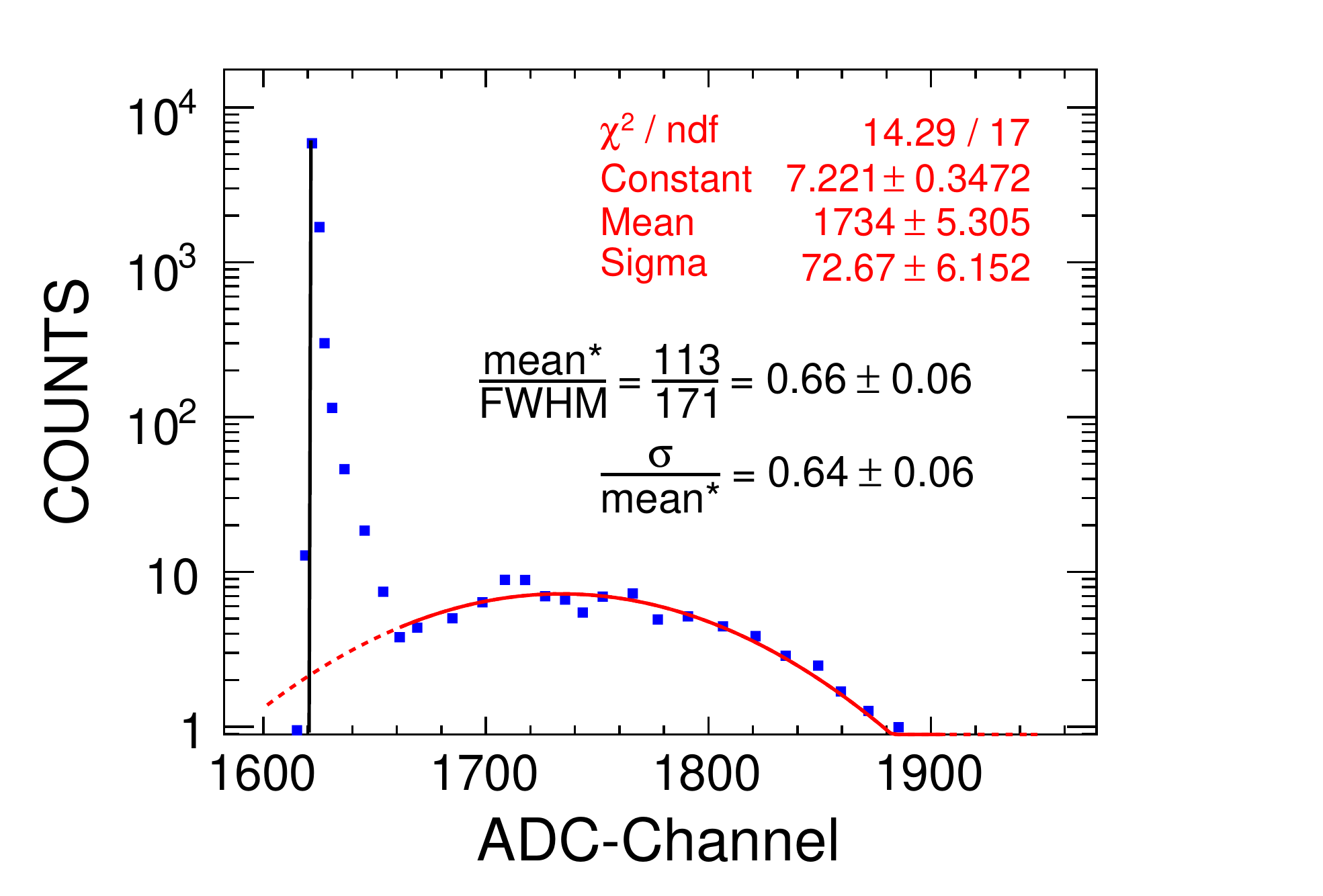}

\caption{Single photoelectron resolution for the MA-PMT H8500/H9500. The left plot taken is from Ref.~\cite{ref_dele}. The data points in the right plot have been extracted from the left figure and fit with a Gaussian function (solid line). The figures are discussed in the text.}
\label{fig:H8500peres}
\end{figure}

Each channel of six fibres is readout by a channel of a Multi-anode photomultipler tube (MA-PMT) 
(Hamamatsu H8500) which is a flat plane of 8 by 8 = 64 channels. The PM was operated at HV = 1050 V which corresponds to a gain of $2\times{}10^{6}$ (according to the data sheet) and with an expected quantum efficiency of approx 20\%. 
This class of MA-PMT has been characterized in the past~\cite{ref_dele}: in this context we want to concentrate on its single photon response (see Fig.~\ref{fig:H8500peres}) as this parameter will be used in the simulation to determine the number of photoelectrons in the cosmic data.

Ref.~\cite{ref_dele} has measured the single photoelectron response of the  MA-PMT H8500/H9500 with the spectrum shown in Fig.~\ref{fig:H8500peres}(left). Extracting values from the histogram by hand and fitting the values above pedestal with a Gaussian distribution, the width of the single photoelectron peak was extracted. The value of mean/FWHM = 0.66 means that it is not possible to separate even the first two photoelectron peaks with this photodetector, as expected. The value of $\sigma{}$/mean is a characteristic of the photodetector that will be used in the Monte Carlo described later.

\begin{figure}[htbp]
\centering
\includegraphics[angle=0,width=0.7\textwidth]{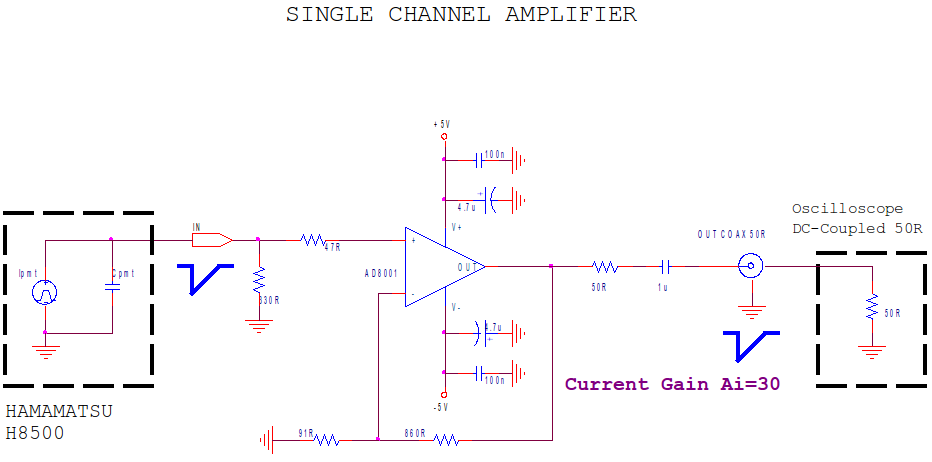} 
\includegraphics[angle=0,width=0.29\textwidth]{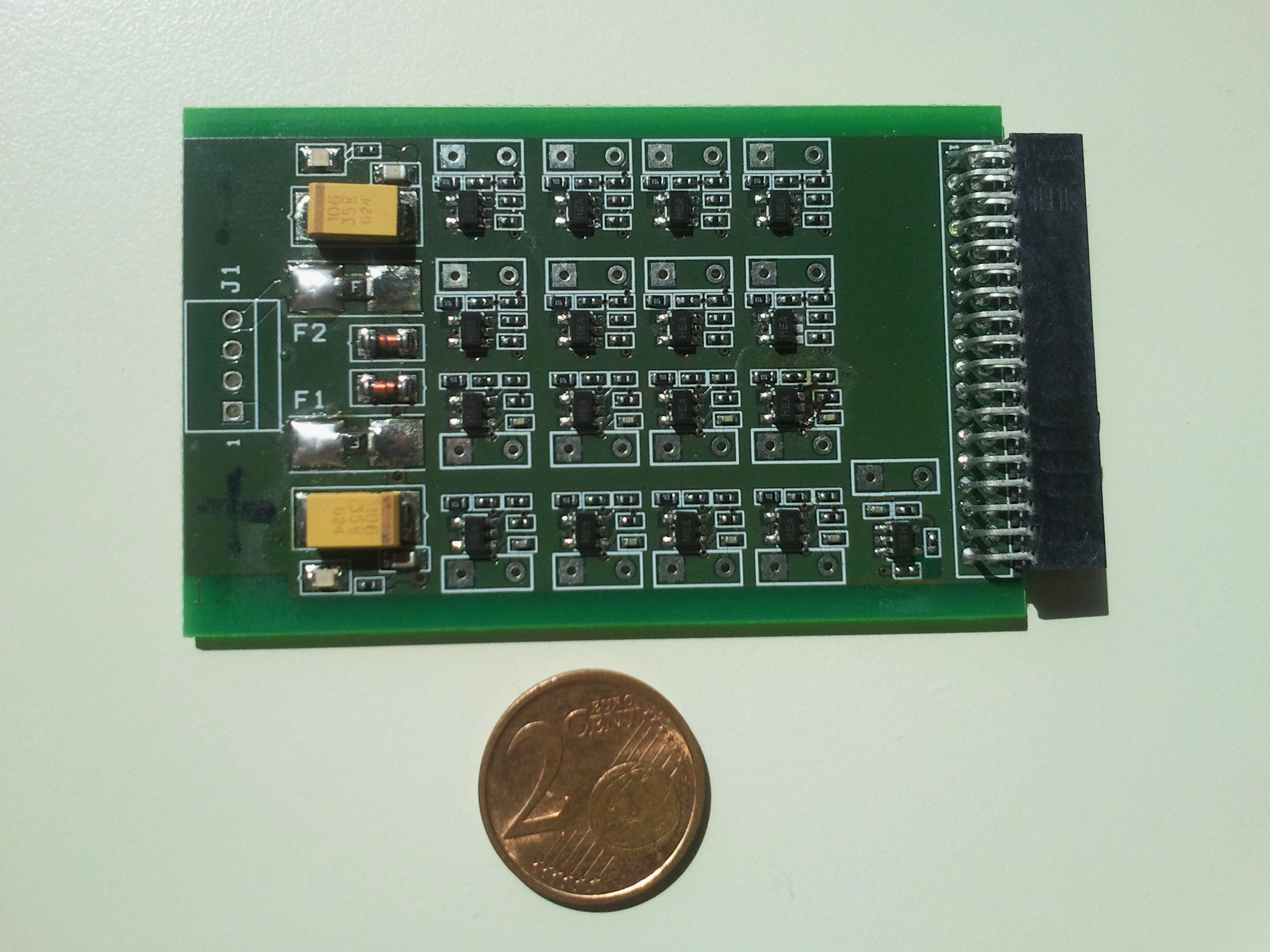} 
\caption{Amplifier schematic (left) and a photo of a single 16 channel board (right).}
\label{fig:amplifier}
\end{figure}

The H8500 Multi-Anode phototube is connected to four 16-channel FEE amplifier boards built with discrete components based on the op-amp AD8001 (the circuit scheme is shown in Fig.~\ref{fig:amplifier} powered by +/- 5 V. 
The current gain factor of the amplifier is 30, with a bandwidth of 200 MHz, which matches the readout needs of the subsequent charge readout circuitry. 
The expected output noise of the amplifier is 0.3 mV.
A photo of one board containing 16 channels is also shown in Fig.~\ref{fig:amplifier}.

\begin{figure}[htbp]
\centering
 \includegraphics[width=0.4\textwidth]{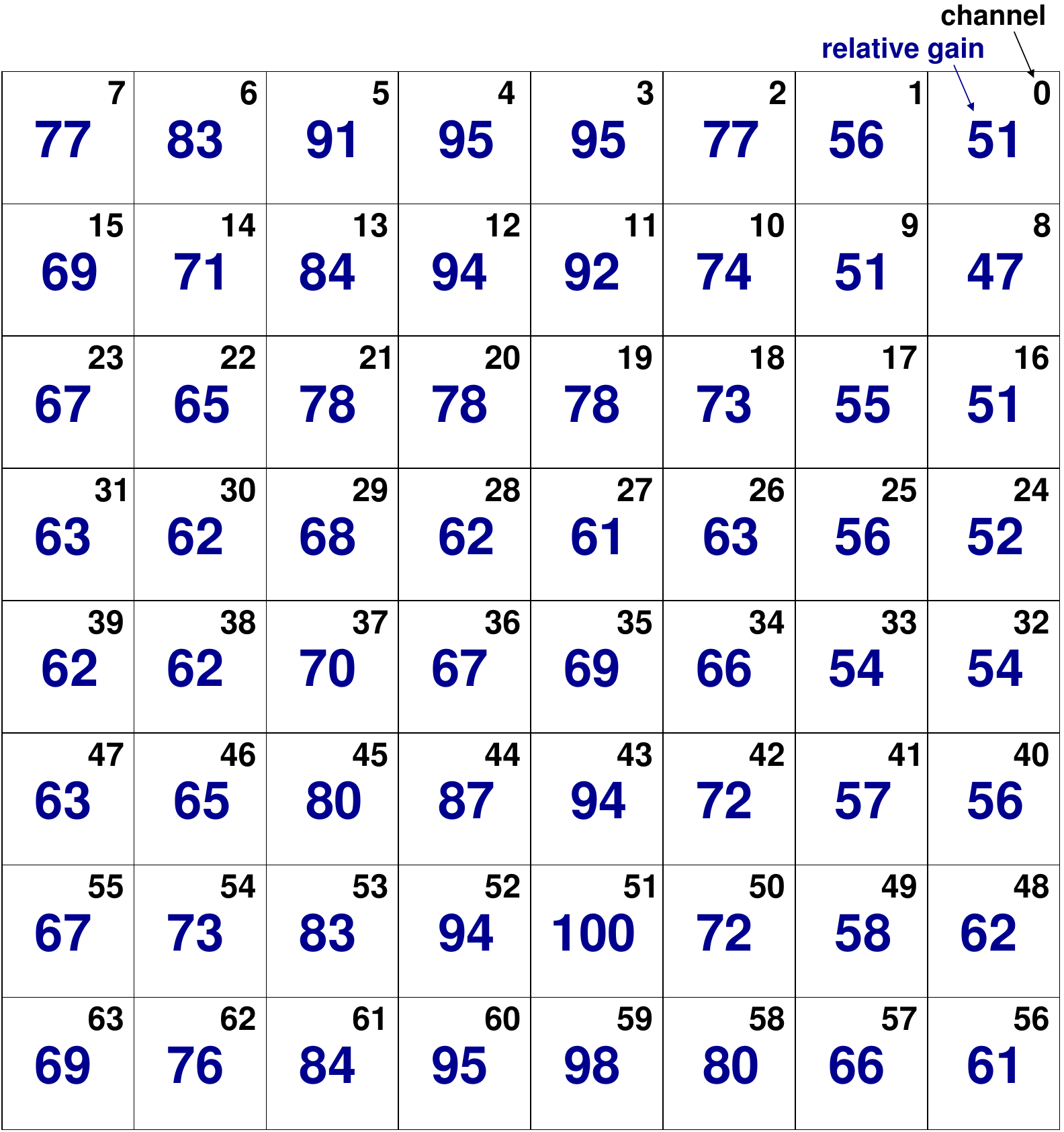}

\caption{The relative gain response of each anode pixel provided by the manufacturer for the H8500 Multi-Anode PMT used in this test setup.}
\label{fig:anoderesponse}
\end{figure}

The manufacturer has supplied the relative gain response of each channel in the MA-PMT, as shown in Fig.~\ref{fig:anoderesponse}. These values are used to equalize the signals in order for each channel to provide a signal of equal amplitude for a  the same light yield on each channel. A secondary correction was applied offline to bring the amplitudes of each channel within 5\% of the mean amplitude to correct for light yield variations.

\section{Tracking Resolutions}

The drift times measured in the muon drift tubes are used to reconstruct the track of the charged 
particle with good precision, such that the fibre plane resolution can be extracted.  Data were taken in 4 different positions of the triggering device along the fibres: at 60, 90, 150 and 230~cm from the MA-PMT. The geometry of the muon drift tubes and the experimental setup did not allow for positions below 60~cm  or above 230~cm.

For an event to be analyzed, it must pass the following two conditions: 1) there must be one hit in each of the 4 layers of drift tubes to produce the track  and 2) the mean drift time of the bottom three layers of tubes must be within $250 ns < t_{mean} < 600 ns$ which will reject tracks that may have produced delta electrons inside the tube. See Appendix A for an explanation of the mean drift time.

\subsection{Drift Tube results}
\begin{figure}[h!tbp]
 \includegraphics[angle=90,width=0.5\textwidth]{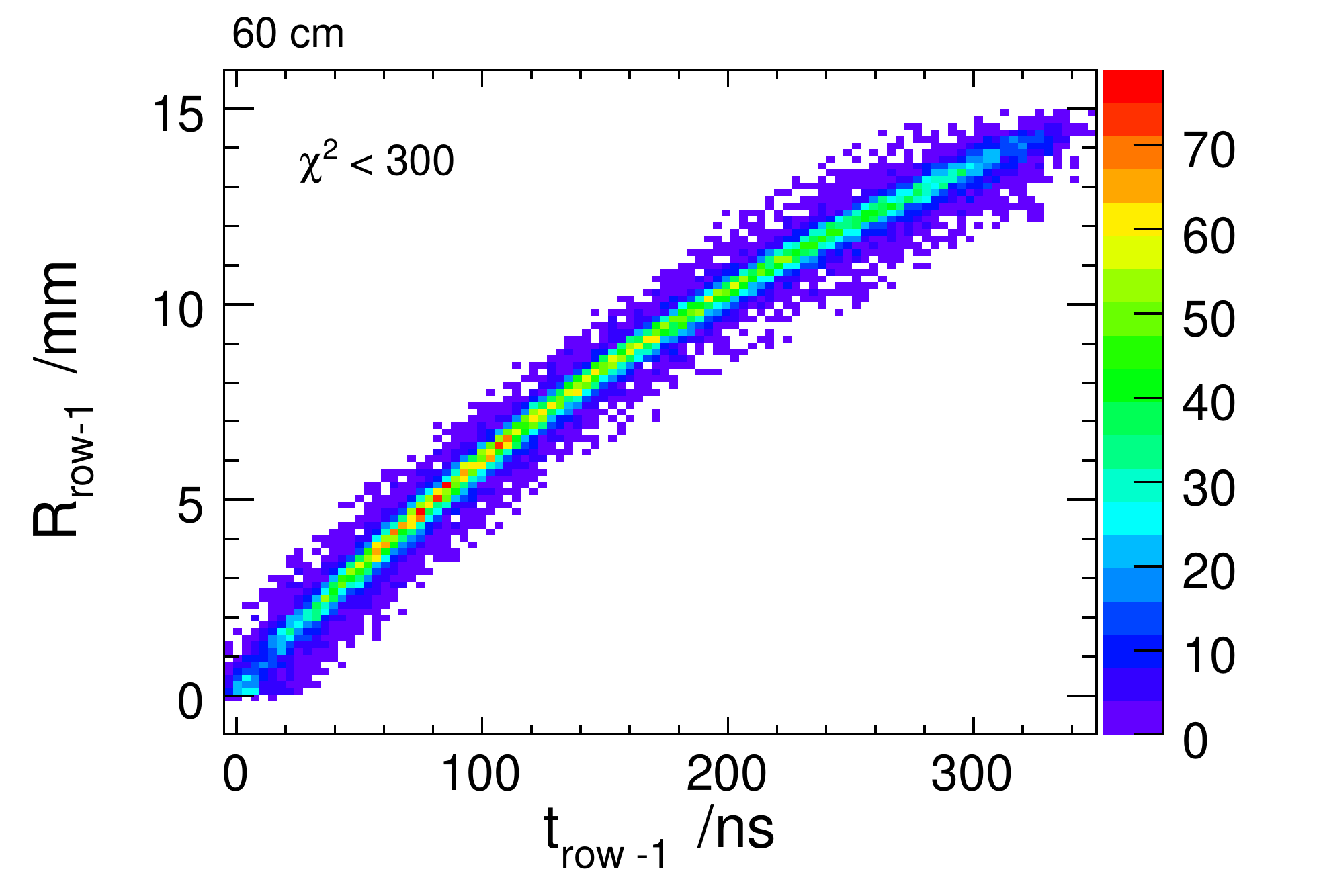}
  \includegraphics[angle=90,width=0.5\textwidth]{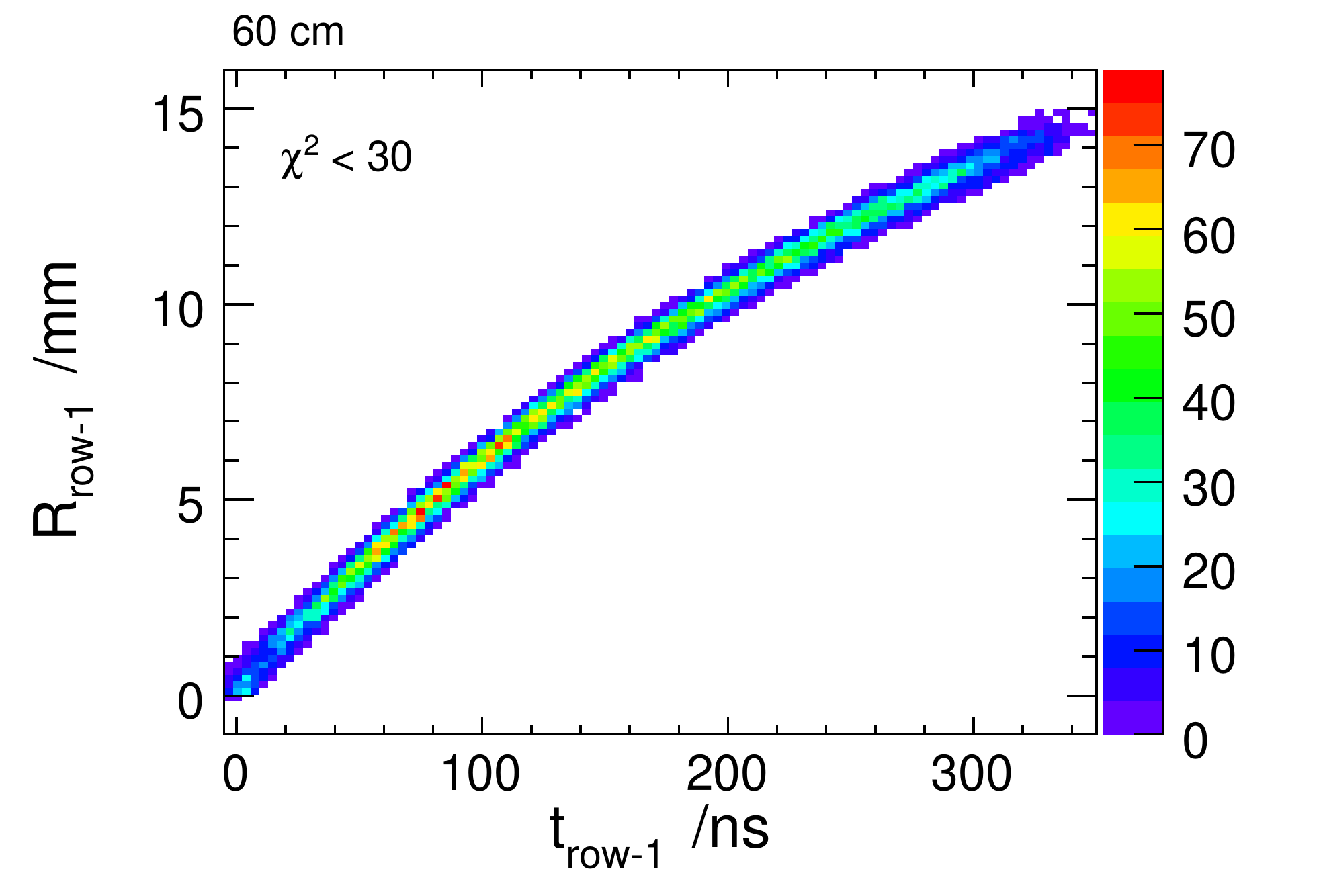}

\caption{Track radius, $R$, versus drift time, $t$, for the offset tube row, Row-1. Under the assumption that all tubes behave the same, the $R(t)$ relation for all the tubes in the data set is derived from this. The left (right) plot has a cut applied of $\chi^{2}<300~(30)$ from the track fit.}
 \label{fig:rtrelation1}
\end{figure}

\begin{figure}[h!tbp]
 \includegraphics[angle=90,width=0.5\textwidth]{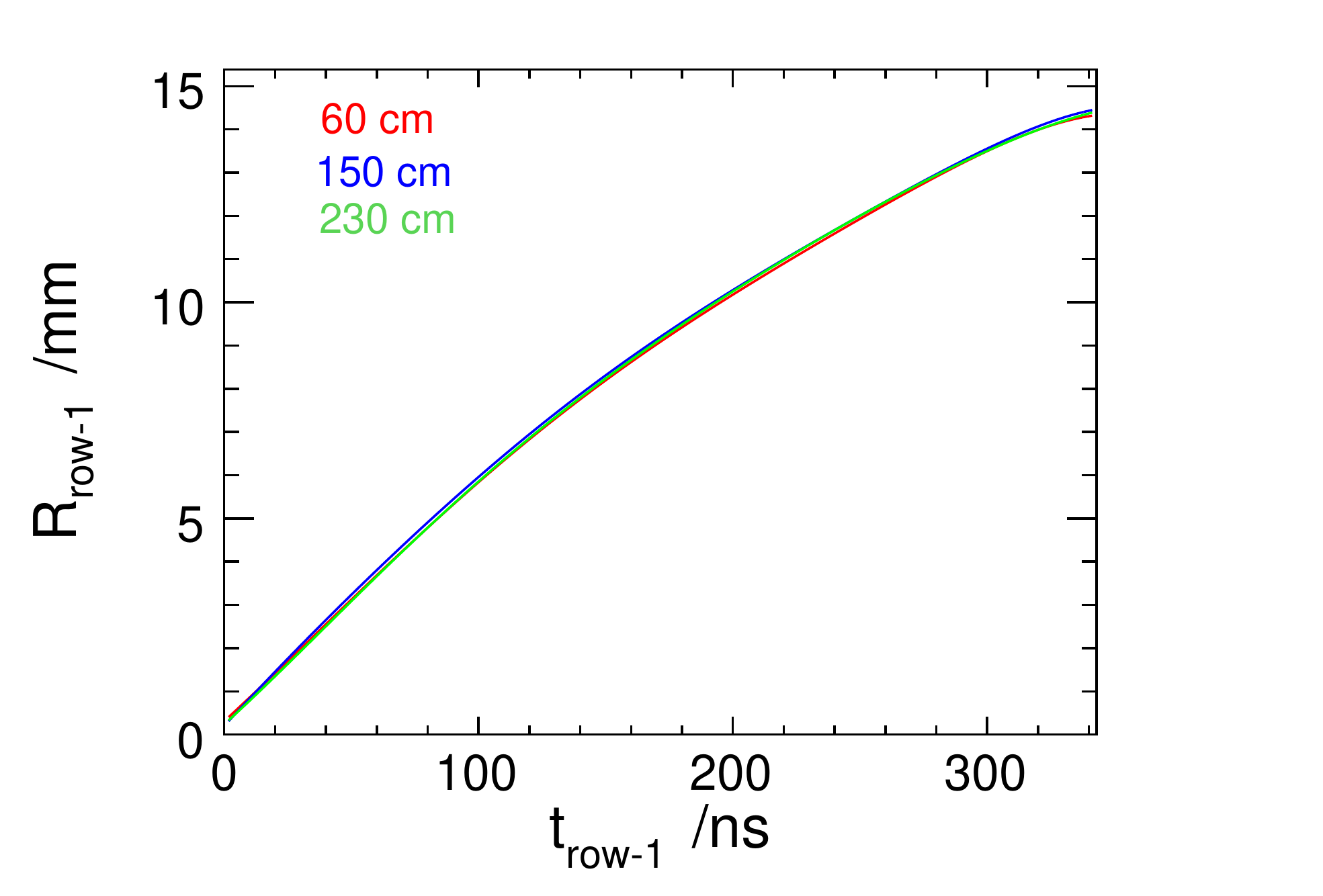}
  \includegraphics[angle=90,width=0.5\textwidth]{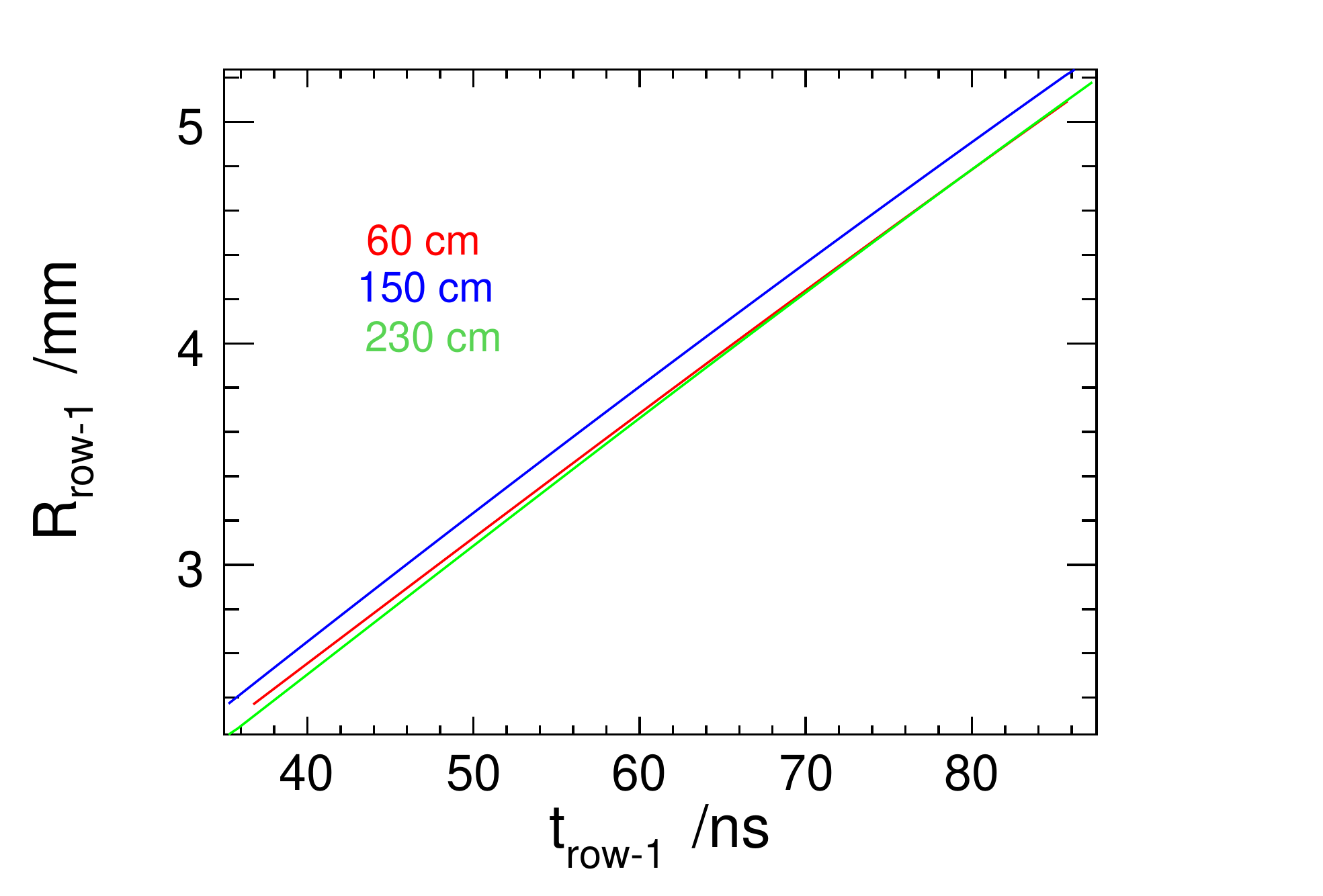}

\caption{The lines are results of constrained 6th order Chebyshev polynomials fits to radius, $R$, as a function of drift time, $t$,  for 60~cm, 150~cm and 230~cm trigger positions.  The right plot is a close up of the fits showing the differences in track radius for a given drift time in the different data sets.}
 \label{fig:rtrelation2}
\end{figure}

Assuming that the eleven tubes are nearly uniform in gas content, construction and electric field shape over the period of data taking, the relationship between drift radius is found through a self calibrating process, by minimizing the $\chi^{2}$ of the radius function and track radius of each tube. The calibration process is described further in Appendix A. The largest uncertainties in the drift radius will result from systematic errors in $t_{0}$ for the tubes, differences in the electric field within each tube as well as changing drift velocities due to the gas tanks being exposed to outdoor temperatures. An example plot of the radius of the wire from the fitted track as a function of drift time is seen in Fig.~\ref{fig:rtrelation1}. Fit functions with constrained sixth order Chebyshev polynomials to the drift data are seen in Fig.~\ref{fig:rtrelation2}. The plot on the right highlights the differences (up to 0.2~mm) in radius for the same drift time for three different data sets, where a resolution of 0.1 to 0.2~mm is needed from the tubes.

\begin{figure}[h!tbp]
 \includegraphics[angle=90,width=0.5\textwidth]{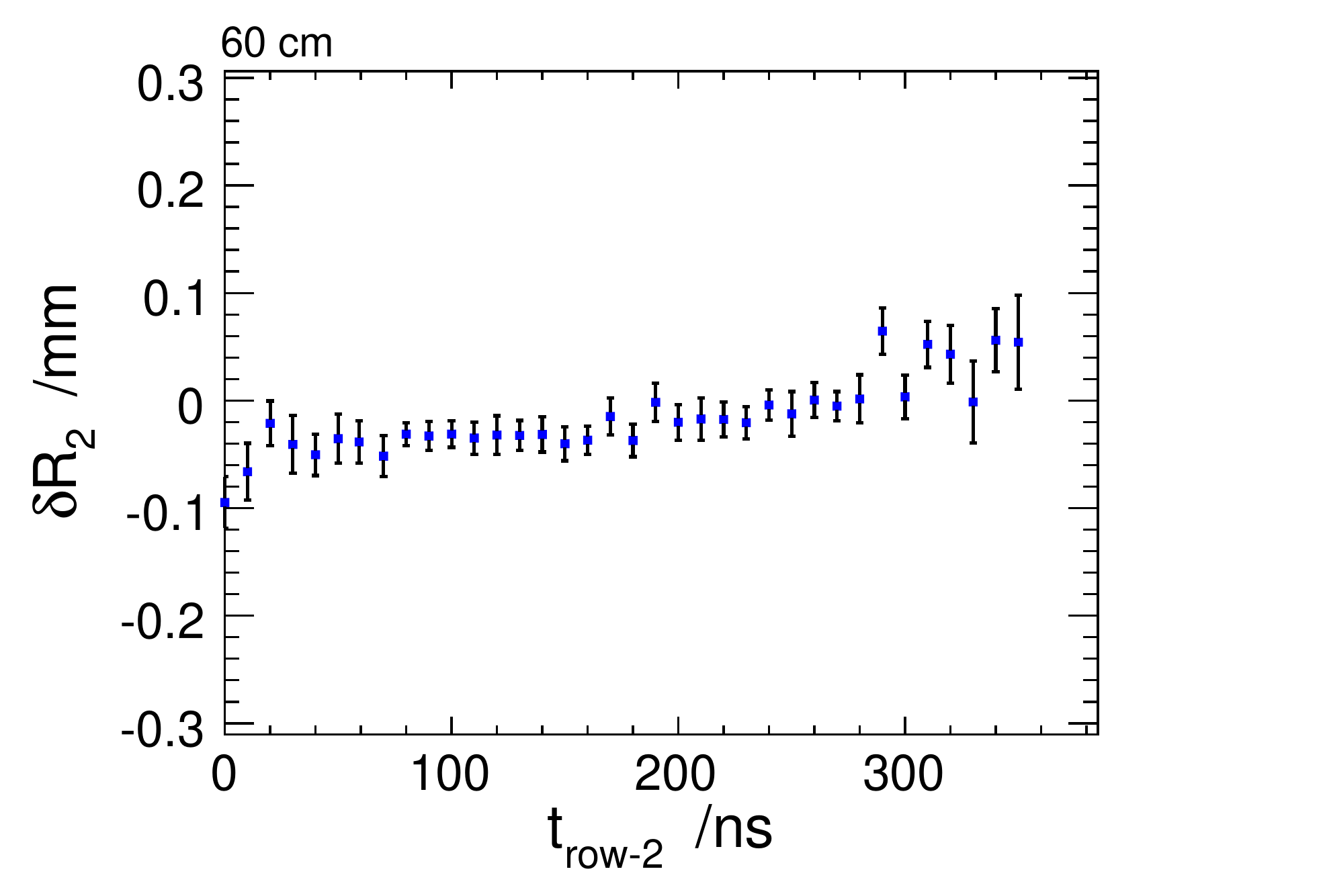}
  \includegraphics[angle=90,width=0.5\textwidth]{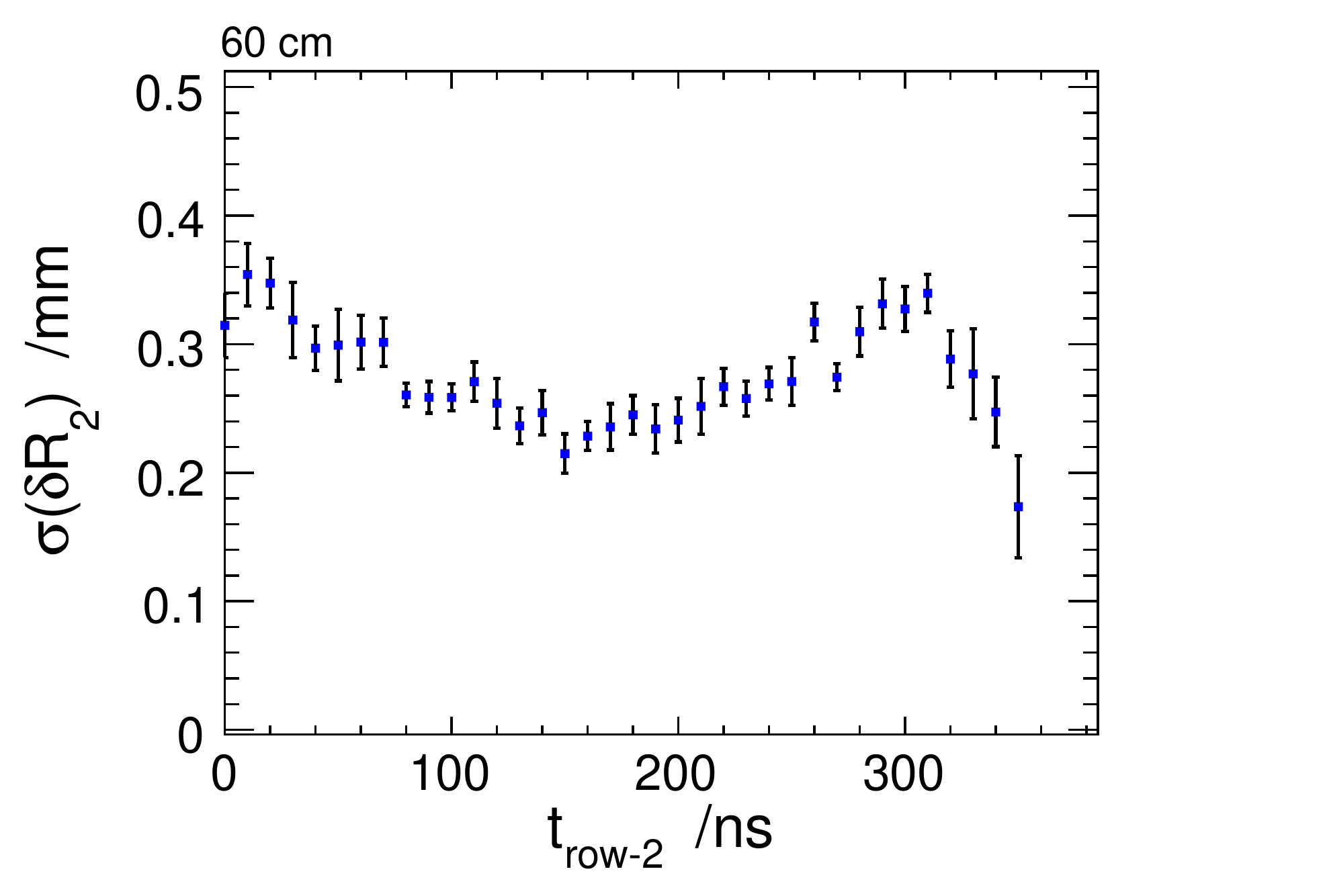}

\caption{\textbf{60~cm.} (left) Mean value of $\delta{}R$ as a function  of time, $t$, for Row-2 tubes, for the 60~cm trigger position. (right) The standard deviation of the $\delta{}R$ over $t$. }
 \label{fig:drtrow260}
\end{figure}

 The difference between the fitted track radius and radius calculated from the drift time, $\delta{}R$, over drift time, $t$, for Row-2, is shown in Fig.~\ref{fig:drtrow260}. The mean value of $\delta{R}$ provides the accuracy with which the drift relation has been determined.  Assuming tube rows 0, 2 and 3  are properly aligned, this distribution should be centred on zero for all values of $t$.  As well, the standard deviation of $\delta{}R$, as shown on the right of Fig.~\ref{fig:drtrow260}, is the combined resolution of the fitted track radius and the drift time radius.

\begin{figure}[h!tbp]
 \includegraphics[angle=90,width=0.5\textwidth]{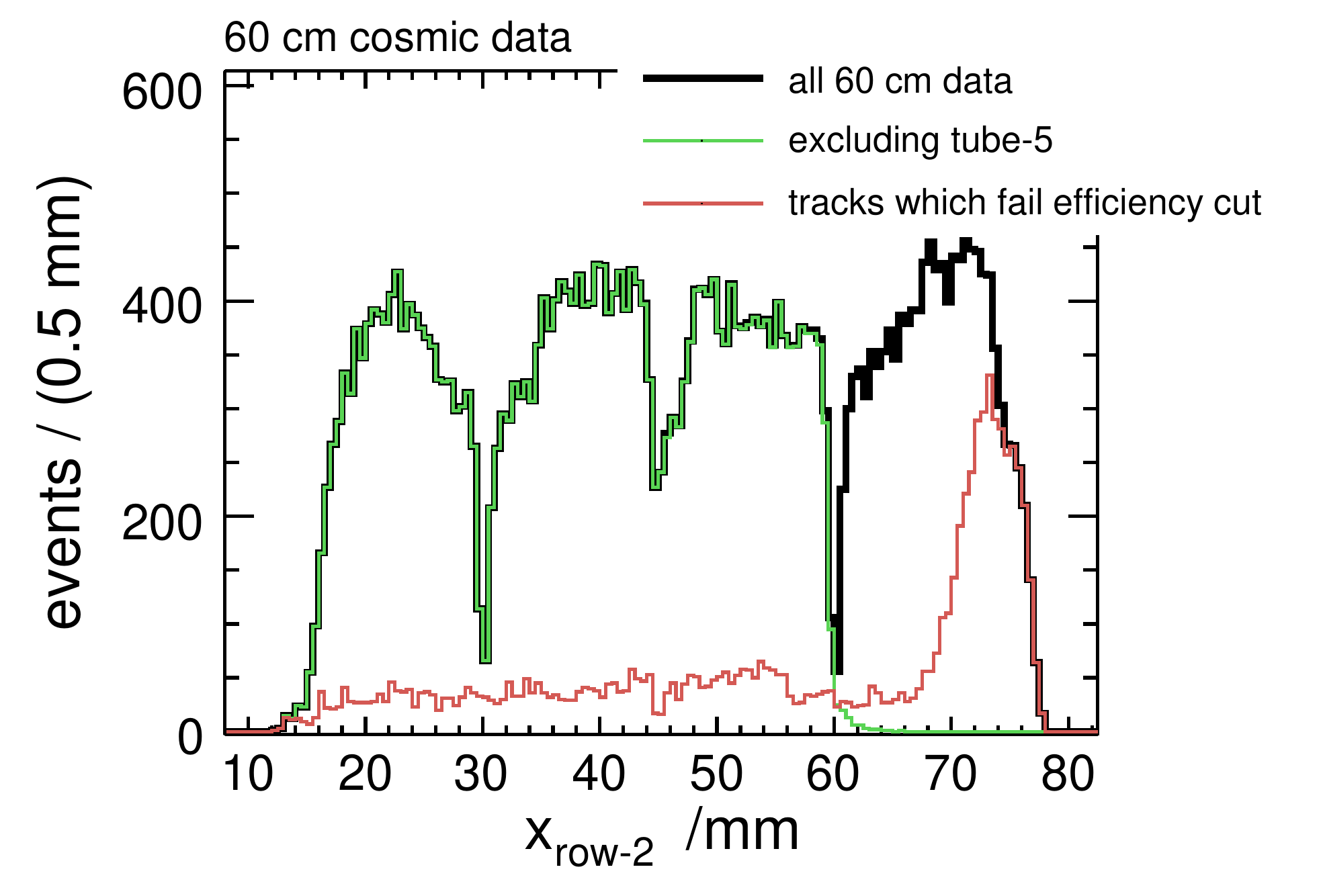}
  \includegraphics[angle=90,width=0.5\textwidth]{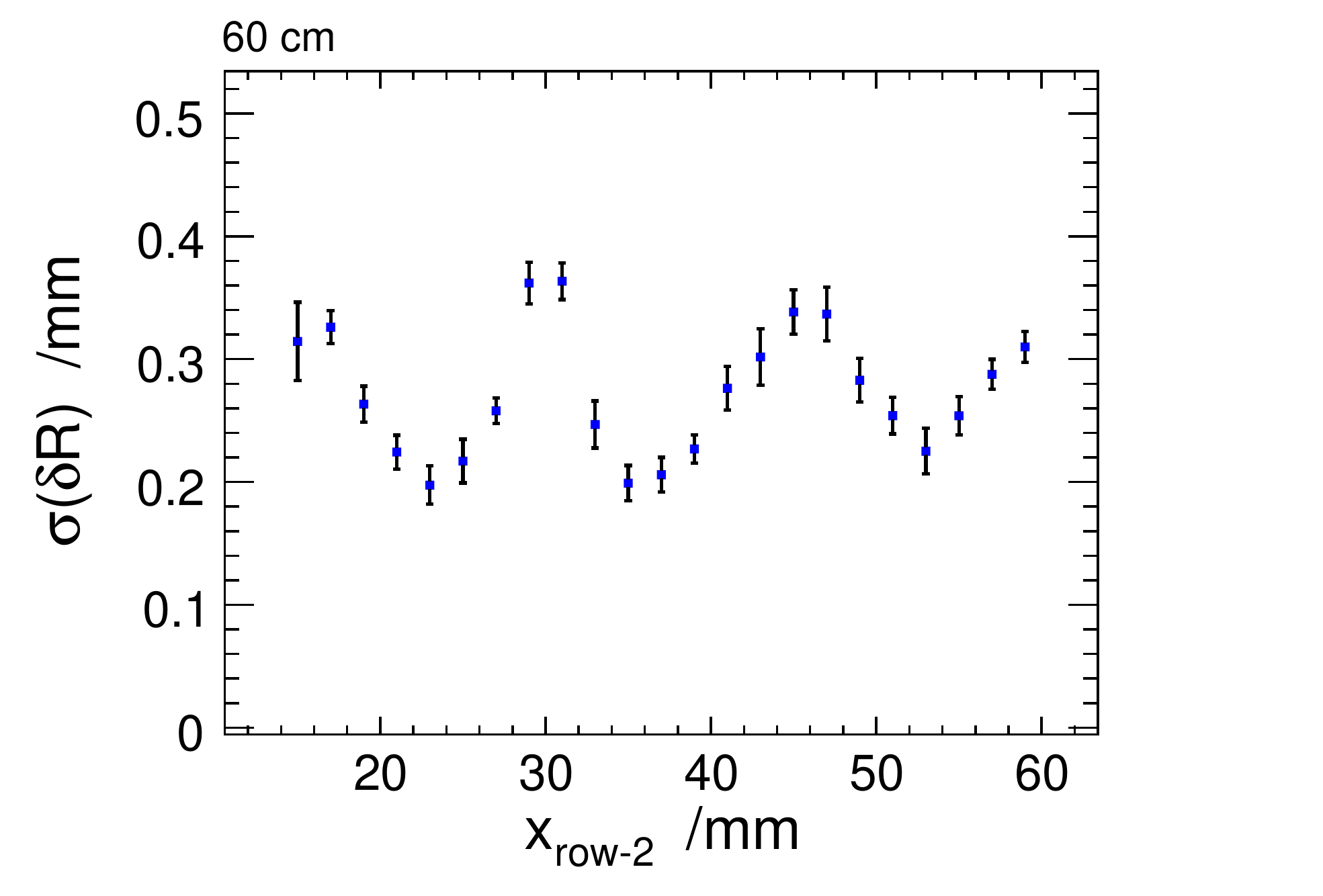}

\caption{\textbf{60~cm.} (left) A histogram of the number of events as a function of the position, $x$, along the plane of the Row-2 wires. The details are explained in the text. (right) The standard deviation of $\delta{}R$ as a function of position, $x$, in the tube plane excluding Tube-5.}
 \label{fig:drXrow260}
\end{figure}

\begin{figure}[h!tbp]
\centering
 \includegraphics[angle=0,width=0.5\textwidth]{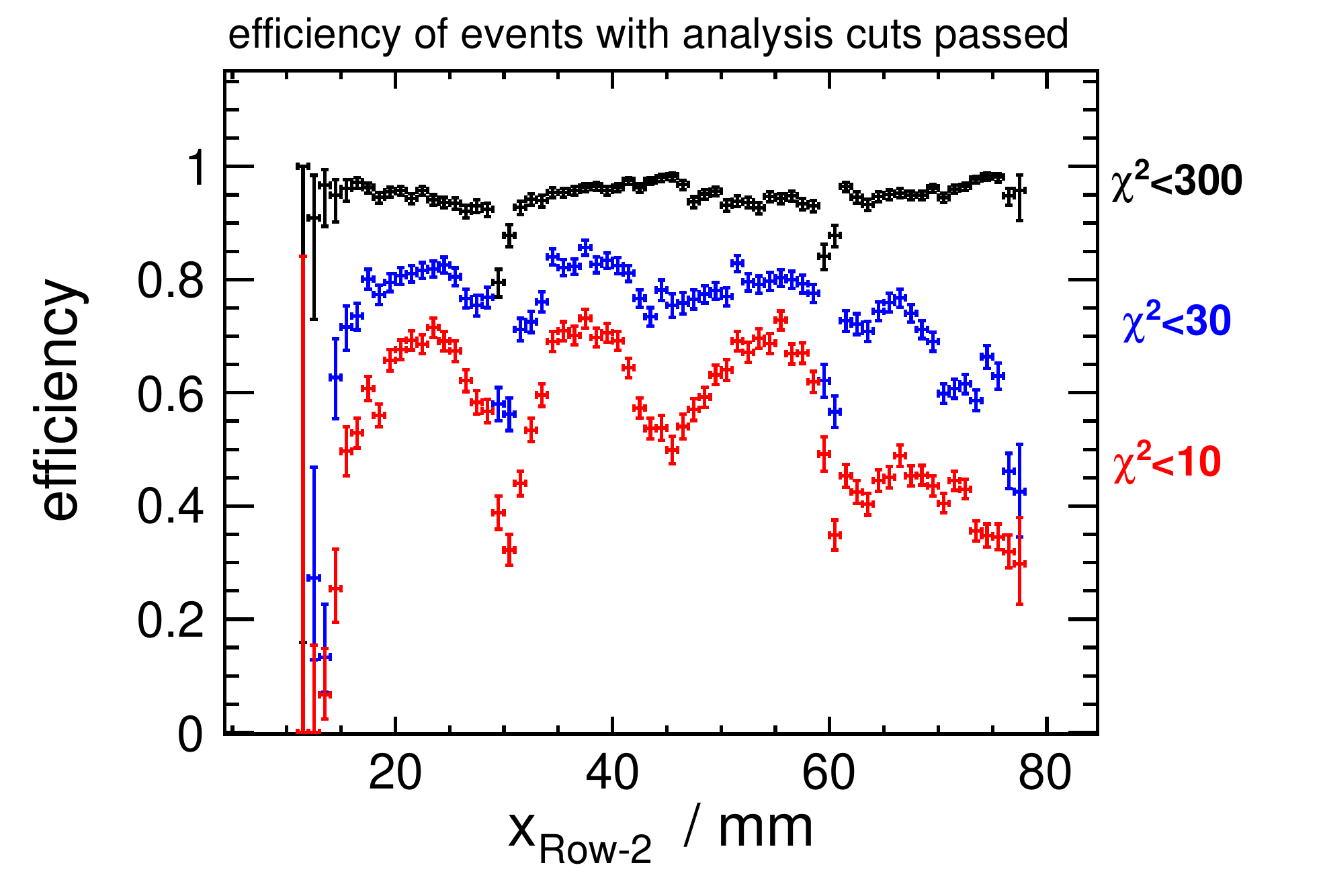}

\caption{\textbf{60~cm.} The efficiency or fraction of the total number of tracks with $\chi^{2} \leq$~10, 30 and 300 from the track fitter, as a function of Row-2 track position. The lower efficiency from 60 to 80~mm is a symptom of the poor time spectra in Tube-5.}
 \label{fig:effrow260}
\end{figure}

One of the consequences of the symmetry of the placement of the drift tubes is that with the resolution of each individual tube being worse near the wire and near the outer edges, where consecutive parts of the detector with poor resolution overlap, creating zones of poor efficiency and resolution, as seen in Fig.~\ref{fig:drXrow260}. The black line Fig.~\ref{fig:drXrow260}(left) are all events with at least four drift tube hits and pass the mean time cut, and cover the 15~mm to 75~mm range. Tube-5 has a wider tdc spectrum than the other 10 tubes which also produces many tracks with a large $\chi^{2}$. Cutting out tracks with hits in this tube produces the green line spectra in Fig.~\ref{fig:drXrow260}(left). The red line are all the drift tube tracks which do not have a corresponding hit in the fibre tracker within 2.7~mm. The peak in the red data past 65~mm is a result of the fibre tracker not extending this far.  The fraction of events which pass a $\chi^{2}$ cut of 10, 30 or 300 as a function of track position in Row-2 is shown in Fig.~\ref{fig:effrow260}. Comparing the track resolution at Row-2, found in Fig.~\ref{fig:drXrow260}(right), to Monte Carlo, one can expect a tube track resolution extrapolated to the fibre plane of 0.175$\pm$0.15~mm. 

\subsection{Fibre tracker results }

\begin{figure}[h!tbp]
 \includegraphics[angle=90,width=0.5\textwidth]{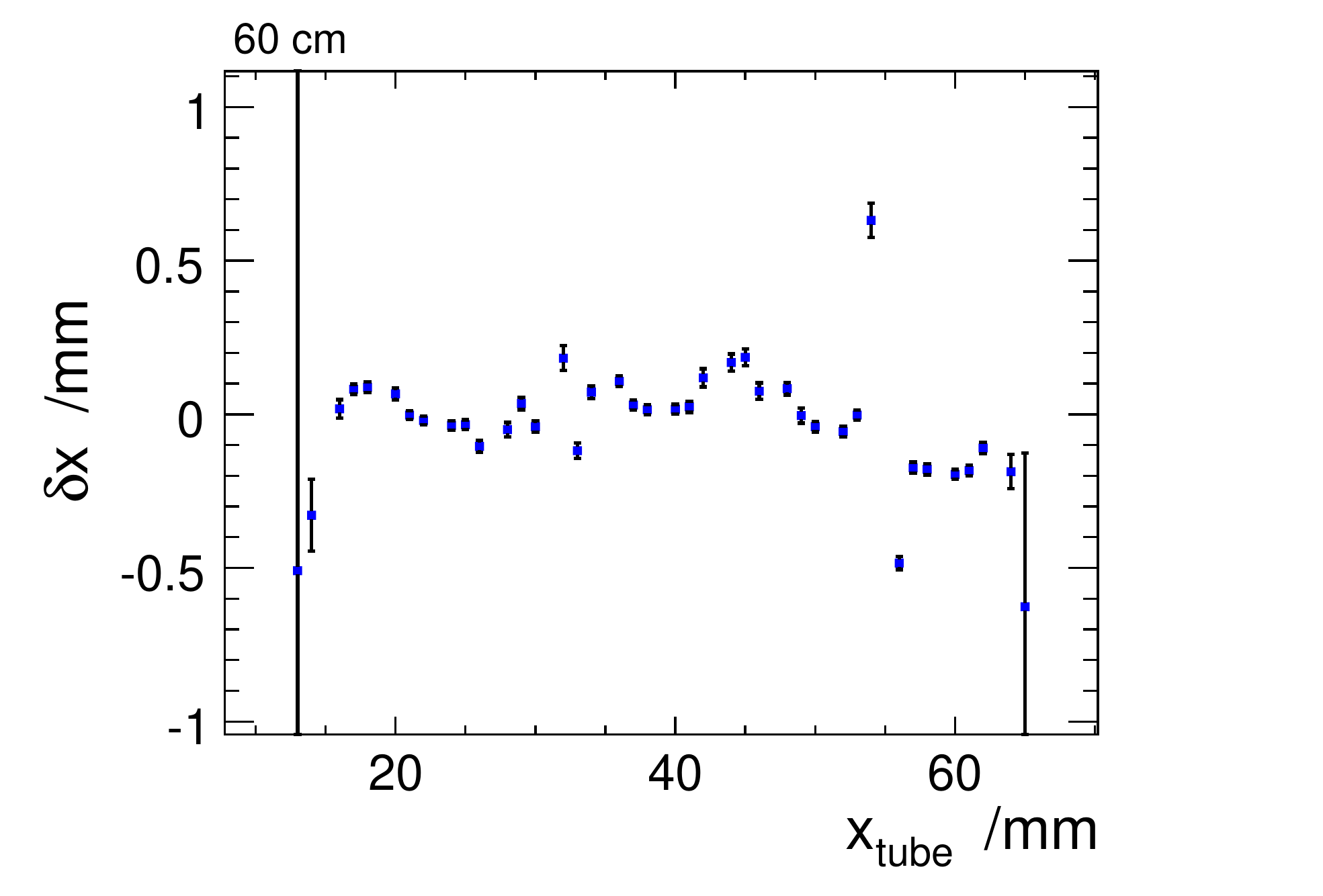}
  \includegraphics[angle=90,width=0.5\textwidth]{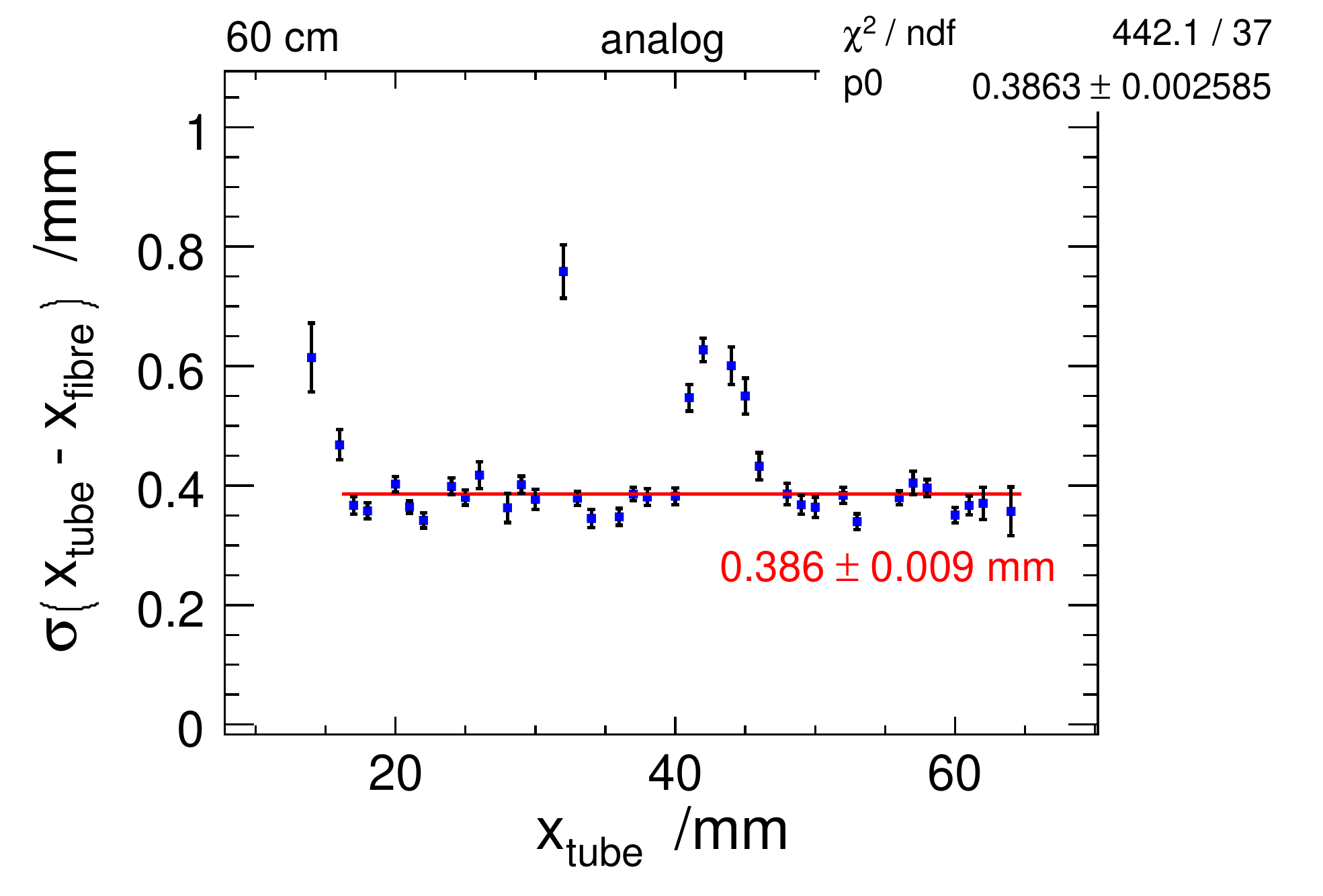}
 
\caption{\textbf{60~cm.} (left) The mean of $\delta{}x$ as a function of $x_{tube}$.  (right) The standard deviation of $\delta{}x$ as a function of $x_{tube}$.  Details are discussed in the text. Error bars in both plots are the statistical errors from Gaussian fits to the data. }
 \label{fig:dxxtube60}
\end{figure}

\begin{figure}[htbp]
 \includegraphics[angle=90,width=0.5\textwidth]{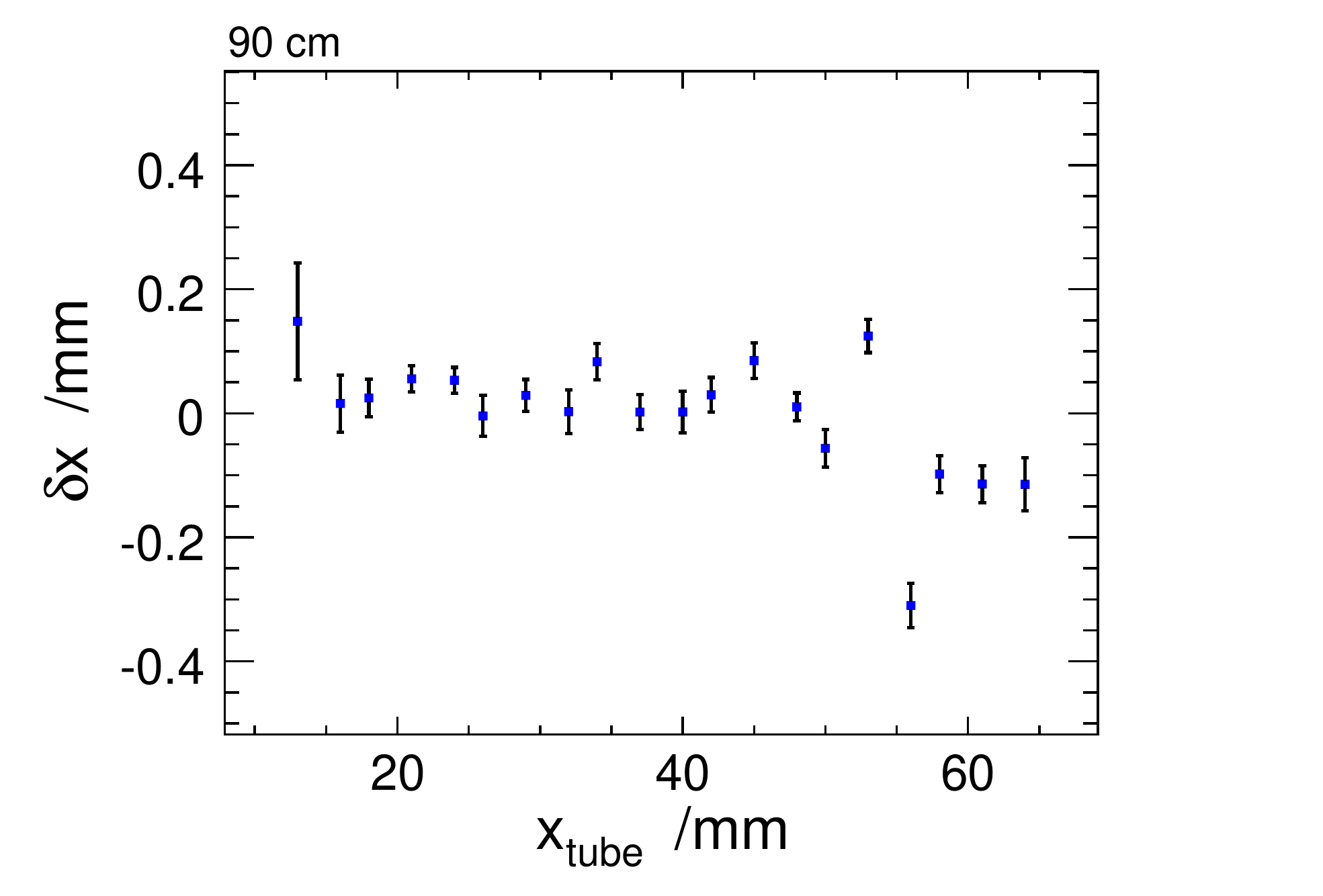}
  \includegraphics[angle=90,width=0.5\textwidth]{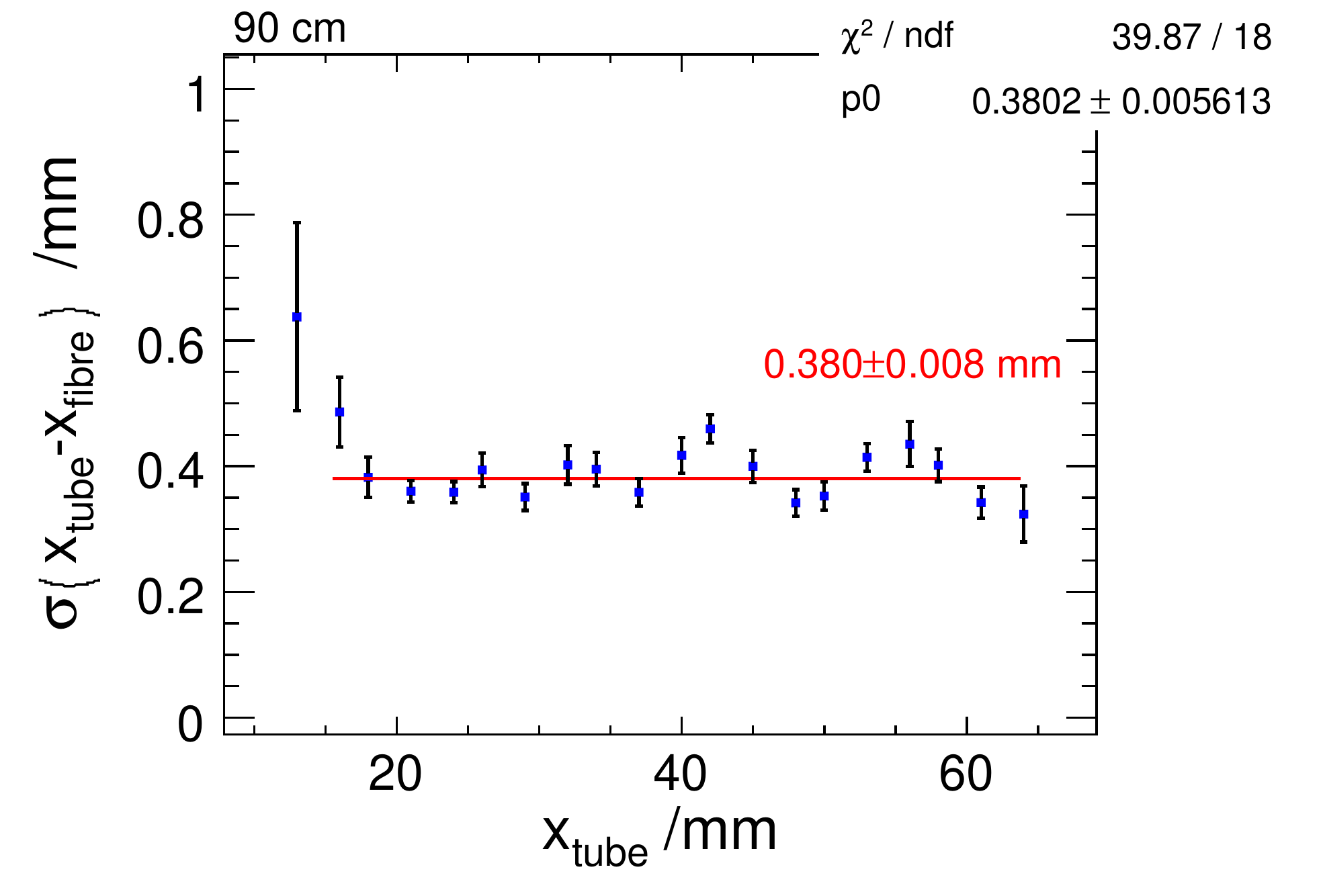}
  
  \includegraphics[angle=90,width=0.5\textwidth]{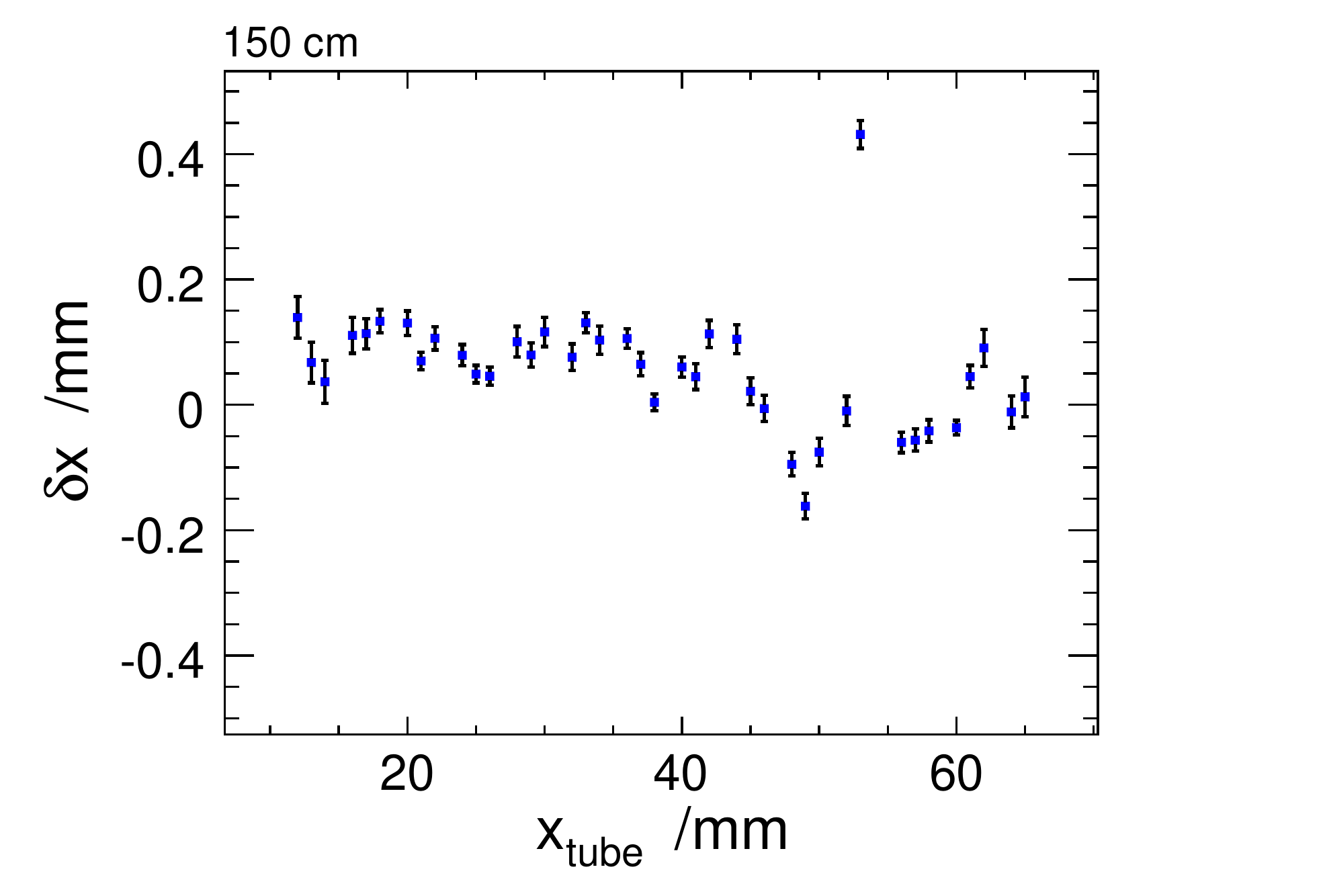}
  \includegraphics[angle=90,width=0.5\textwidth]{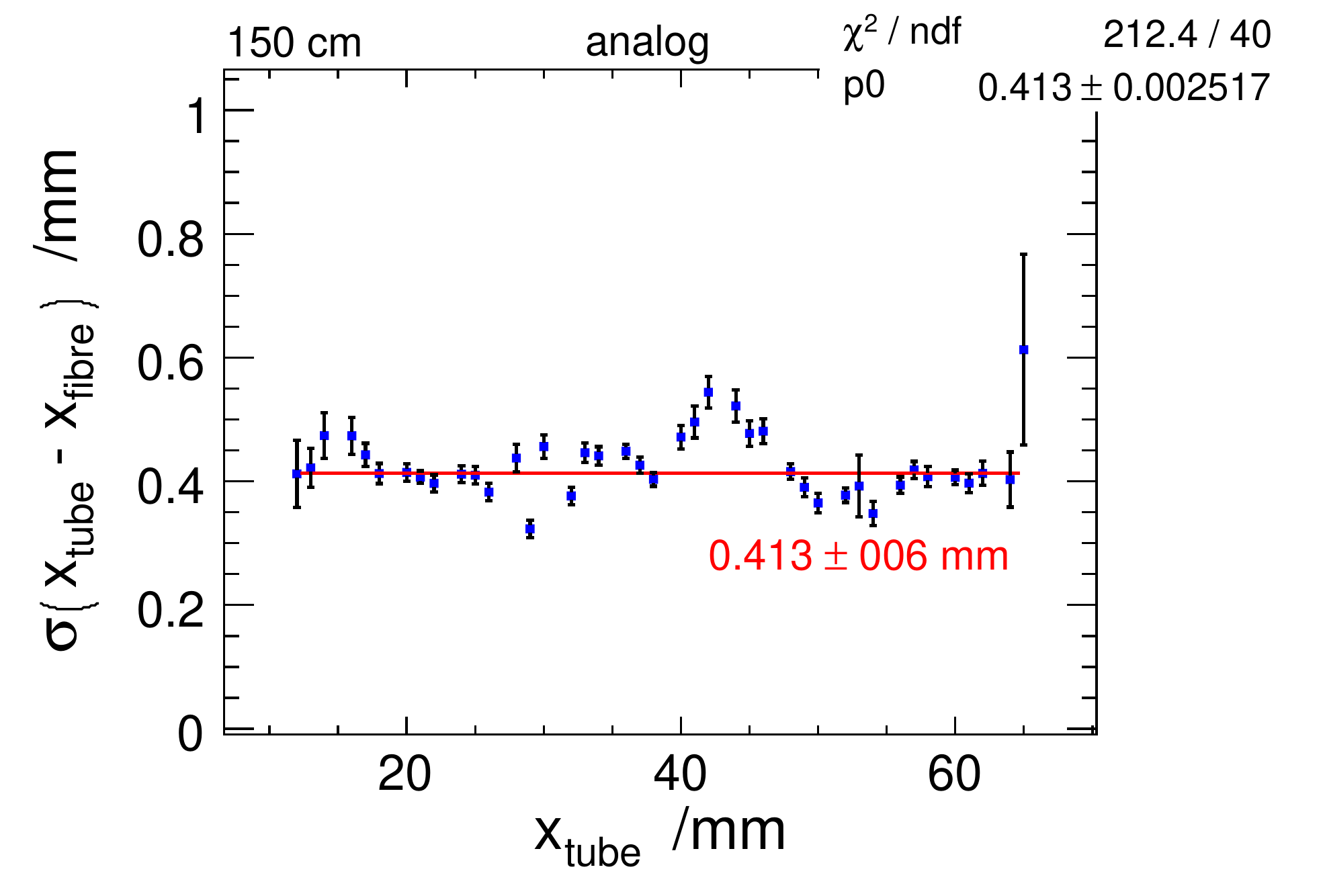}
  
  \includegraphics[angle=90,width=0.5\textwidth]{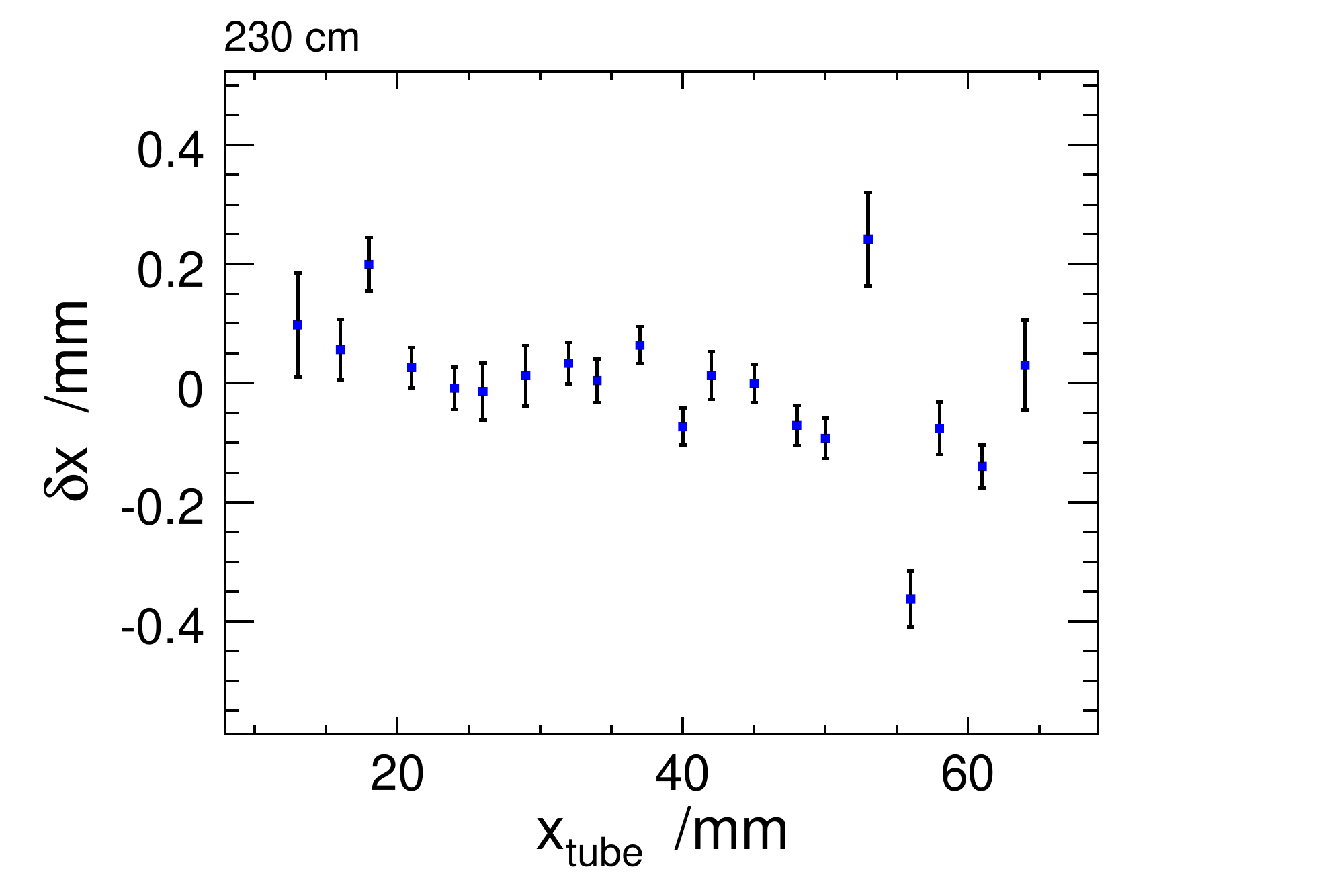}
  \includegraphics[angle=90,width=0.5\textwidth]{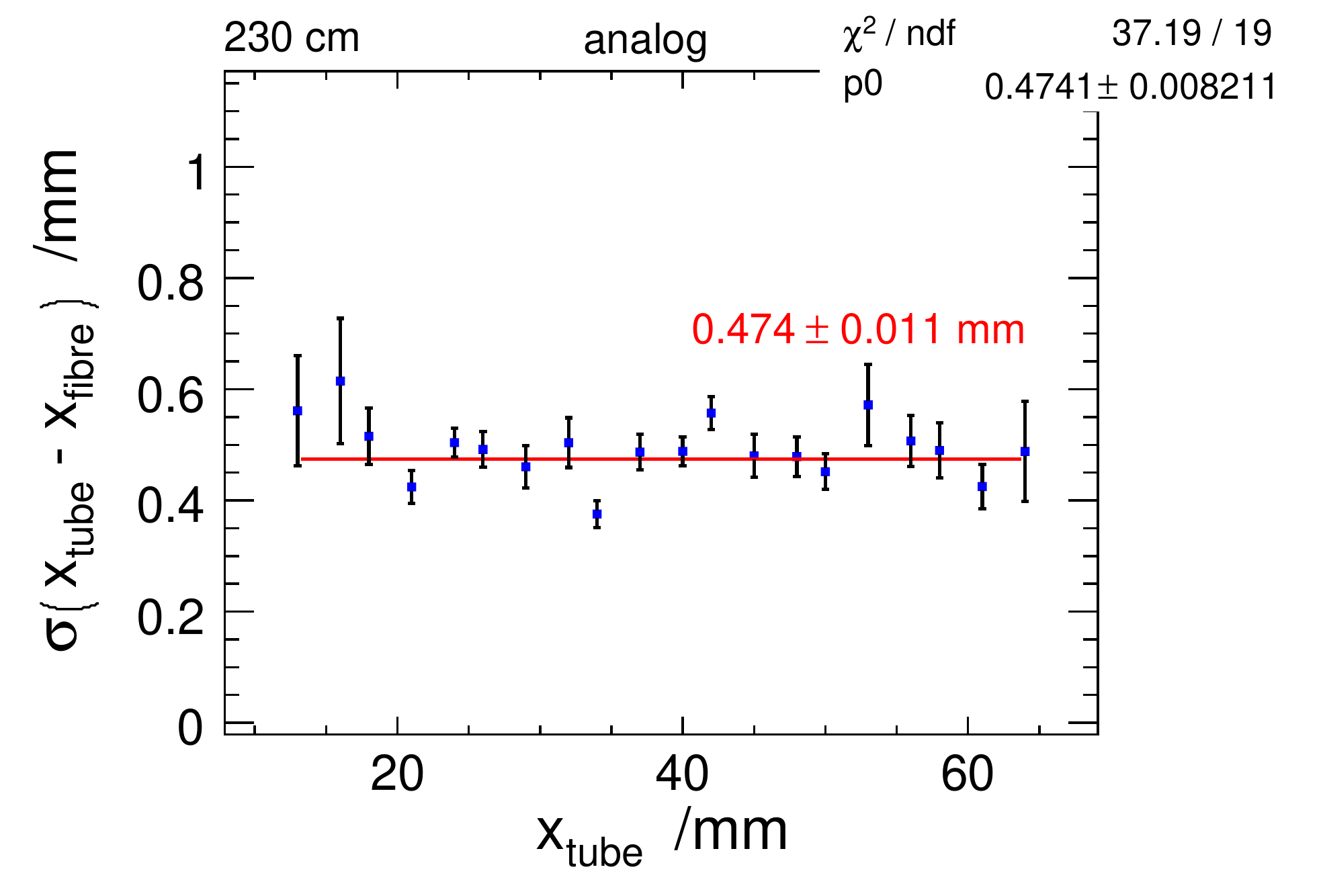} 
\caption{\textbf{90, 150 and 230~cm.} (left) The mean value of $\delta{}x$ as a function of $x_{tube}$. (right) The standard deviation of $\delta{}x$ as a function of $x_{tube}$. This is the combined resolution of the extrapolated drift tube track and the fibre plane. Error bars in both plots are the statistical errors from a Gaussian fit to the data. }
 \label{fig:dxxtubeall}
\end{figure}

The difference, $\delta{}x$, in the position between the extrapolated track position, $x_{tube}$  and charge-weighted hit position from channels which exceed 10~pC, $x_{fibre}$, is shown in Fig.~\ref{fig:dxxtube60} for 60~cm trigger position data as a function of position. The threshold, after subsequent analysis, is equivalent to approximately 0.5 photoelectrons.    The addition of the extrapolated tube track resolution in quadrature with the fibre tracker resolution will equal the standard deviation of $\delta{}x$ shown in Fig.~\ref{fig:dxxtube60}. The expected resolution of only the extrapolated tube track at the fibre plane is 0.175$\pm$0.015~mm. The left plot in Fig.~\ref{fig:dxxtube60} is the mean of $\delta{}x$.  The deviations at 15, 30 and 45~mm from zero are symptomatic of the poor resolution and efficiency in this region due to the overlapping edges and wires of the four rows of drift tubes. As well, the drift radius and time relation may not be very accurate here as a consequence of the low efficiency.  The deviations in the mean from zero from 55~mm to 65~mm are likely a result of an overall shift in the fibre placement in this region of the fibre tracker. There is also a dead fibre channel at 55~mm. The right plot in Fig.~\ref{fig:dxxtube60} shows the standard deviation of $\delta{}x$, which is the combined resolution of the extrapolated tube track and the fibre tracker, at the fibre plane, with a mean value $\sigma{}(\delta{}x) = 0.39\pm{}0.01$~mm.  Results for 90, 150 and 230~cm trigger position data are shown in Fig.~\ref{fig:dxxtubeall} and show similar characteristics as found in the 60~cm data.

\begin{figure}[htbp]
\centering
\includegraphics[angle=0,width=0.49\textwidth]{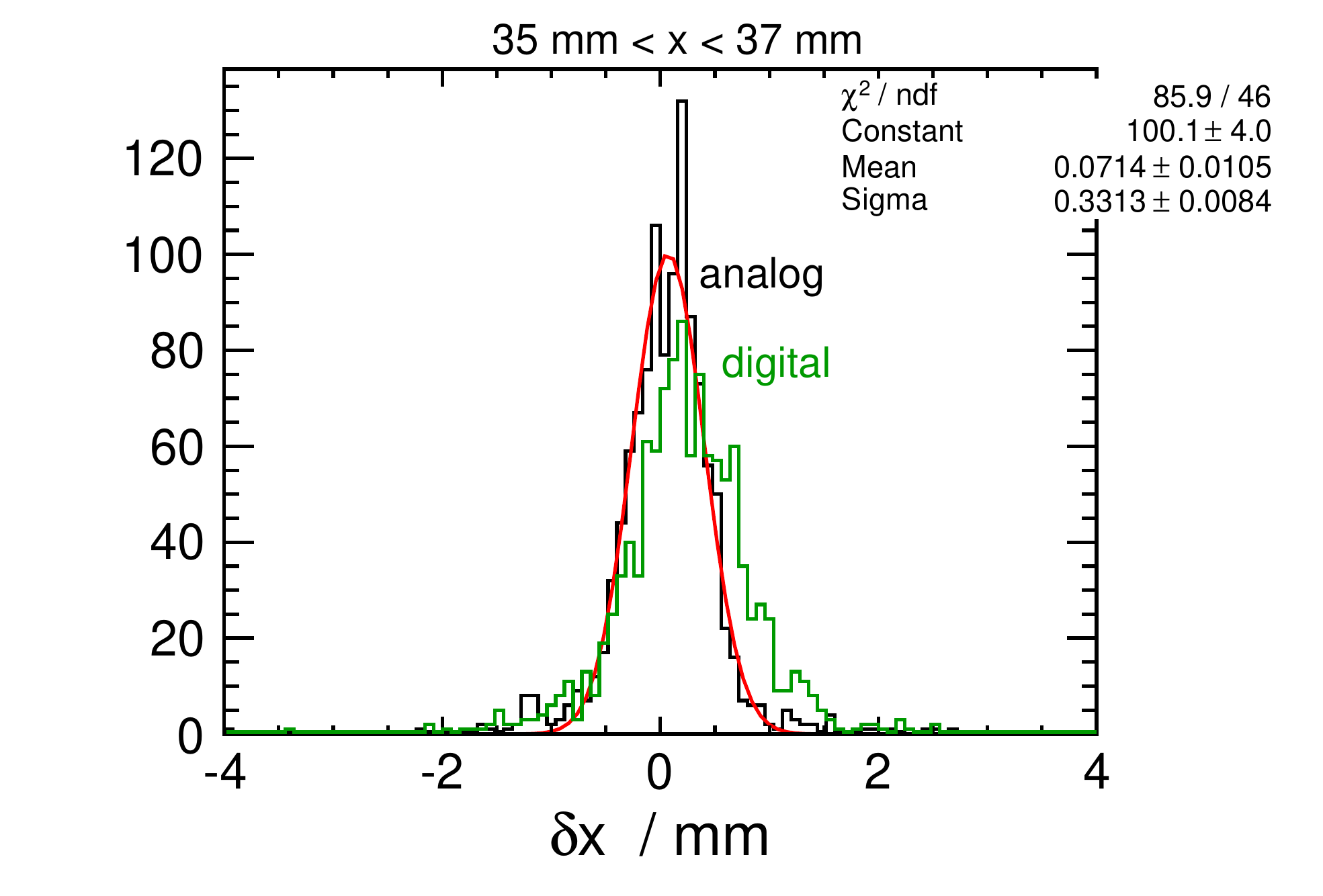} 
\caption{The difference in position, $\delta{}x$, for the region $35 < x_{tube} < 37$~mm. The analog data is in black, with a Gaussian fit overlaid in red. $\delta{}x$ found using a digital method from the fibre plane is overlaid in green.}
\label{fig:deltax3537}
\end{figure}

The $\delta{}x$ distribution for 2 neighbouring data points of Fig.~\ref{fig:dxxtube60} is shown in Fig.~\ref{fig:deltax3537}. The charge-weighted data is shown in black, with the digital data overlaid in green. A Gaussian function is fit to the charge-weighted data with the fit parameters shown in the figure. The digital data typically has a wider distribution from the larger impact of crosstalk in adjacent channels. The relatively low amplitude (1-2 p.e.) of the crosstalk has a smaller impact on the charge-weighted method.

\begin{figure}[htbp]
\includegraphics[angle=90,width=0.49\textwidth]{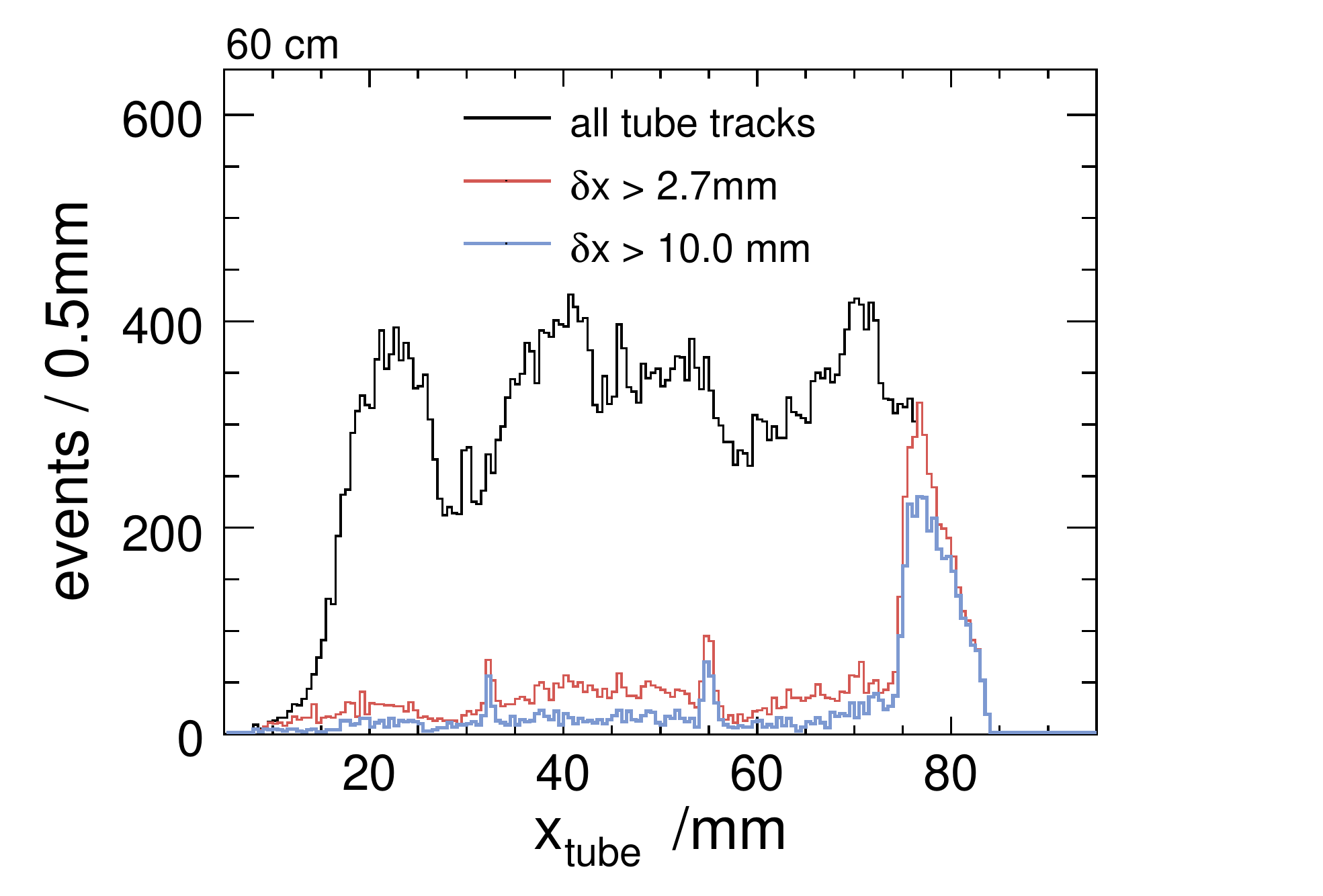} 
\includegraphics[angle=90,width=0.49\textwidth]{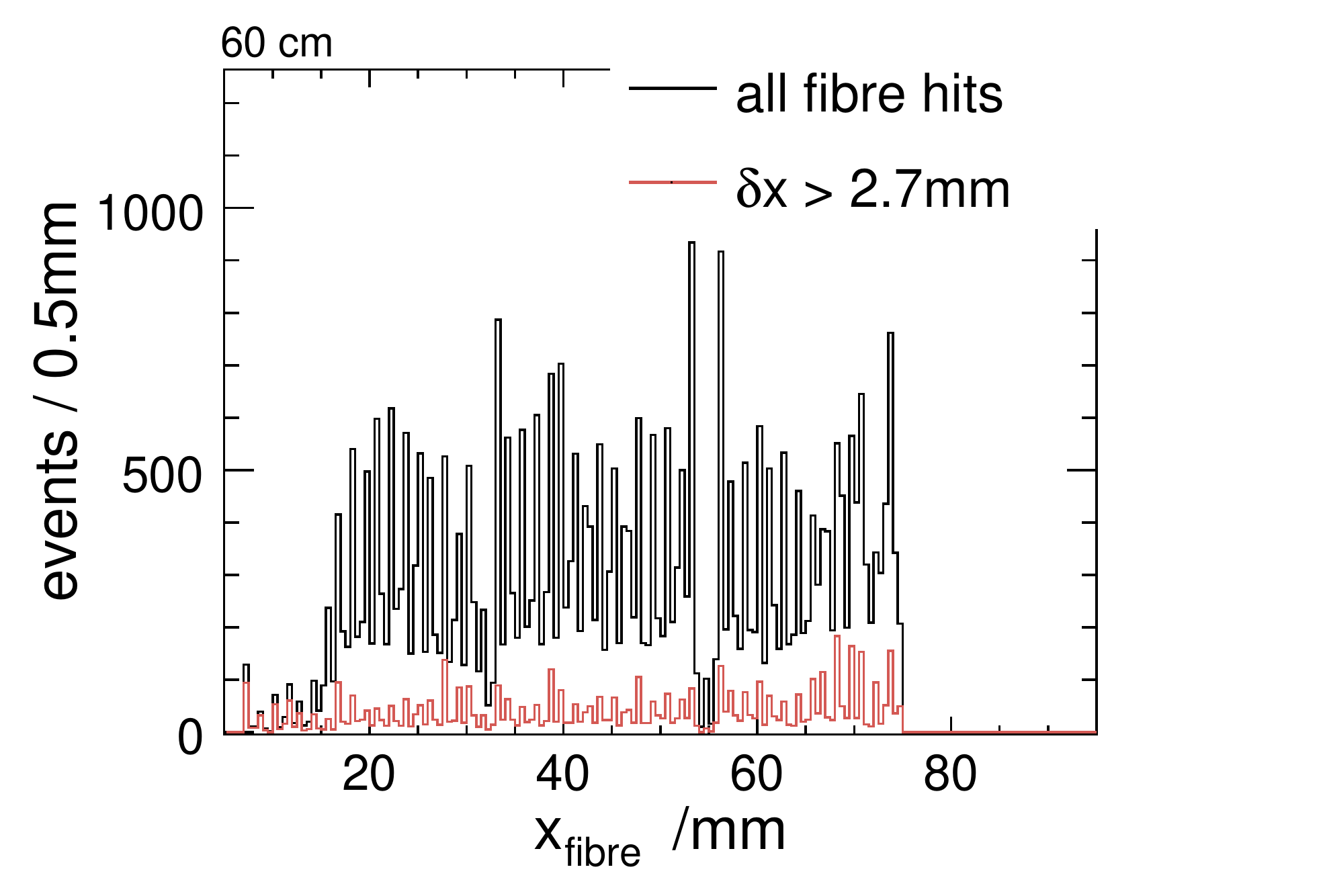} 

\caption{\textit{All events} which have tracks in the drift tubes as well as a hit in the fibre tracker are seen.  (left) A histogram of the extrapolated track position data. (right) A histogram of the charge weighted fibre tracker position data .  No $\chi^{2}$ cut has been applied to remove poor tube tracks in these plots.}
\label{fig:tubefibrex}
\end{figure}

Histograms of the extrapolated position from the drift tubes, as well as the position found from hits in the fibre tracker, are  seen in Fig.~\ref{fig:tubefibrex}. All events which have tracks in the drift tubes as well as a hit in the fibre tracker are seen without cuts on the $\chi^{2}$ value of the track. Events which have a large \textit{disagreement} between the fibre tracker and tube track also also reported, in order to highlight regions with poor efficiency. The extrapolated track position, shown on the right of Fig.~\ref{fig:tubefibrex},  covers the range from 15 to 85~mm. The requirement of having four tube-layer hits  reduces the left/right ambiguity of the drift radius, but also reduces the effective coverage of the drift tubes.  The spikes in the fibre hit position in the right plot  are due to many events having only a single fibre hit.  The fibre tracker spans the position range from 5 to 75~mm. Noisy fibres at the extreme edge were excluded due to a possible light leak, narrowing the effective area of the tracker.

A summary of the resolution and efficiency for the four different trigger positions are shown in Table.~\ref{tab:summary1}. The efficiencies are found from the total number of events which pass the required cuts, not the fits shown previously. The digital resolution is found using the same analysis method as described above, but the position in the fibre plane is found by taking the center of the channels which only exceed a threshold in the largest (by charge) cluster. As will be shown in the Monte Carlo section below, the crosstalk has a larger negative impact on the digital resolution than on the energy weighted method. The fibre resolution is determined by subtracting the expected contribution from the combined resolution, in quadrature. The efficiency is determined by counting the fraction of \textit{normalized events}  that fall within 2.7~mm of the extrapolated track position, determined from the largest charge-weighted cluster. normalized events require hits in four layers of drift tubes and a mean drift time in the bottom layers to be larger than 250~ns and a $\chi^{2}$ from the track fitter.

\begin{figure}[htbp]
\includegraphics[angle=0,width=0.49\textwidth]{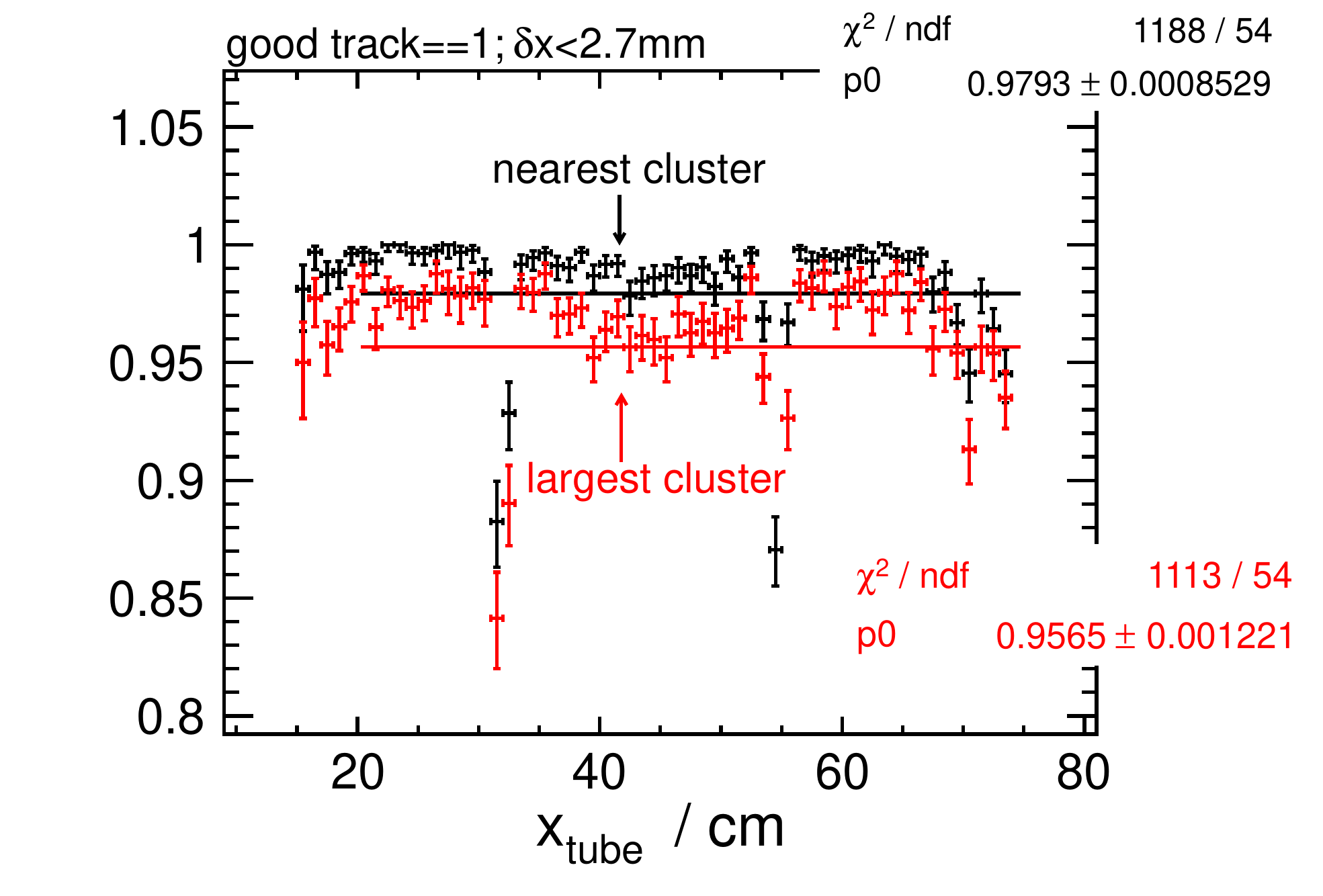}
\includegraphics[angle=0,width=0.49\textwidth]{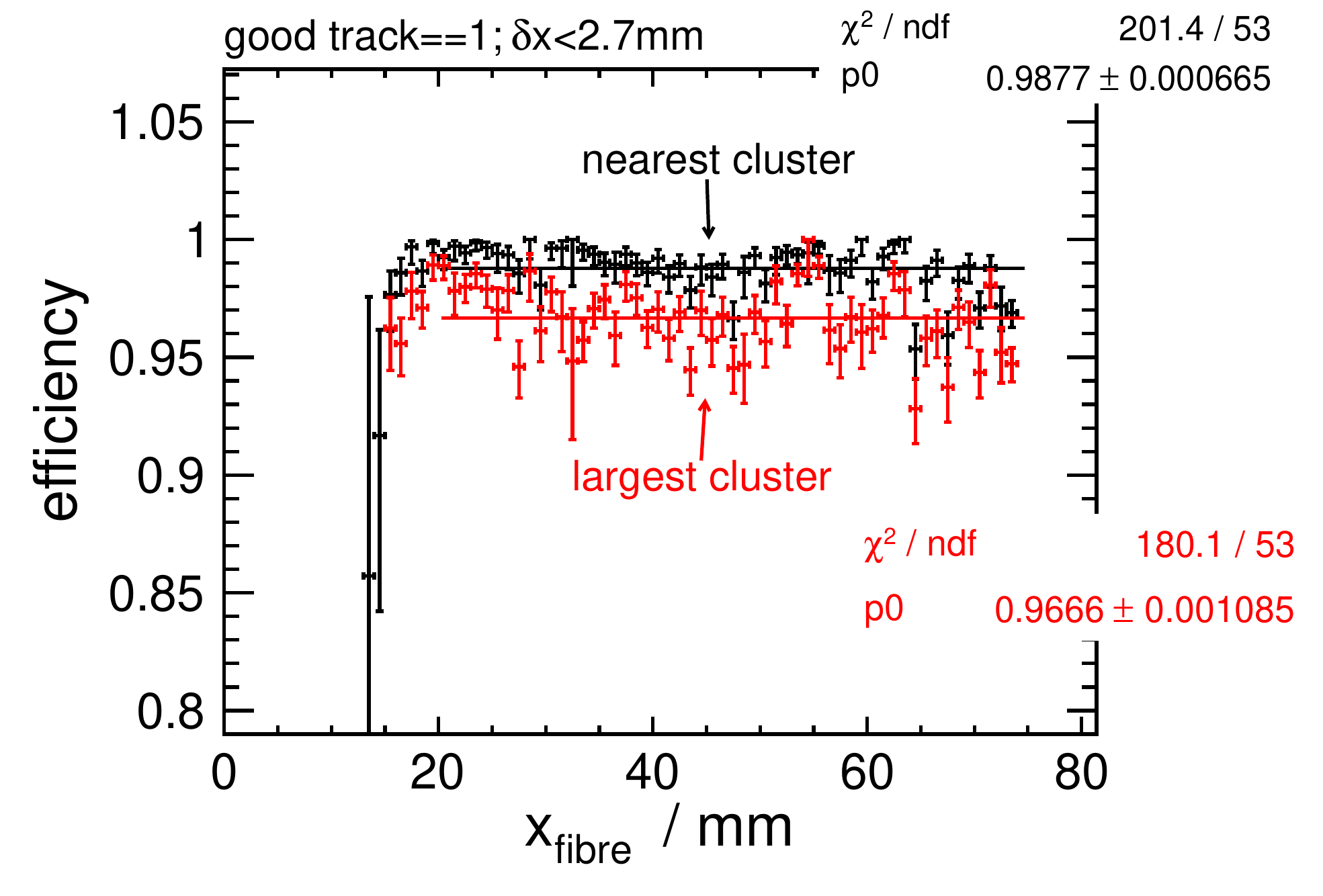}
\caption{When clusters are chosen by the largest total amplitude, the efficiency, $\epsilon_{2.7}$ (fraction of \textit{normalized events} with $|\delta{}x|$ less than 2.7~mm) seen at the position found from the tube (fibre) track is seen in the left (right) plot.  From the difference in efficiency these two plots, it can be seen that at 60~cm approximately 2\% of efficiency is lost from a lack of signal above threshold in the fibre tracker and approximately 2.5\% of efficiency is lost to a noise or crosstalk signal exceeding the cosmic track signal. Choosing the cluster nearest to the drift tube track instead, shows a 2\% higher efficiency than the largest cluster method, indicating again that noise cluster amplitudes exceed the track cluster amplitudes approximately 2\% of the time.}
\label{fig:efftubefibrex}
\end{figure}

Losses in efficiency are attributed to downward fluctuations in the light yield and signal response to amplitudes below the 10~MeV threshold. If the largest cluster is chosen, clusters with signals that fall below large crosstalk and electronics noise amplitudes are misidentified. The efficiency over  detector position can be seen in Fig.~\ref{fig:efftubefibrex} for 60~cm cosmic data. These plots agree within error with Table~\ref{tab:summary1}.  Approximately 2.1\% of efficiency is lost due to a lack of signal above threshold in the fibre tracker and approximately 2.5\% of efficiency is lost due to  noise or crosstalk signal exceeding the cosmic track signal. Choosing the cluster nearest the track, regains the lost inefficiency due to larger amplitude noise and crosstalk clusters.

\begin{table}[htbp]
\centering
\caption{Cosmic data summary of the combined experimental drift tube and fibre tracker resolutions with statistical errors, resolutions with the expected drift tube contribution (see Fig.~\ref{fig:drXrow260}) at the fibre plane subtracted in quadrature and efficiency of matching normalized tracks (also having fibre track hits).}

\begin{tabular}{|c|c|c|c|c|}
\hline trigger pos. & 60~cm & 90~cm & 150~cm & 230cm \\  \hline
\hline combined res.(ana)/mm & 0.386$\pm$0.009 & 0.380$\pm$0.008   & 0.413$\pm$0.006   & 0.474$\pm$0.011   \\ 
\hline combined res.(dig)/mm &0.427$\pm$0.011   &0.427$\pm$0.008   & 0.453$\pm$0.005  & 0.508 $\pm$ 0.011  \\ \hline
\hline fibre res.(ana)/mm &0.355$\pm$0.029 & 0.349$\pm$0.029   &0.385 $\pm$ 0.032  &0.450 $\pm$ 0.037  \\ 
\hline fibre res.(dig)/mm &0.400$\pm$  0.033 &0.400$\pm$0.033  &0.427 $\pm$ 0.036 &0.485  $\pm$ 0.040 \\ \hline
\hline efficiency $\epsilon_{2.7}$ & 0.960  &0.953  & 0.942  & 0.895  \\ 
\hline 

\end{tabular} 
\label{tab:summary1}
\end{table}

\section{Monte Carlo Simulation of Fibre Tracker}

\begin{figure}[htbp]
\includegraphics[width=0.49\textwidth]{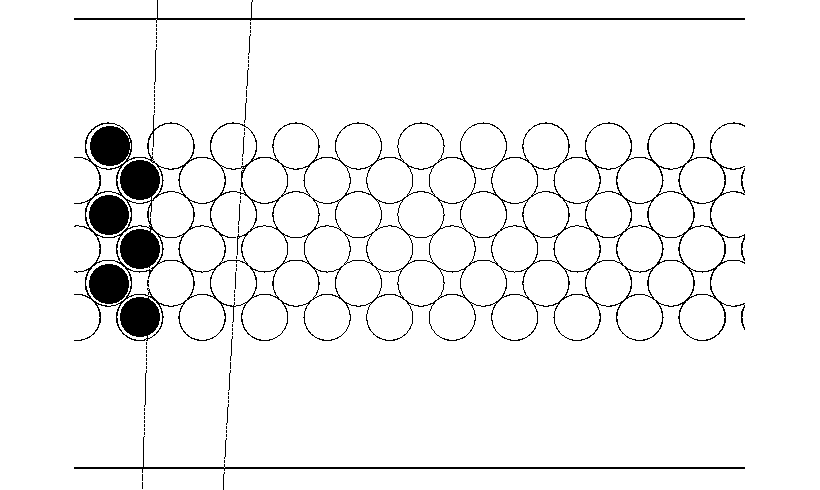} 
\includegraphics[angle=90,width=0.49\textwidth]{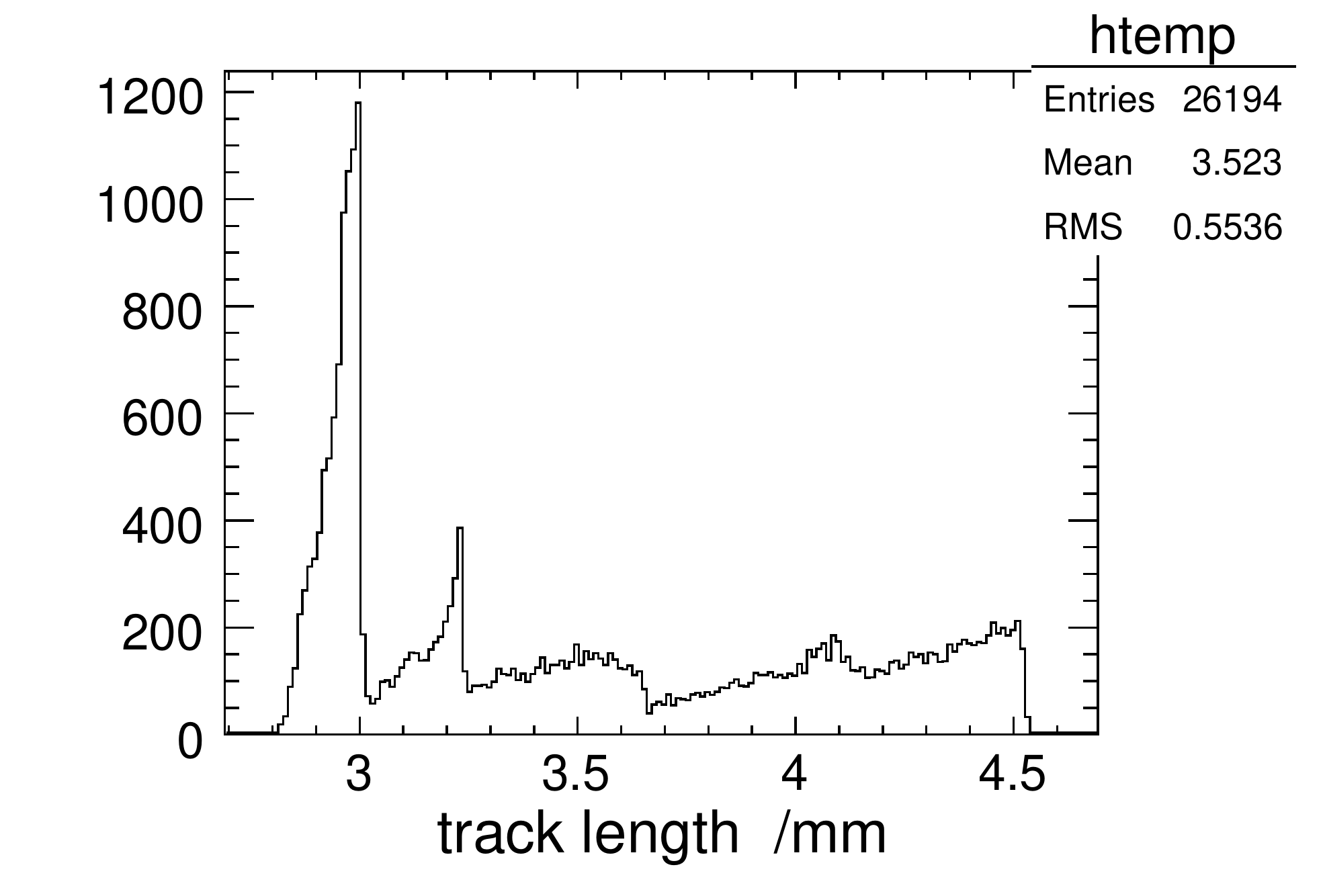} 
\caption{\textbf{1.35mm} (left) A schematic of the fibre layout in Monte Carlo. The 6 shaded fibres represent  a single channel. (right) The total track length in the fibre detector.}
\label{fig:tracktotal_135mm}
\end{figure}

A GEANT4~\cite{ref_geant4} detector simulation was used to simulate the energy deposition of 4~GeV muons in plastic scintillator. The simulated muons had an angular distribution flat in cos$^{2}(\theta)$, similar to cosmic rays, spread uniformly over the surface area of the top trigger paddle. Events which deposited energy in both the top and bottom trigger scintillator paddles were written to file. The 1~mm fibres were composed of a material with a density of 1.05~g/cm$^{2}$, with a chemical composition C$_{8}$H$_{8}$. Polystyrene is (C$_{8}$H$_{8}$)$_{n}$. To simulate the saturation in the scintillation process by reducing the energy deposited per mm, Birk's constant is set to 0.126~mm/MeV for plastic scintillator. No glue or lead material was added to the simulated fibre detector. This would not have affected  the energy deposition in the fibres, but may have reduced multiple scattering in the overall detector. The fibres have a uniform placement as seen in Fig.~\ref{fig:tracktotal_135mm}(left). The centre of each channel (a column of six fibres) was uniformly separated by 1.35~mm. The horizontal pitch of the fibres was 0.675~mm  with a vertical pitch of 0.74~mm. Unlike the fibre detector in the experimental setup seen in the Fig.~\ref{fig:fibreplanephoto}, the simulated detector is perfectly uniform in fibre placement. Variation in the pitch would have the effect of increasing the overall position resolution of the detector, but this variation has not been measured in the actual detector.

\begin{figure}[htbp]
\includegraphics[angle=90,width=0.49\textwidth]{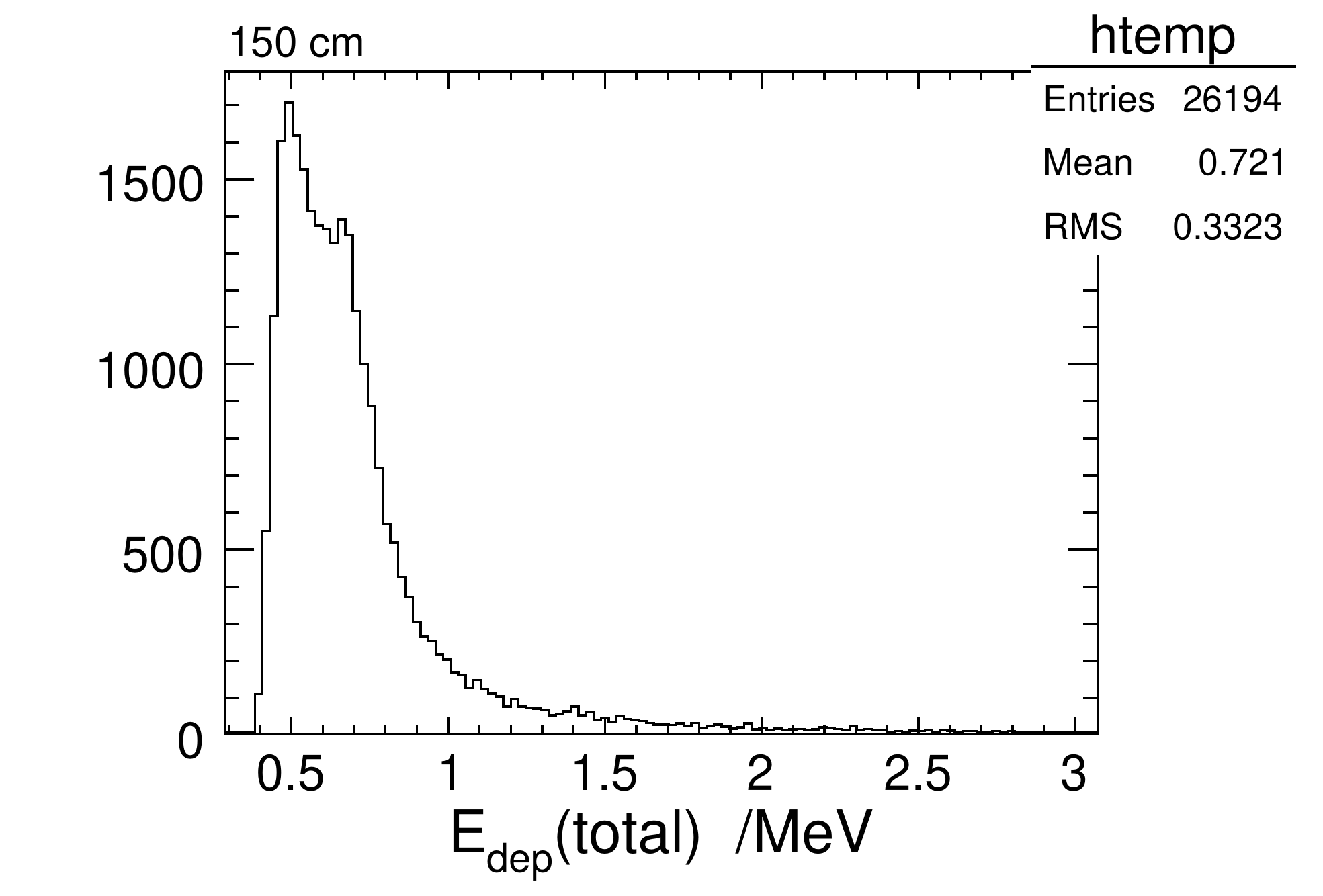}
\includegraphics[angle=90,width=0.49\textwidth]{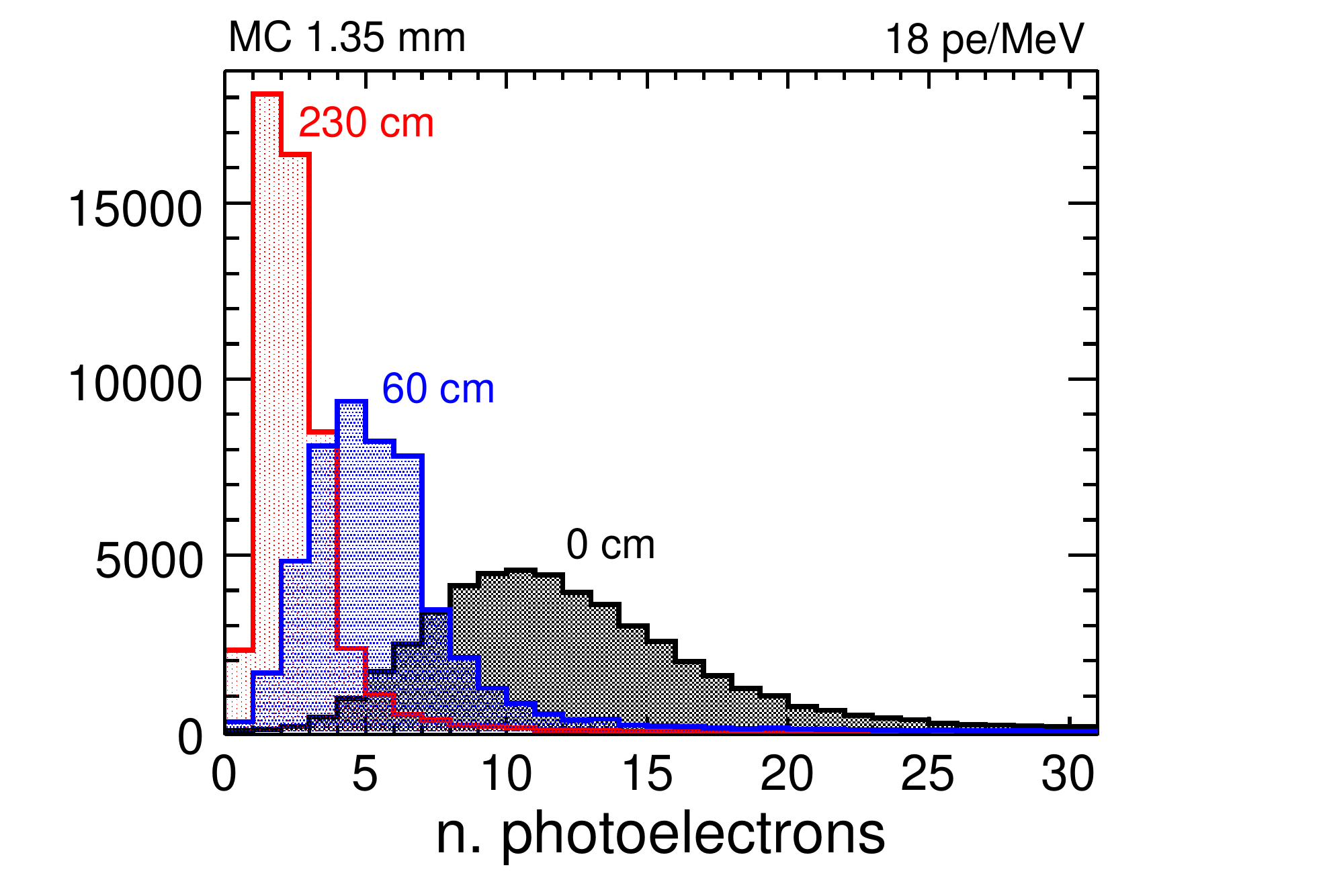}

\caption{\textbf{1.35mm} (left) The total energy deposited in the fibre tracker. (right) The photoelectron distribution as seen at the source (0~cm), at 60~cm and 230~cm. }
\label{fig:edep_135mm}
\end{figure}

The total track length through the fibres is shown in Fig.~\ref{fig:tracktotal_135mm}(right), while the total energy deposited in the fibres shown in Fig.~\ref{fig:edep_135mm}(left). The large variation in energy deposition and track length will prevent an analog signal analysis from providing more reconstruction  information than a digital analysis would.

\subsection{Number of Photoelectrons and Charge}
 The charge response of the detector is modelled in the following way:

\begin{itemize}
\item The simulated number of photoelectrons seen by the photodetector, from light produced by a MIP depositing energy in the fibre core, follows a Poisson distribution of integer values:
\begin{equation}
npe(0~cm) = Poisson\left(npe / MeV \cdot{} E_{dep}(MeV)\right)
\end{equation}
\item The probability that a photon is transmitted to the end of the fibre is described by an exponential function, such that the photoelectron yield at the detector is
\begin{equation}
npe(x) = Integer\left[npe(0~cm)\cdot{}\left(A_{1}e^{-x/\lambda_{1}} + A_{2}e^{-x/\lambda_{2}} \right)\right]
\end{equation} 
\item The charge response of the photodetector is described by a Gaussian probability distribution where the mean is $npe(x)\cdot{}G_{1pe}$ where $G_{1pe}$ is the charge amplitude of a single photoelectron. The standard deviation, $\sigma$, of the charge response is a quadratic sum of the pedestal width, $\sigma_{ped}$ and individual photoelectron peak, $\sigma_{1pe}\cdot{}\sqrt{npe(x)}$. The overall charge is then
\begin{equation}
C = G_{1pe}\cdot{}Gauss\left(npe(x),\sqrt{npe(x)\sigma_{1pe}^{2}+\sigma_{ped}^{2}}\right)\,.
\end{equation}
\item electronics noise in individual channels and crosstalk between adjacent MA-PMT channels are modelled with rates and distributions seen in the experimental Cosmic data.

\end{itemize}

The mean number of photoelectrons/MeV in the  Monte Carlo was calibrated by matching charge distributions to  create similar shapes to the Cosmic data. The width of the distribution is dominated by the number of photoelectrons, and the efficiency (npe greater than zero) determines the normalization.

\begin{table}
\centering
\begin{tabular}{|l|l|}
\hline Parameter & Value \\ 
\hline $npe / MeV$ & 18 \\ 
\hline $\lambda_{1},~\lambda_{2}$ (cm) & 210.7,~0 \\ 
\hline $A_{1},~A_{2}$ (pC) & 1,~0 \\ 
\hline $\sigma_{1pe}$ & 0.6 \\ 
\hline $\sigma_{ped}$ & 0.02 \\ 
\hline $G_{1pe}$ (pC)  & 18.6 \\ 
\hline 
\end{tabular} 

\caption{Parameters and their values input to the fibre detector response model.}
\label{tab:MCparameters}
\end{table}

\begin{figure}[htbp]
 \includegraphics[angle=90,width=0.5\textwidth]{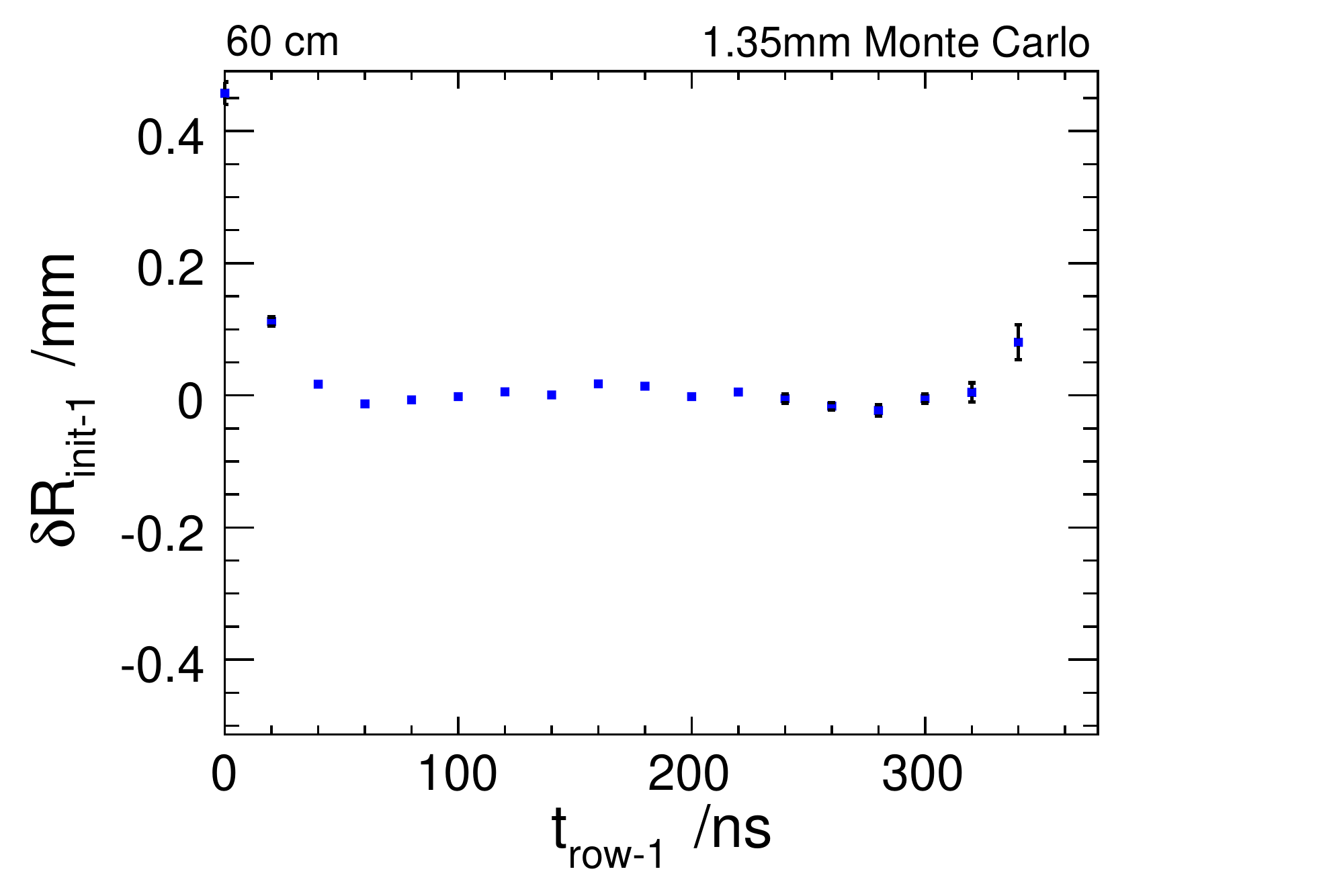}
  \includegraphics[angle=90,width=0.5\textwidth]{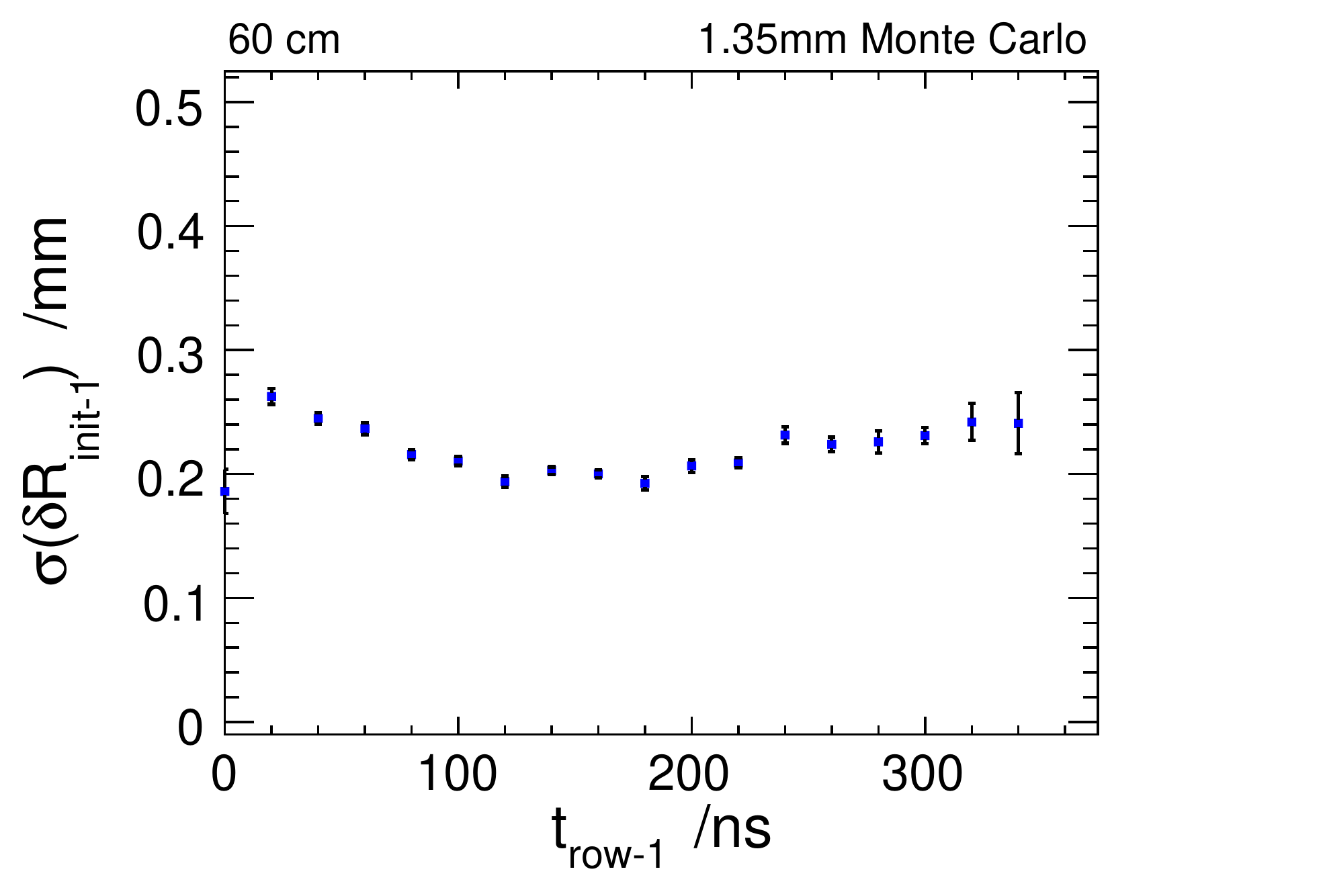}

\caption{\textbf{MC60~cm.}(left plot) The mean value of $\delta{R}$ as a function of $t, $before a $\chi^2$-minimization, for Row-1 tubes, for the 60~cm trigger position. Assuming Row-0, Row-2 and Row-3 tubes have their correct R(t) relation, this distribution should be centred on zero. (right plot) The standard deviation of $\delta{}R$ as a function of drift time, $t$.}
 \label{fig:drtrow1MC60}
\end{figure}

\begin{figure}[htbp]
 \includegraphics[angle=90,width=0.5\textwidth]{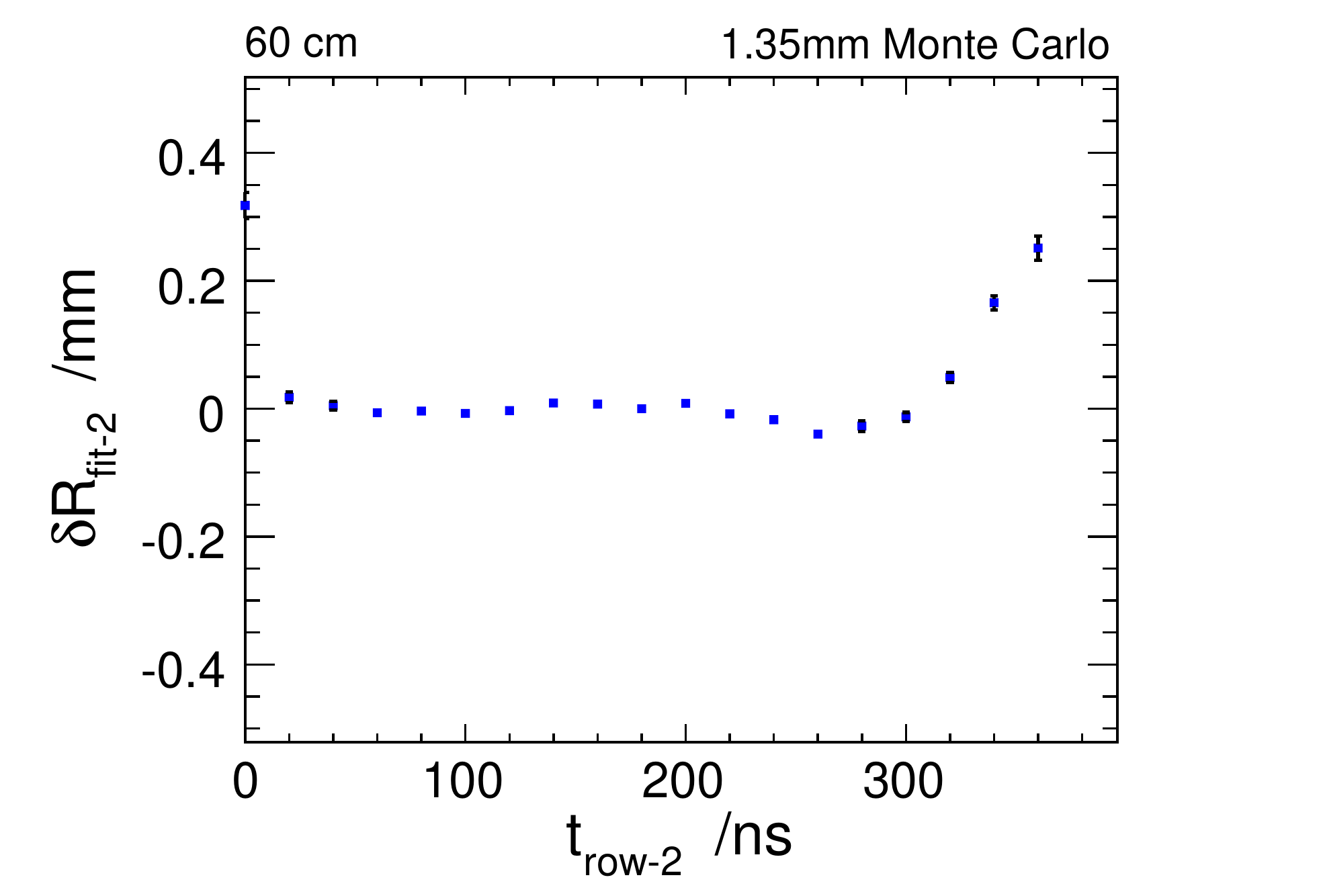}
  \includegraphics[angle=90,width=0.5\textwidth]{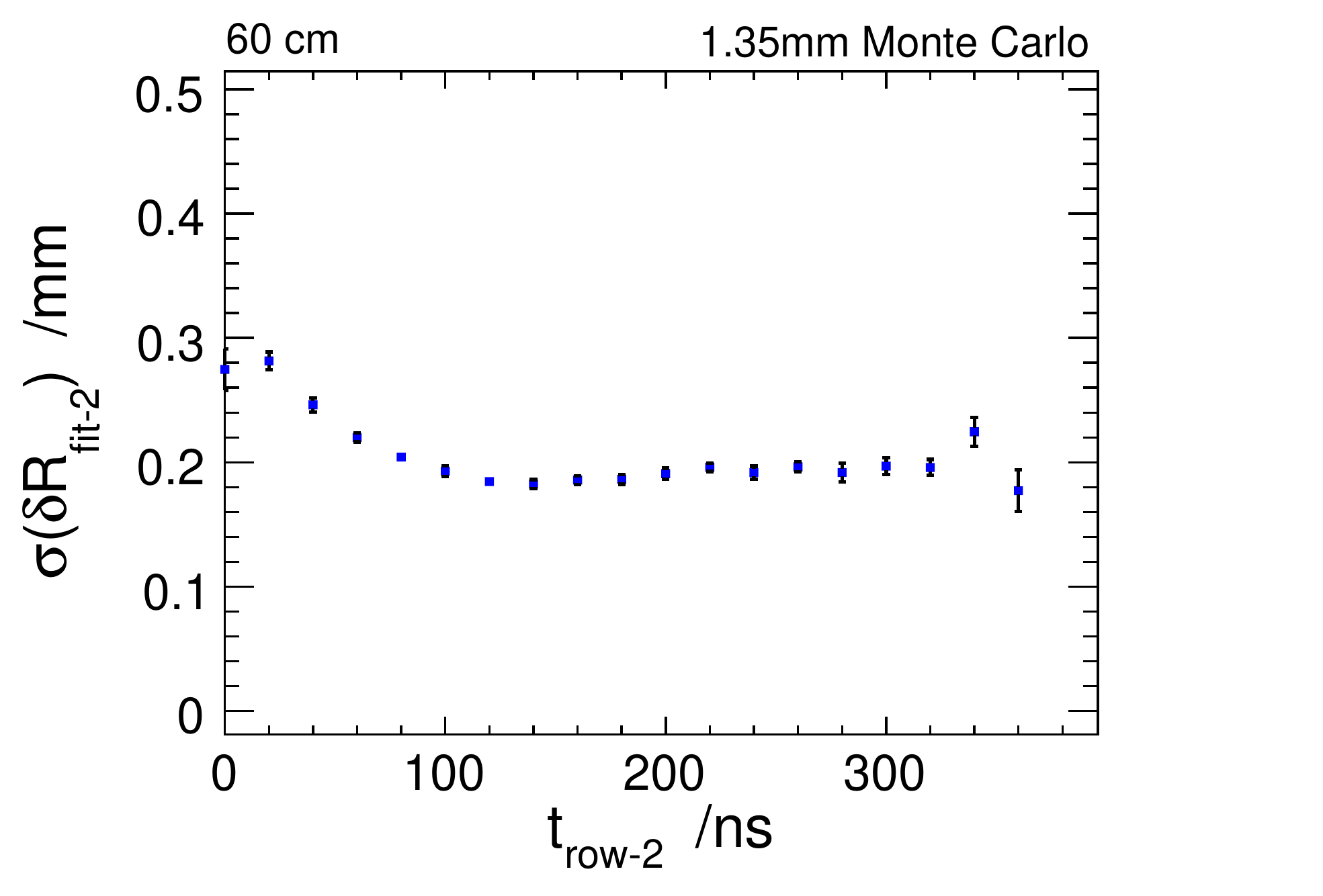}
 
\caption{\textbf{MC60~cm.}(left plot) The mean  value of $\delta{R}$  as a function of $t$, for Row-2 tubes, after a $\chi^2$-fit, with Row-2 excluded from the fit.  (right plot) The standard deviation of $\delta{R}$ as a function  of $t$.}
 \label{fig:drtrow2MC60}
\end{figure}

The Monte Carlo muon tracks through the simulated detector have their radii smeared from the resolution found in the data and given a time from the inverse of the R(t) relation found from the data. The Monte Carlo data was reconstructed with the same method applied to the experimental data, re-establishing an R(t) relation for the tubes and using the same analysis methods to determine the drift tube resolution. Figs.~\ref{fig:drtrow1MC60}~and~\ref{fig:drtrow2MC60} show the $\delta{}R$ precision and resolution  of the reconstructed Monte Carlo tracks for Row-1 tubes before the minimization fit is done, and Row-2, after the minimization fit from which it was excluded.

\begin{figure}[htbp]
 \includegraphics[angle=90,width=0.5\textwidth]{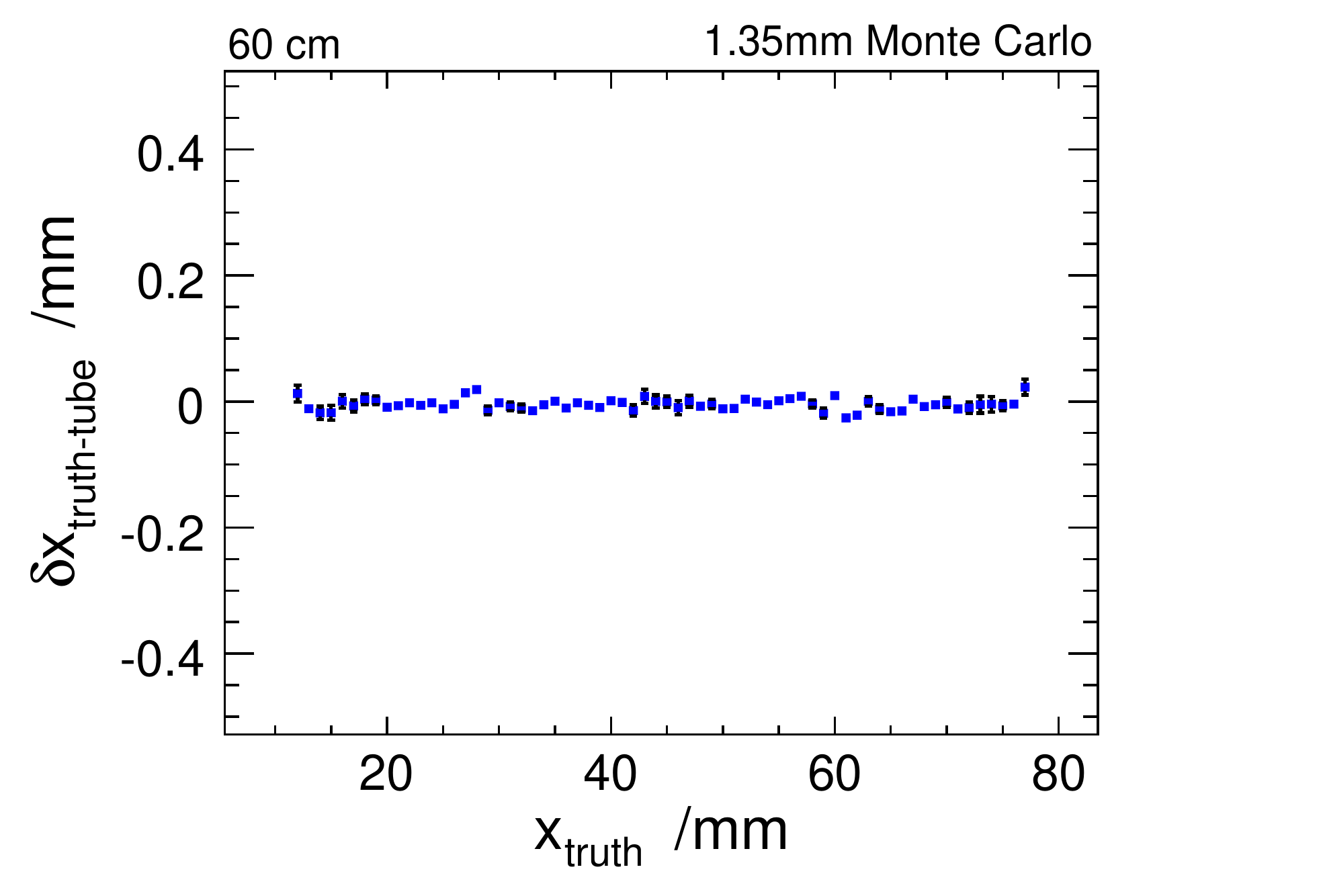}
  \includegraphics[angle=90,width=0.5\textwidth]{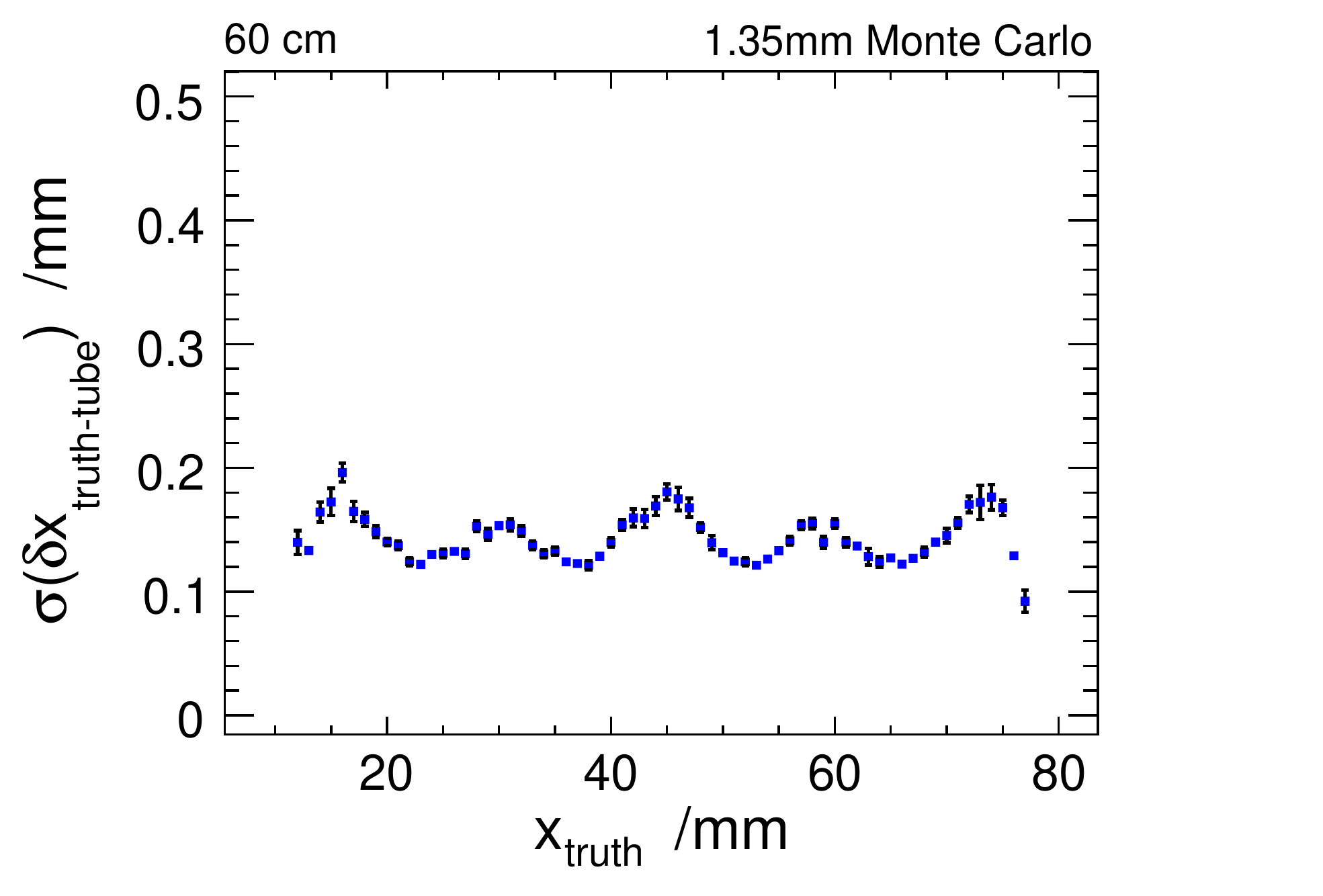}

\caption{\textbf{MC60~cm.} The mean value of  the difference between the true Monte Carlo track position and the  extrapolated track position as a function of $x$ from the Monte Carlo.  (right plot) The standard deviation of $\delta{}x_{(truth-tube)}$ over $x$.}
 \label{fig:dxtruthtubeMC60}
 \end{figure}

With the simulated data, it is possible to extract the contributions  of the extrapolated drift tube track and the reconstructed fibre position. The difference between the extrapolated track position at the fibre plane and the true position in the Monte Carlo simulation and its standard deviation is shown in Fig.~\ref{fig:dxtruthtubeMC60}. This is the contribution of the tube track to the total combined resolution measured at the fibre plane. Since the resolution is modelled after the experimental data, the same features appear near the tube wire and edge positions. However, the efficiency in these regions is still high compared to the data.

\begin{figure}[htbp]
 \includegraphics[angle=0,width=0.5\textwidth]{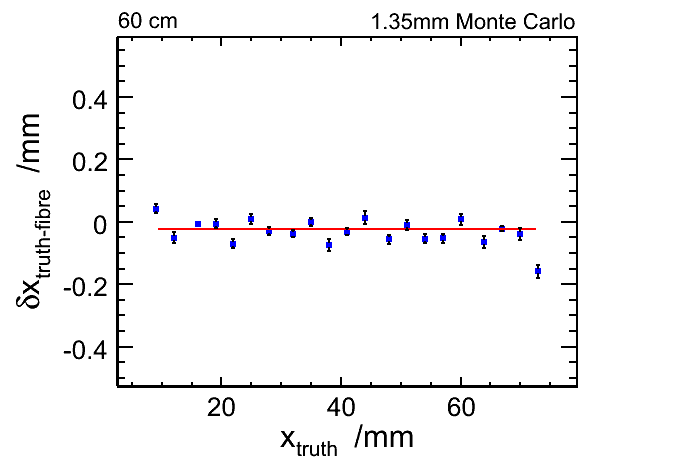} 
  \includegraphics[angle=0,width=0.5\textwidth]{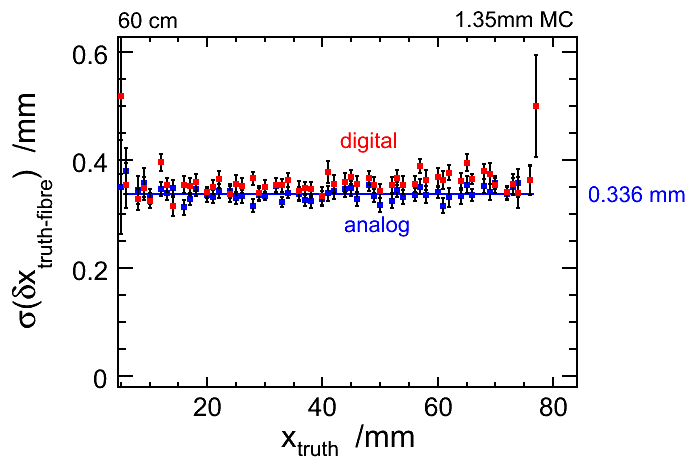}

\caption{\textbf{MC60~cm.} The mean value of  the difference between the true Monte Carlo track position and the  reconstructed fibre position as a function of $x$ from the Monte Carlo.  (right plot) The standard deviation  $\delta{}x_{(truth-fibre)}$ over $x$. Both digital and analog results are shown.  This is the contribution of the fibre plane resolution to the overall resolution, which is what can be measured in the experimental setup.  }
 \label{fig:dxtruthfibreMC60}
 \end{figure}

The difference between the reconstructed fibre position  and the true position in the Monte Carlo simulation and its standard deviation is shown in Fig.~\ref{fig:dxtruthfibreMC60}. This is the contribution of the fibre plane to the total combined resolution measured. Unlike the data, there are no features due to irregular spacing or dead fibres aside from the same spikes due to predominant single channel hits.

\begin{figure}[htbp]
 \includegraphics[angle=0,width=0.5\textwidth]{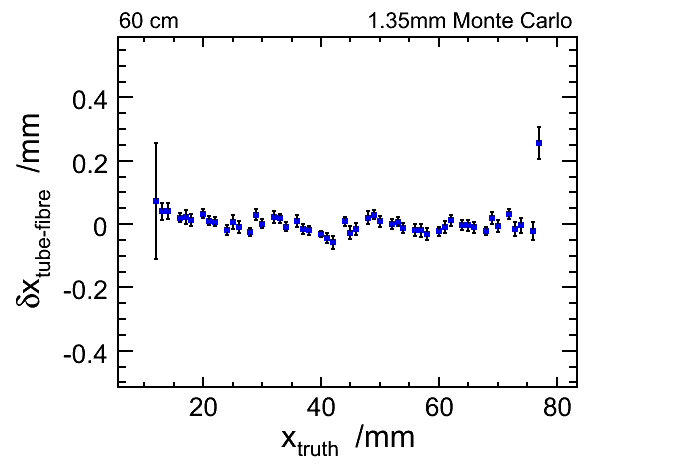}
  \includegraphics[angle=0,width=0.5\textwidth]{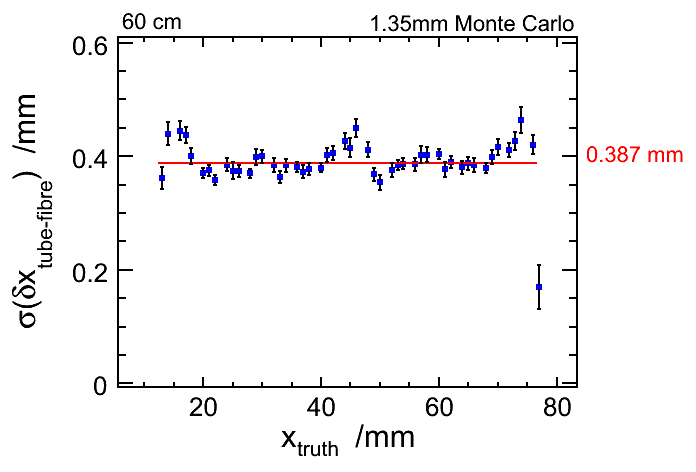}
 
\caption{\textbf{MC60~cm.} (left plot) The mean value of the difference between the extrapolated track position and the reconstructed fibre position as a function of $x$ from the Monte Carlo.   (right plot) The standard deviation of $\delta{}x_{(tube-fibre)}$ over $x$ is the combination of the track and fibre plane resolutions.}
 \label{fig:dxxtubeMC60}
\end{figure}

With the correct parameters in the Monte Carlo, the total combined resolution measured should be equal or slightly better than that found in the experimental setup. The improved resolution would be due to effects the regular fibre arrangement in the Monte Carlo, as well as improved track finding and efficiency in the simulated drift tubes. The  difference between the reconstructed fibre position and extrapolated track position in the Monte Carlo simulation and its standard deviation are shown in Fig.~\ref{fig:dxxtubeMC60}. This is the total combined resolution that should be compared to the combined resolution found in the experimental data. A summary of the measured fibre plane resolutions and efficiencies  for the four trigger positions is shown in Table~\ref{tab:summarymc135mm}, which are in good agreement with the experimental data in Table.~\ref{tab:summary1}.

\begin{figure}[htbp]
 \includegraphics[angle=90,width=0.5\textwidth]{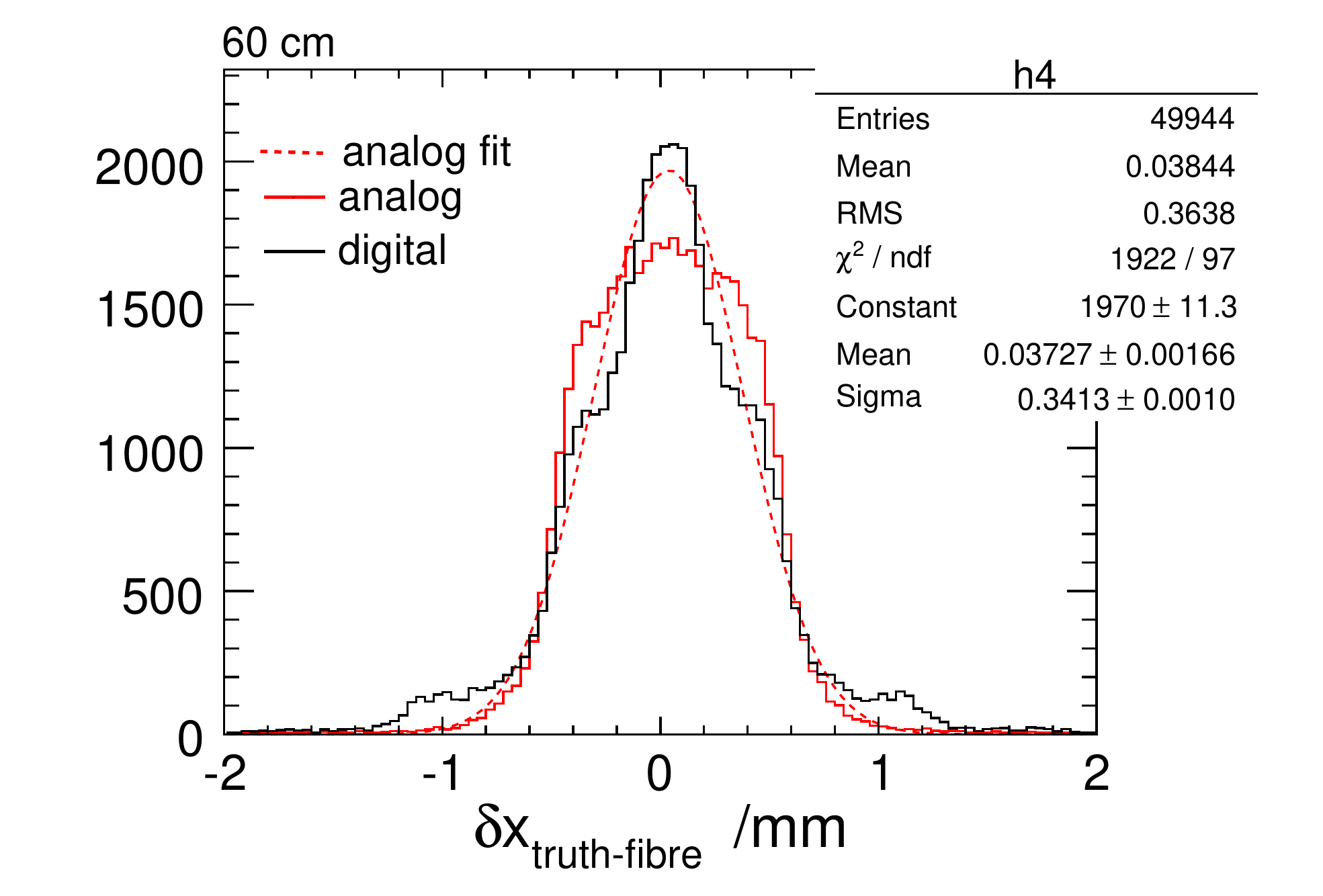} 
 \includegraphics[angle=90,width=0.5\textwidth]{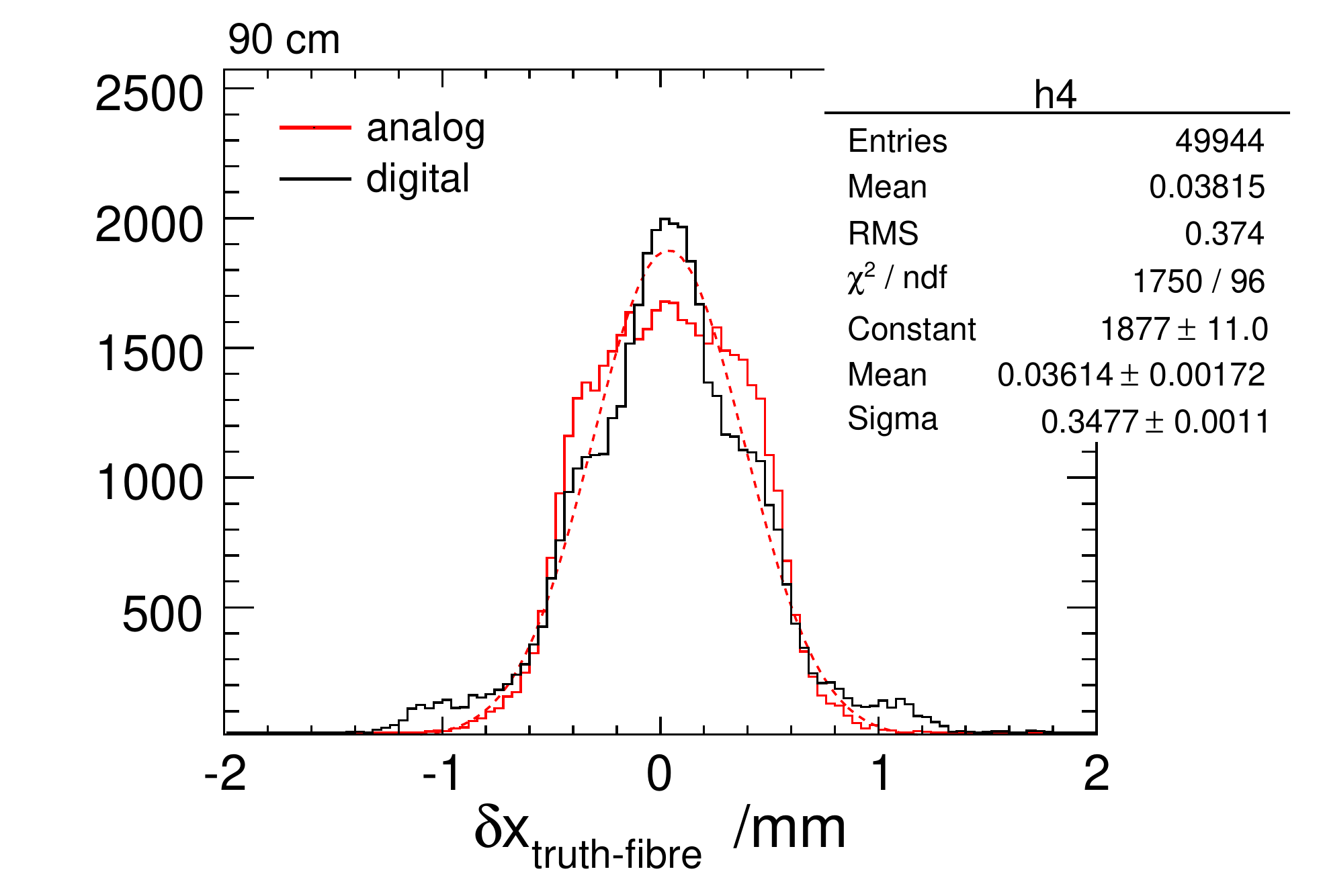} 
  \includegraphics[angle=90,width=0.5\textwidth]{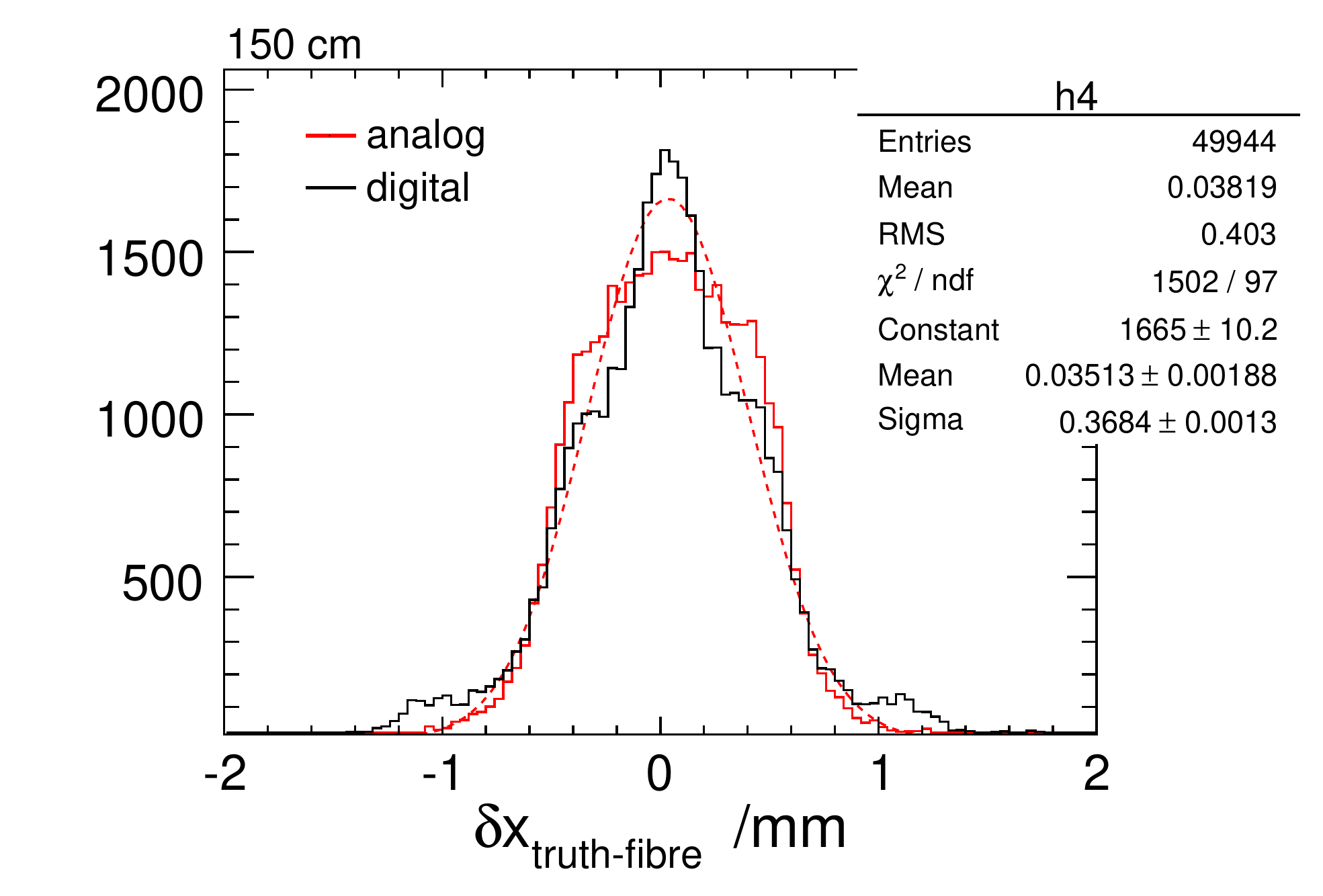} 
\includegraphics[angle=90,width=0.5\textwidth]{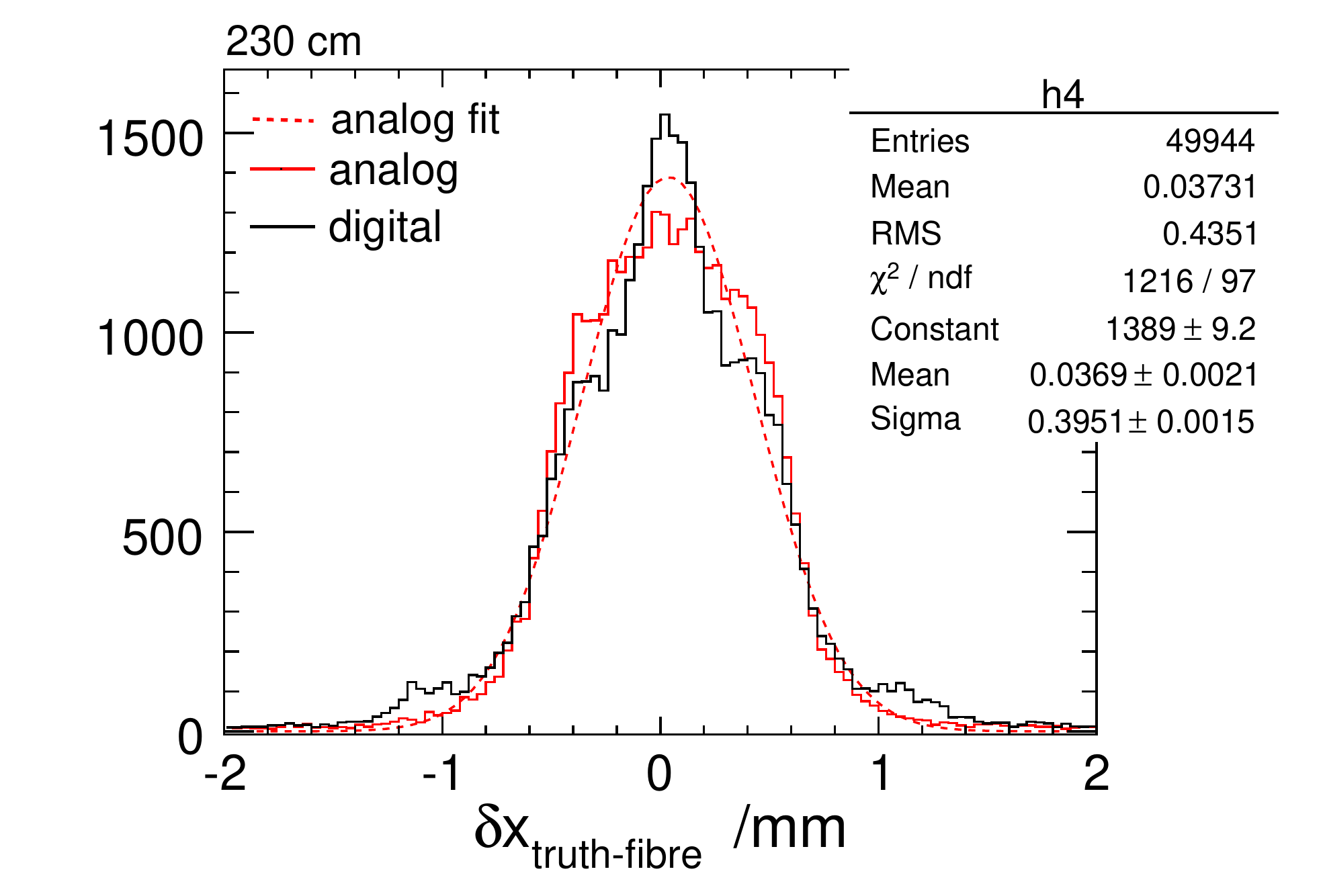} 

\caption{ A histogram of the difference in truth and fibre plane position for 60, 90, 150 and 230~cm data sets. The width of the distributions can be seen to increase as the trigger position becomes farther from the photodetector.}
 \label{fig:dxtruthfibreMC90}
 \end{figure}

Histograms of $\delta{}x_{(truth-fibre)}$ for Monte Carlo data are shown in Fig.~\ref{fig:dxtruthfibreMC90}. Naively, one would expect a flat distribution with a resolution of $width/\sqrt{12}$. The shaping of the analog reconstructed position with a peak near the center is an indication of the improvement in position resolution  from measuring the charge in clusters with multiple channel hits and overlapping channels. The charge in a single channel cluster does not provide any extra position information and the $\delta{}x_{(truth-fibre)}$ distribution of only single channel hits exhibits a flat shape as expected. Tails in the distributions of both the analog and digital reconstructions are a product of false-positive hits from crosstalk and noise in channels adjacent to those with actual energy deposit from the track. As well, false-negative hits contribute to the tails where the signal from the light yield falls below the 10 pC threshold in one of the shared track channels. The narrower part of the peak near zero, especially in the digital spectra, is a result of the small distance of overlap of adjacent channels (see Figs.~\ref{fig:fib1}b and~\ref{fig:thickpos}d). If there is a hit in this region, and the signals exceed threshold in both channels, the resolution is significantly better, as expected, than the much wider single channel hit regions.

\begin{table}[hbp]
	\centering	
	\caption{Summary of the resolution and efficiency for Monte Carlo data for four trigger positions. The efficiency, $\epsilon_{2.7}$,  is the fraction of events with extrapolated track position from normalized track and a $\chi^{2}\leq{}30$ and fibre hit position within 2.7~mm. .}
\begin{tabular}{|c|c|c|c|c|}
\hline MC1.35mm & 60~cm & 90~cm & 150~cm & 230cm \\  
\hline resolution (analog) /mm & 0.340  & 0.348  & 0.368  & 0.395  \\ 
\hline resolution (digital) /mm & 0.390  &   &   &   \\ 
\hline efficiency $\epsilon_{2.7}$  & 0.981  & 0.975   & 0.953  & 0.927 \\
\hline 
\end{tabular} 
\label{tab:summarymc135mm}
\end{table}

\section{Fibre Tracker Performance}

The high voltage of the MA-PMT is set to reach a gain of approximately $1.5- 2\times{}10^{6}$ where one photoelectron corresponds to 20 to 30 pC. Correcting for the relative gain of each channel of the MA-PMT, the few photoelectrons expected from the light yield of the scintillating fibres should produce signals between 20 and 200~pC, depending on the trigger position. If the width of the signal is dependent only on photoelectron statistics, the mean number of photoelectrons can be extracted. Two methods to extract these values are discussed below. The cosmic data has had secondary corrections applied to produce mean peak channel distributions with an  RMS of the peak means of 5--10\%. The \textit{peak channel} refers to the single channel in the fibre plane with the largest observed charge. The Monte Carlo data has had their channel amplitude means smeared by 5\%. As seen from Table~\ref{tab:moyalnpe} below, the width of the charge distributions  by channel from cosmic data vary with an RMS of ~15--20\%. Since the width goes with the square root of the number of photoelectrons, this indicates a variation in light yield on the order of 25--40\% between channels. The measured charged signals and the corresponding Monte Carlo detector charge model are discussed below.

\subsection{Charge Distribution and Attenuation Length}

\begin{figure}[htbp]
\centering
\includegraphics[angle=90,width=0.7\textwidth]{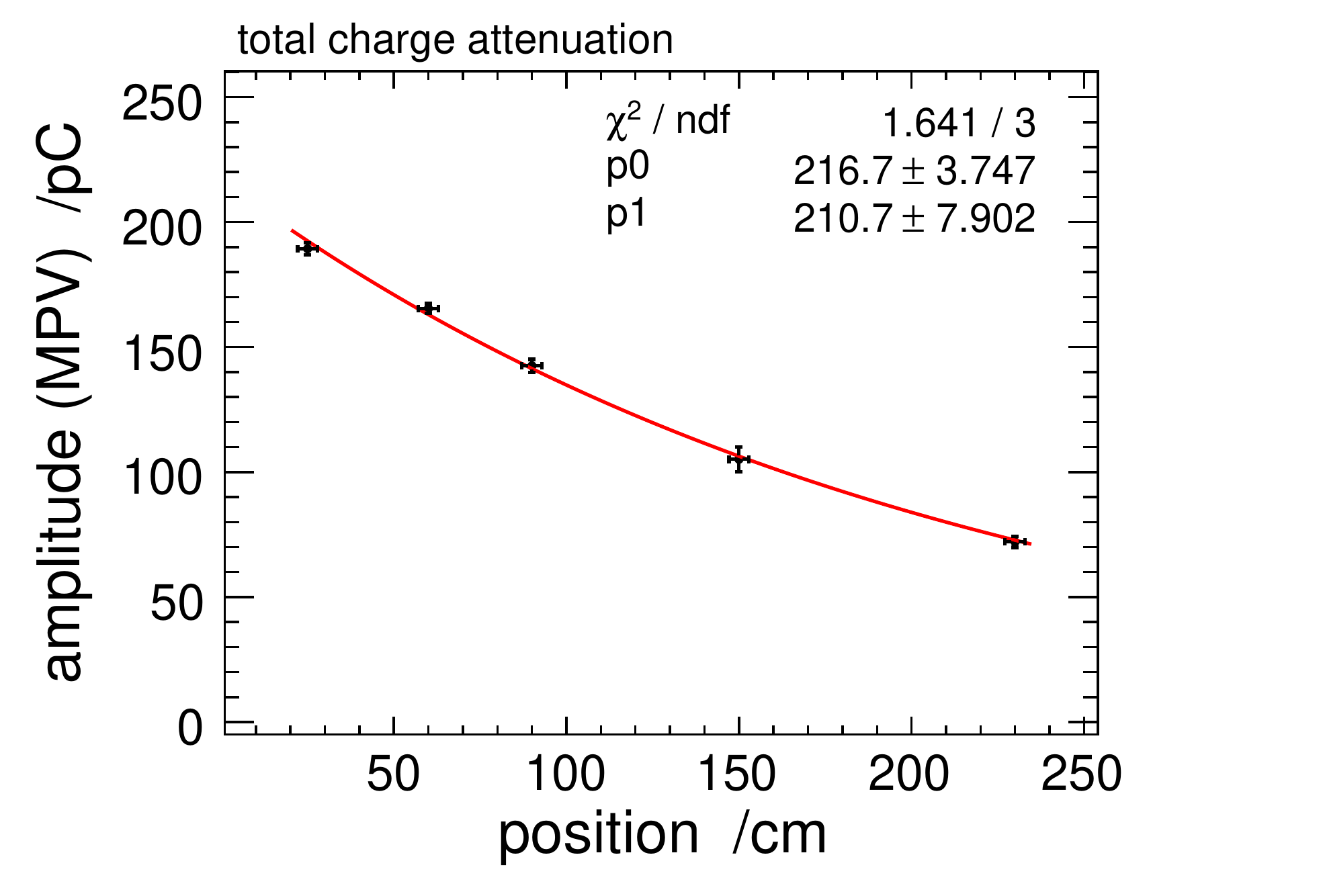} 
\caption{\textbf{attenuation} The mean of the total charge distributions as a function of trigger position. The data are fit to a single  exponential function with a decay length of 210~cm. }
\label{fig:atten}
\end{figure}

The mean value of total collected charge as a function of trigger position is shown in Fig.~\ref{fig:atten}. Normally, the attenuation of the light yield in a scintillating fibre follows a double exponential decay, with short ($\sim$30~cm) and long component ($\sim$400~cm), but with only five data points, the data fits well to a single exponential function over the positions measured, with a bulk attenuation length of 210~cm when measured from 25~cm to 230~cm. Quoted attenuation length from manufacturers are typically measured from 1~m to 3~m. Differences in the attenuation length are likely a consequence of the wavelength dependence of the attenuation, quantum efficiency of the photodetector and wavelength spectra of the scintillator as well as the quality of polish on the far ends of the fibres.

\begin{figure}[htbp]
\includegraphics[angle=90,width=0.49\textwidth]{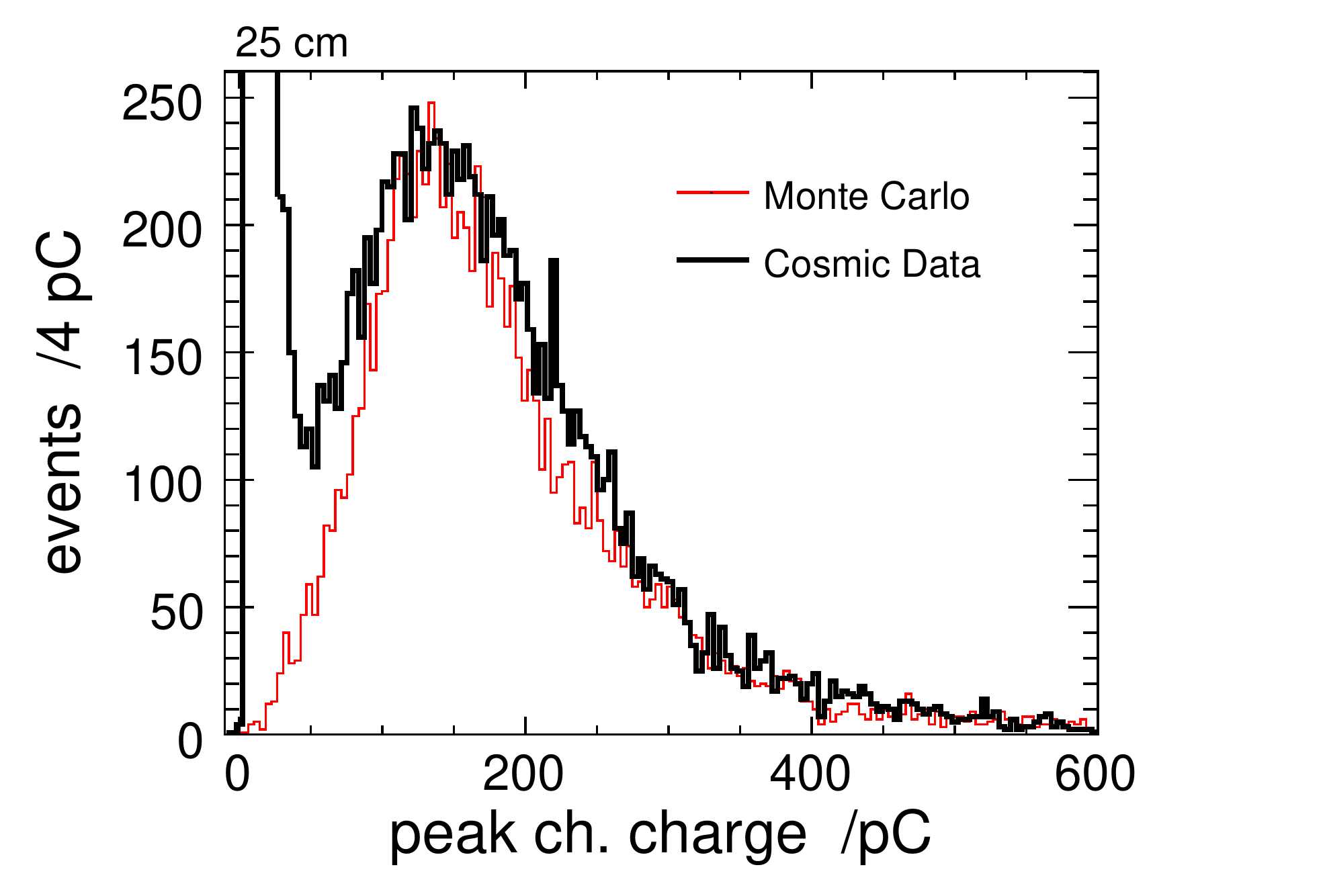} 
\includegraphics[angle=90,width=0.49\textwidth]{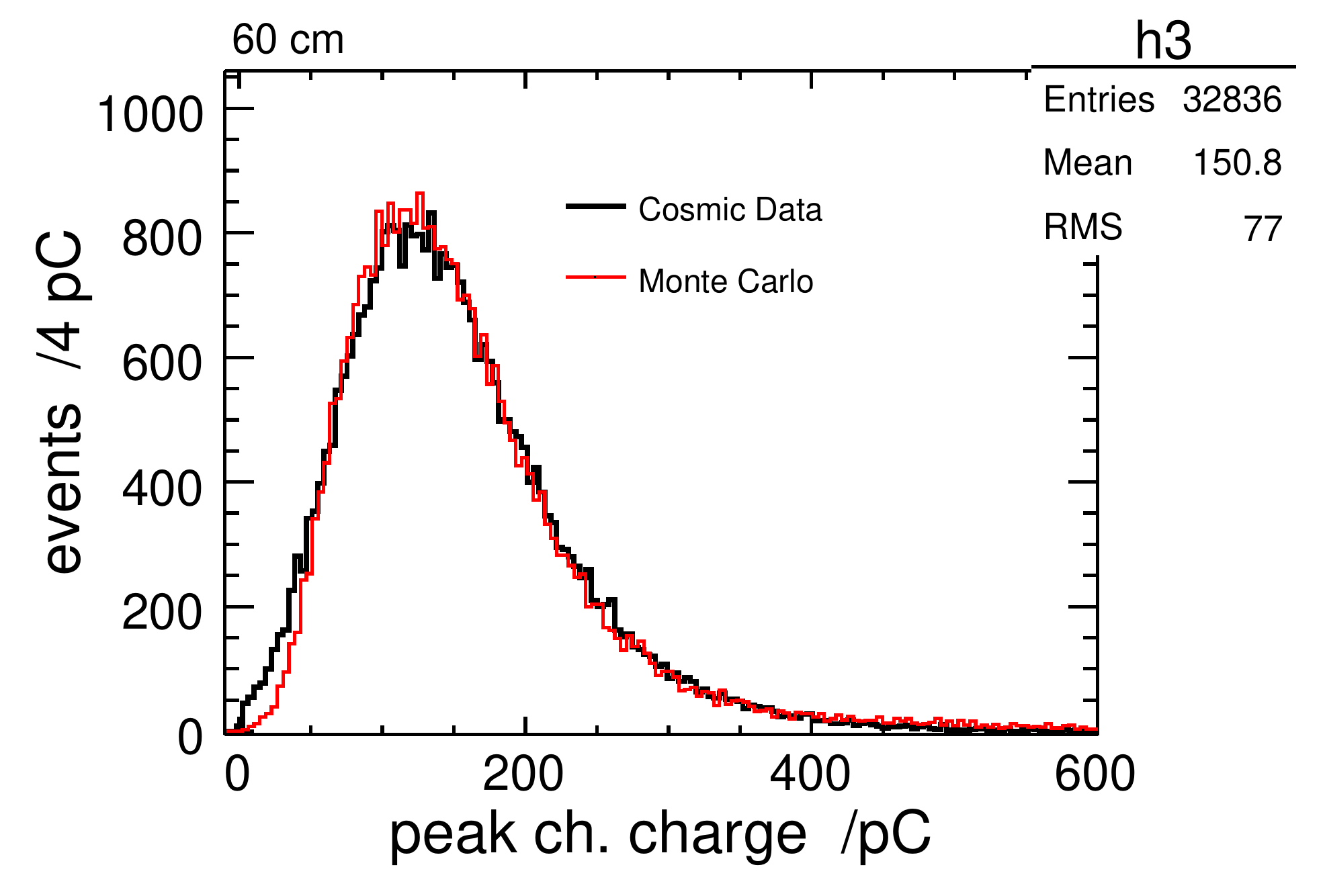} 
\includegraphics[angle=90,width=0.49\textwidth]{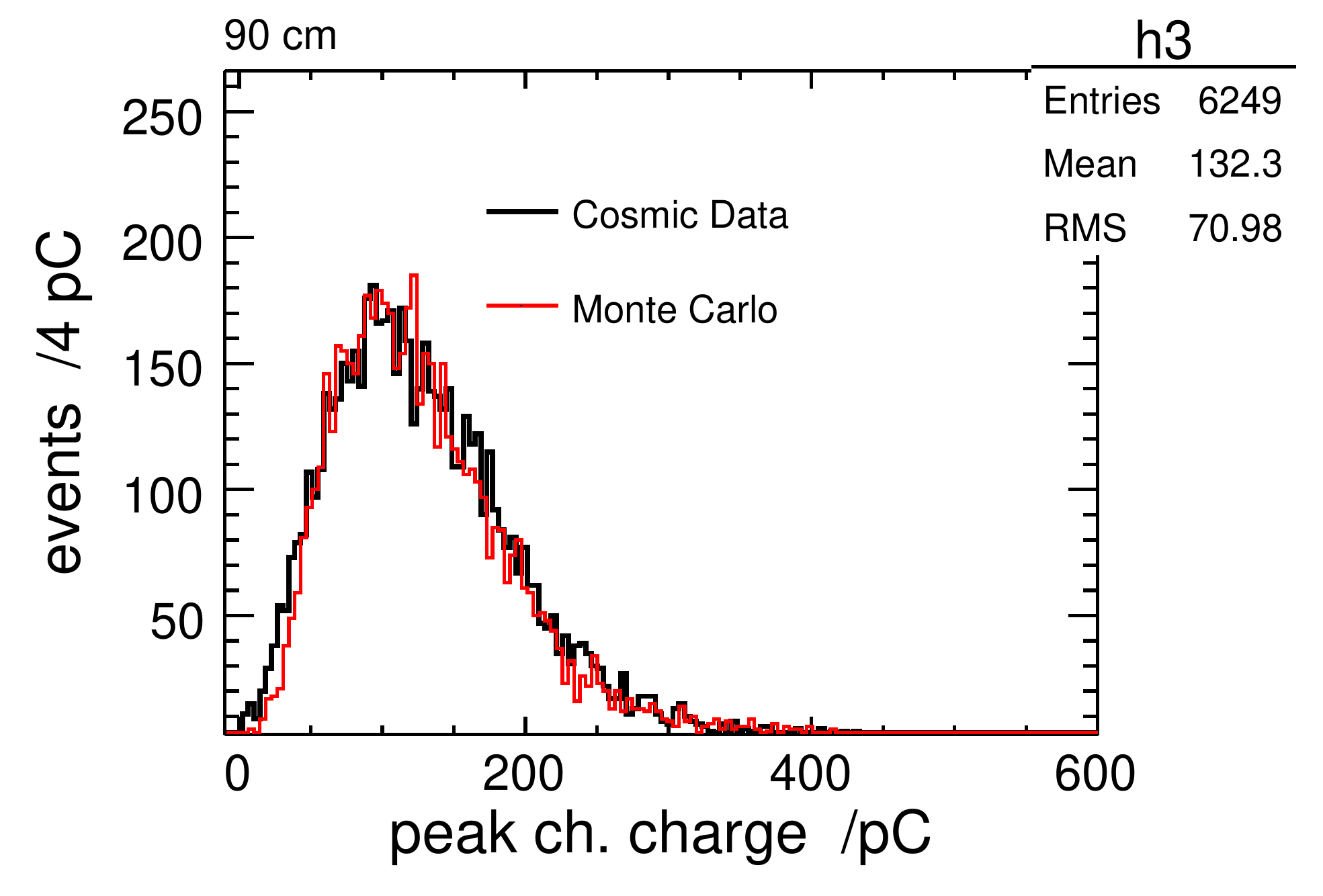} 
\includegraphics[angle=90,width=0.49\textwidth]{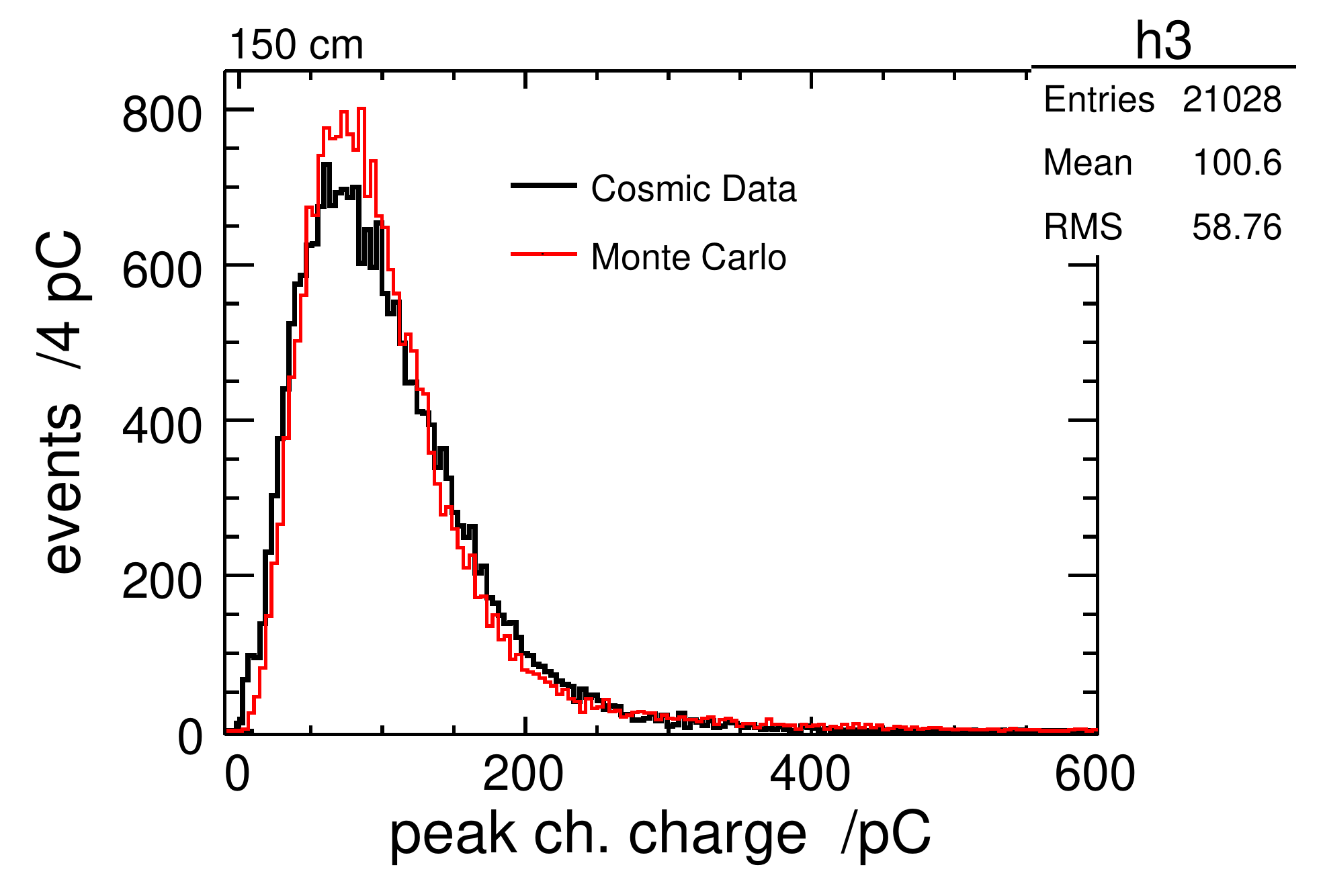} 
\includegraphics[angle=90,width=0.49\textwidth]{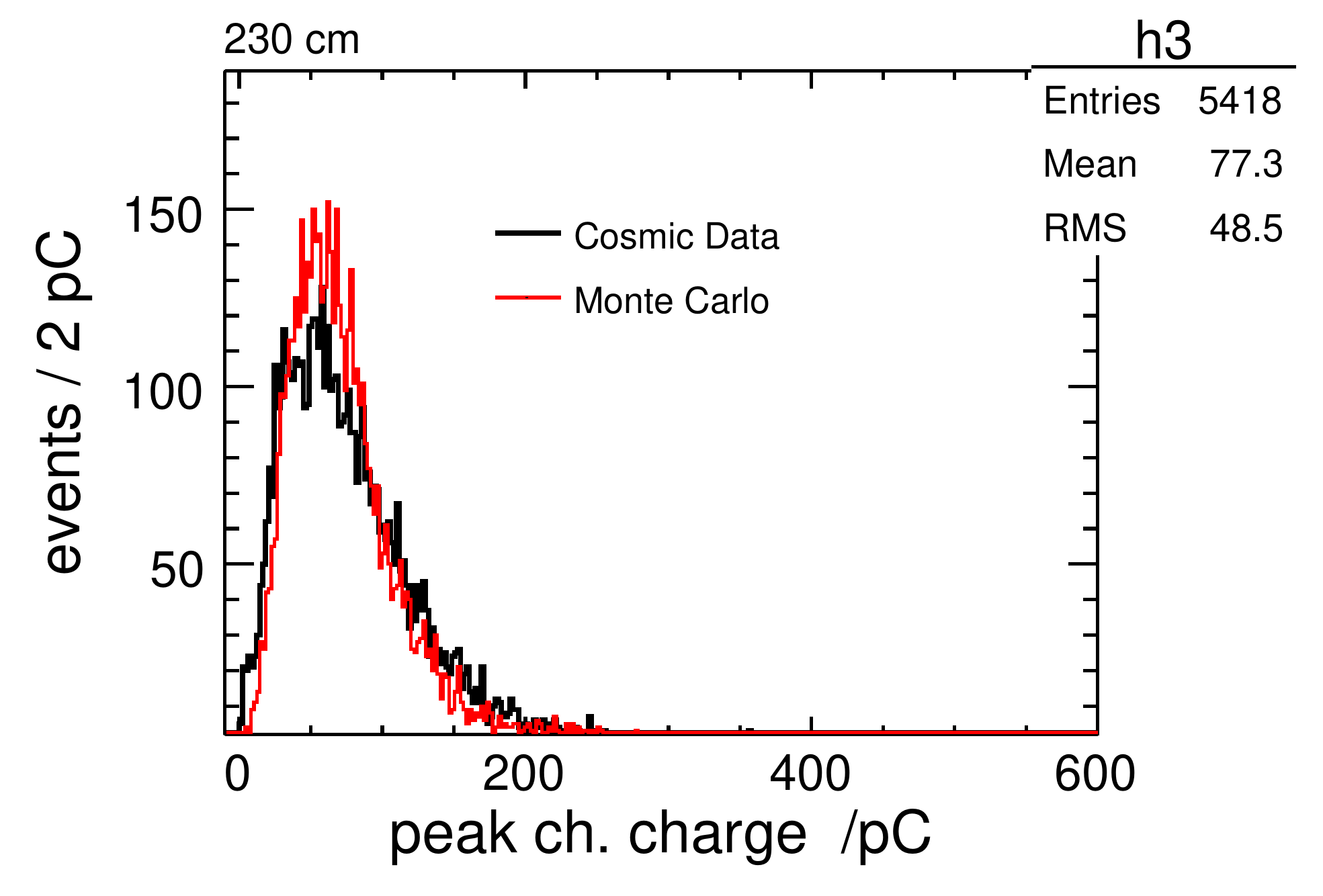} 
\caption{\textbf{peak channel charge} The charge in the peak channel from cosmic data and Monte Carlo for 25, 60, 90, 150 and 230~cm trigger positions.  The 25~cm Cosmic data has no drift tube information to provide good cluster identification. }
\label{fig:chargepeak_135mm}
\end{figure}

\begin{figure}[htbp]
\includegraphics[angle=90,width=0.49\textwidth]{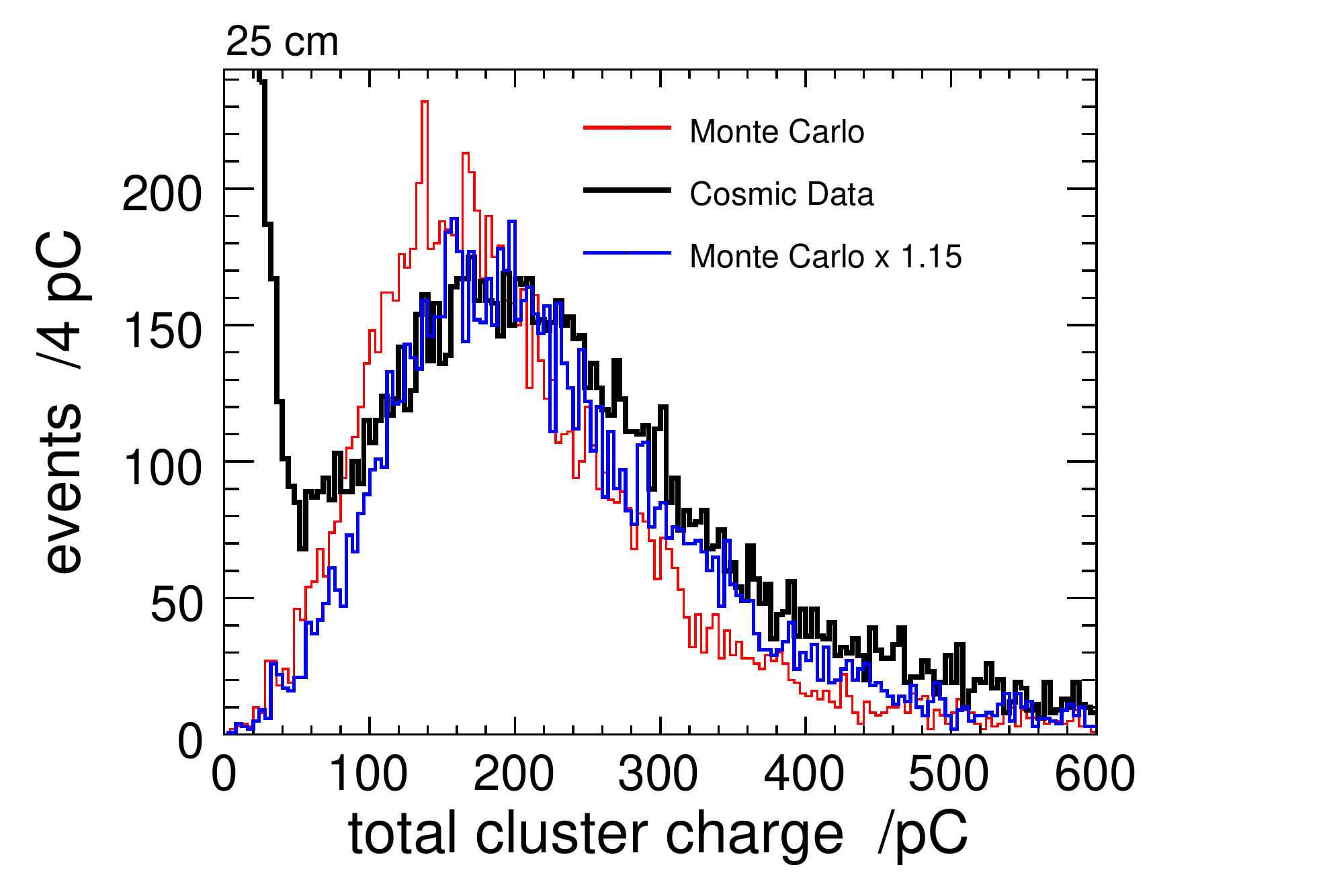} 
\includegraphics[angle=90,width=0.49\textwidth]{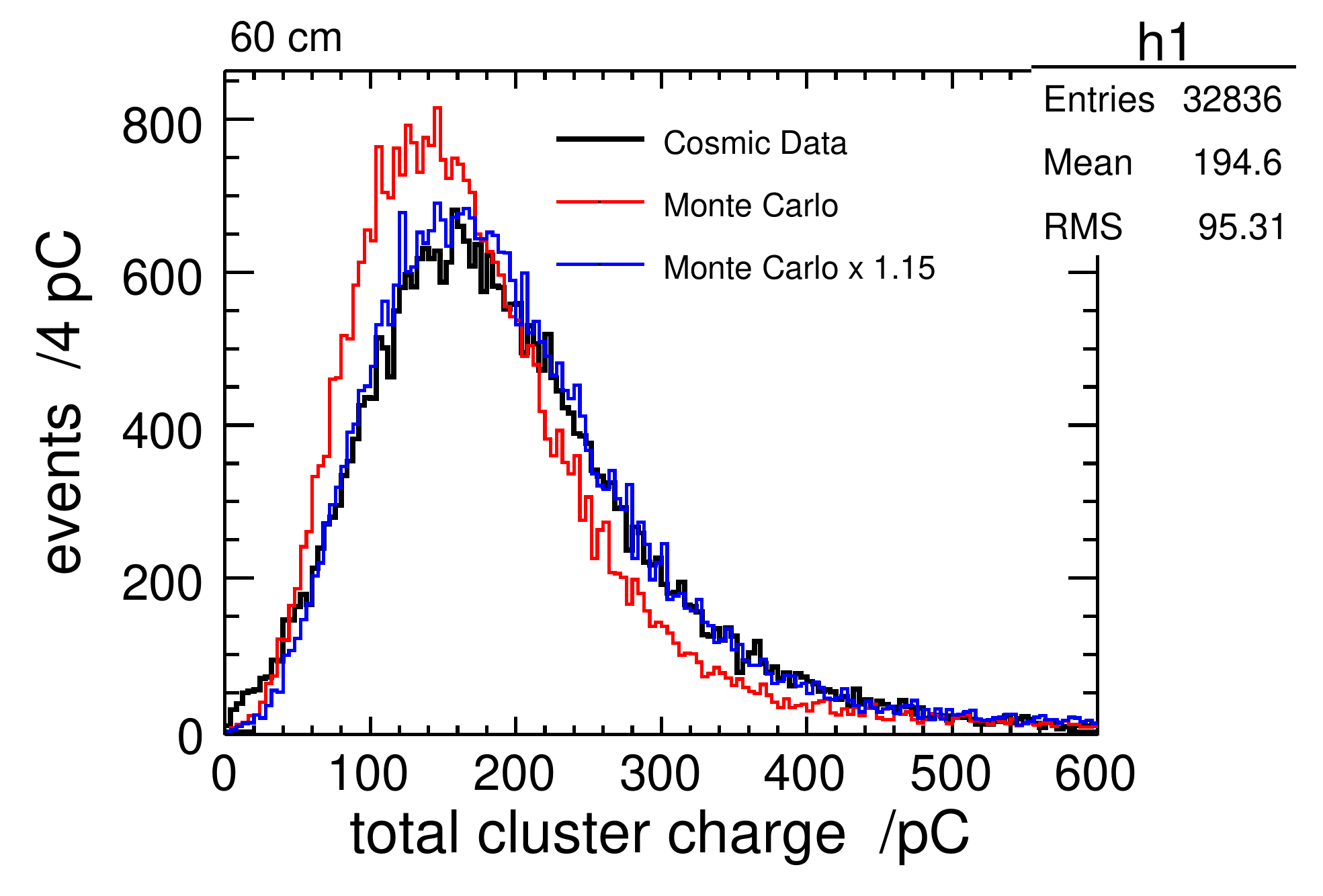} 
\includegraphics[angle=90,width=0.49\textwidth]{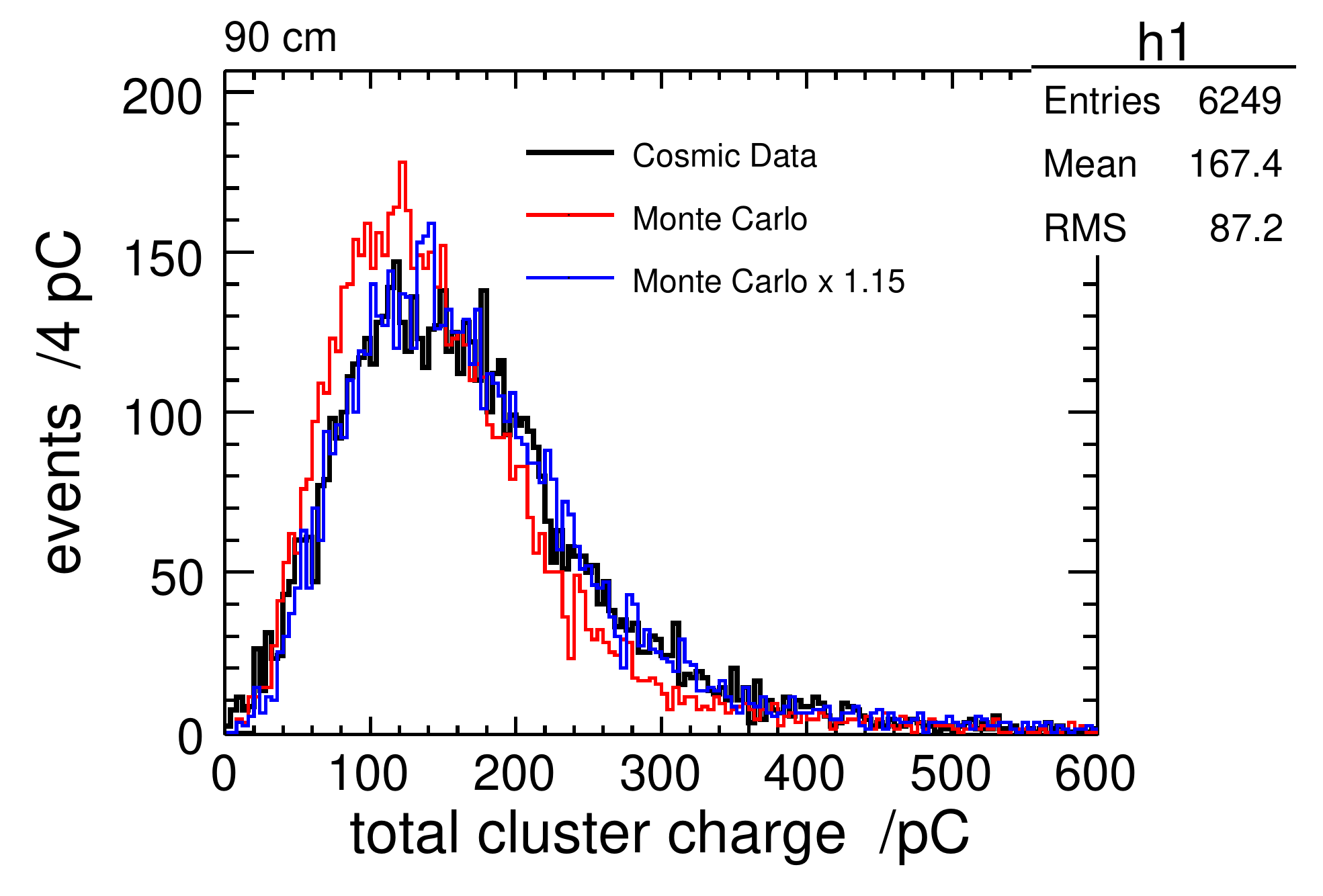} 
\includegraphics[angle=90,width=0.49\textwidth]{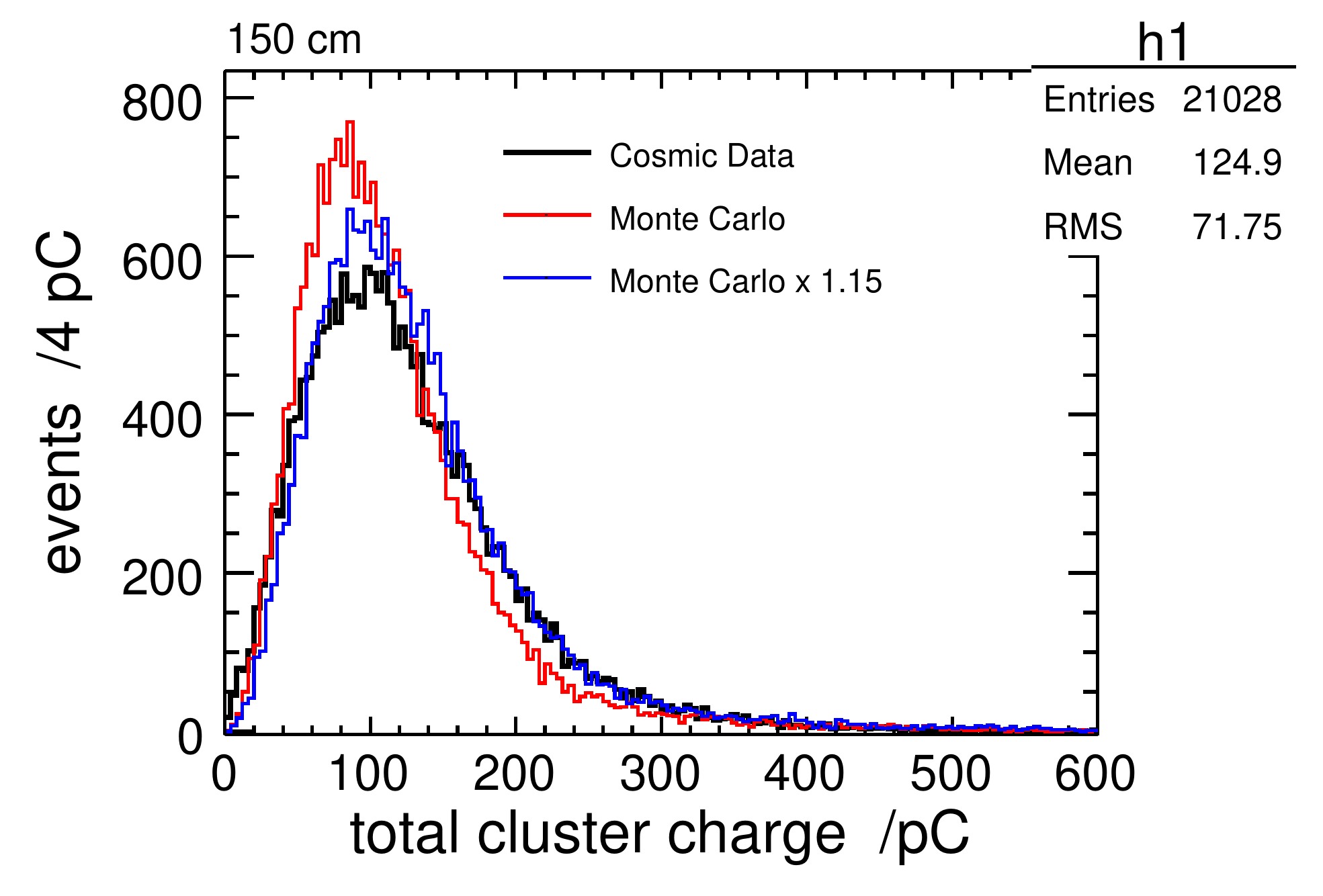} 
\includegraphics[angle=90,width=0.49\textwidth]{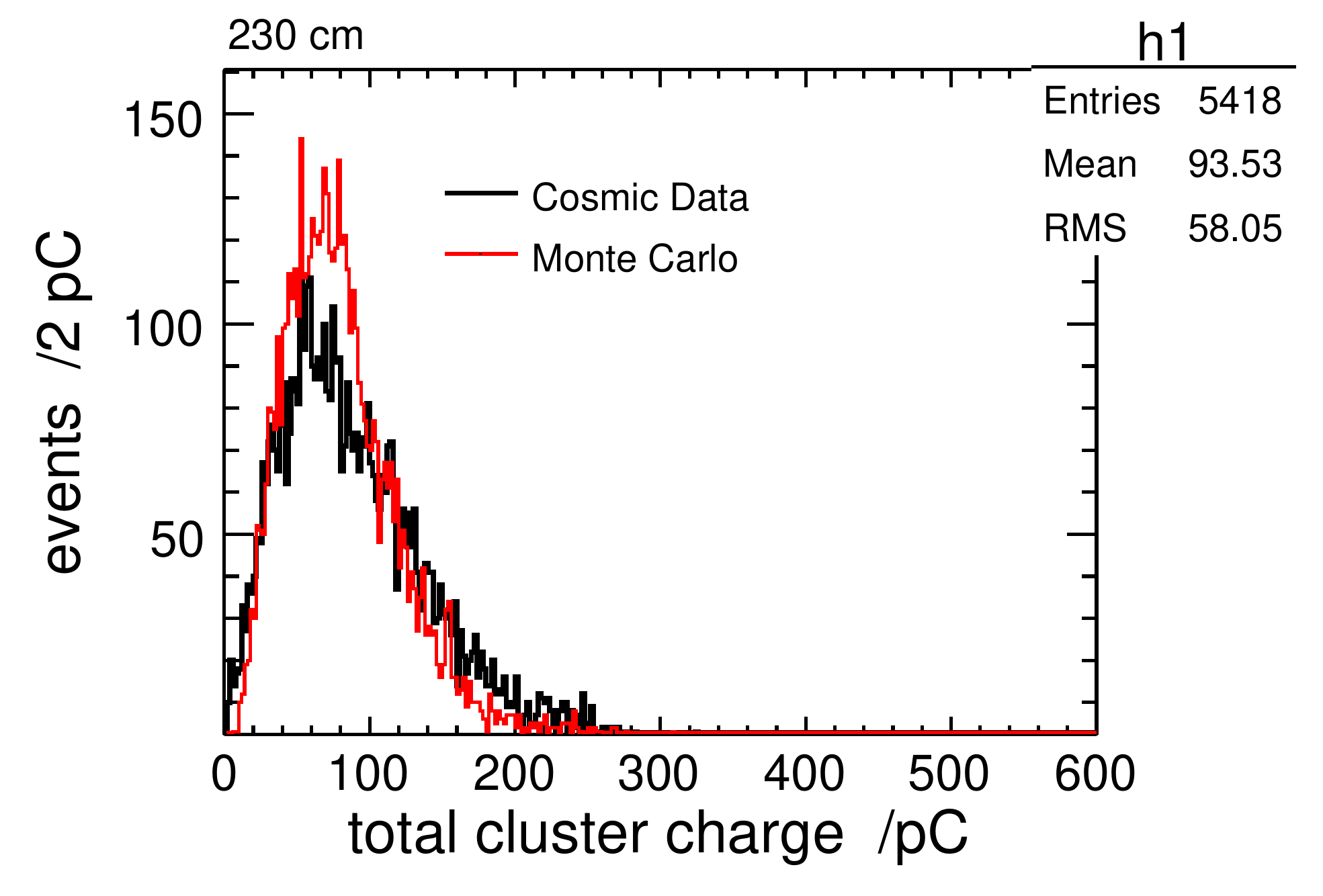} 
\caption{\textbf{total cluster charge} The total charge in hit cluster channels from  cosmic data and Monte Carlo for 25, 60, 90, 150 and 230~cm trigger positions. The 25~cm Cosmic data has no drift tube information to provide good cluster identification. }
\label{fig:chargetotal_135mm}
\end{figure}

The charge distributions observed in the fibre detector from Cosmic rays for the different trigger positions are shown in Fig.~\ref{fig:chargepeak_135mm} (the peak channel only) and Fig.~\ref{fig:chargetotal_135mm} (the sum of the charge in the hit cluster channels).  Only events with a position within 2.7~mm of the extrapolated drift tube track are shown (except for the 25~cm position where no track from the drift tubes was possible due to geometrical constraints). The Monte Carlo charge distributions, based on the model described in the previous section, are also shown in the figures, and explained in further detail below.

\subsection{Number of photoelectrons}

Despite the fact that the H8500 is not a photosensor with photon counting capabilities, it was still possible to estimate the number of photoelectrons from the charge distributions. Given that the width of the $i$th photoelectron peak goes as
\begin{equation}
\sigma_{i} = G_{1pe}\cdot\sqrt{\left(i\sigma_{1pe}^{2} + \sigma_{other}^{2} \right)}
\end{equation}
  where $\sigma_{1pe} = 0.6$ for the H8500 photodetecto, the width of the spectrum will be dominated by the photodetector resolution and photoelectron statistics. $G_{1pe}$ is the single photoelectron charge gain. Other contributions to the total width, $\sigma_{other}$, will be a result of errors in the gain corrections for each MA-PMT channel, as well as variations in the uniformity of the total energy deposited in each channel. However, beyond the first peak, $\sigma_{other}$  contributes little.
  
  Two methods have been used to estimate the mean number of photoelectrons seen at each trigger position:
\subsubsection*{First method}  
  
   The first method uses the Moyal function, Eq.~\ref{eqn:moyal}, which represents a simple analytic approximation of the detector response to a minimum ionizing particle passing through the detector:

\begin{equation}
\centering
M ( x ) = A \cdot exp \left( -\frac{1}{2} \left(( \frac{ x -\bar{x} } {\sigma}) + exp(- \frac{ x - \bar{x}}  {\sigma})\right) \right)  
\label{eqn:moyal}
\end{equation}

   The structure is simply a Gaussian distribution with a tail, approximating the behaviour of the tails in the energy deposit of the cosmic ray energy deposition. If the width of the distribution is purely Poisson in nature, then the fractional width of the distribution will give us an estimate of the number of photoelectrons:
   
   \begin{equation}
   \frac{\bar{x}^{2}}{\sigma_{x}^{2}} \sim \left(\frac{npe}{\sqrt{npe}}\right)^{2} = npe
\end{equation}

\begin{figure}[htbp]
\centering
\includegraphics[angle=0,width=0.49\textwidth]{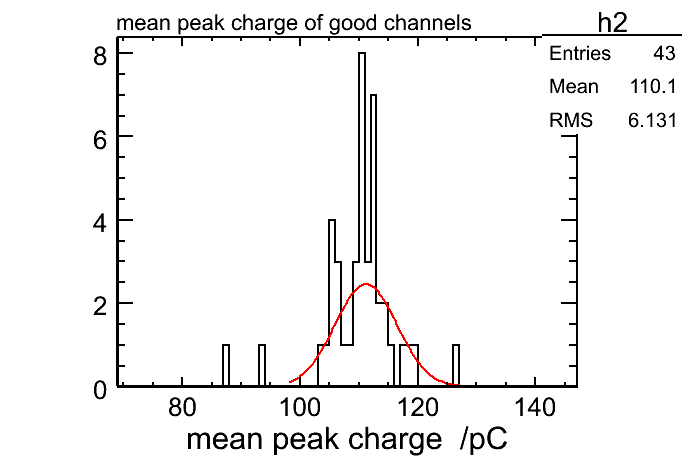} 
\includegraphics[angle=0,width=0.49\textwidth]{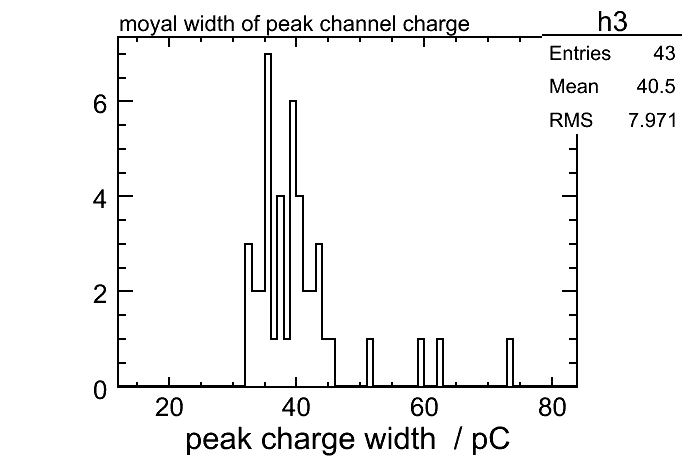}
\caption{Histograms of the mean, $\bar{x}$ (left) and width, $\sigma_{x}$, (right) of Moyal fits (Eq.~\ref{eqn:moyal}) to the charge spectra of each fibre channel with the largest signal in each event for 60~cm cosmic data. }
\label{fig:moyalhist}
\end{figure}

Examining each peak fibre channel   individually, and fitting with the Moyal function, a histogram of the mean and width of the peak charge distributions is shown in Fig.~\ref{fig:moyalhist}. This method seems to work better for the closer trigger positions and larger number of photoelectrons where the Poisson distribution is more similar to a Gaussian distribution such that the Moyal fit is a better approximation.

\begin{table}[htbp]
\centering
\caption{Estimate of the number of photoelectrons from Moyal fits (Eq.~\ref{eqn:moyal}) to peak charge  distributions. The uncertainties are the RMS values of the distributions.}
\begin{tabular}{|c|c c c c|}
\hline trigger pos.  & \textbf{60~cm} & \textbf{90~cm} & \textbf{150~cm} & \textbf{230~cm}   \\ 
\hline \hline   \multicolumn{5}{|c|}{\textbf{Monte Carlo}} \\  

\hline $\bar{x}$ & $ 108\pm5 $ & $ 89\pm5 $& $62 \pm 3$& $ 45\pm 18$\\ 
\hline $\sigma_{x}$ &  $ 38\pm 2$& $ 33\pm 3$& $ 25\pm 2$& $ 18\pm 2$\\ 
\hline npe &  $8.1 \pm0.9 $& $7.2 \pm1.1 $& $6.1 \pm0.8 $& $ 6.1\pm 1.1$ \\ 

\hline \hline \multicolumn{5}{|c|}{\textbf{Cosmic Data}}\\
\hline $\bar{x}$ & $110 \pm 6$ & $98 \pm 17$ & $68 \pm 7$ & $48 \pm 7$\\
\hline $\sigma_{x}$ & $41 \pm 8$  & $36 \pm 6$ & $30 \pm 5$ & $24 \pm 5$ \\
\hline npe & $8.0 \pm 2.6$ & $7.5 \pm 2.7$ & $5.5 \pm 1.2$ & $4.7 \pm 2.3$\\

\hline 
\end{tabular} 
\label{tab:moyalnpe}
\end{table}

\subsubsection*{Second method}
The second method of estimating the number of photoelectrons involves determining the single photoelectron gain and then fitting the charge spectrum with a series of Gaussian functions. Certain peak channels along the edge of the MA-PMT (7,15,23,\dots) will have adjacent fibre channels far away, guaranteeing that there is no crosstalk in these channels and the spectra is a result of light yield, hence the photoelectrons are result of a Poisson process. If the width of each peak is described by the number of photoelectrons and the resolution of the photodetector,  the charge spectra (the intensity, $I$, at charge, $x$) can be described as follows:

\begin{equation}
I(x)= A\cdot\sum_{i}Poisson(i,\overline{npe})\cdot Gauss(x,i\cdot{}G_{1pe},\sqrt{i}\cdot\sigma_{1pe}) \,,
\label{eqn:npespectra}
\end{equation}

where $A$ is some normalization factor; $Poisson(i,\overline{npe})$ guarantees the relative Poisson distribution of the $i$th photoelectron with a distribution mean, $\overline{npe}$;  $Gauss(x,i\cdot{}G_{1pe},\sqrt{i}\cdot\sigma_{1pe})$ is the Gaussian  photodetector response with a mean $i\cdot{}G_{1pe}$ and standard deviation $\sqrt{i}\cdot\sigma_{1pe}$. 

\begin{figure}[htbp]
\centering
\includegraphics[angle=0,width=0.49\textwidth]{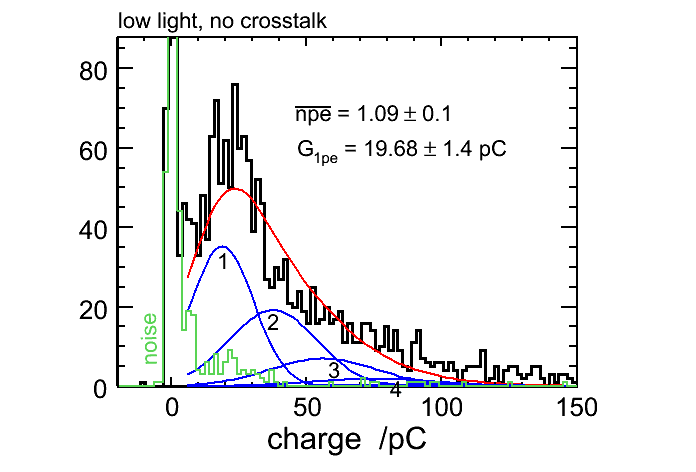} 
\includegraphics[angle=0,width=0.49\textwidth]{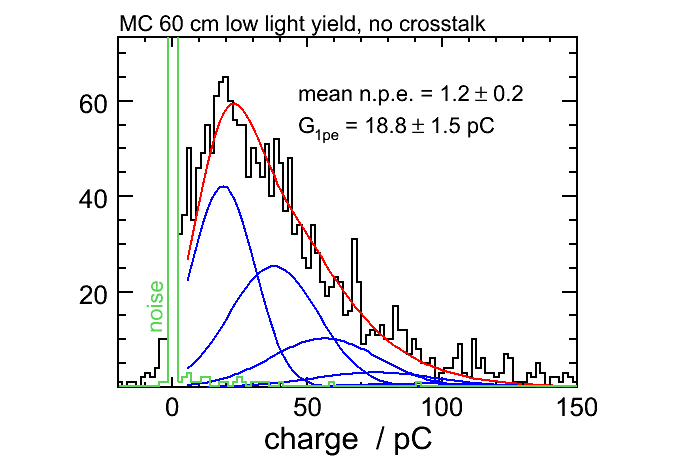} 
\caption{Charge spectra for low light yield channels with no electronic crosstalk, fit to Eq.~\ref{eqn:npespectra}.  Cosmic data at 60~cm are shown in the left plot, Monte Carlo data are shown in the right plot. The first four individual photoelectron peaks are shown with blue lines and labelled by number.}
\label{fig:mod7plus1}
\end{figure}
 
Spectra dominated by single photoelectrons and a narrow pedestal with little noise are shown in Fig.~\ref{fig:mod7plus1}. In the fit to the data, $A$, $\overline{npe}$ and $G_{1pe}$ are free parameters. The photodetector resolution is fixed at $\sigma_{1pe} = 0.6$. The result of the fit is shown as a red line in Fig.~\ref{fig:mod7plus1}  with a value for the mean number of photoelectrons  of $\overline{npe} = 1.09\pm0.14$, and the single photoelectron gain equal to $G_{1pe} = 19.7\pm1.3$~pC for cosmic data. Applying the same method to the Monte Carlo data returns a value of $G_{1pe} = 18.8\pm1.5$~pC, in agreement with, but found independently from, the input parameter for the Monte Carlo. Assuming the gain is consistent for all the cosmic ray data, fixing this value along with the detector resolution would allow for an accurate estimate of the mean number of photoelectrons seen at each trigger position.

\begin{figure}[htbp]
\centering
\includegraphics[angle=90,width=0.49\textwidth]{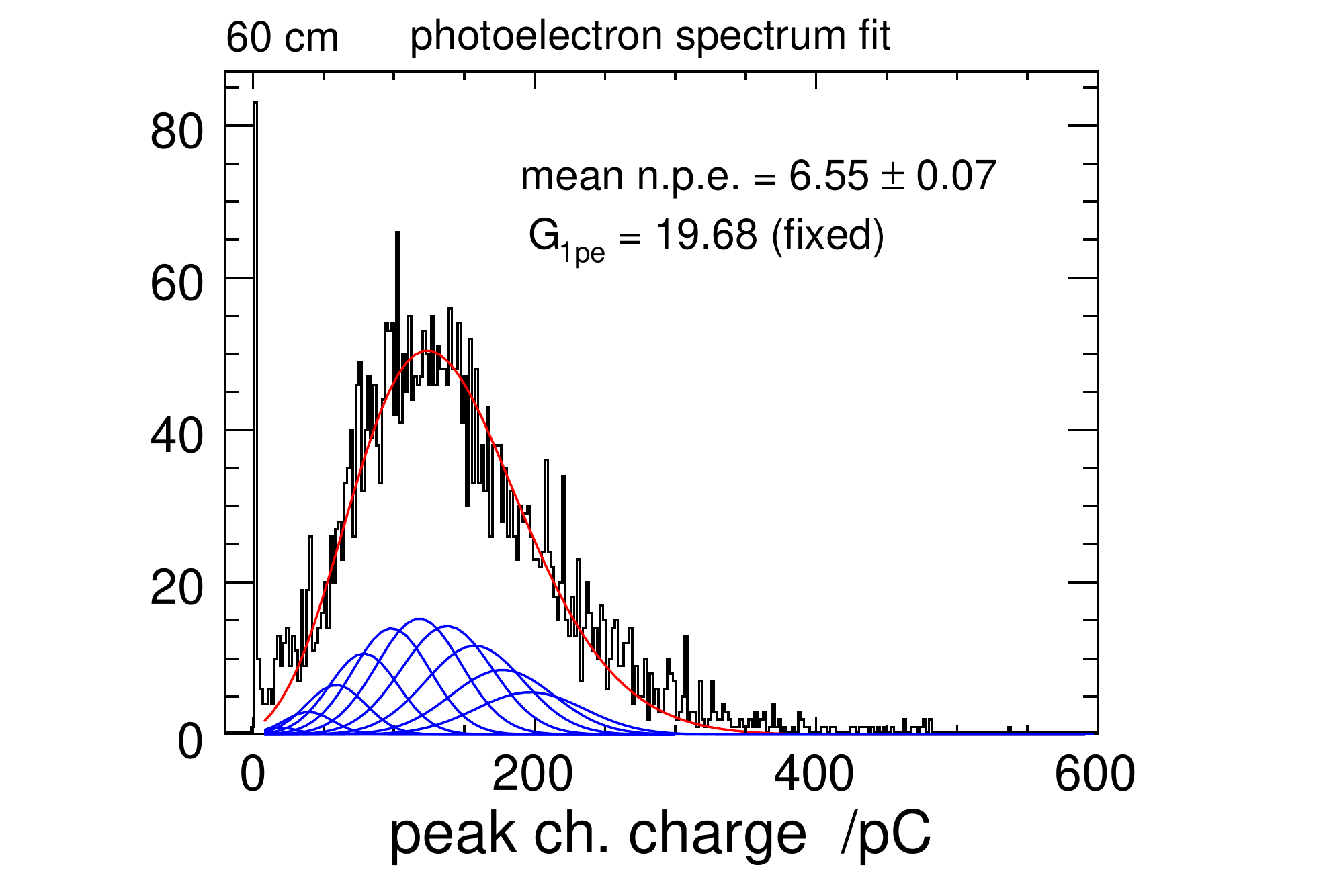} 
\includegraphics[angle=90,width=0.49\textwidth]{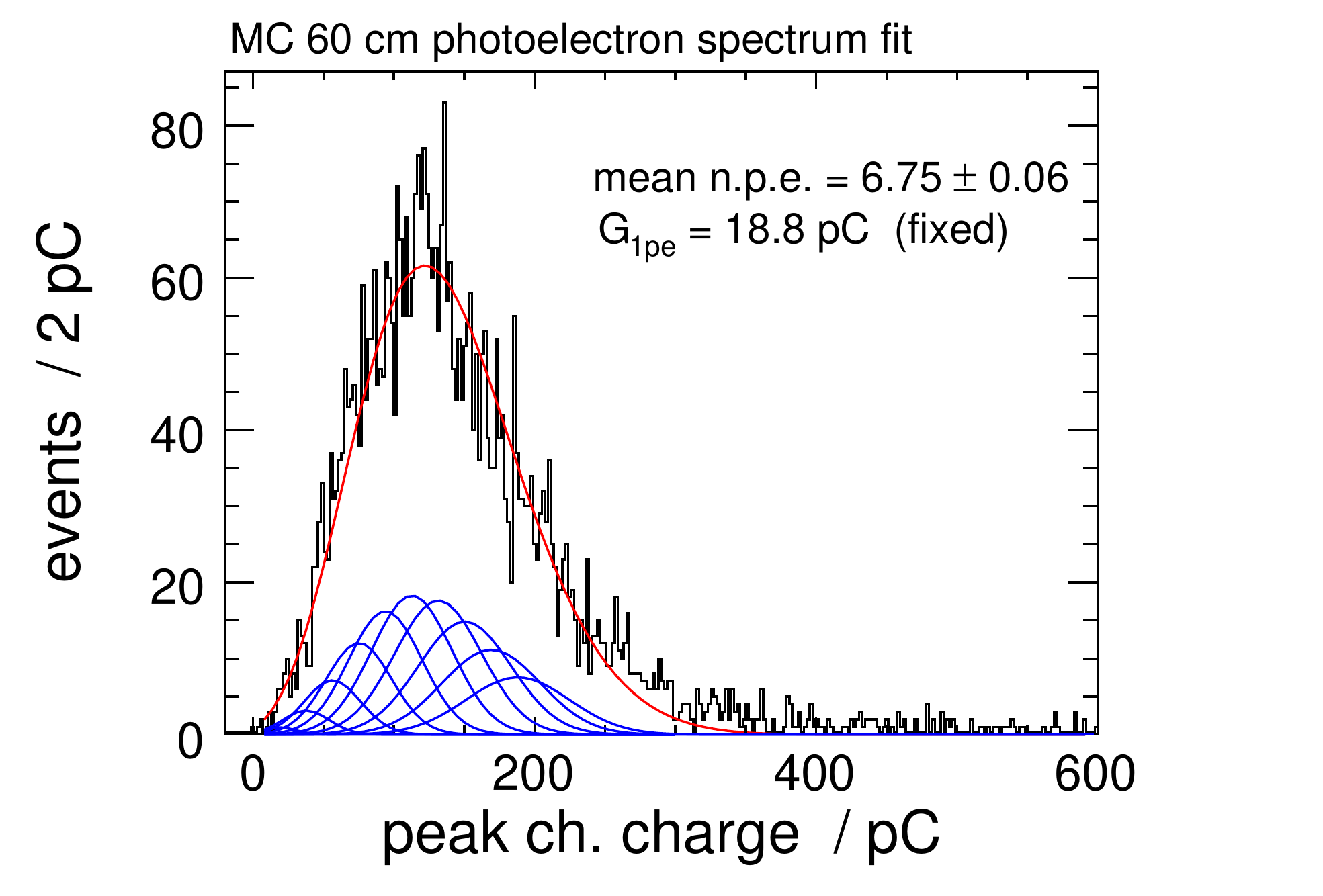} 
\includegraphics[angle=90,width=0.49\textwidth]{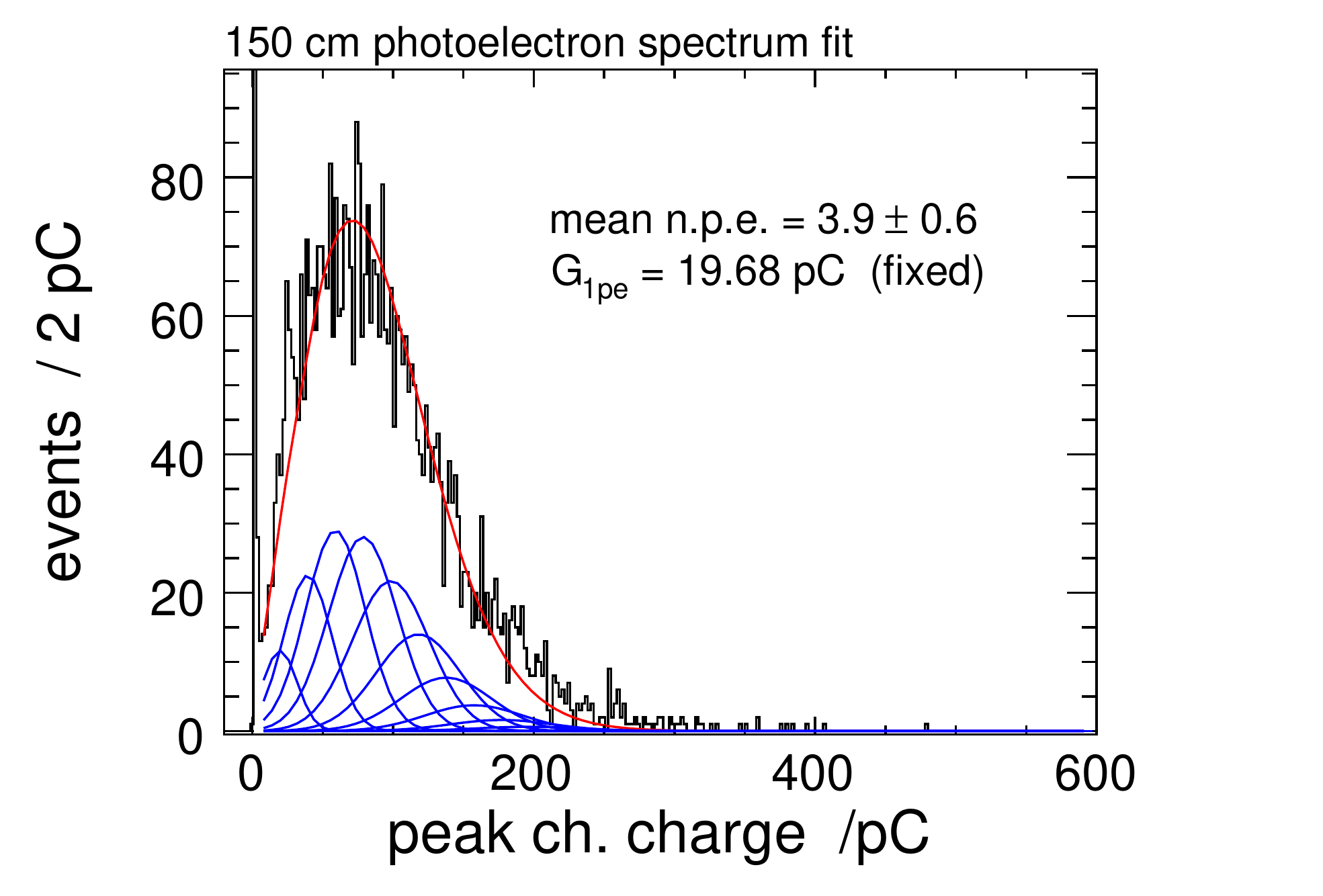} 
\includegraphics[angle=90,width=0.49\textwidth]{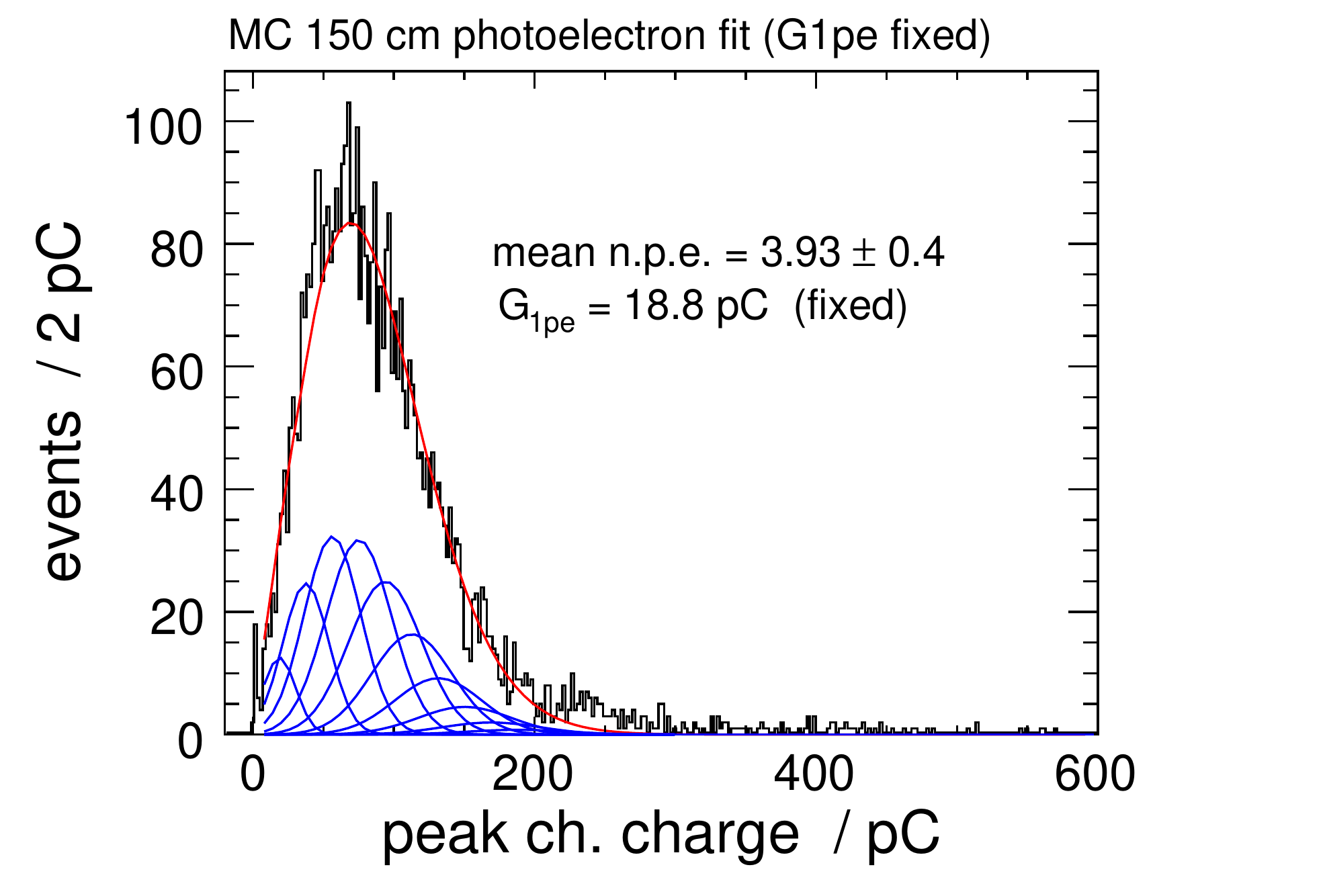} 
\caption{The red line indicates the fit to the  peak channel charge distributions with  Eq.~\ref{eqn:npespectra} for 60~cm (top) and 150~cm (bottom)  trigger positions from Cosmic data (left) and Monte Carlo data (right). The  individual photoelectron peaks  are shown with blue lines.}
\label{fig:npespectrafit}
\end{figure}

From Fig.~\ref{fig:npespectrafit}, the mean number of photoelectrons at each trigger are seen in Table~\ref{tab:spectranpe} for cosmic data and Monte Carlo generated data. The fit to the data with Eq.~\ref{eqn:npespectra} was done twice: once with $G_{1pe}$ fixed to the value found from Fig.~\ref{fig:mod7plus1}, and second time allowing $G_{1pe}$ to float free in the fit in order to see how the mean of the Poisson distribution changes with an unconstrained $G_{1pe}$. In Table~\ref{tab:spectranpe}, $\overline{npe}$(truth) is the mean of a single Poisson fit to the generated photoelectron distribution.  Fitting with Eq.~\ref{eqn:npespectra} and a fixed $G_{1pe}$ appears to underestimate the mean number of photoelectrons by $\sim10\%$ but the uncertainties are about equal to the differences. With the attenuation length found, this produces an average of $9.9 \pm 0.2$ photoelectrons in the peak channel at the cosmic track source in the Monte Carlo which is consistent with the 18 pe/MeV used as an input parameter. The underestimation of the number of photoelectrons in the free and fixed $G_{1pe}$ cases  are most likely a result of extra contributions to the resolution over and above the photoelectron statistics. Extra width in the spectra would appear as a few number of photoelectrons in the fit.

Comparing the $\overline{npe}$ found in the cosmic data to the Monte Carlo, shows a good agreement, especially for fixed-$G_{1pe}$ fits. The cosmic data tends to have $\overline{npe}$ slightly lower than Monte Carlo values,  likely due to extra width contributions, such as variations in the  track length and light yield in each channel of the fibre tracker, that were not be modelled in the simulation. 

\begin{table}[htbp]
\centering
\caption{Estimates of the number of photoelectrons seen by the MA-PMT for cosmic ray  and Monte Carlo data for four trigger positions where the single pe gain parameter $G_{1pe}$ has either been fixed ($G_{1pe}$-fixed) or allowed to float freely ($G_{1pe}$-free) in the fit to Eq.~\ref{eqn:npespectra}. The mean of the generated photoelectrons in the Monte Carlo, $\overline{npe}$(truth),  is also shown for comparison.}
\begin{tabular}{|c|c c c c|}
\hline trigger pos. & \textbf{60~cm} & \textbf{90~cm} & \textbf{150~cm} & \textbf{230~cm} \\
\hline \hline  \multicolumn{5}{|c|}{\textbf{Monte Carlo}} \\ 
 
\hline $\overline{npe} (G_{1pe}$-fixed)&$6.8 \pm 0.1 $ &$5.6 \pm 0.1 $&$ 3.9 \pm 0.4 $&$2.9 \pm 0.1 $ \\
 \hline $\overline{npe}(G_{1pe}$-free)&$ 6.0 \pm0.2 $&$4.8 \pm 0.4 $&$4.0 \pm 0.1 $&$ 4.0 \pm 0.3 $ \\
  $G_{1pe} / pC$  &$ 21.4\pm 0.6 $&$ 22.0\pm 1.4 $&$ 18.7  \pm 0.6 $&$14 \pm 1 $\\
\hline $\overline{npe}$(truth)& $7.2 \pm 0.2$& $6.1\pm 0.2$ & $4.4\pm 0.1$ & $3.0 \pm 0.2$\\

\hline \hline \multicolumn{5}{|c|}{\textbf{Cosmic Data}}  \\  
\hline $\overline{npe} (G_{1pe}$-fixed) & $6.6 \pm 0.1$ & $5.9 \pm 0.2$ & $3.9 \pm 0.6 $&$ 2.9 \pm 0.1$ \\ 
\hline $\overline{npe}(G_{1pe}$-free) & $5.3 \pm 0.2$ & $3.3 \pm 0.4$ & $2.8 \pm 0.2 $&$ 2.2 \pm 0.3$ \\ 
 $G_{1pe} / pC$ &$24\pm1$ &$34\pm4$ &$26\pm1$ & $25\pm3$\\ \hline

\hline 
\end{tabular} 
\label{tab:spectranpe}
\end{table}


\subsection{Efficiency}
If the  efficiency of the fibre tracker, $\epsilon_{2.7}$, is defined as how many events seen in the fibre tracker are within 2.7~mm of the extrapolated track from the drift tubes, the efficiency is somewhat entangled with the track reconstruction efficiency and resolution of the drift tubes.  To define what events  are included in the  \textit{normalized} events, the drift tubes must have one hit in each layer, the mean drift time in the bottom three layers must be greater than 250~ns, and any events with a hit in tube-5 are excluded (which also excludes the region which the fibre plane does not cover) due to some anomalies. A summary of the data yield for the four data sets at 60, 90, 150 and 230~cm is shown in Table~\ref{tab:datasets}.

\begin{table}[htbp]
\centering
\caption{Summary of data sets detailing the raw number of events in each data file and the number of normalized events with well defined tracks found in the raw file. The percentage of normalized events which also contain at least one channel that exceeds a 10~pC threshold in the fibre plane is also shown.}
\begin{tabular}{|c|c c c c|}
\hline
 trigger pos. & \textbf{60~cm} & \textbf{90~cm} & \textbf{150~cm} & \textbf{230~cm} \\ 

\hline \hline   \multicolumn{5}{|c|}{\textbf{Monte Carlo}} \\  
 
\hline trig. events & 50000&12000  &50000 &12000  \\
\hline norm. events &36128 &8638 &36128 &8638\\
\hline w/$\ge$1 fibre hit & 99.6\%&99.8\% &99.4\% &99.1\% \\
 \hline \hline  \multicolumn{5}{|c|}{\textbf{Cosmic Data}} \\
 \hline trig. events &198435 &40664 &174515 &37851\\
 \hline norm. events &30712 &6438 &33915 & 5705\\
 \hline w/$\ge$1 fibre hit &98.5\% &97.8\% &96.8\% &93.8\%\\
\hline 


\end{tabular} 
\label{tab:datasets}
\end{table}

\begin{figure}[htbp]
\includegraphics[angle=90,width=0.49\textwidth]{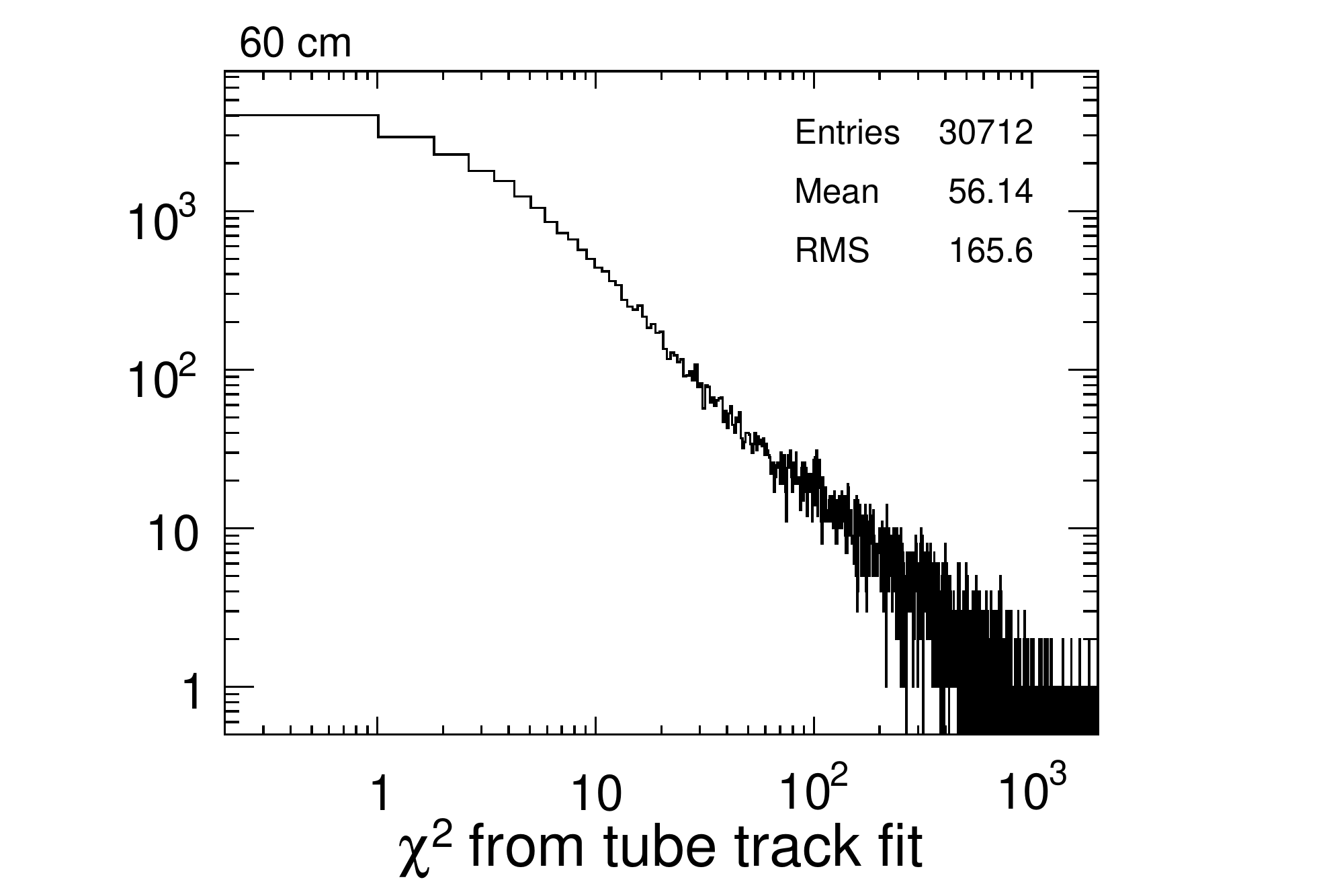} 
\includegraphics[angle=90,width=0.49\textwidth]{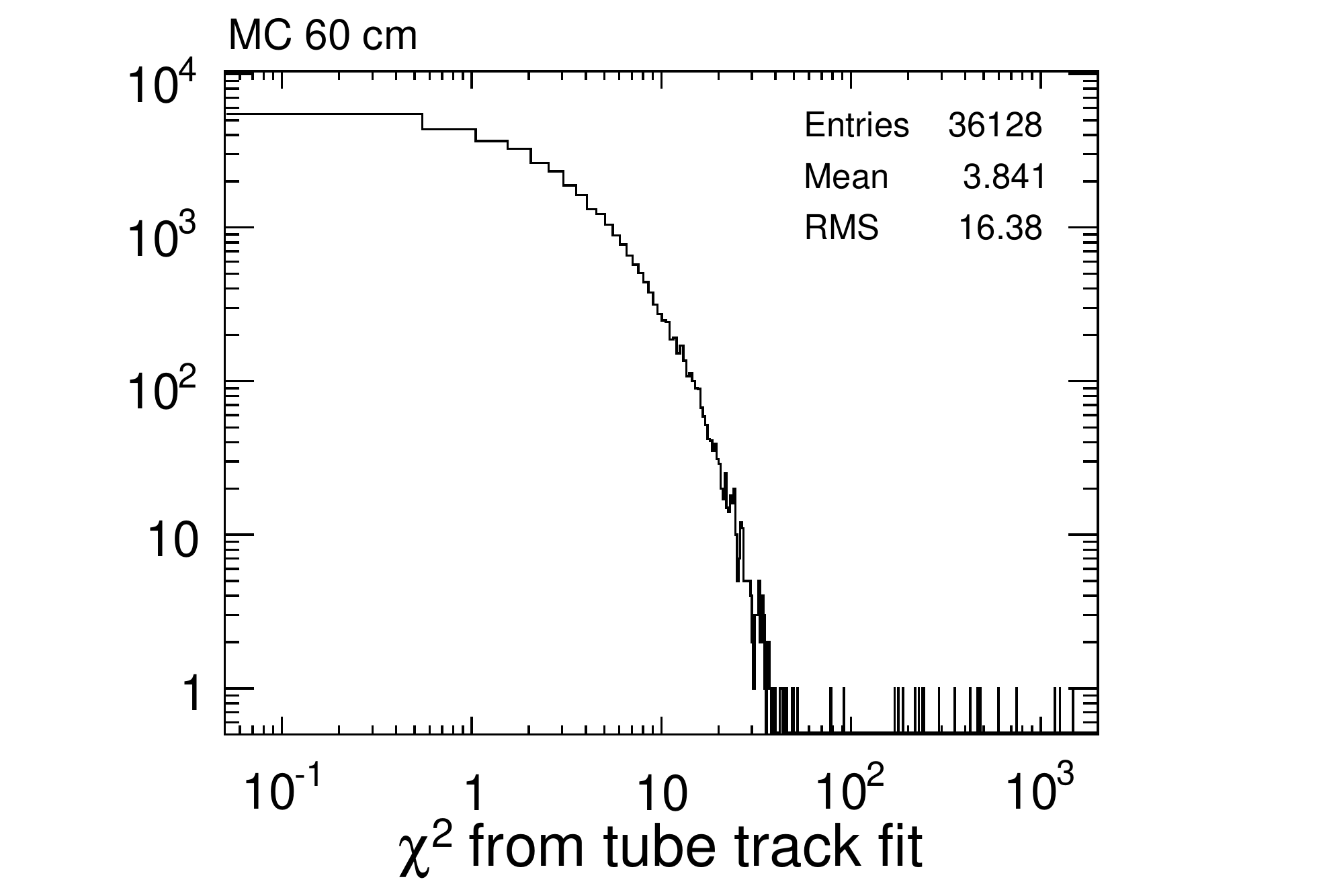} 
\caption{The $\chi^{2}$ of linear track fits to the drift radii from cosmic data (left) and Monte Carlo (right). }
\label{fig:chi2eff}
\end{figure}

Of the normalized events, the accuracy of a linear track fit to the drift times of the muon drift tubes is quantified by its $\chi{}^{2}$ value. Histograms of the $\chi^{2}$ for normalized tracks in the cosmic data and simulation are shown in Fig.~\ref{fig:chi2eff}. Events with a large  $\chi{}^{2}$ indicate a large extrapolated position uncertainty for the track. Successively smaller cuts on  the $\chi{}^{2}$ value should remove poorly defined tracks and exclude events which may degrade the measured position resolution and efficiency. The cosmic data appears to have some poorly defined tracks which the Monte Carlo does not simulate.

Table~\ref{tab:eff1}  lists the number of normalized events , and fraction of events which are within 2.7~mm  (2 channels) or 1.35~mm  (1 channel) of the track position in the fibre plane, $\epsilon_{2.7}$ or $\epsilon_{1.35}$ for the four cosmic ray  and Monte Carlo data sets. This will be the efficiency for the fibre tracker. It can be seen that the efficiency converges for the cosmic data as the track $\chi^{2}$ cut is more restrictive. The Monte Carlo efficiency appears unaffected by $\chi^{2}$ cut, but this is a result of the tracks in Monte Carlo only being smeared with their resolution, and other effects that may occur in the cosmic data, such as delta ray production, are not involved. The Monte Carlo efficiency is only affected by charge, crosstalk and noise effects.

\begin{table}[htbp]
\centering
\caption{The efficiency for tracks at a given trigger position  which fall within 2.7~mm (2 channels) of the extrapolated tube track, $\epsilon_{2.7}$, and within 1.35~mm (1 channel), $\epsilon_{1.35}$. }

\begin{tabular}{| c | c c c | c c c |}
\hline & \multicolumn{3}{c|}{\textbf{Cosmic Data}} & \multicolumn{3}{c|}{\textbf{Monte Carlo}} \\  
\hline $\chi^{2}~\textless $ & n. events & $\epsilon_{2.7}$ & $\epsilon_{1.35}$ & n. events & $\epsilon_{2.7}$ & $\epsilon_{1.35}$   \\

\hline \multicolumn{7}{|c|}{\textbf{60 cm}}\\
\hline 10 & 19363 & 0.966 &  0.932 & 33614 & 0.981 &0.978\\
\hline 30 & 24094 & 0.960  & 0.918 & 36054& 0.981 &0.977\\
\hline 60 & 25877 & 0.953  & 0.903 & 36099& 0.981 & 0.977\\
\hline 150 & 27814 & 0.937  & 0.875 &- &-& -\\
\hline 

\multicolumn{7}{|c|}{\textbf{90 cm}}\\
\hline 10 & 4281 & 0.956  &  0.927 & 8051&0.975  &0.971\\
\hline 30 & 5022 & 0.953  & 0.913 & 8619& 0.975  &0.971\\
\hline 60 & 5373 & 0.945  & 0.895 &8628 & 0.975 & 0.971\\
\hline 150 & 5760 & 0.930 & 0.867 & - & - &-\\
\hline 

\multicolumn{7}{|c|}{\textbf{150 cm}}\\
\hline 10 & 21325 & 0.943 &  0.916 &33614 &0.953&0.948\\
\hline 30 & 27274 & 0.942 & 0.905 &36054 & 0.953&0.947\\
\hline 60 & 29243 & 0.933  & 0.892  & 36099& 0.953&0.947\\
\hline 150 & 31133 & 0.920  & 0.868&  -& -&-\\
\hline 

\multicolumn{7}{|c|}{\textbf{230 cm}}\\
\hline 10 &3617  & 0.898 & 0.868 &8051 &0.927  &0.919\\
\hline 30 & 4432 & 0.895  & 0.856 & 8619 &0.927  &0.919\\
\hline 60 & 4765 & 0.889  & 0.841 & 8628& 0.927 & 0.919\\
\hline 150 & 5139 & 0.873  & 0.812&  -& -&-\\
\hline 

\end{tabular} 
\label{tab:eff1}
\end{table}

\subsection{Crosstalk and noise evaluation}

\begin{figure}[htbp]
\includegraphics[angle=90,width=0.49\textwidth]{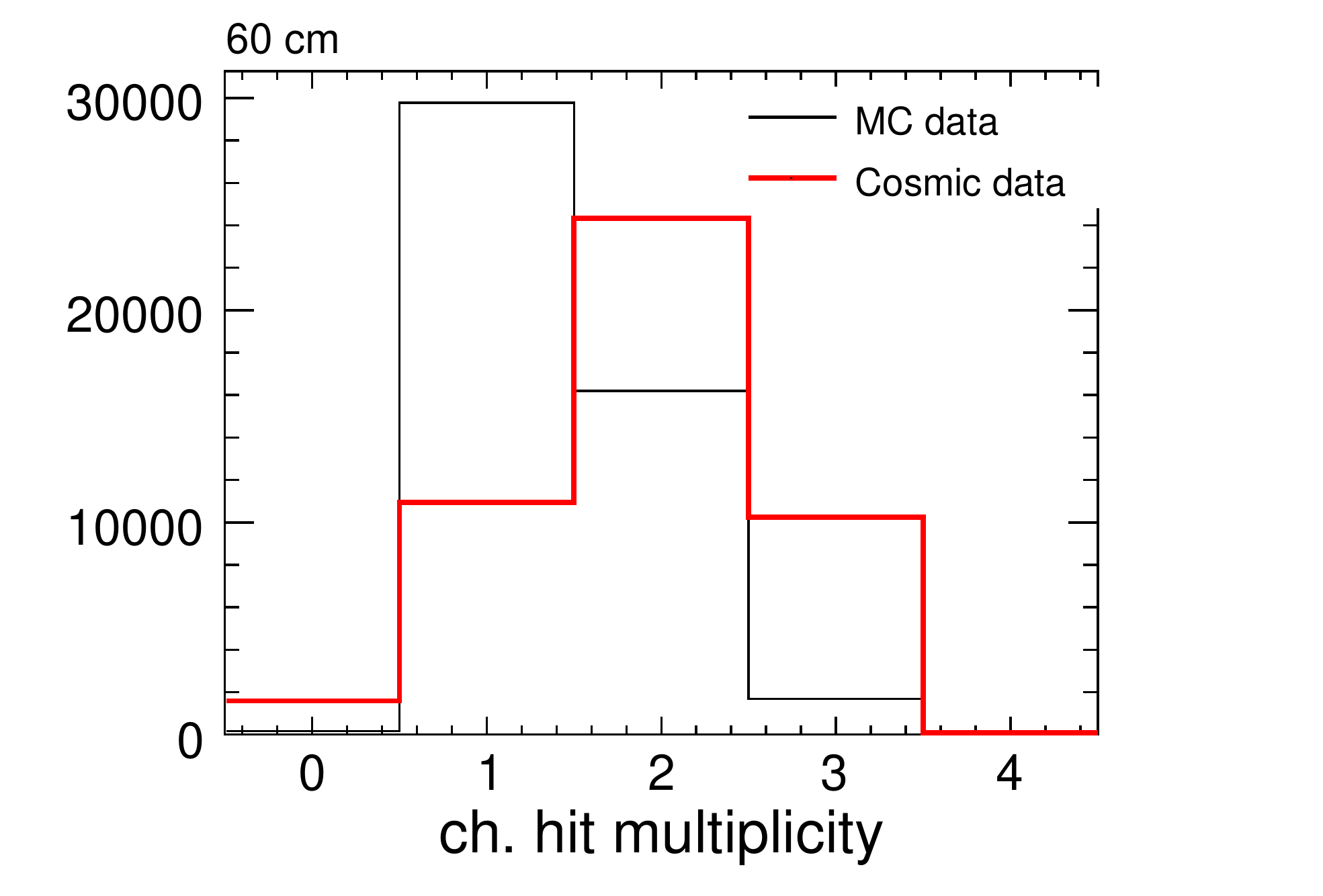} 
\includegraphics[angle=90,width=0.49\textwidth]{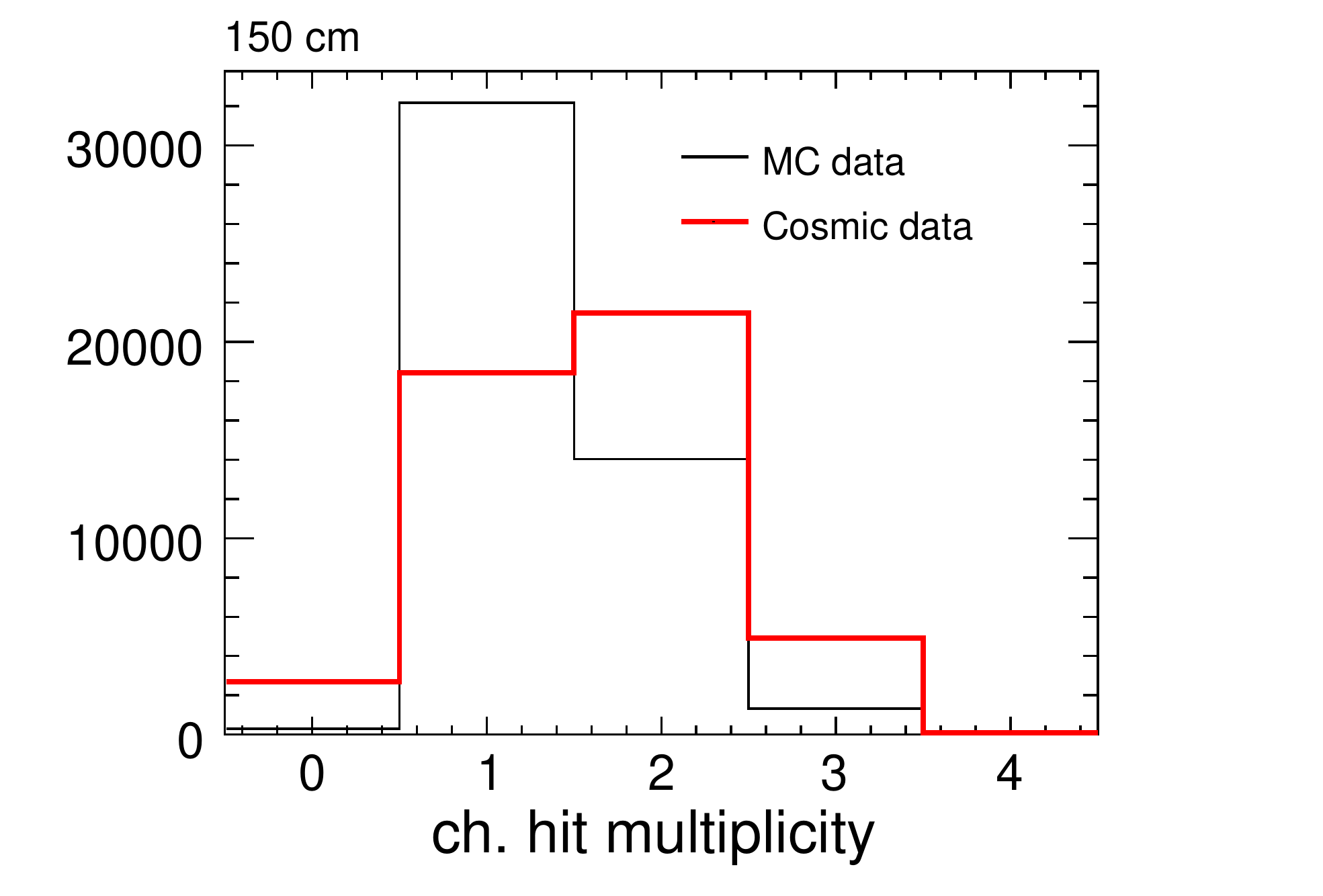} 

\caption{\textbf{multiplicity} MC and Cosmic data channel hit multiplicity for the 60 and 150~cm trigger positions. Experimental equivalent electronic noise and MA-PMT crosstalk are included in the MC described here. }
\label{fig:multiplicity}
\end{figure}

\begin{figure}[htbp]
\centering
\includegraphics[width=0.4\textwidth]{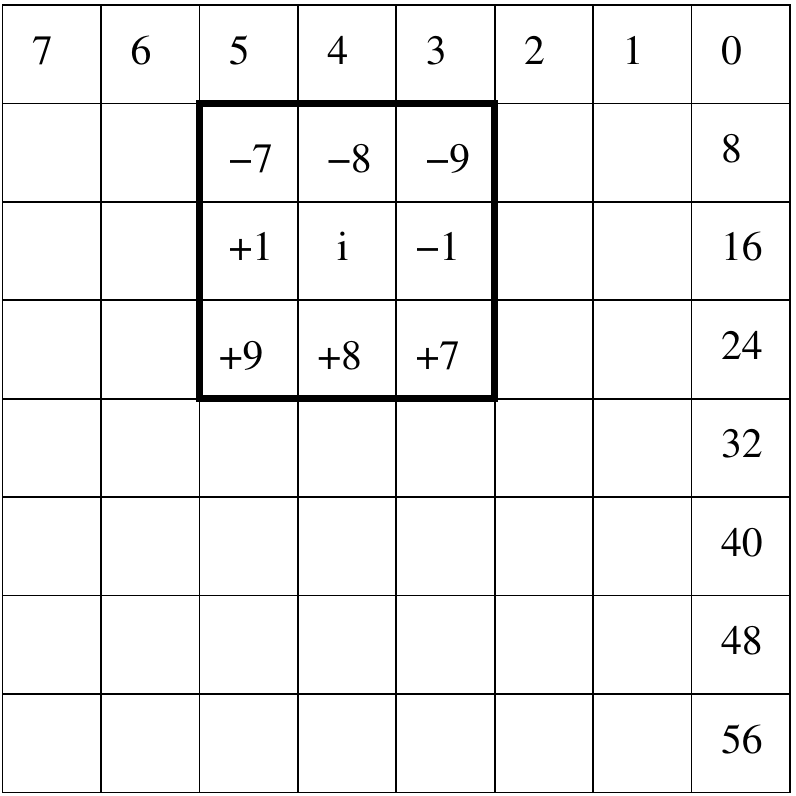} 
\caption{A sketch of the multi-anode PMT channels, where $i$ is the channel with the maximum charge seen. }
\label{fig:crosstalk}
\end{figure}

Since the larger fraction of the area of the fibre plane is covered by a single channel, one would expect the channel multiplicity in the clusters formed to be peaked at one, with the fraction of the area covered by two overlapped channels to be smaller. However, as is shown in Fig.~\ref{fig:multiplicity}, the multiplicity distribution is peaked at two channels with a significant number of events having three or more channels above threshold in a cluster. Three or more channels should be geometrically impossible, but the crosstalk on the MA-PMT will frequently create a low amplitude signal that, due to the unfortunate arrangement of the fibre channels on the MA-PMT,  will add the crosstalk channel to the fibre channel hit cluster as an extra fibre hit. Fibre channels are placed  on the MA-PMT with a 1-to-1 mapping by channel number as shown in Fig.~\ref{fig:crosstalk} where $i$ is the peak charge channel. 

\subsubsection*{Clusters}
\begin{figure}[htbp]
\includegraphics[angle=90,width=0.49\textwidth]{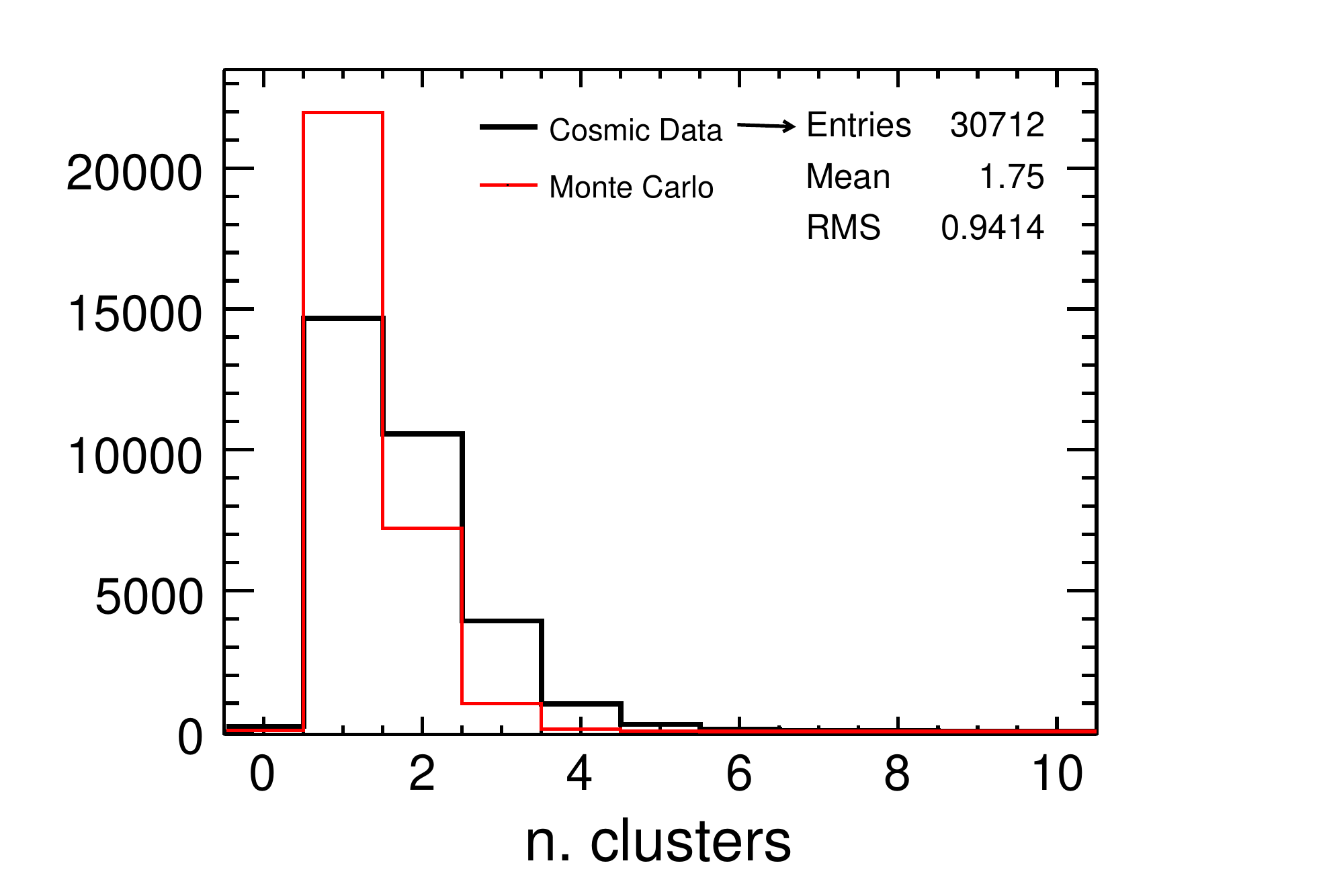} 
\includegraphics[angle=90,width=0.49\textwidth]{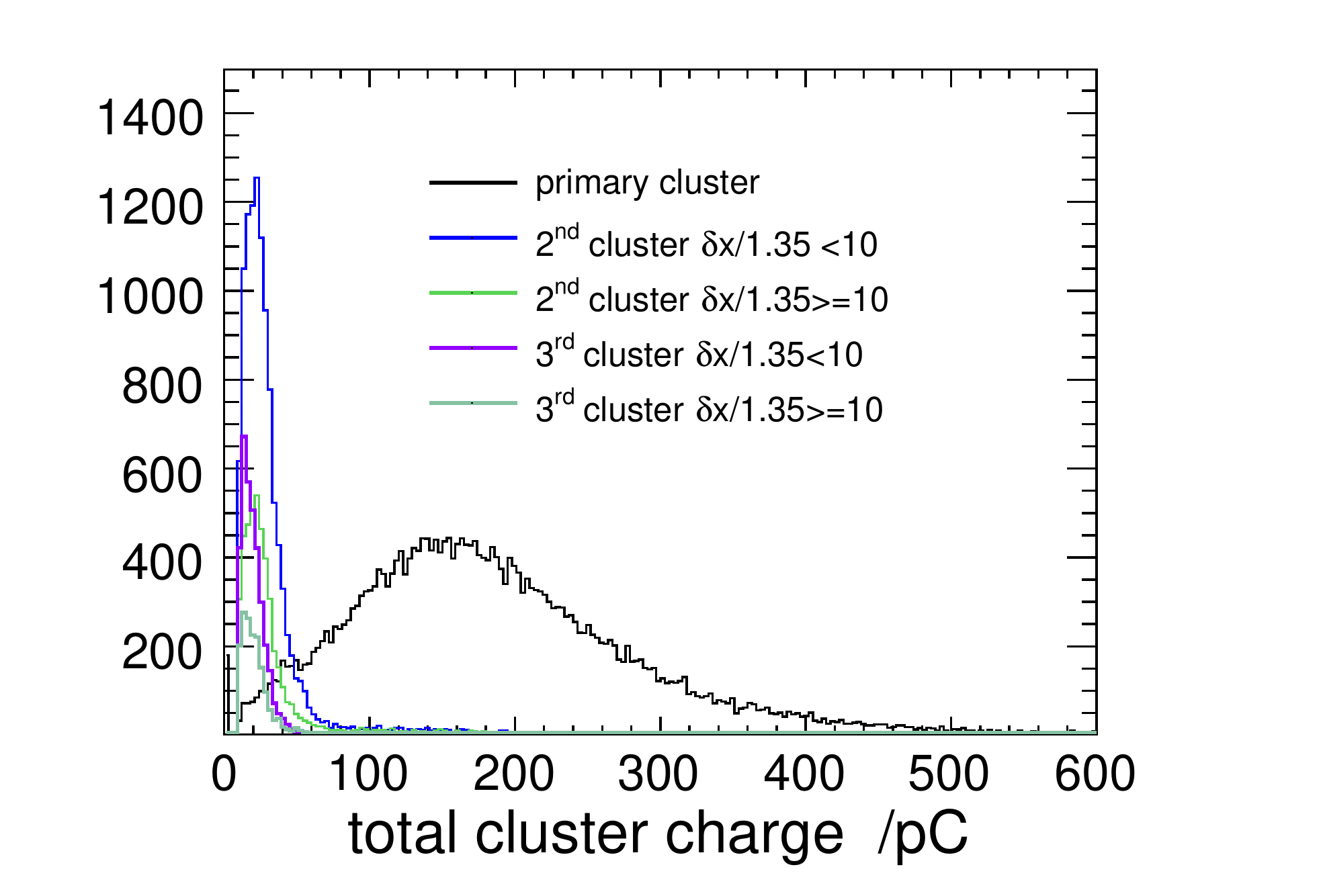} 
\caption{(left) The number of clusters seen in the fibre tracker for each cosmic track event. (right) The total charge in the primary, secondary and tertiary clusters.}
\label{fig:cluster1}
\end{figure}

In a separate analysis of the fibre tracker, nearby channels which exceeded a threshold of 10~pC were grouped into clusters, and then sorted in order of total cluster charge. The cluster with the largest total charge was assumed to be the primary cluster corresponding to a cosmic ray track. One could also take the nearest cluster to the track, depending on whether you assume the fibre tracker is a stand-alone detector or used in conjunction with other trackers, but we will assume the largest total charge cluster here.  The number of clusters found in the fibre plane is shown in Fig.~\ref{fig:cluster1}(left). The  60~cm fibre plane cosmic data has approximately 1.75 clusters in every cosmic events. The total charge for primary clusters, secondary and tertiary clusters which are most likely crosstalk ($\delta{}x/1.35mm < 10$) as well as secondary and tertiary clusters which are most likely noise ($\delta{}x/1.35mm \geq 10$) are shown in Fig.~\ref{fig:cluster1}(right).

\begin{figure}[htbp]
\includegraphics[angle=90,width=0.49\textwidth]{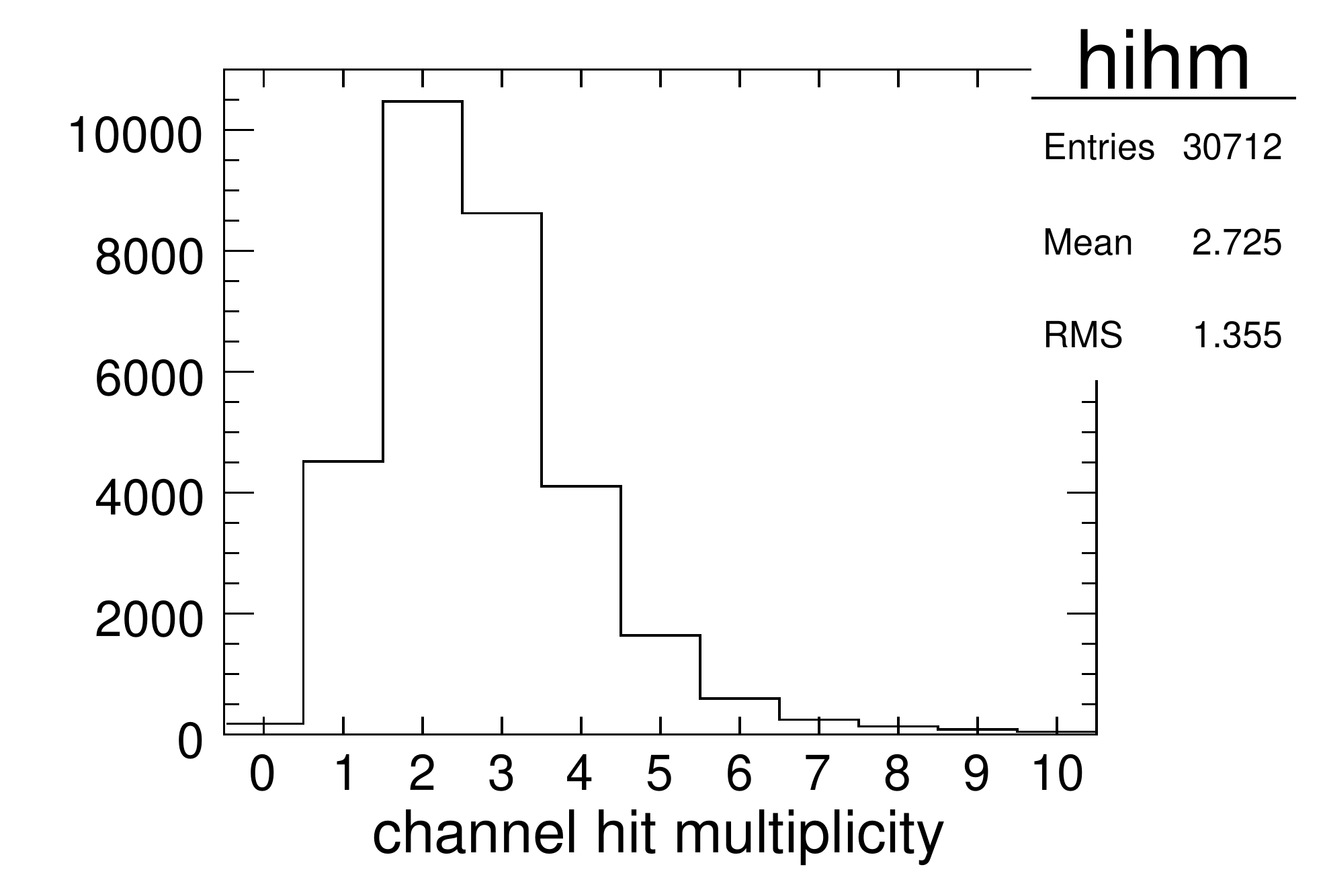} 
\includegraphics[angle=90,width=0.49\textwidth]{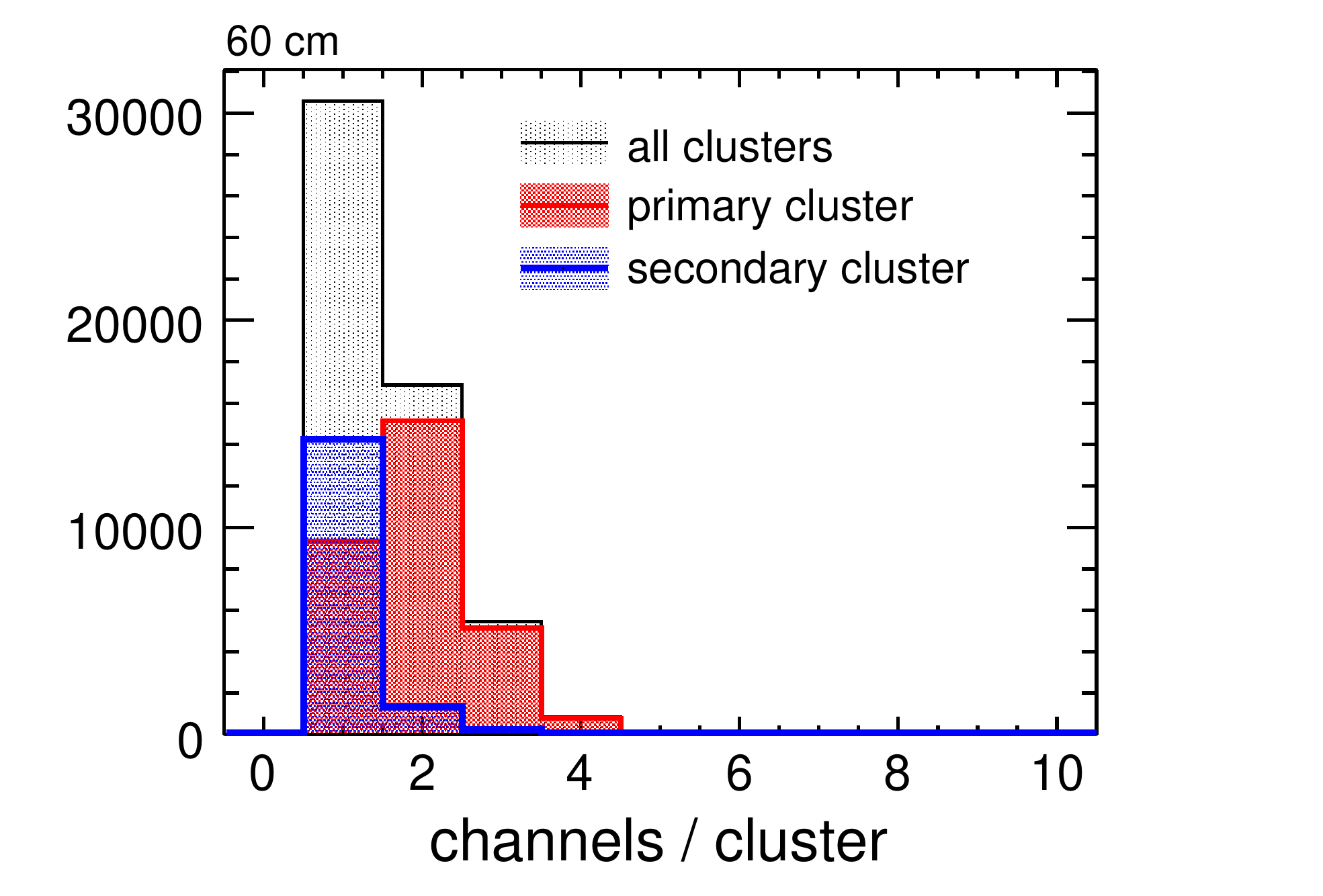} 
\caption{(left) The number of channels in the fibre plane which exceed a threshold of 10~pC during one event. (right) The number of channels per cluster for primary and secondary type clusters.}
\label{fig:cluster2}
\end{figure}

At 60~cm, there are on average 2.75 channels per event that exceed a threshold of 10~pC, as shown in Fig. Fig.~\ref{fig:cluster2}(left). Each cluster has an average of 1.6 channels, with primary clusters having 1.9 channels and secondary clusters normally having 1.1 channels included, as shown in Fig.~\ref{fig:cluster2}(right).

\begin{figure}[htbp]
\includegraphics[angle=90,width=0.49\textwidth]{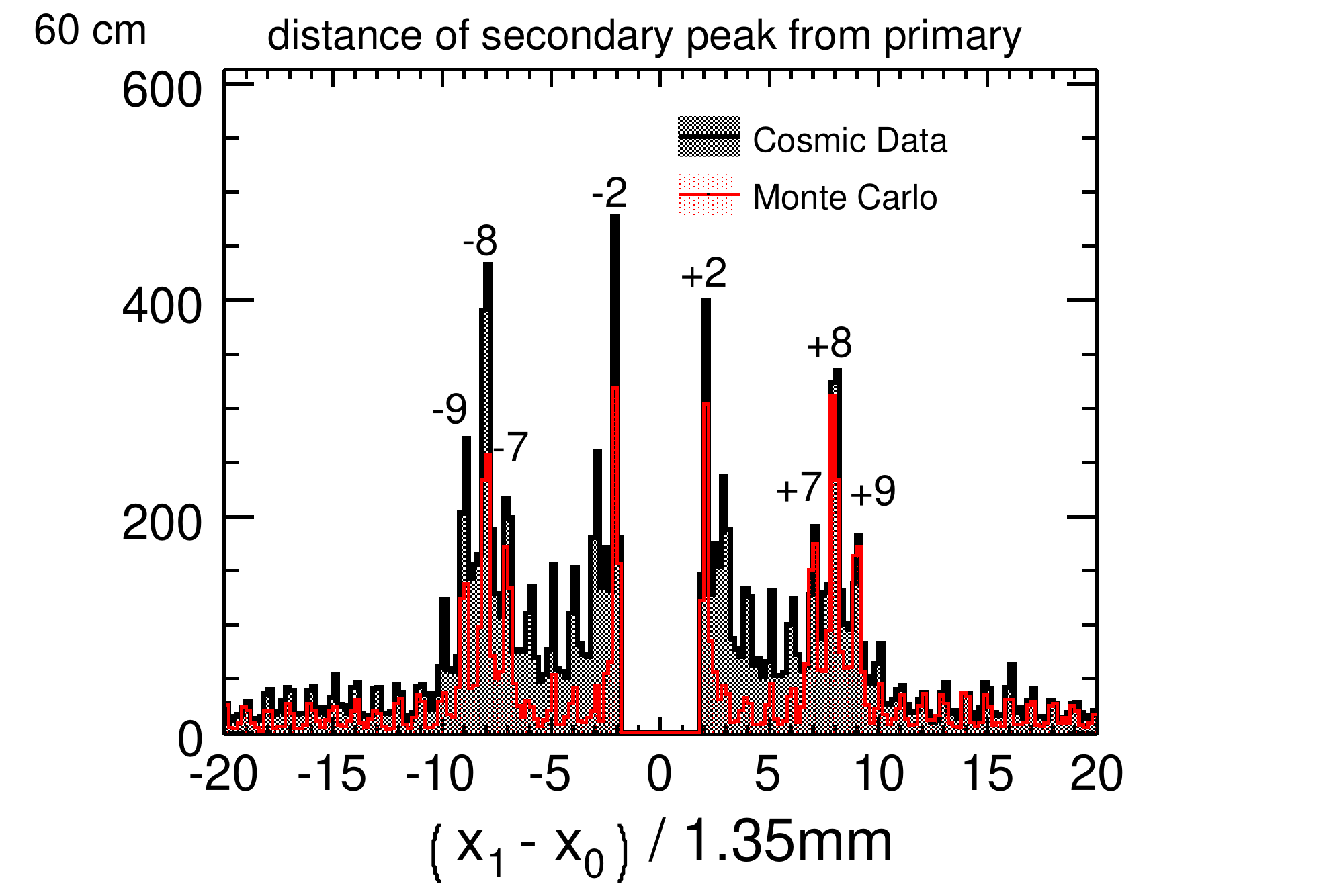} 
\includegraphics[angle=90,width=0.49\textwidth]{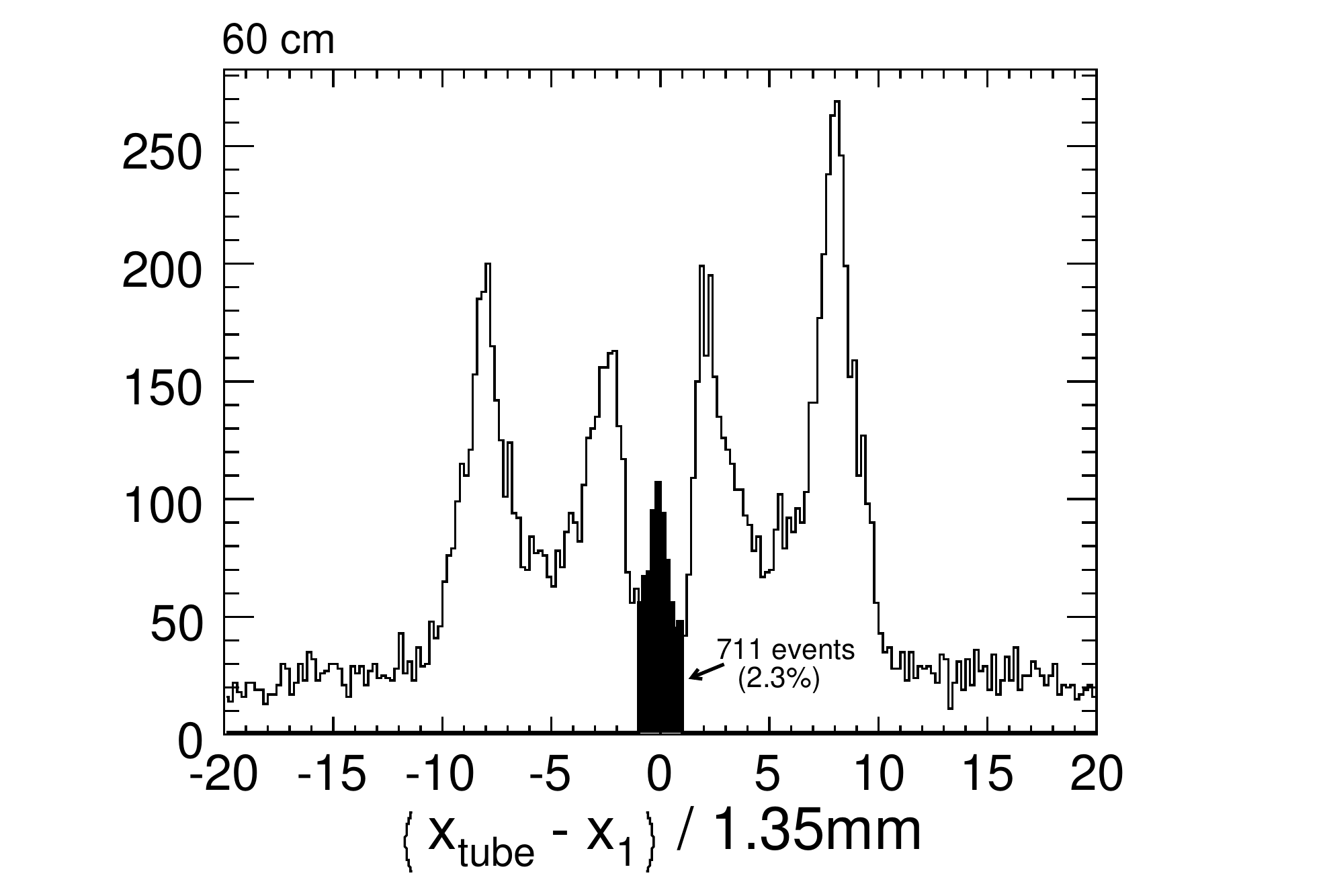} 
\caption{(left) The difference in channel centroid position of the primary and secondary clusters. (right) The difference in channel fibre channel centroid position of the tube track and the secondary fibre cluster.}
\label{fig:cluster3}
\end{figure}

The majority of extra clusters appear to be generated from crosstalk on the MA-PMT, as shown in Fig.~\ref{fig:cluster3}(left), where the distance in channel spacing of the secondary cluster from the primary cluster is reported. The largest peaks are seen at $\pm2, 7, 8$ and 9 channels widths from the primary cluster position. Any electronic crosstalk at $\pm1$ would be included in the primary cluster charge integration (a good reason to separate adjacent fibre channels from adjacent MA-PMT channels). An enhancement of the peaks in channels 3 through 6 above noise are likely a result of optical crosstalk, additional secondary electronic crosstalk and channel (position) resolution.  Fig.~\ref{fig:cluster3}(right) shows the difference in  tube track position and secondary fibre cluster position. The peak at zero indicates that the secondary peak is most likely the cluster due to the cosmic ray and the primary peak was a large noise or crosstalk signal. The shaded area in the figure contains 711 events, or 2.3\% of the total tracks.

\begin{figure}[htbp]
\includegraphics[angle=90,width=0.49\textwidth]{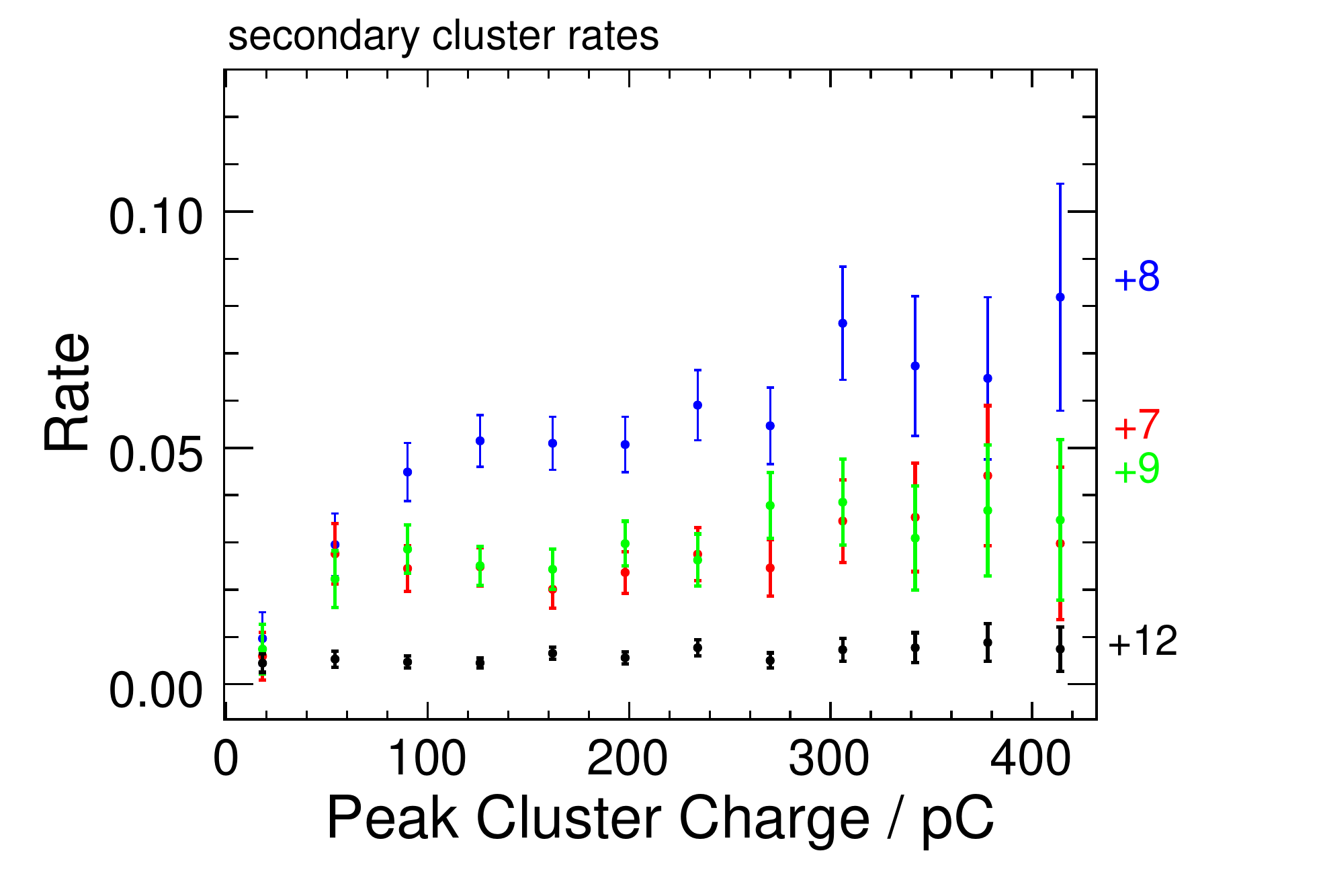} 
\includegraphics[angle=0,width=0.49\textwidth]{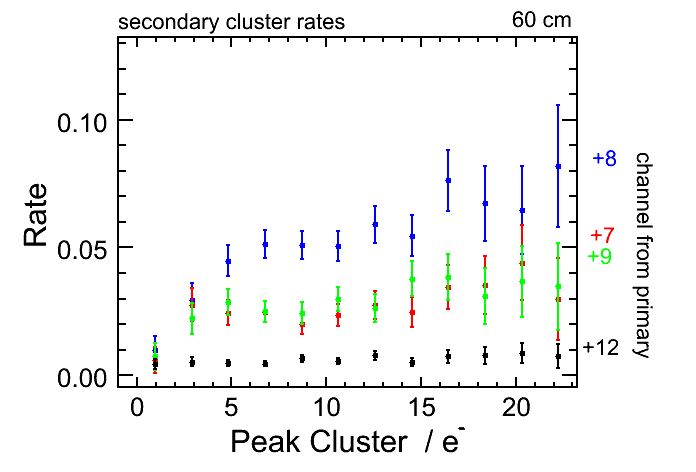} 
\caption{(left) The crosstalk rate for channels +7,+8, +9 and +12 from the cluster channel center as a function of primary cluster amplitude. (right) The crosstalk rate for clusters as a function of photoelectrons, assuming the Monte Carlo single photoelectron amplitude of 18.6 pC.}
\label{fig:ctrate}
\end{figure}

Ref.~\cite{ref_anab} has shown previously that there is a crosstalk rate dependence on the signal amplitude of the peak channel for the H9500 MA-PMT. The amplitude of the crosstalk, found from this work and reported  in Fig.~\ref{fig:cluster1} as secondary and tertiary clusters, is of the order of one to two photoelectron level. The crosstalk rates (fraction of events in which secondary clusters are seen nearby on the MA-PMT) as a function of the peak (primary) channel charge amplitude for channels +7, +8, +9 and +12 channels from the primary cluster are reported in the Fig.~\ref{fig:ctrate}. The plot on the right of Fig.~\ref{fig:ctrate} has a horizontal axis assuming the Monte Carlo single photoelectron amplitude of 18.6~pC. If channel~+12 is a good indication of the nominal noise rate, there is a 0.5\% probability of seeing a secondary cluster in each channel purely from electronics noise or accidentals.   

For channels +8 from the peak cluster, from zero to 100~pC ($\sim$5~photoelectrons), there is a 1\% probability  of seeing crosstalk in that channel per  photoelectron in the peak cluster, and 0.5\% per photoelectron for channels +7 and +9. After 100~pC, the probability of crosstalk increases slowly with increasing peak cluster charge, at a rate of about 0.1\% per photoelectron for both channels +7, +8 and +9, though the uncertainty for this is quite large. Channel +12 remains constant, as expected. Results for -7, -8, -9 and -12 are nearly identical. Assuming the crosstalk in +1 and -1 is similar to +8 and -8, there is an average 10\% probability of seeing crosstalk in an adjacent  fibre channel with the current arrangement, and a 30\% probability of crosstalk in one adjacent MA-PMT channel.

\subsection{Hit Threshold}

\begin{figure}[htbp]
\includegraphics[angle=0,width=0.49\textwidth]{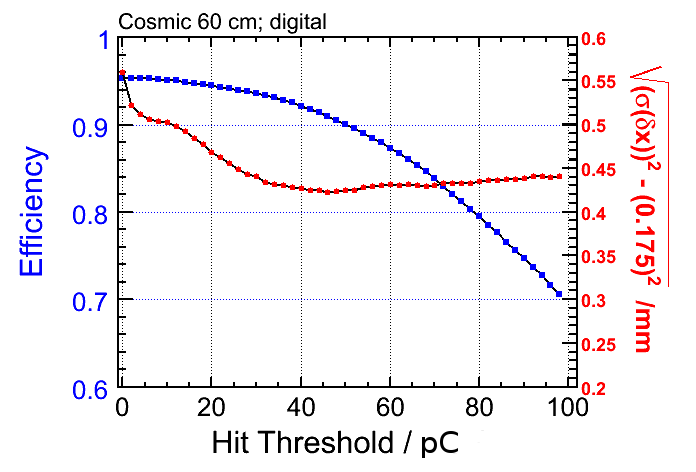} 
\includegraphics[angle=0,width=0.49\textwidth]{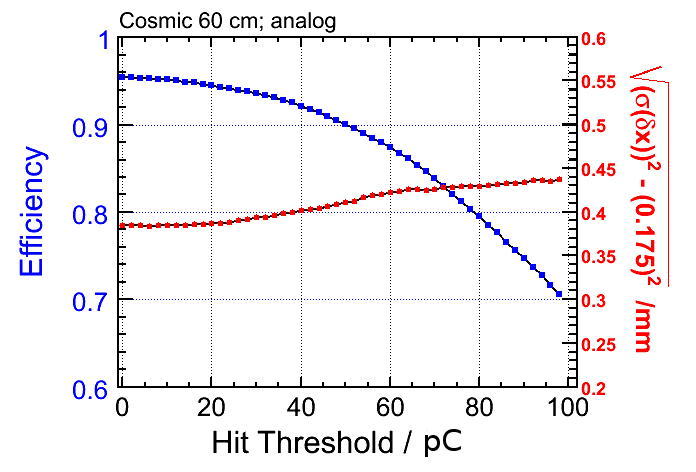} 
\caption{A plot of standard deviation of the difference in tube track position and fibre tracker position (with the expected tube contribution subtracted)  for both digital (left) and analog (right) determined positions. The efficiency for finding the track as a function of threshold value is also shown.}
\label{fig:effthresh}
\end{figure}

The choice of whether to include a channel hit in determining the cosmic ray position in the fibre tracker is entirely based on whether or not it exceeds a certain  threshold. This is a simple method to exclude the smaller noise/crosstalk signals which degrade the resolution, but will also remove small signals that are from energy deposited in that channel's fibres which could be used to improve the resolution. A plot of standard deviation of the difference between tube track position and fibre tracker position (with the expected tube contribution subtracted) are seen in Fig.~\ref{fig:effthresh} for both digital and analog determined positions.  The larger standard deviation than the previously quoted resolution is due to some remaining contribution to the resolution from systematic offsets of the individual fibre channels that is present in the faster method used to determine the resolution. The efficiency for finding the track as a function of threshold value is also seen in Fig.~\ref{fig:effthresh}. 
Fig.~\ref{fig:effthresh} illustrates the degrading effect that the crosstalk has on the position resolution, which can be overcome with a higher threshold value, but the efficiency is sacrificed. The analog resolution is as effected by the crosstalk and benefits from a lower threshold value. A threshold of 10~pC  is consistently used as a threshold in the analysis in other sections.

\begin{figure}[htbp]
 \includegraphics[angle=0,width=0.5\textwidth]{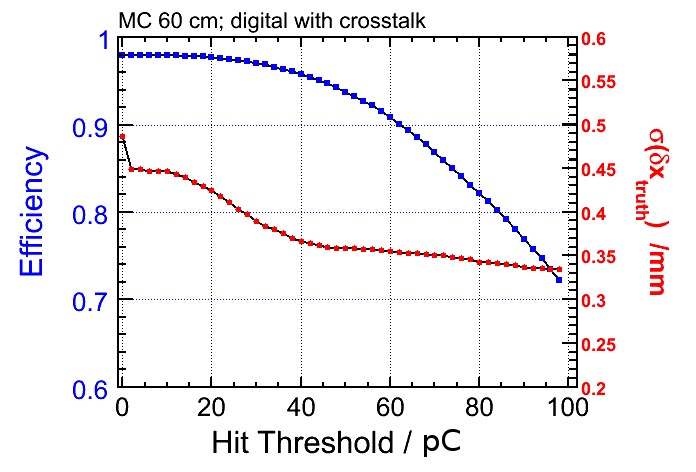} 
\includegraphics[angle=0,width=0.5\textwidth]{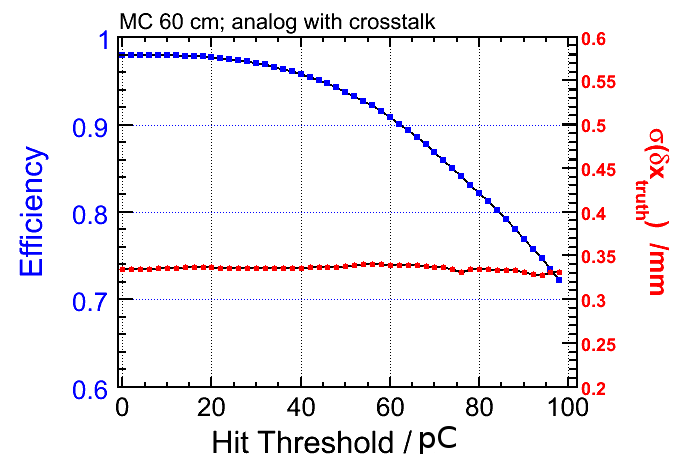} 
\includegraphics[angle=0,width=0.5\textwidth]{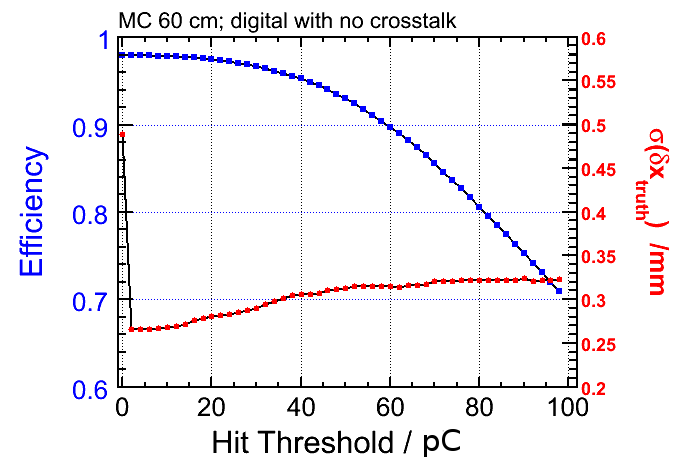}
 \includegraphics[angle=0,width=0.5\textwidth]{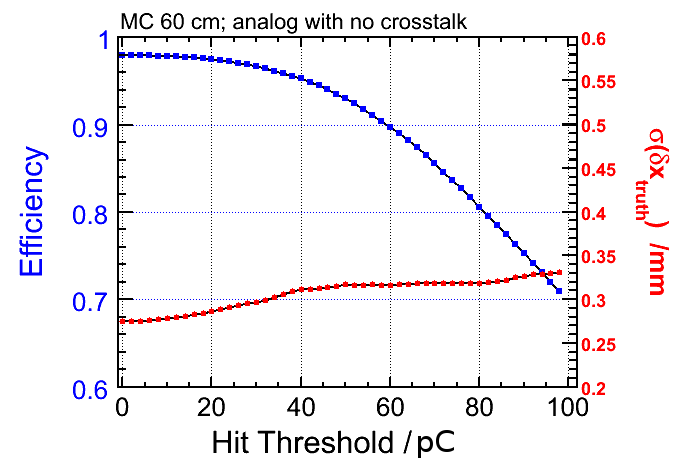}
 \caption{The efficiency and resolution for digital (left) and analog (right) modes where the detector Monte Carlo includes (top) and excludes (bottom) simulated crosstalk.}
 \label{fig:MCeffthresh}

 \end{figure}

The same method was used to look at efficiency and resolution in the Monte Carlo detector simulation of the fibre tracker using similar crosstalk amplitudes and frequencies to mimic the experimental cosmic ray setup. The data here is much cleaner because the fibre tracker and tubes have perfect alignment without missing channels. The calculated fibre tracker position is compared to the known truth information for the track in the Monte Carlo. The efficiency and resolution for the track are shown in Fig.~\ref{fig:MCeffthresh}. One can infer from the MC plots that crosstalk and threshold are the driving factors for the resolutions at low threshold. At low threshold more crosstalk channels are included in the position calculation degrading the resolution but more channels with true energy deposition are retained improving the resolution. The crosstalk has a much larger effect on the digital position resolution even though it is of small amplitude, it is weighted equally with the peak energy deposit channel when determining the hit centroid.

\section{Monte Carlo - new designs}

Taking  the knowledge and experience gained from this first fibre tracker prototype, and implementing this into a Monte Carlo detector simulation that reliably approximates the detector response, improved designs with other fibre configurations and geometries were tested.

The goals of the overall design are to have an improved position resolution which meets an experimental requirement of less than 0.200~mm, while minimizing material and monetary costs. An improved light yield, which increases the number of detected photoelectrons, will improve the resolution. This can be achieved by a greater thickness of scintillator but this also introduces more material into the detector.  Double-clad fibres that have a circular cross section (\textit{rnd}) can also be used to increase the trapping efficiency by a factor of 5.3/3.5 (7.3/3.5 in the case of fibres with a square (\textit{sqr}) cross section ). This will increases the light yield without increasing the material thickness, but also increase the cost per fibre.

The finite channel width also affects the resolution greatly. A smaller channel width will also improve the resolution, in a digital/binary readout mode. The single photoelectron resolution and efficiency of the photodetector also contributes to this. False-negative and false-positives significantly affect the digital readout resolution, if a track shared between adjacent channels is not registered in one channel, or a false-positive includes an adjacent channel where none should be included. For the results below, it is assumed that crosstalk has been minimized in a possible experimental scenario, and therefore has not been included in the resolution and efficiency results. The electronic noise is the same as that for the 1.35~mm detector simulation described in the previous section.

For the simulations below, the single photoelectron resolution, $\sigma_{1pe}/G_{1pe}$, of the photodetector is set to  a value of 0.4.   This should be equivalent to the resolution of the super-bialkali  H7546A-100 MA-PMT which is expected to have a quantum efficiency(QE) of 34\%, a large improvement over the 22\% QE of the H8500. The ultra-bialkali  H7546A-200 MA-PMT has an even larger QE of 40\%. The single photoelectron gain, $G_{1pe}$, is set to 18.6~pC/pe. The simulations also assume that crosstalk is not an issue as adjacent fibre channels are no longer placed next to each other on the MA-PMT array. The photoelectron yield per energy deposited differs in each of the different scenarios described below, depending on the fibre type used.

\subsection{4 layers of round 1~mm fibres, 1mm horizontal spacing\label{sec:4rnd1mm}}

\begin{figure}[hp]
\includegraphics[width=0.49\textwidth]{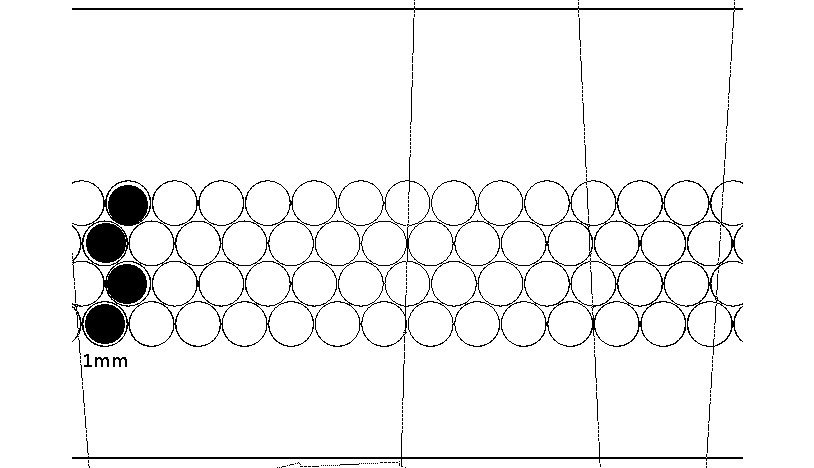} 
\includegraphics[angle=90,width=0.49\textwidth]{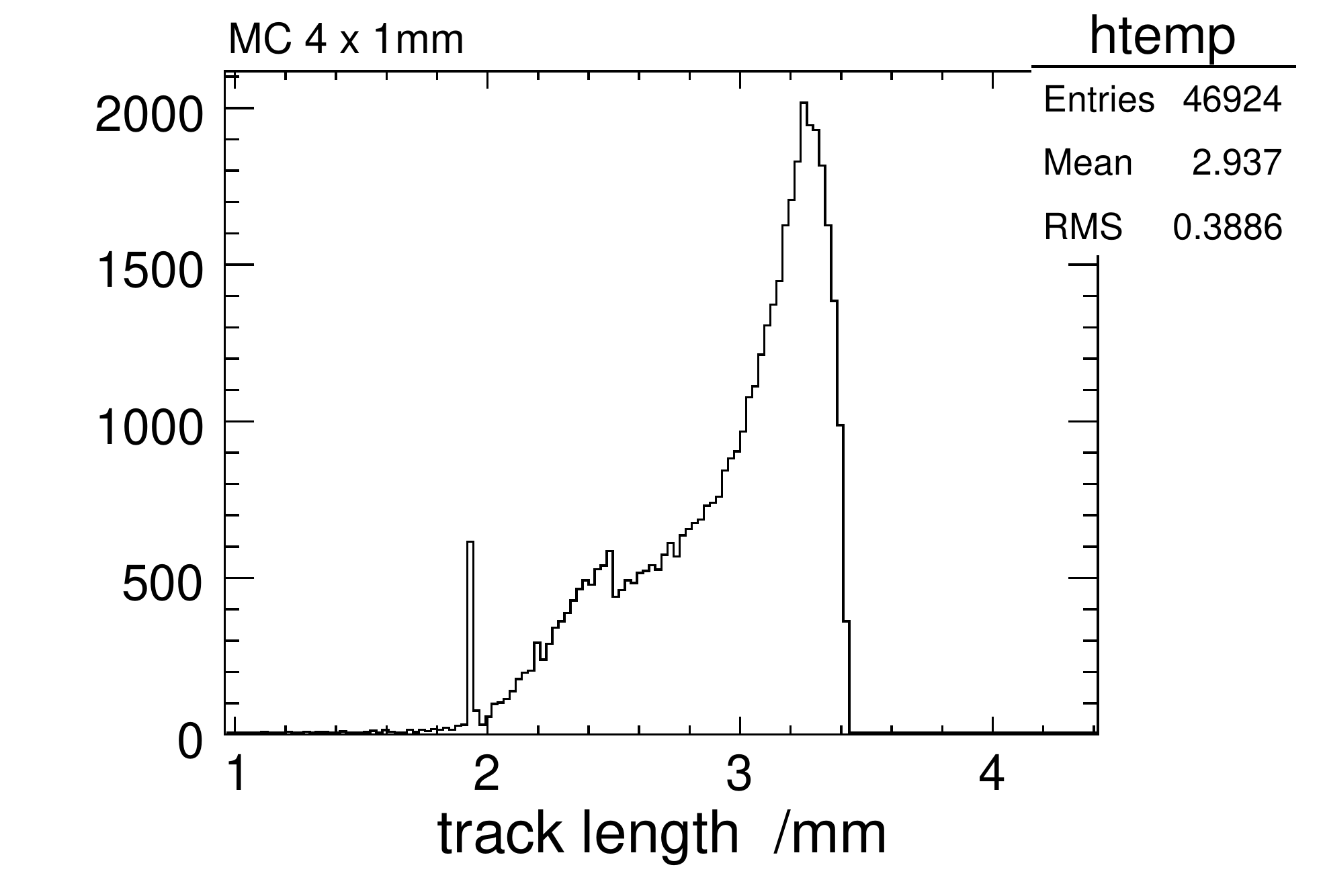} 
\caption{\textbf{4rnd1mm} (left) A schematic of the fibre layout in Monte Carlo. The 4 shaded fibres form  a single channel. (right) The total track length in the fibre detector.}
\label{fig:tracktotal_C41mm}
\end{figure}

\begin{figure}[htbp]
\centering
\includegraphics[angle=90,width=0.49\textwidth]{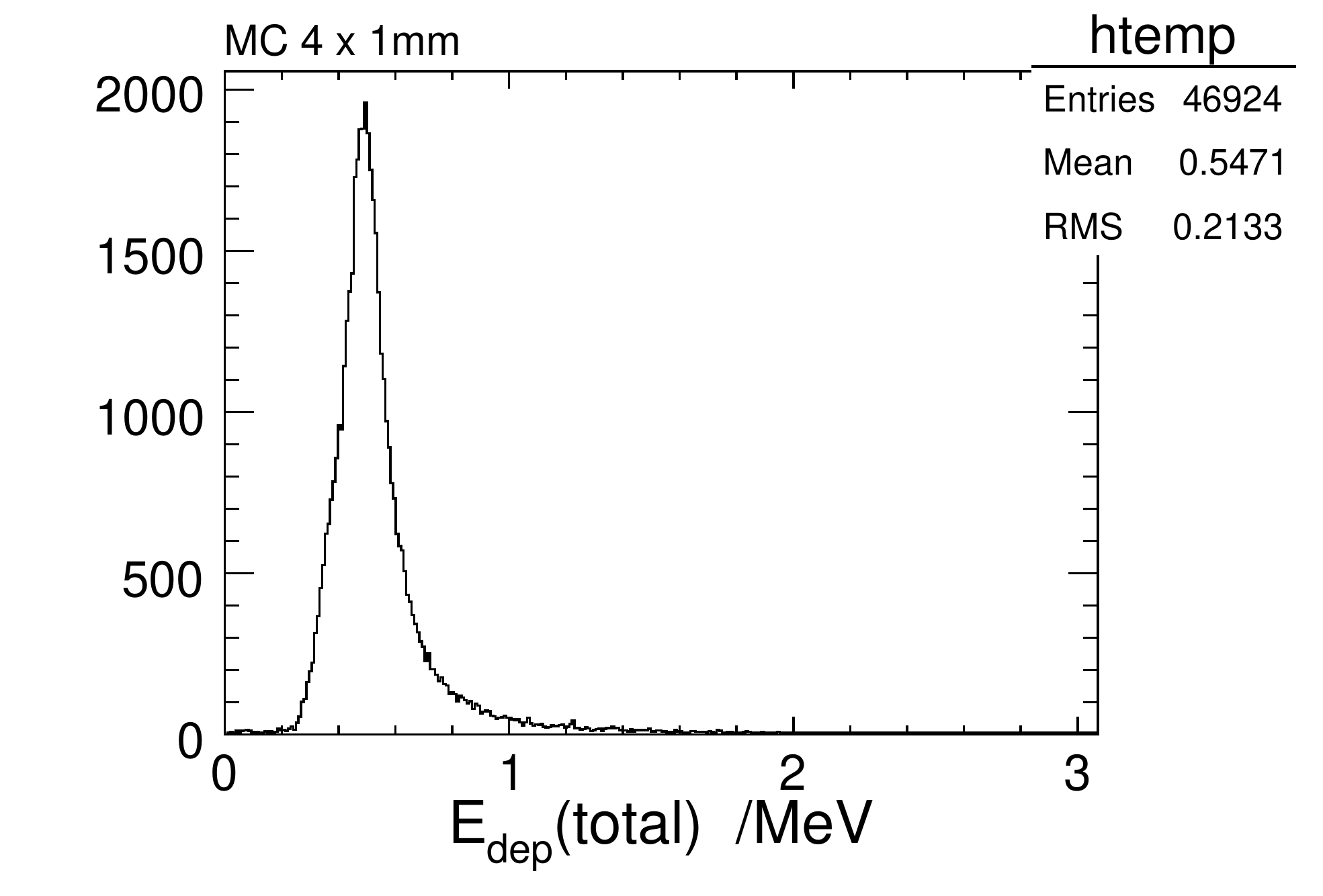}
\includegraphics[angle=90,width=0.49\textwidth]{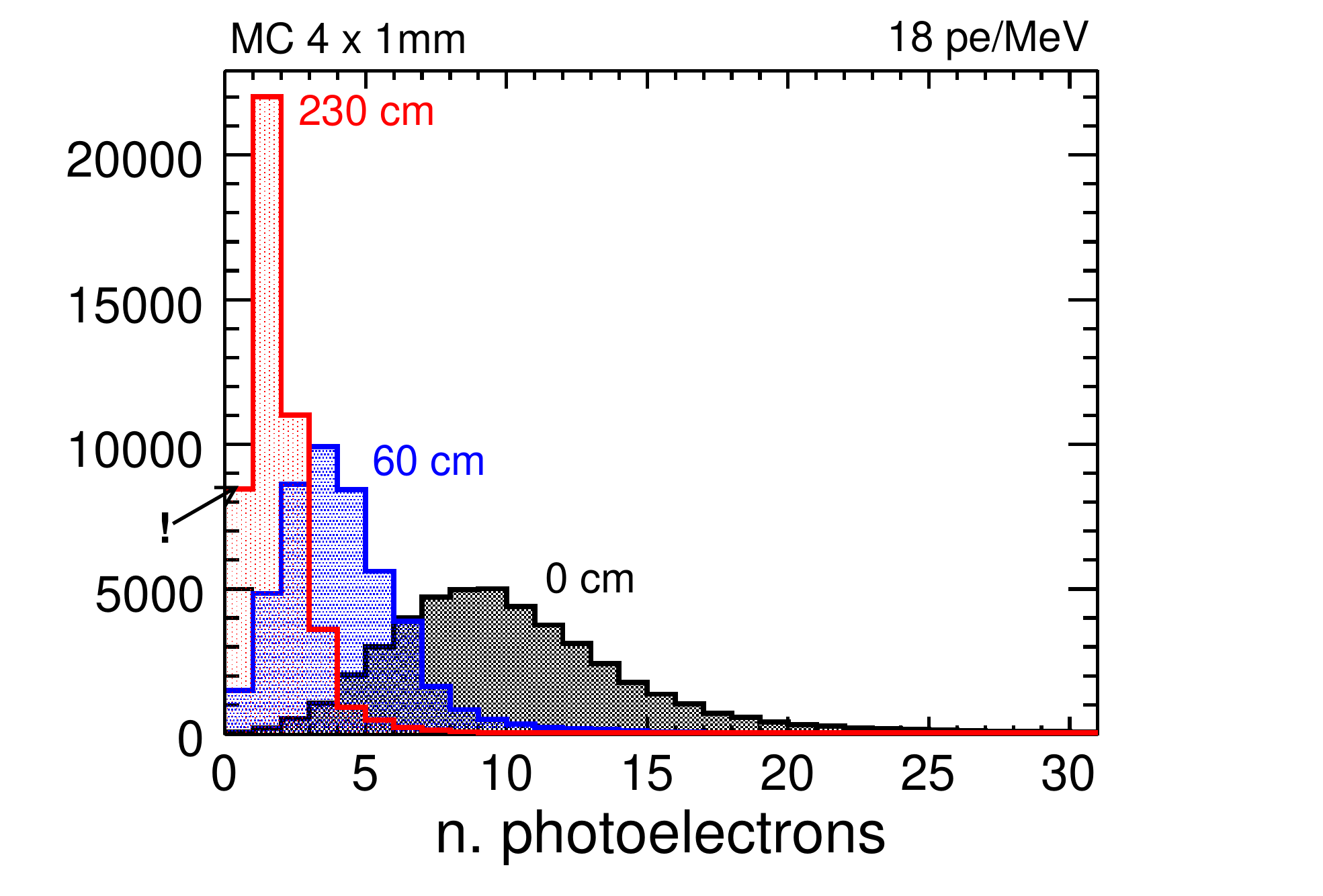}

\caption{\textbf{4rnd1mm} (left) The total energy deposited in the fibre tracker. (right) The photoelectron distribution as seen at the source (0~cm), at 60~cm and 230~cm. }
\label{fig:edep_C41mm}
\end{figure}

\begin{figure}[htbp]
\centering
\includegraphics[angle=90,width=0.49\textwidth]{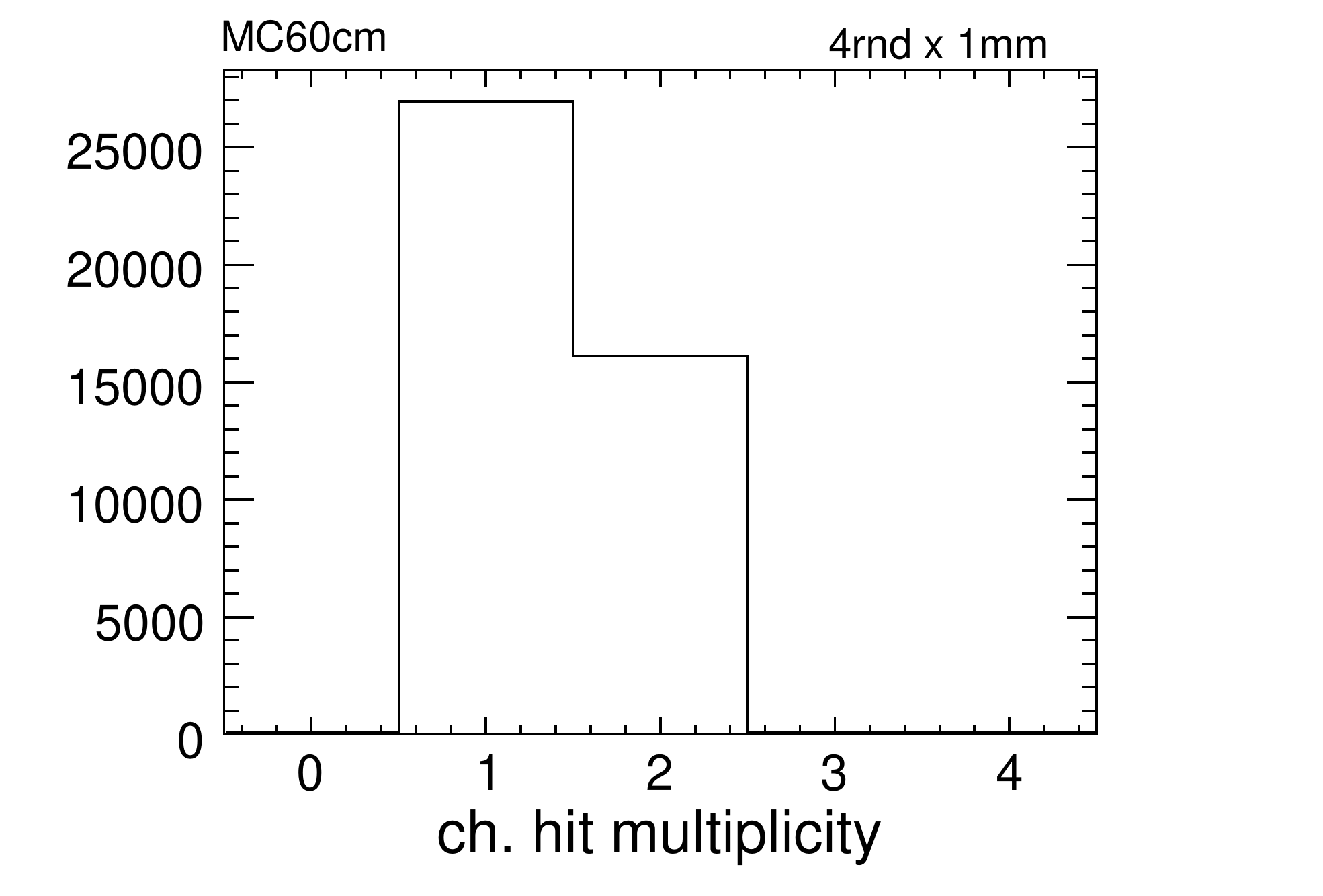}
\includegraphics[angle=90,width=0.49\textwidth]{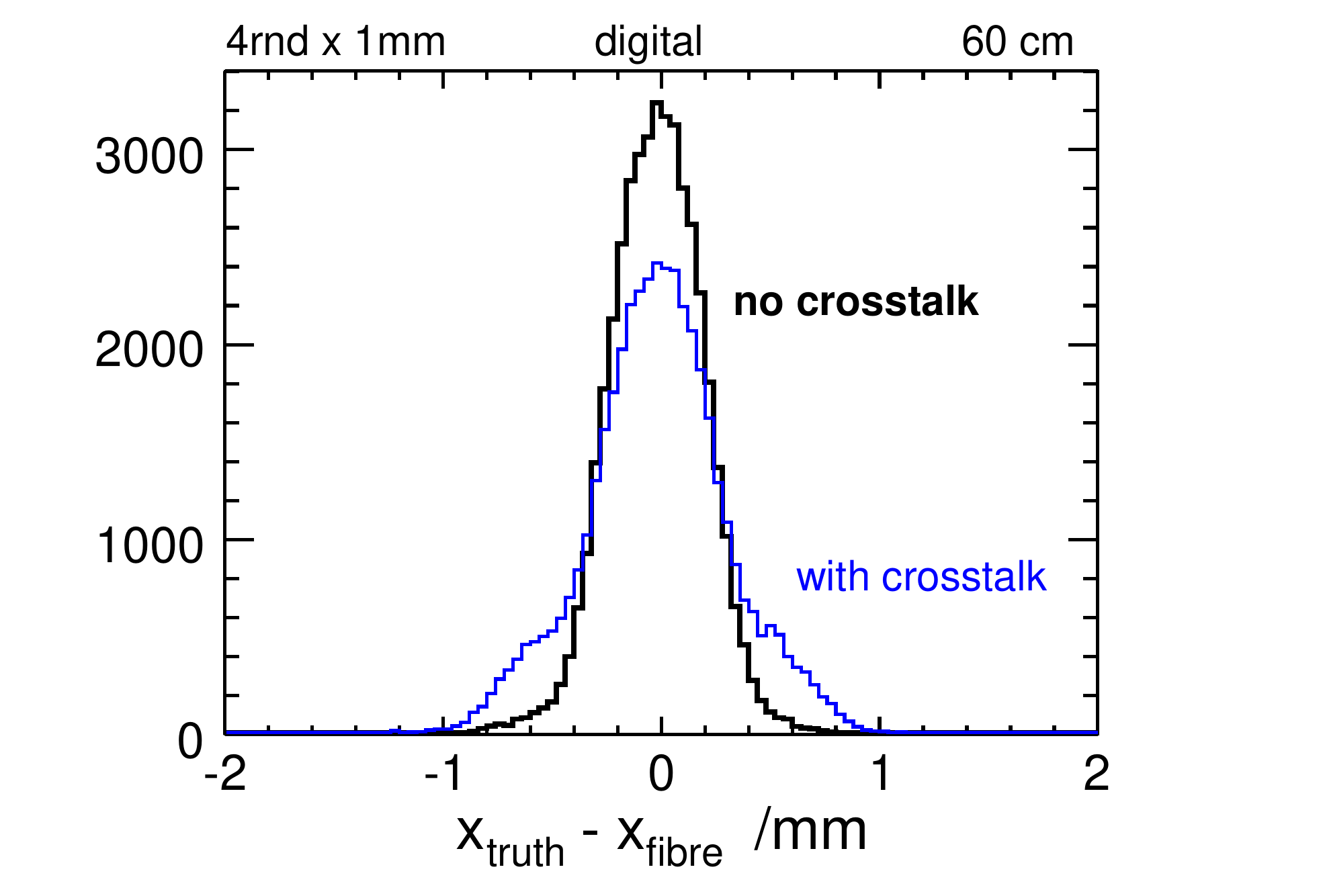}

\caption{\textbf{4rnd1mm} (left) Channel hit multiplicity at 60~cm with no crosstalk. (right) The expected resolution at 60 cm with no crosstalk (black) and if  crosstalk similar to the experimental prototype is included (blue). }
\label{fig:mult_C41mm}
\end{figure}

Here, it is assumed that the fibres are single clad, producing 18 photoelectrons per MeV, as in the first prototype detector setup. The fibres are packed together as closely  as possible, with 1 fibre diameter separating channel centres, in such a way that each channel overlaps the one in a consecutive layer by half a fibre diameter, as in Fig.~\ref{fig:tracktotal_C41mm}(left).  The path length of the tracks through the fibres is shown in Fig.~\ref{fig:tracktotal_C41mm}(right). The energy deposited in the fibres in this configuration along with the expected number of photoelectrons are shown in Fig.~\ref{fig:edep_C41mm}. The hit multiplicity, shown in Fig.~\ref{fig:mult_C41mm}, will indicate whether the track passed though the center of the channel or the region shared with the adjacent channel, improving the effective resolution to that of half the fibre width, $\sim{}0.5/\sqrt{12}$~mm. However, the resolution of the photodetector, the reduction in fibre material away from the channel centre and photoelectron statistics will widen the resolution to somewhat greater than this. A summary of the resolutions and efficiency are shown in Table.~\ref{tab:summary4rnd} for this fibre tracker configuration.

\begin{table}[h!b]
\centering
\caption{Summary of the four layer, 1~mm diameter, single clad fibre design from Monte Carlo.}
\begin{tabular}{|l|c|c|c|c|}
\hline 4rnd x 1mm & 60~cm & 90~cm & 150~cm & 230cm \\ 
\hline resolution (digital) /mm & 0.204 & 0.205 & 0.236 & 0.236 \\ 
\hline resolution (analog) /mm & 0.224 & 0.231 & 0.259 & 0.260 \\ 
\hline efficiency(2.7~mm) &  0.972 &0.962  & 0.936 & 0.920 \\ 
\hline 
\end{tabular} 
\label{tab:summary4rnd}
\end{table}

\pagebreak
\subsection{6 layers of round 1~mm fibres, 1mm horizontal spacing}
\label{sec:6rnd1mm}

\begin{figure}[htbp]
\centering

\includegraphics[width=0.49\textwidth]{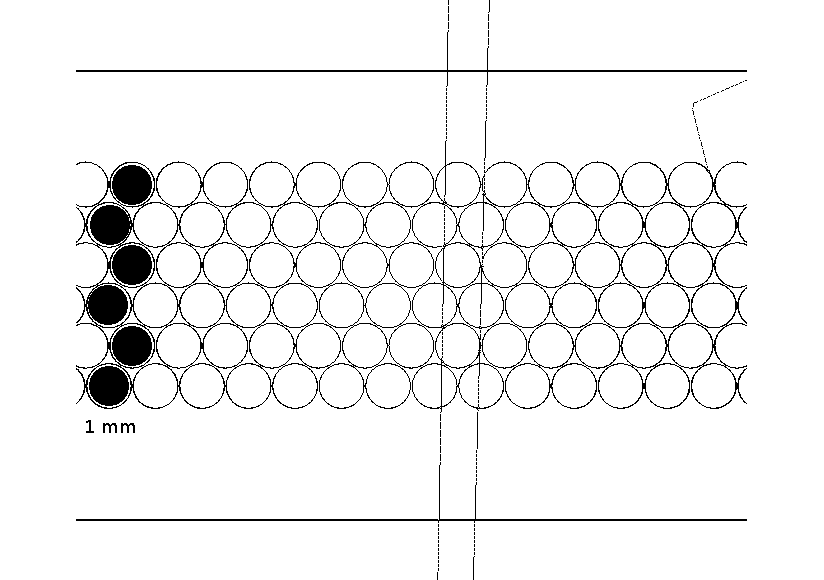} 
\includegraphics[angle=90,width=0.49\textwidth]{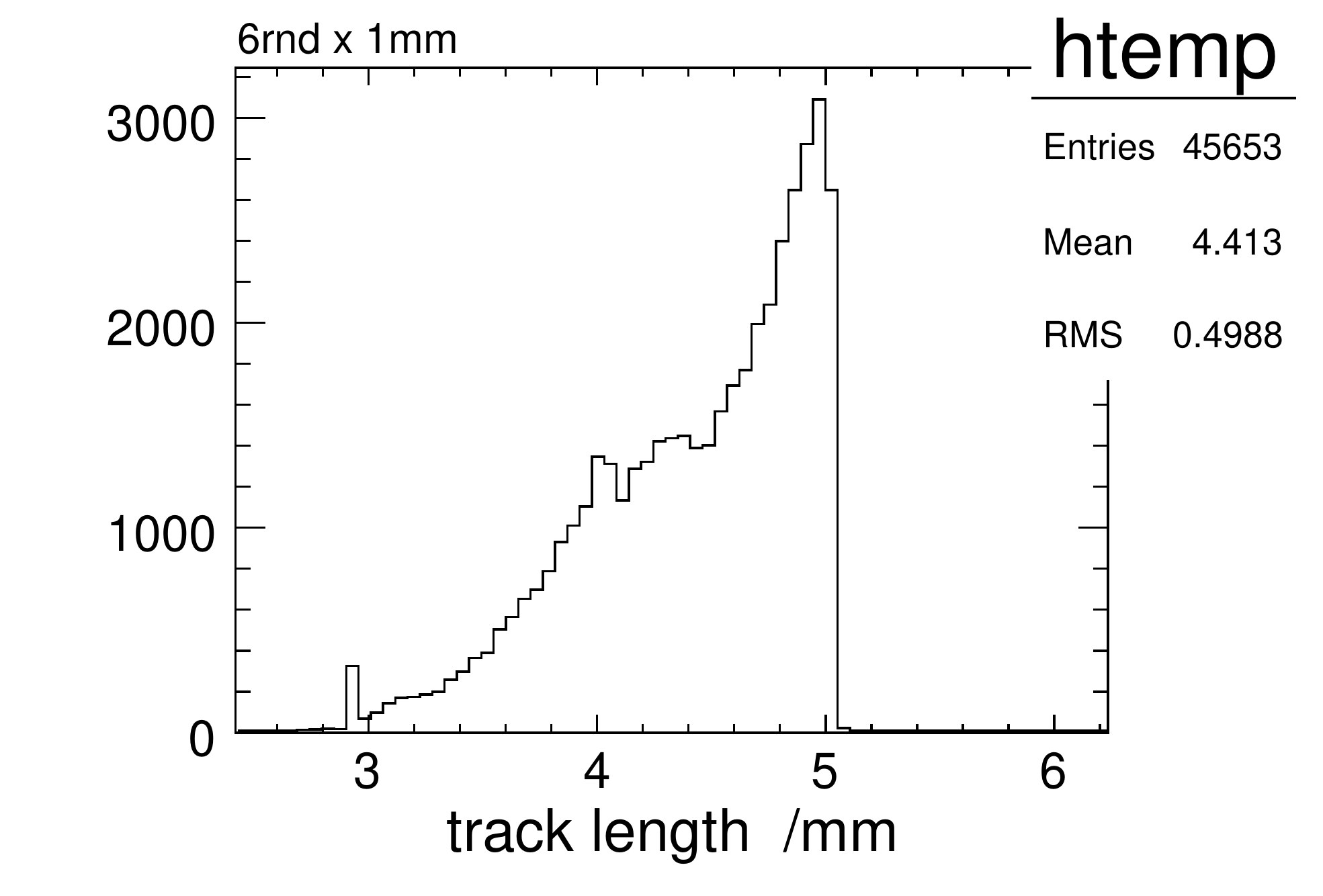} 
\caption{\textbf{6rnd1mm} (left) A schematic of the fibre layout in Monte Carlo. The 6 shaded fibres are  a single channel. (right) The total track length in the fibre detector.}
\label{fig:tracktotal_C61mm}
\end{figure}

\begin{figure}[htbp]
\centering
\includegraphics[angle=90,width=0.49\textwidth]{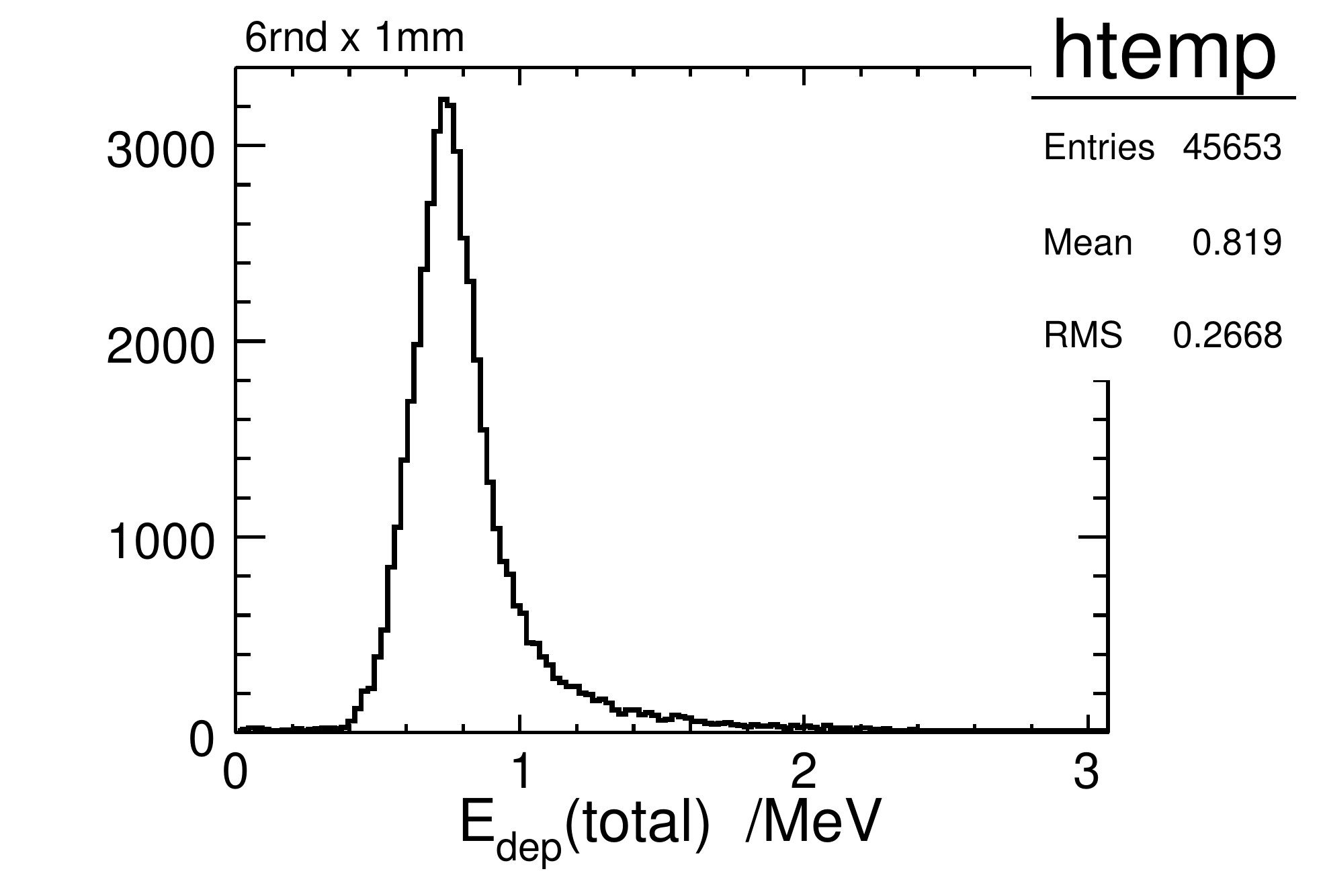}
\includegraphics[angle=90,width=0.49\textwidth]{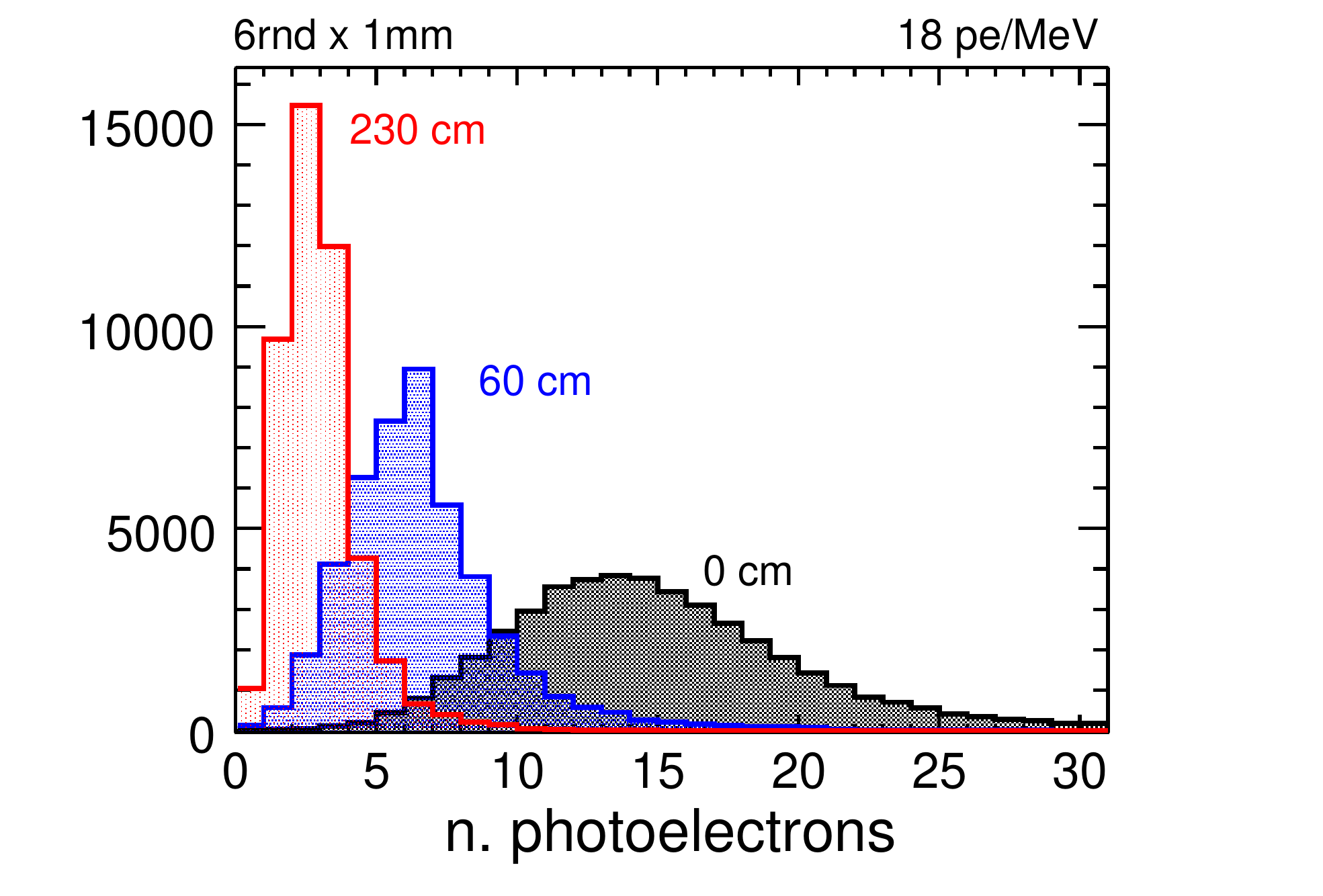}

\caption{\textbf{6rnd1mm} (left) The total energy deposited in the fibre tracker. (right) The photoelectron distribution as seen at the source (0~cm), at 60~cm and 230~cm. }
\label{fig:edep_C61mm}
\end{figure}

\begin{figure}[htbp]
\centering
\includegraphics[angle=90,width=0.49\textwidth]{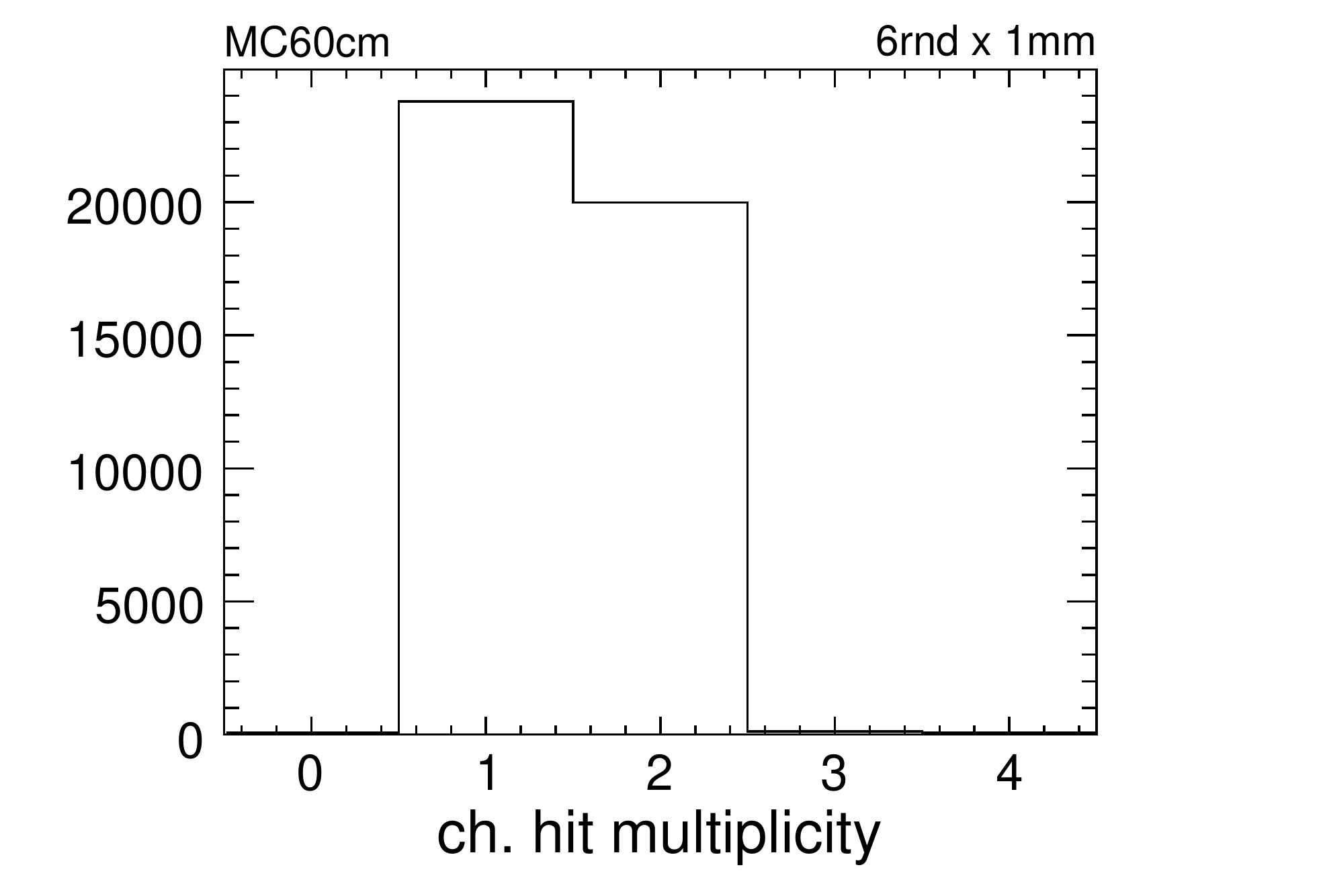}
\includegraphics[angle=90,width=0.49\textwidth]{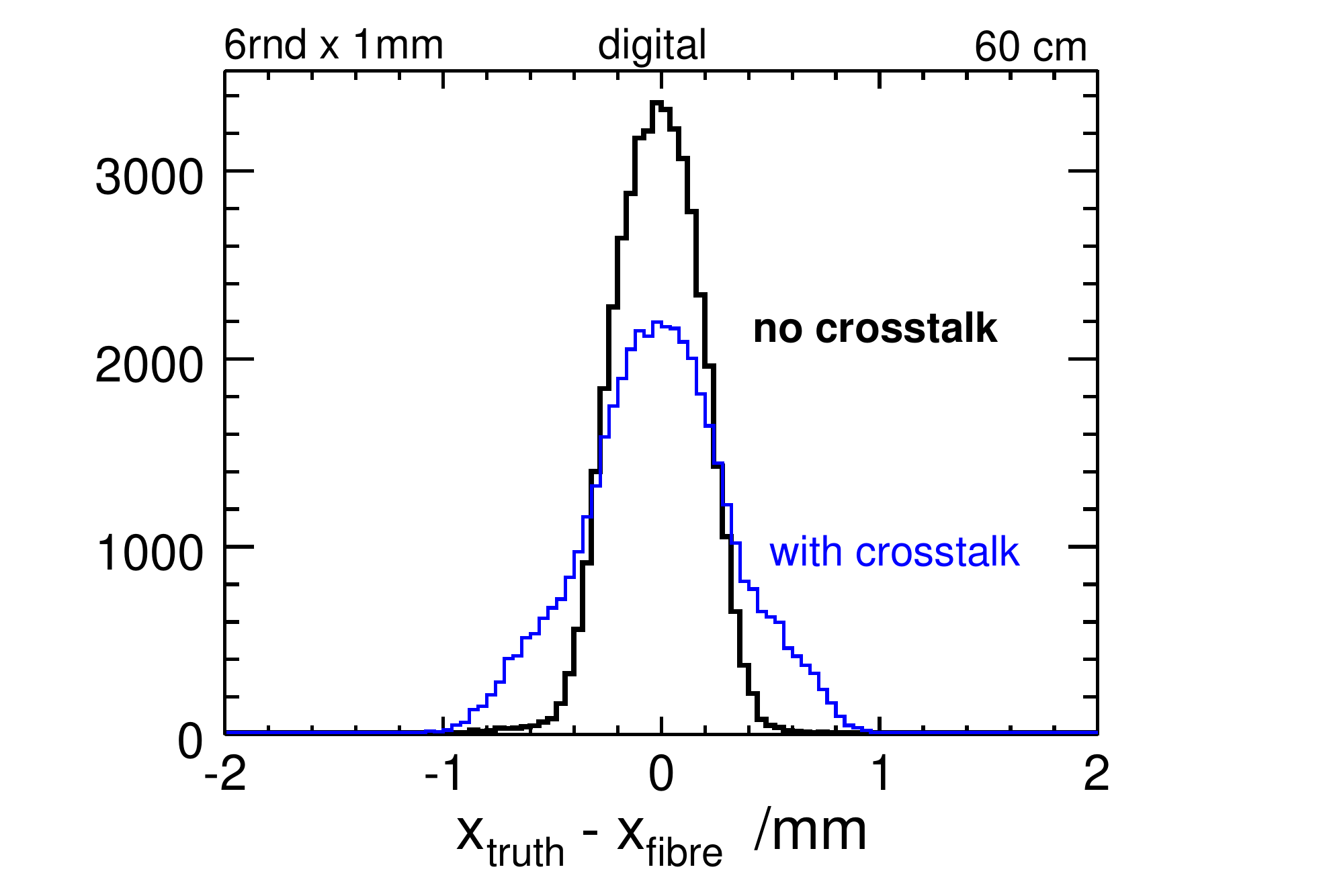}

\caption{\textbf{6rnd1mm} (left) Channel hit multiplicity at 60~cm with no crosstalk. (right) The expected resolution at 60 cm with no crosstalk (black) and if similar crosstalk to experiment is included (blue). }
\label{fig:mult_C61mm}
\end{figure}

Here, it is assumed that the fibres are single clad, producing 18 photoelectrons per MeV, as in the first prototype detector setup. The fibres are packed together as closely  as possible, with 1 fibre diameter separating channel centres, in such a way that each channel overlaps the one in a consecutive layer by half a fibre diameter, as in Fig.~\ref{fig:tracktotal_C61mm}(left).  The longer path length of the tracks through the six layers fibres is shown in Fig.~\ref{fig:tracktotal_C61mm}(right). The energy deposited in the fibres in this configuration along with the expected number of photoelectrons are shown in Fig.~\ref{fig:edep_C61mm}. The hit multiplicity, shown in Fig.~\ref{fig:mult_C61mm}, indicates that more channels are sharing hits which exceed threshold than in Sec.~\ref{sec:4rnd1mm}, as a result of the angle and longer path length of the tracks.  A summary of the resolution and efficiency are shown in Table.~\ref{tab:summaryC61mm} for this fibre tracker configuration. Only the digital resolution is shown as this would be the likely readout method for this fibre configuration. The resolution can be seen to be improved with the increase in material and more light, over the four layers in the previous section.

\begin{table}[h!]
\centering
\caption{Summary of the six layer, 1~mm diameter, single clad fibre tracker resolutions and efficiencies.}
\begin{tabular}{|l|c|c|c|c|}
\hline 6rnd x 1mm & 60~cm & 90~cm & 150~cm & 230cm \\ 

\hline resolution (digital) /mm & 0.189 & 0.189 & 0.202 & 0.203 \\ 
\hline efficiency(2.7~mm)  & 0.986 & 0.983 &0.974 &  0.963\\ 
\hline 
\end{tabular} 
\label{tab:summaryC61mm}
\end{table}

\pagebreak

\subsection{6 layers of round 0.7~mm, 0.7 mm horizontal spacing}

\begin{figure}[htbp]
\centering

\includegraphics[width=0.49\textwidth]{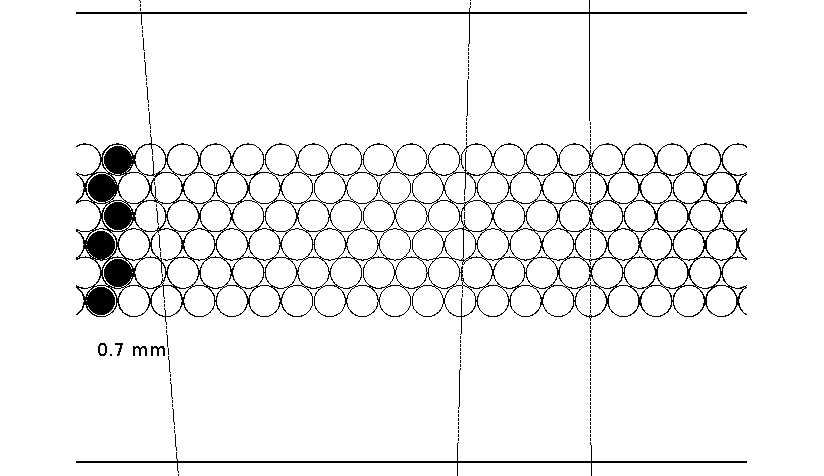} 
\includegraphics[angle=90,width=0.49\textwidth]{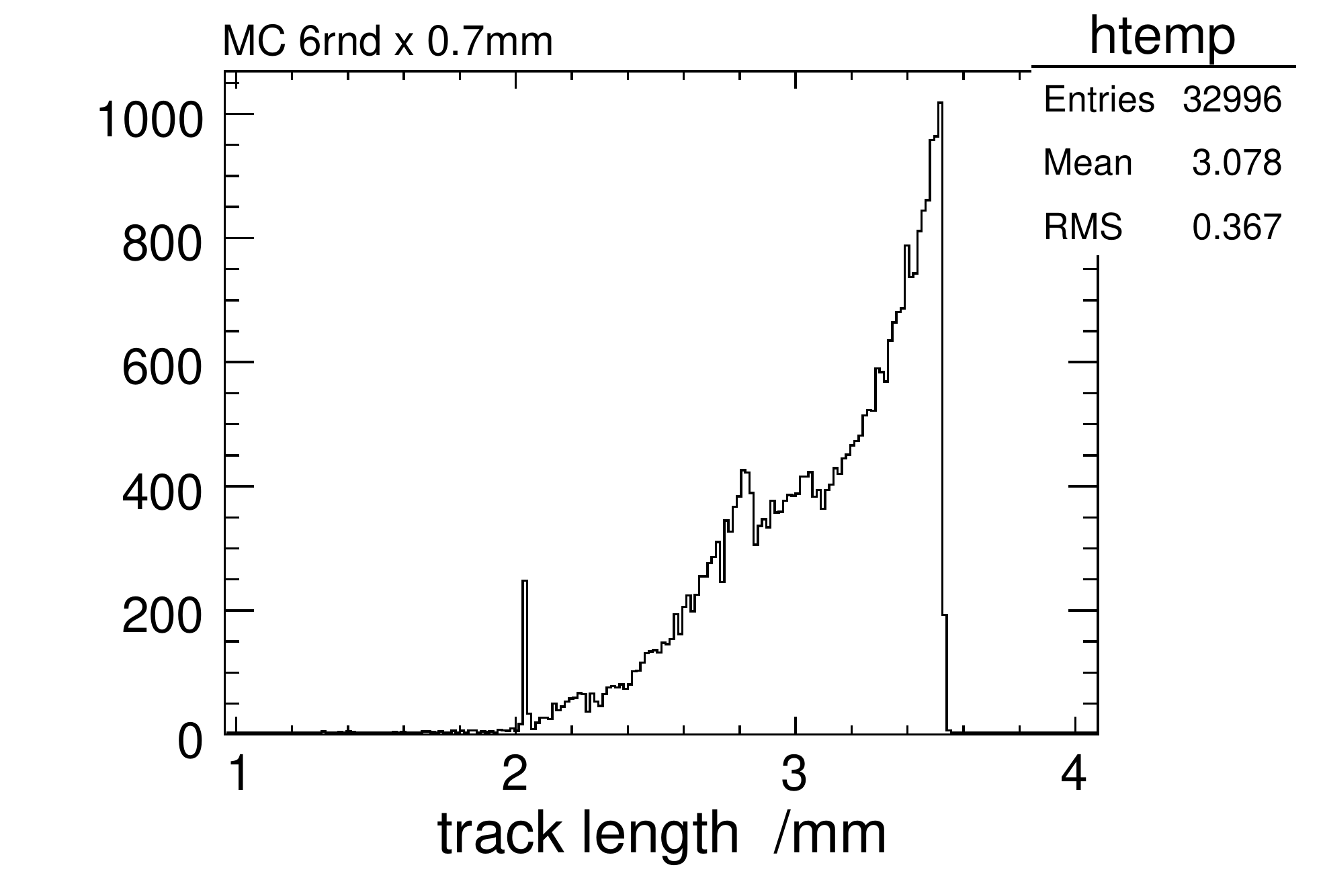} 
\caption{\textbf{6rnd700} (left) A schematic of the fibre layout in Monte Carlo. The 6 shaded fibres are  a single channel. (right) The total track length in the fibre detector.}
\label{fig:tracktotal_6rnd700}
\end{figure}

\begin{figure}[htbp]
\centering
\includegraphics[angle=90,width=0.49\textwidth]{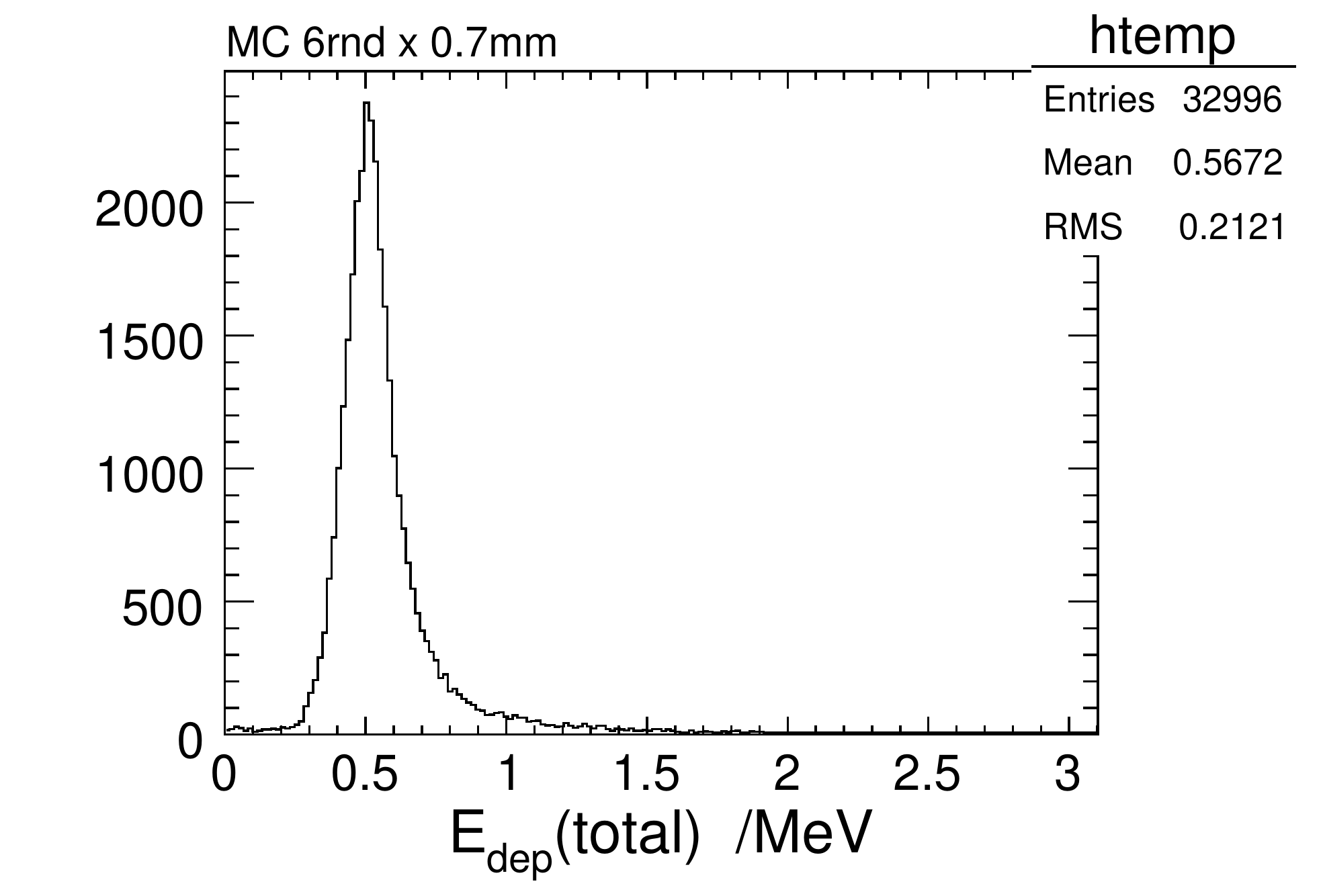}
\includegraphics[angle=90,width=0.49\textwidth]{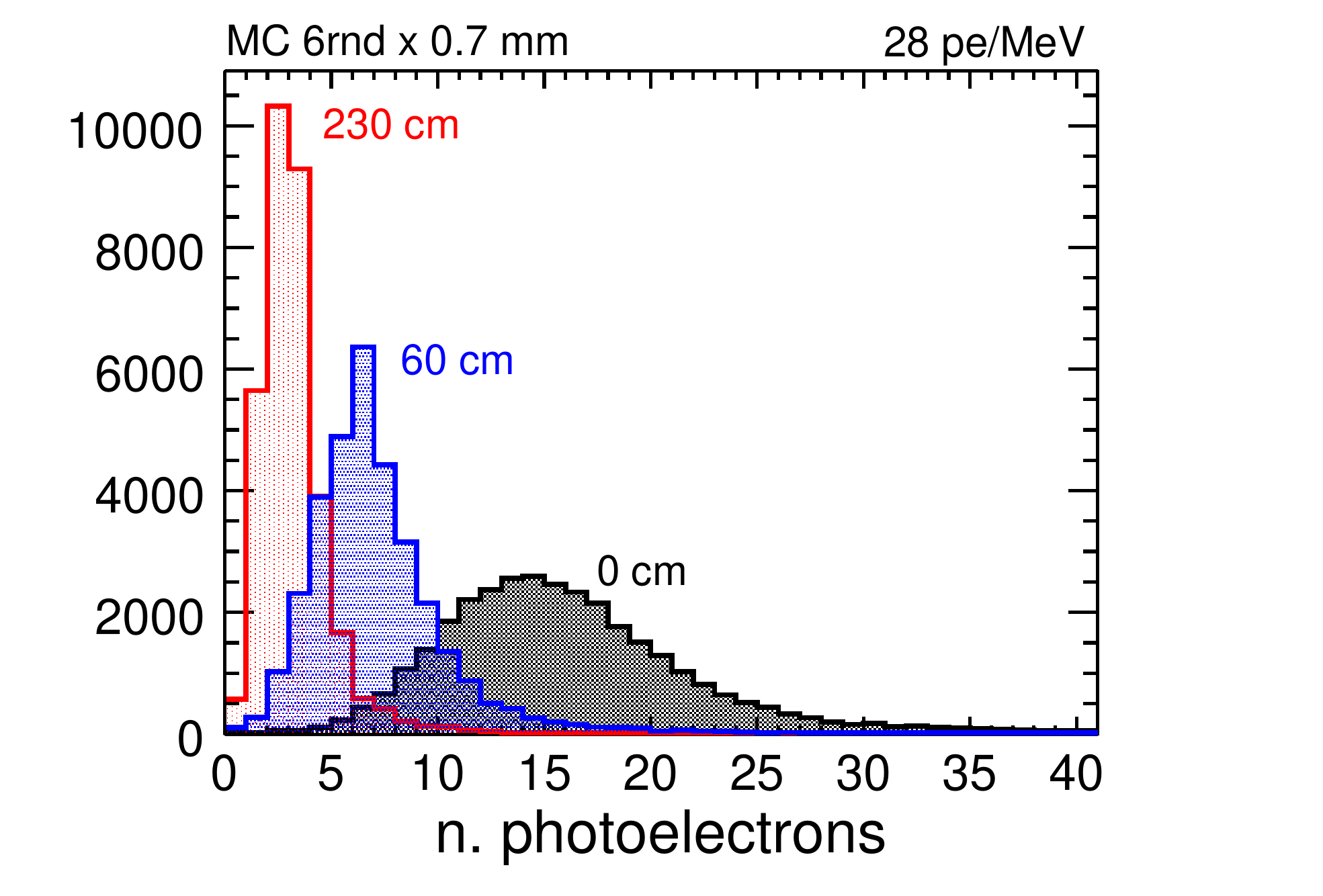}

\caption{\textbf{6rnd700} (left) The total energy deposited in the fibre tracker. (right) The photoelectron distribution as seen at the source (0~cm), at 60~cm and 230~cm. }
\label{fig:edep_6rnd700}
\end{figure}

\begin{figure}[htbp]
\centering
\includegraphics[angle=90,width=0.49\textwidth]{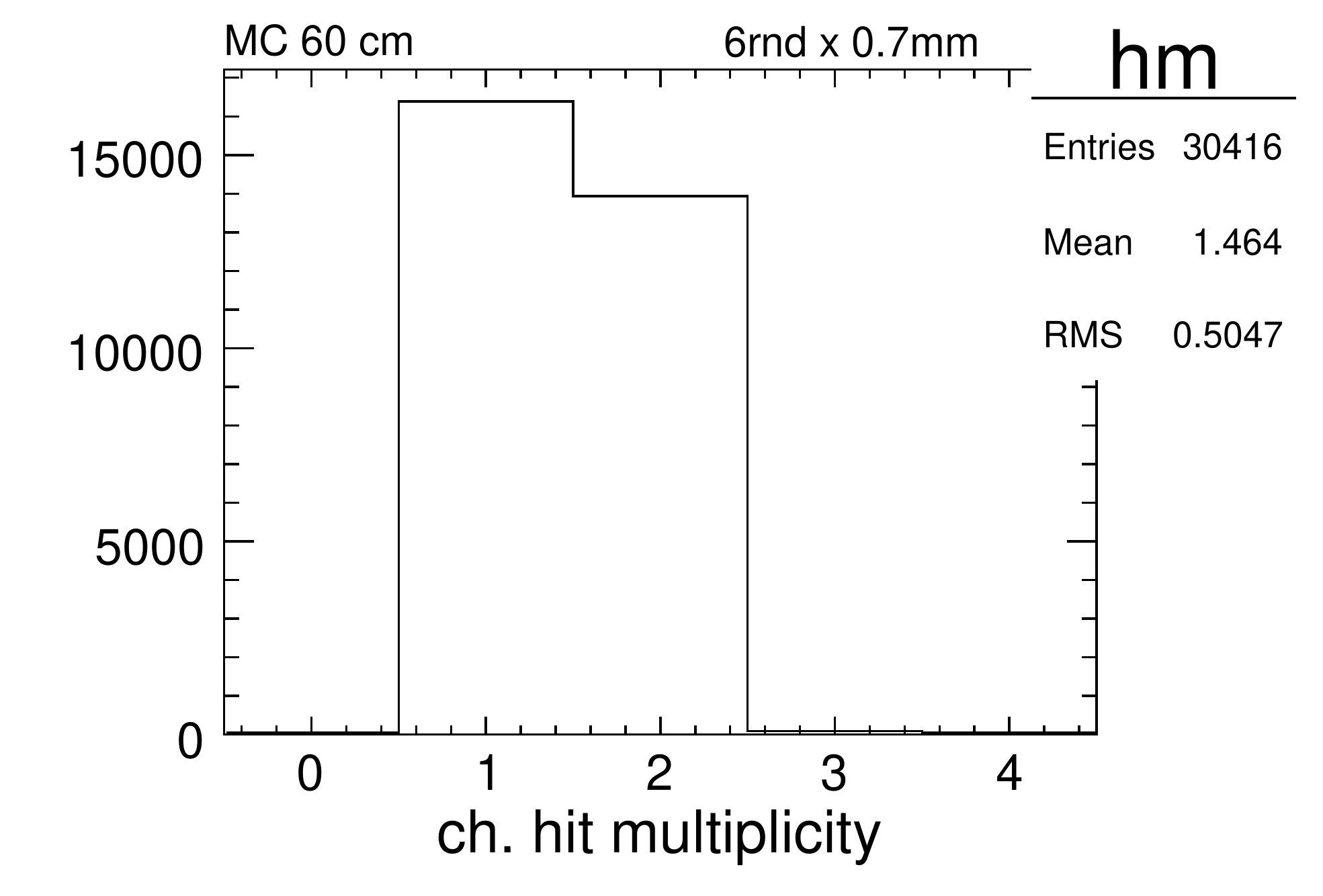}
\includegraphics[angle=90,width=0.49\textwidth]{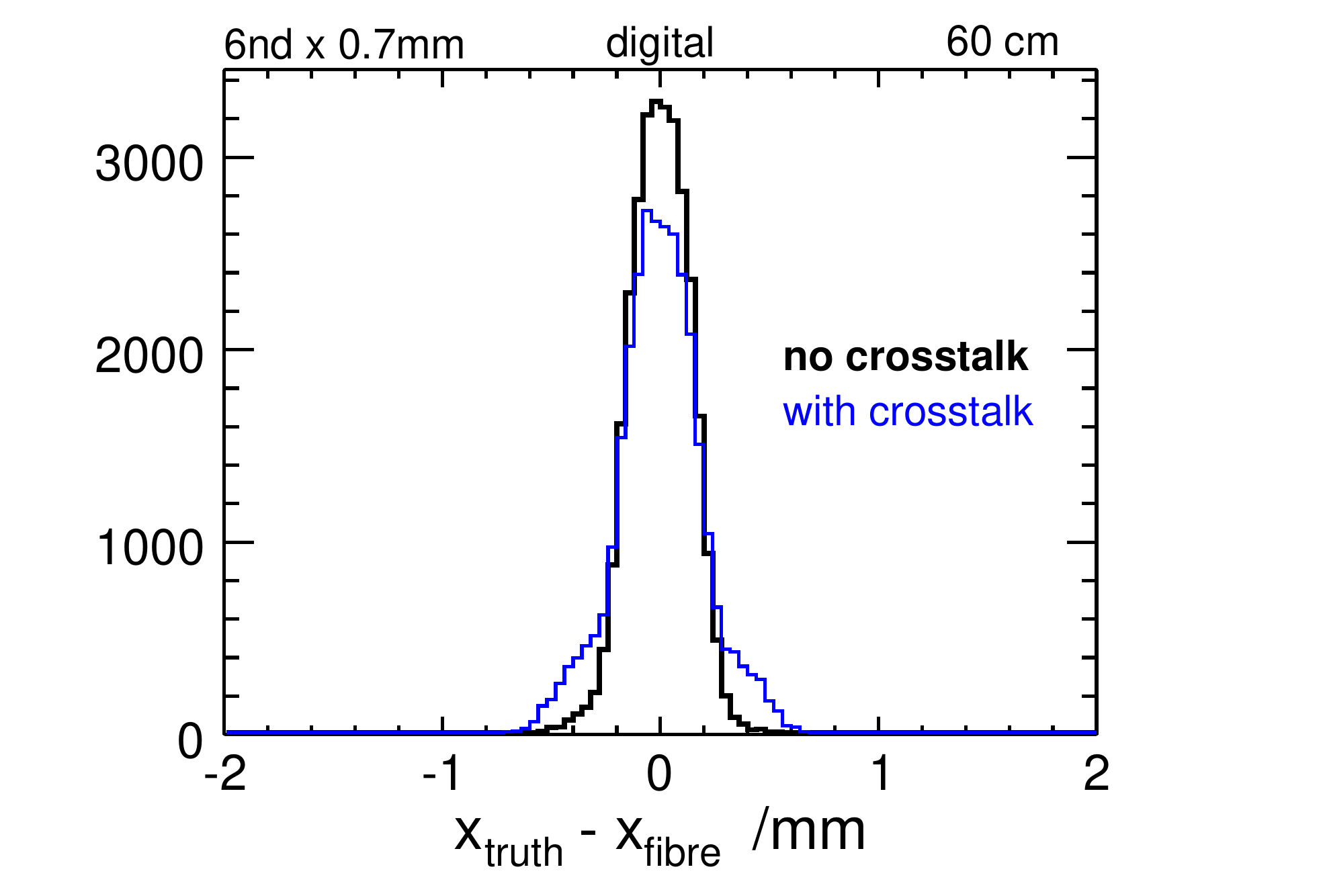}

\caption{\textbf{6rnd700} (left) Channel hit multiplicity at 60~cm with no crosstalk. (right) The expected resolution at 60 cm with no crosstalk (black) and if similar crosstalk to experiment is included (blue). }
\label{fig:mult_6rnd700}
\end{figure}

With respect to the design of Section~\ref{sec:6rnd1mm}, the same fibre packing was used but the fibres have been replaces with 0.7~mm diameter \textit{double} clad fibres, producing 28 photoelectrons per MeV (see Fig.~\ref{fig:tracktotal_6rnd700}). The path length of the tracks through the six layers fibres is shown in Fig.~\ref{fig:tracktotal_6rnd700}(right), is similar to the four layers of 1~mm fibres. The energy deposited in the fibres in this configuration along with the expected number of photoelectrons are shown in Fig.~\ref{fig:edep_6rnd700}. With the same path length as Sec.~\ref{sec:4rnd1mm}, slightly more photoelectrons are seen than in  Sec.~\ref{sec:4rnd1mm}.  The hit multiplicity, shown in Fig.~\ref{fig:mult_6rnd700}, indicates that the channels are sharing hits similar to the previous six layer design in Sec.~\ref{sec:6rnd1mm}.  A summary of the resolution and efficiency are shown in Table.~\ref{tab:summary6rnd700} for this fibre tracker configuration. The resolution can be seen to be significantly improved with the increase in light yield and smaller channel width, compared to the 1~mm single clad fibre trackers in the previous sections.

\begin{table}[h!b]
\centering
\caption{Summary of the six layer, 0.7~mm diameter, double clad fibre tracker resolutions and efficiencies. }
\begin{tabular}{|l|c|c|c|c|}
\hline 6rnd x 0.7mm & 60~cm & 90~cm & 150~cm & 230cm \\ 
\hline resolution (digital) /mm & 0.122 & 0.123 & 0.130 & 0.130 \\ 
\hline resolution (analog) /mm & 0.131 & 0.131 & 0.138 & 0.138 \\ 
\hline efficiency(2.7~mm)  & 0.981 &0.977  & 0.966 & 0.951 \\ 
\hline 
\end{tabular} 
\label{tab:summary6rnd700}
\end{table}

\pagebreak

\subsection{4 layers of square 1~mm fibres (analog)}

\begin{figure}[htbp]
\centering

\includegraphics[width=0.49\textwidth]{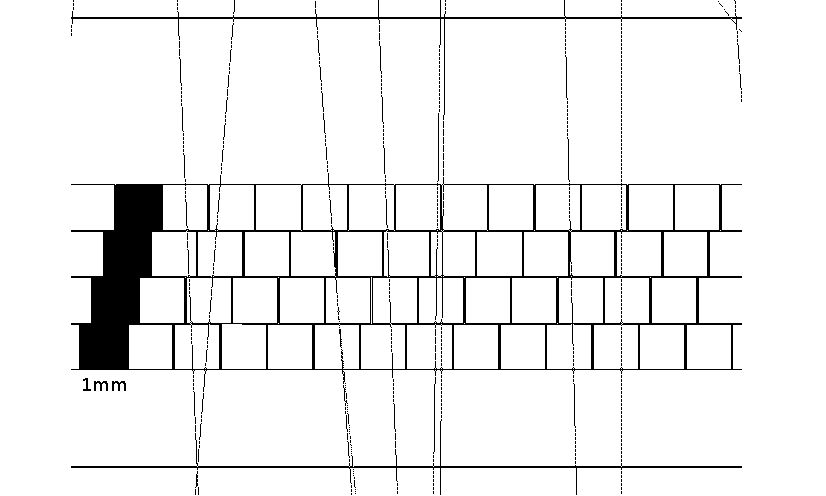} 
\includegraphics[angle=90,width=0.49\textwidth]{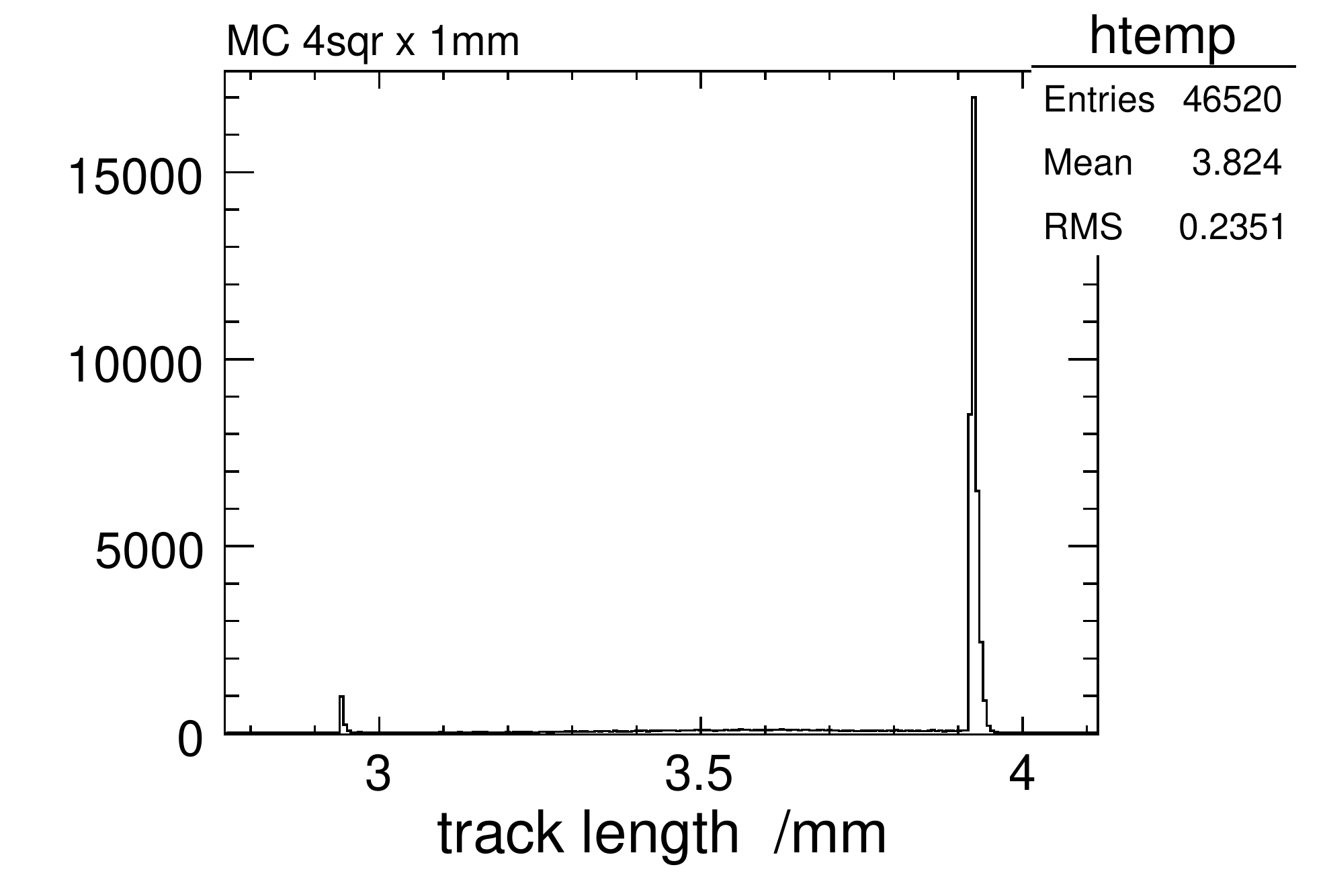} 
\caption{\textbf{4sqr1mm} (left) A schematic of the fibre layout in Monte Carlo. The 4 shaded fibres are  a single channel. (right) The total track length in the fibre detector.}
\label{fig:tracktotal_4sqr}
\end{figure}

\begin{figure}[htbp]
\centering
\includegraphics[angle=90,width=0.49\textwidth]{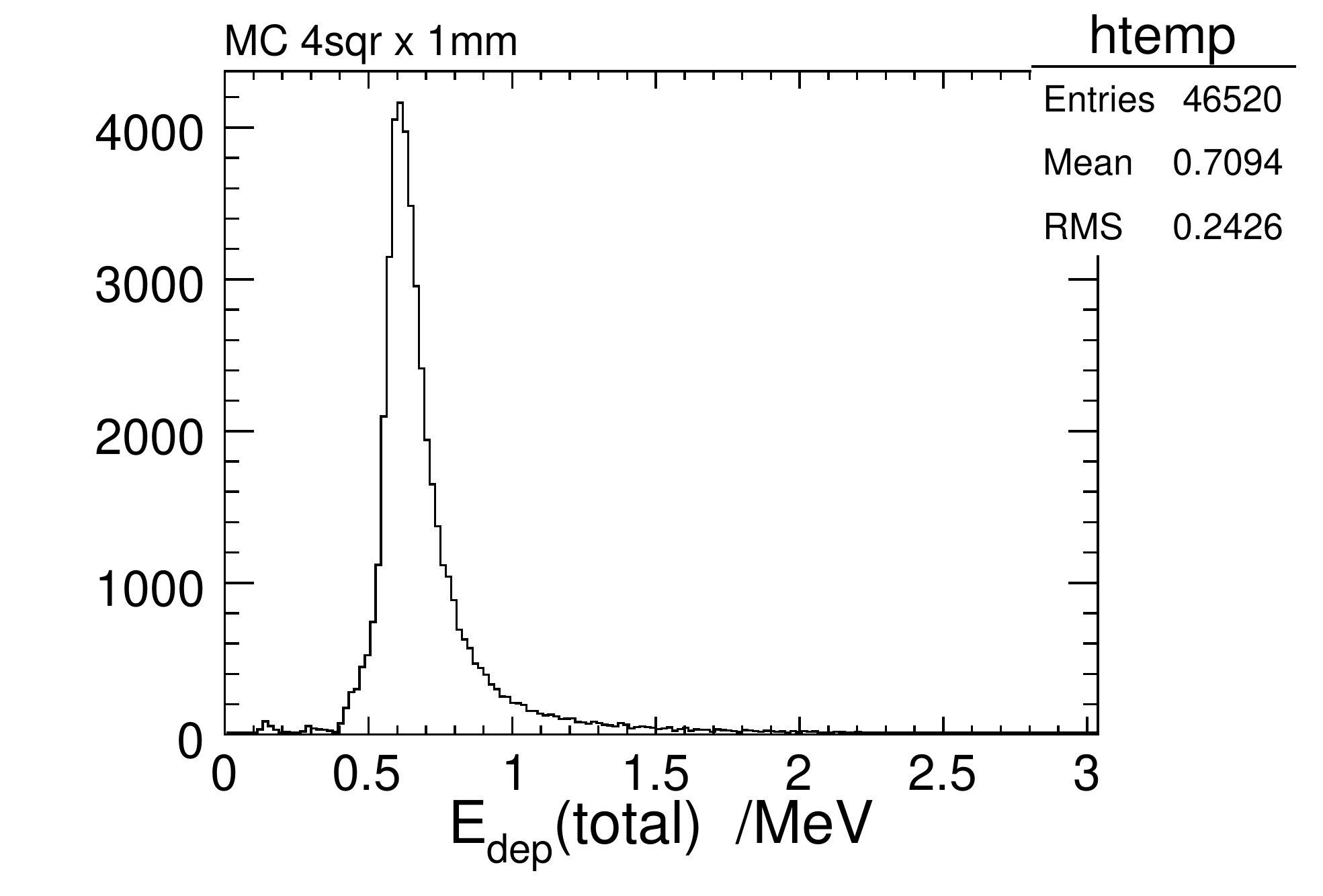}
\includegraphics[angle=90,width=0.49\textwidth]{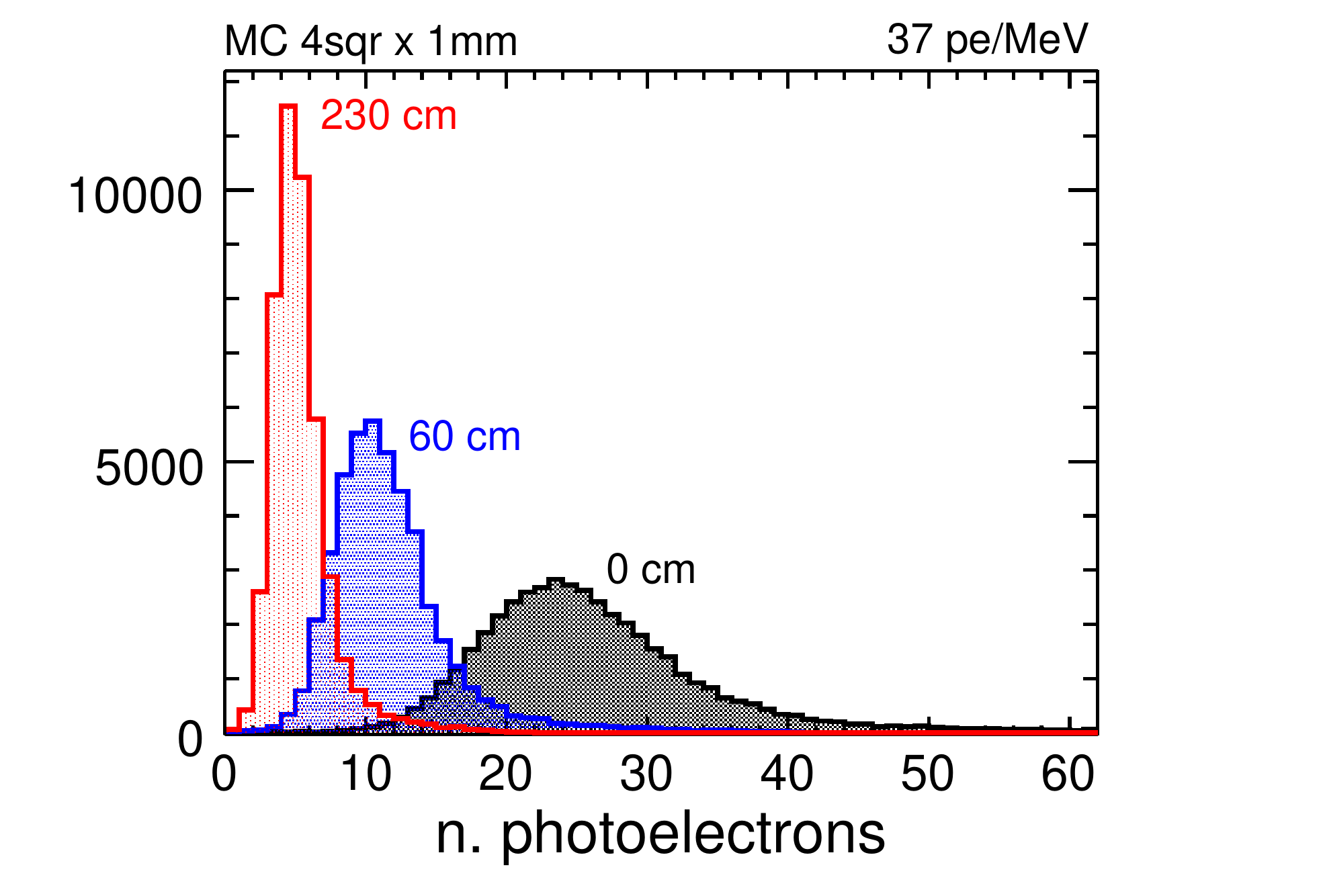}

\caption{\textbf{4sqr1mm} (left) The total energy deposited in the fibre tracker. (right) The photoelectron distribution as seen at the source (0~cm), at 60~cm and 230~cm. }
\label{fig:edep_4sqr}
\end{figure}

\begin{figure}[htbp]
\centering
\includegraphics[angle=90,width=0.49\textwidth]{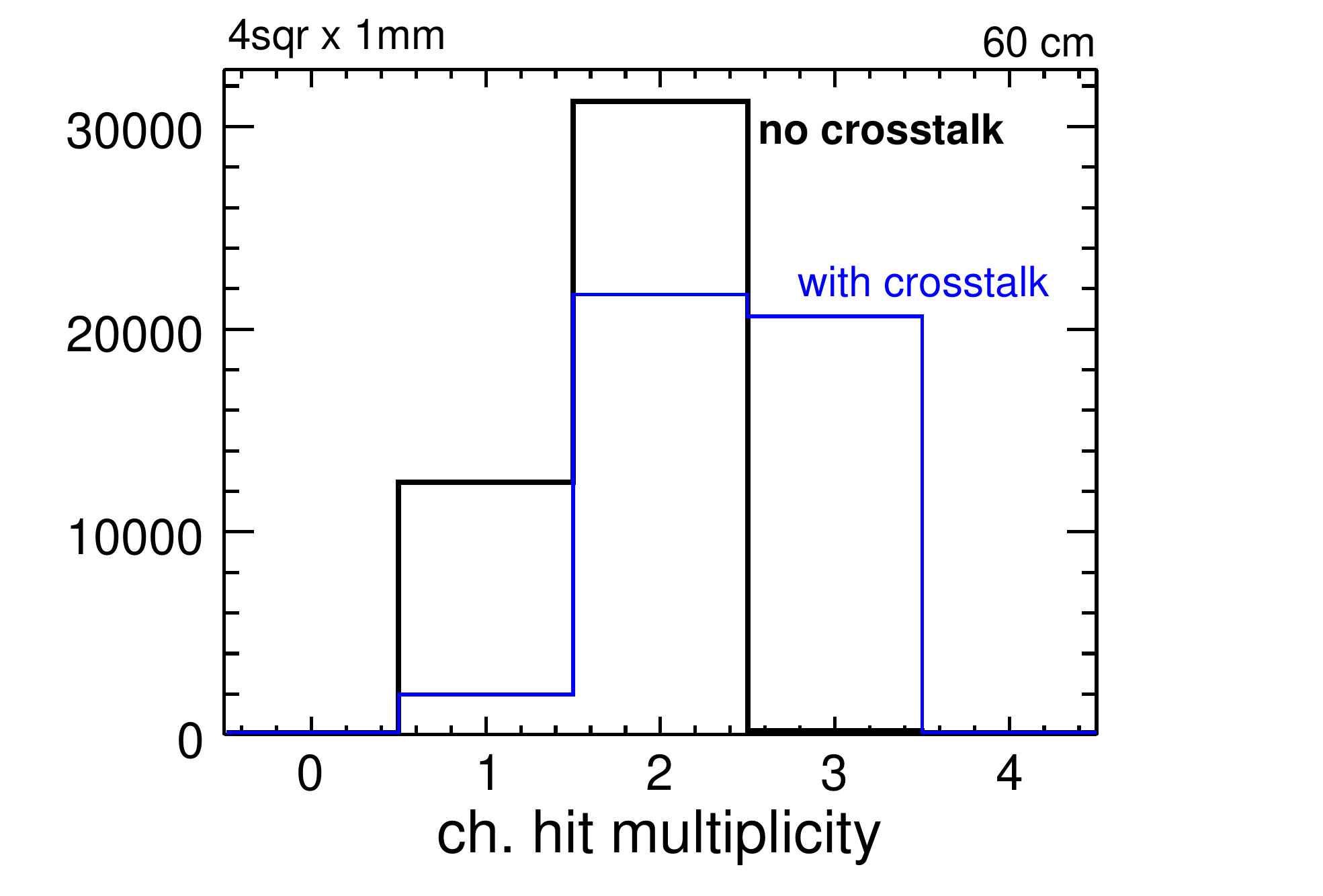}
\includegraphics[angle=90,width=0.49\textwidth]{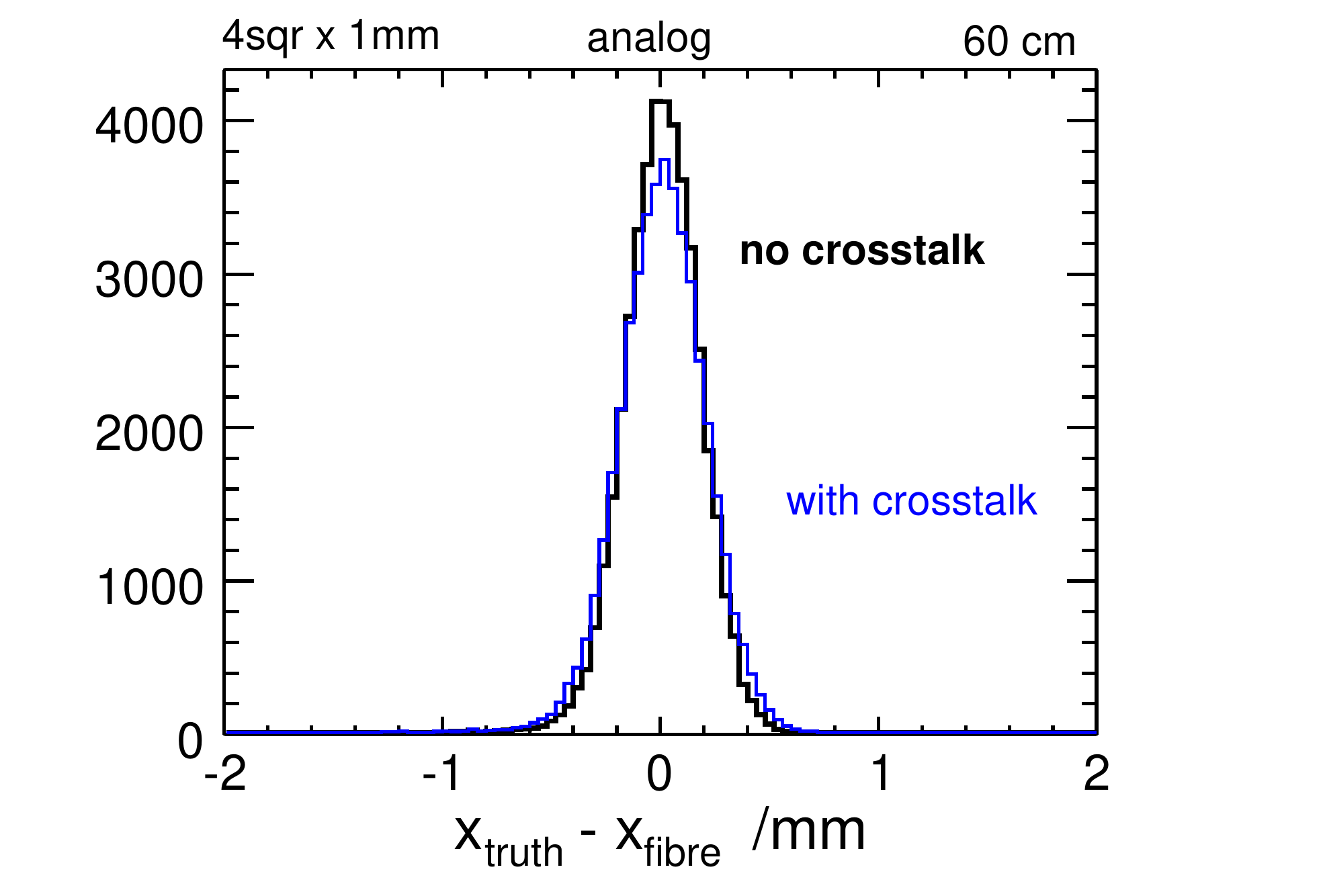}

\caption{\textbf{4sqr1mm} (left) Channel hit multiplicity with no crosstalk. (right) The expected resolution at 60 cm with no crosstalk (black) and if similar crosstalk to experiment is included (blue). }
\label{fig:mult_4sqr}
\end{figure}

This configuration is meant to take advantage of the analog signal readout. All other proposed designs are meant for digital readout. The double-clad, square, 1~mm fibres are arranged such that distance of the track from the channel center should be correlated to the amount of scintillator crossed in that channel's fibres. The maximum signal should be at the channel centre, while an equal distance between adjacent channels should produce equal signals in each channel. The resolution of this configuration will depend strongly on the resolution of the photodetector as well a photoelectron statistics, such that the relative signals in each channel can be measured accurately to determine an accurate position. The benefits of the square fibres are shown in the narrow track length (see Fig.~\ref{fig:tracktotal_4sqr}(right)) and energy deposited (Fig.~\ref{fig:edep_4sqr}(left)). As well, square fibres have a higher light trapping efficiency of 7.3\%.  Particle tracks in this design will produce a consistent signal independent of the part of the tracker they encounter, unlike the round fibre designs. The multiplicity of channels hit per track, shown in (Fig.~\ref{fig:mult_4sqr}(left)), is dominated by two channels, due to the better overlap of channels than the other design, such that the charge sharing with the overlap can be taken advantage of in producing an good analog resolution. One can also see that the crosstalk effect on the resolution in Fig.~\ref{fig:mult_4sqr}(right) is minimal with an analog measurement and large light yield due to the low amplitude of the crosstalk  relative to the photoelectron yield.  The efficiency here is also the highest of the four layer designs. A summary of the analog resolution and efficiency is shown in Table.~\ref{tab:4sqr}.

\begin{table}[h!b]
\centering
\caption{Summary}
\begin{tabular}{|l|c|c|c|c|}
\hline 4sqr x 1mm & 60~cm & 90~cm & 150~cm & 230cm \\ 
\hline resolution (ana) /mm  & 0.170  & 0.174 & 0.182 & 0.182 \\ 
\hline efficiency(2.7~mm)  & 0.985 &0.982  &0.974 & 0.960 \\ 
\hline 
\end{tabular} 
\label{tab:4sqr}
\end{table}

\pagebreak
\subsection{4 layers of square 0.7~mm fibres (digital)}

\begin{figure}[htbp]
\centering

\includegraphics[width=0.49\textwidth]{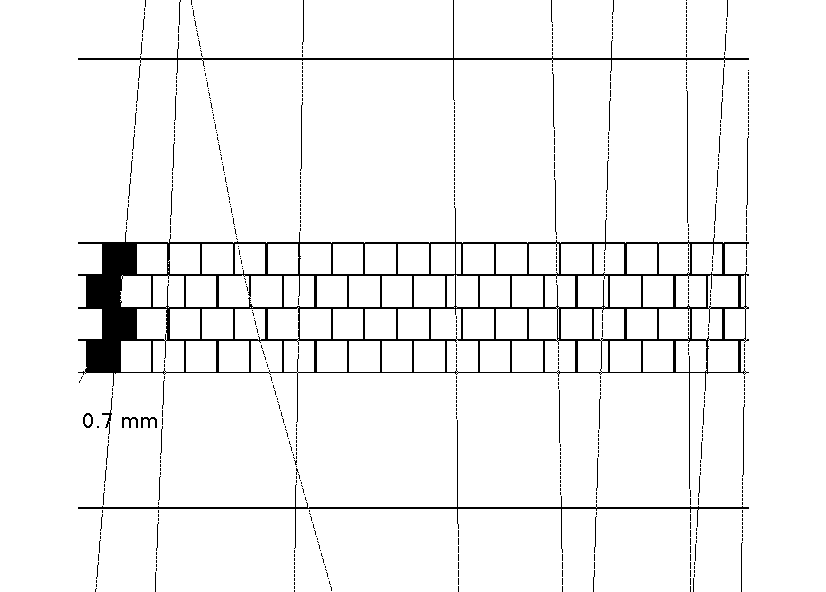} 
\includegraphics[angle=90,width=0.49\textwidth]{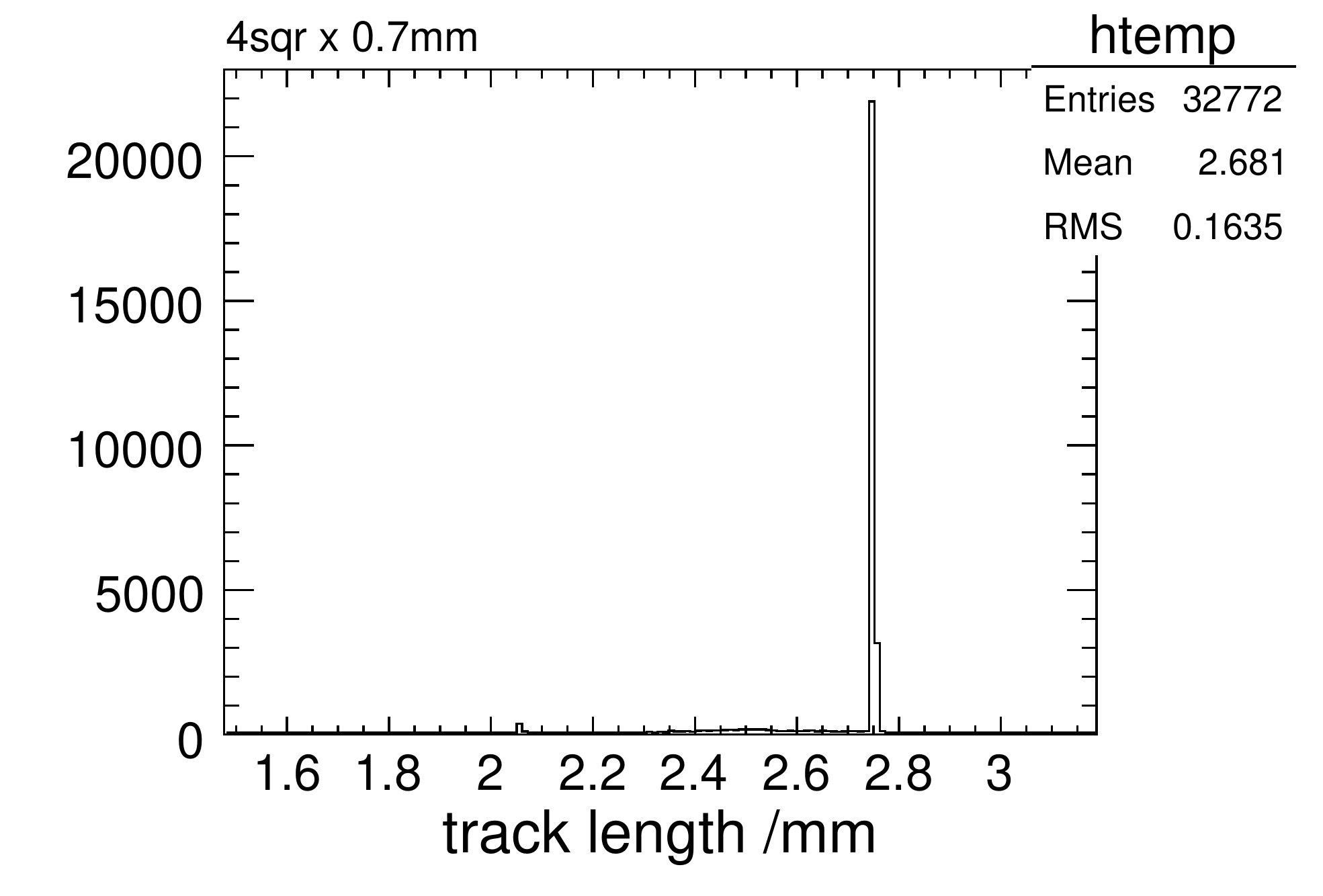} 
\caption{\textbf{4sqr700} (left) A schematic of the fibre layout in Monte Carlo. The 4 shaded fibres are  a single channel.(right) The total track length in the fibre detector.}
\label{fig:tracktotal_4sqr700}
\end{figure}

\begin{figure}[htbp]
\centering
\includegraphics[angle=90,width=0.49\textwidth]{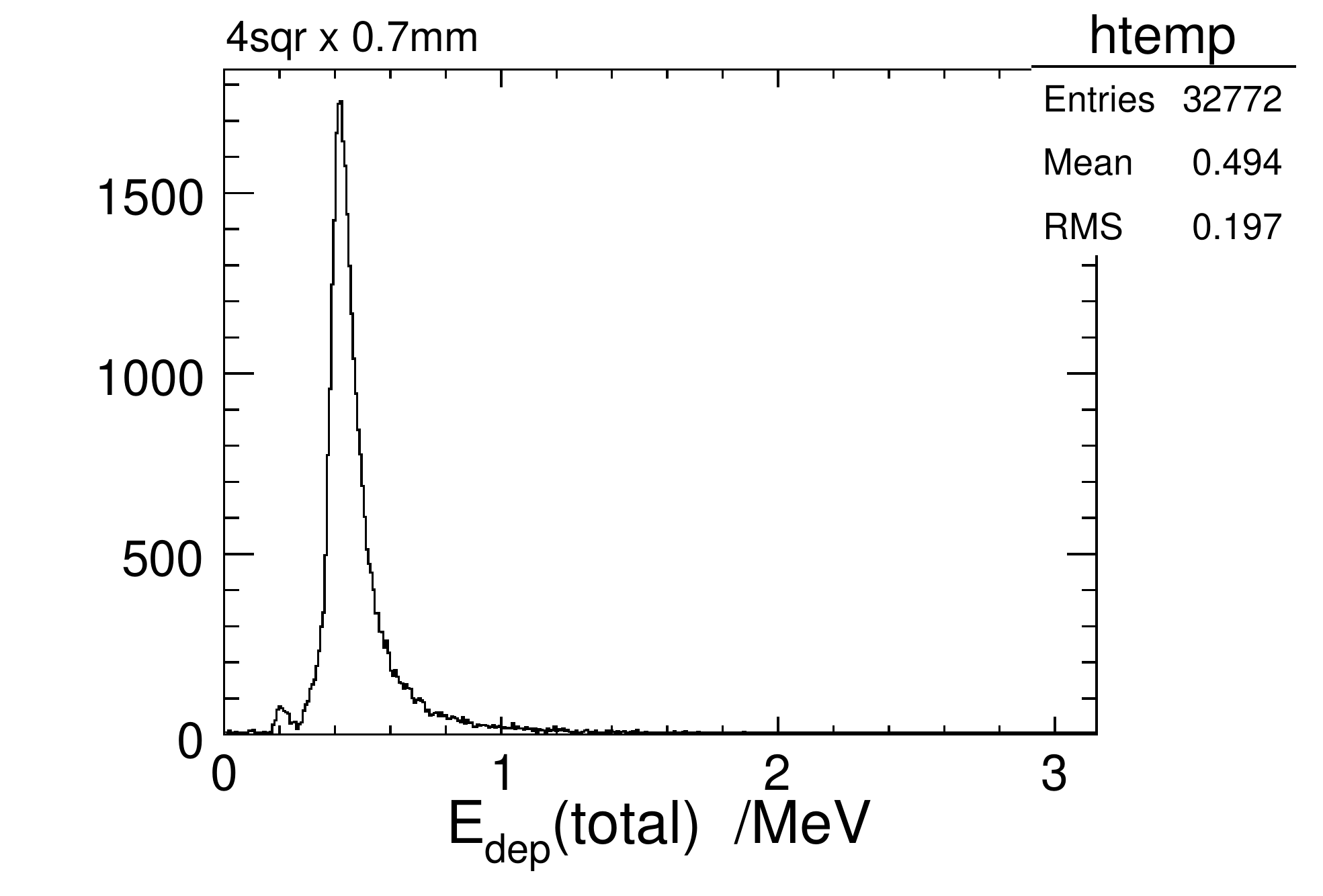}
\includegraphics[angle=90,width=0.49\textwidth]{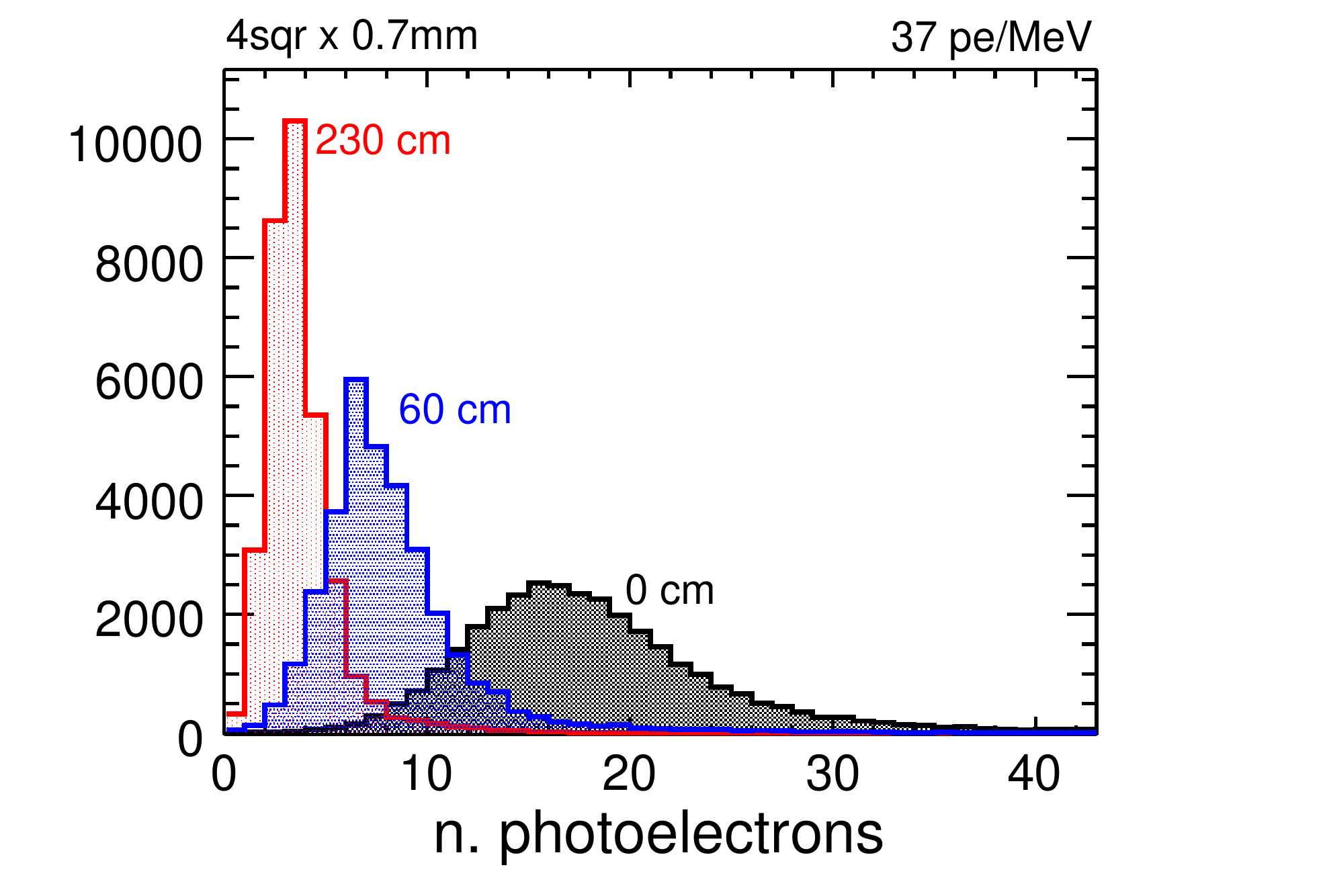}

\caption{\textbf{4sqr700} (left) The total energy deposited in the fibre tracker. (right) The photoelectron distribution as seen at the source (0~cm), at 60~cm and 230~cm. }
\label{fig:edep_4sqr700}
\end{figure}

\begin{figure}[htbp]
\centering
\includegraphics[angle=90,width=0.49\textwidth]{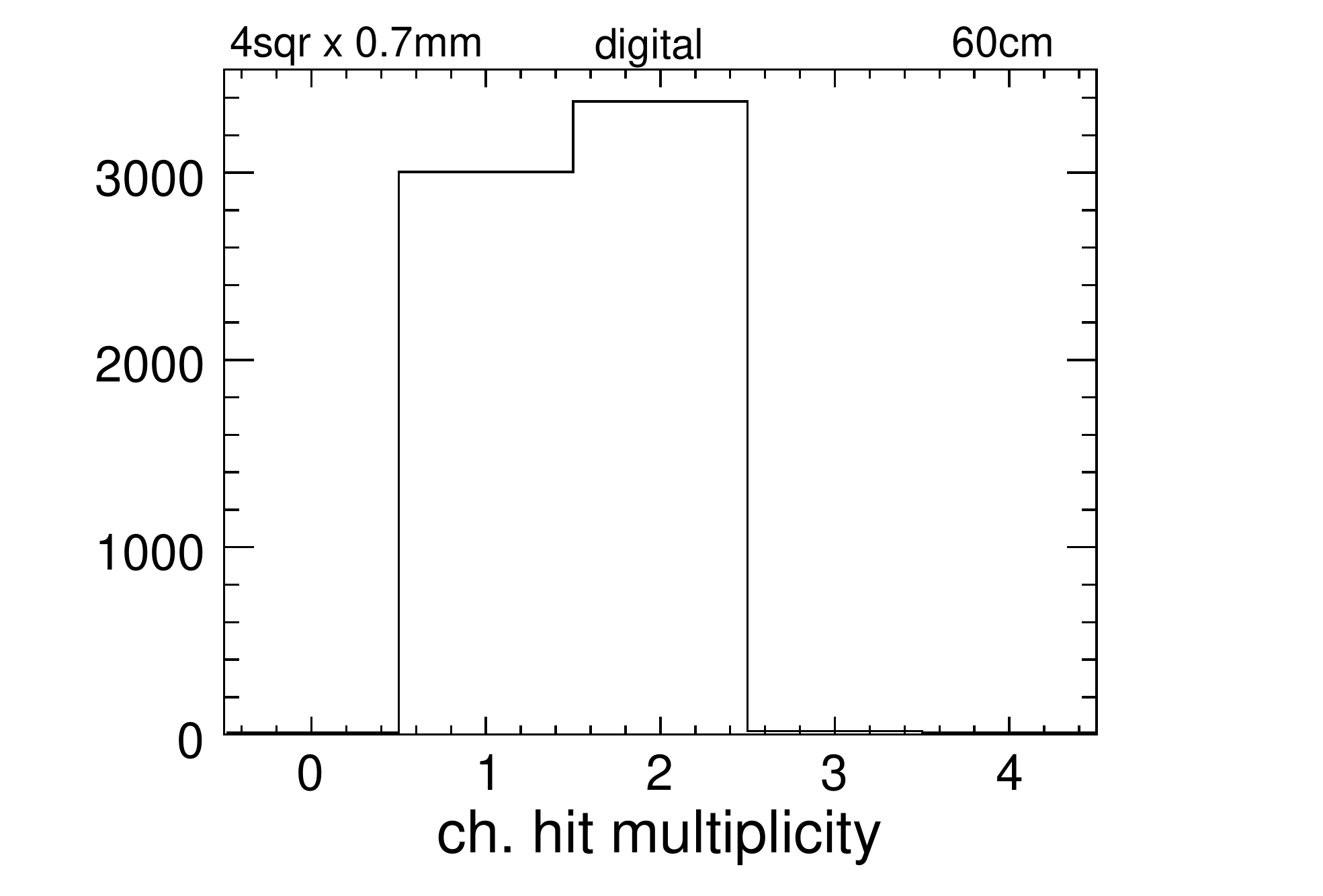}
\includegraphics[angle=90,width=0.49\textwidth]{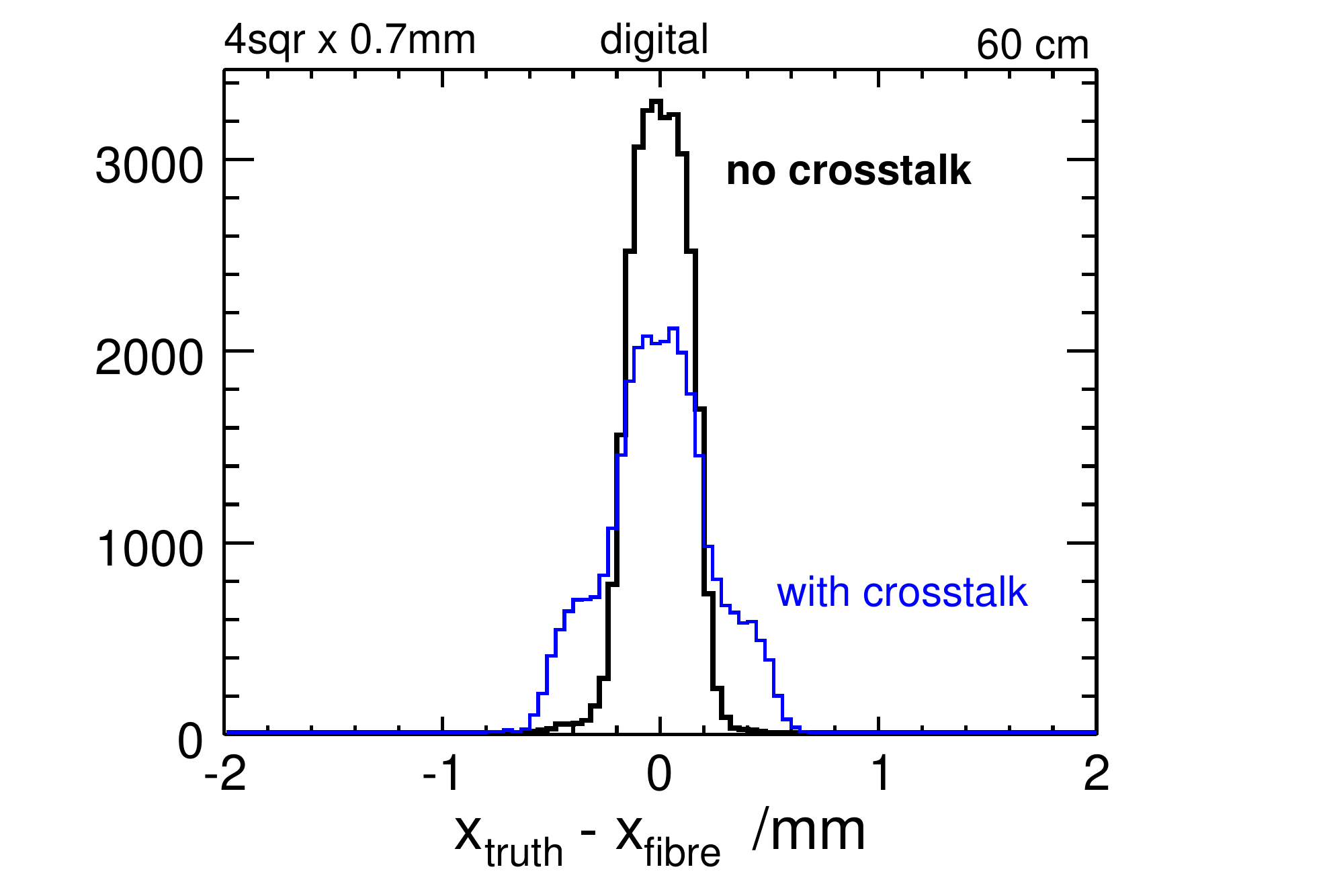}

\caption{\textbf{4sqr700} (left) Channel hit multiplicity at 60~cm with no crosstalk. (right) The expected resolution at 60 cm with no crosstalk (black) and if similar crosstalk to experiment is included (blue). }
\label{fig:mult_4sqr700}
\end{figure}

This configuration is comparable to the 0.7~mm round, double clad fibres. However, square fibres have a higher light trapping efficiency of 7.3\%, producing 37 photoelectrons per MeV. The fibres are packed as close together as possible, with 1 fibre diameter separating channel centres, with the fibre pitches such that each channel overlaps the adjacent one by half a fibre diameter, as in Fig.~\ref{fig:tracktotal_4sqr700}. The benefits of the square fibres are shown in the narrow track length (see Fig.~\ref{fig:tracktotal_4sqr700}(right)) and energy deposited (Fig.~\ref{fig:edep_4sqr700}(left)). Particle tracks in this design will produce a consistent signal independent of the part of the tracker they encounter, unlike the round fibre designs. The multiplicity of channels hit per track, shown in (Fig.~\ref{fig:mult_4sqr700}(left)), is split evenly between single and two channel hits, due to the overlap of channels. One can also see that the crosstalk has a large effect on the resolution in Fig.~\ref{fig:mult_4sqr700}(right) unless one rearranges the channels on the MA-PMT.  The efficiency here is also the highest of the four layer designs. A summary of the analog resolution and efficiency is shown in Table.~\ref{tab:summary4sqr700}.

\begin{table}[h!b]
\centering
\caption{Summary of resolutions and efficiency for 4 layers of square 0.7~mm fibres with 0.7~mm horizontal channel spacing }
\begin{tabular}{|l|c|c|c|c|}
\hline 4sqr x 0.7~mm & 60~cm & 90~cm & 150~cm & 230cm \\ 
\hline resolution (dig) /mm & 0.118 & 0.118 & 0.119 & 0.119 \\ 
\hline resolution (ana) /mm & 0.132 & 0.135 & 0.140 & 0.140 \\ 

\hline efficiency(2.7~mm) &  0.976 & 0.972 &0.959  & 0.941 \\ 
\hline 
\end{tabular} 
\label{tab:summary4sqr700}
\end{table}

\section{Conclusion}
\begin{figure}[htbp]
\centering
\includegraphics[angle=270,width=0.9\textwidth]{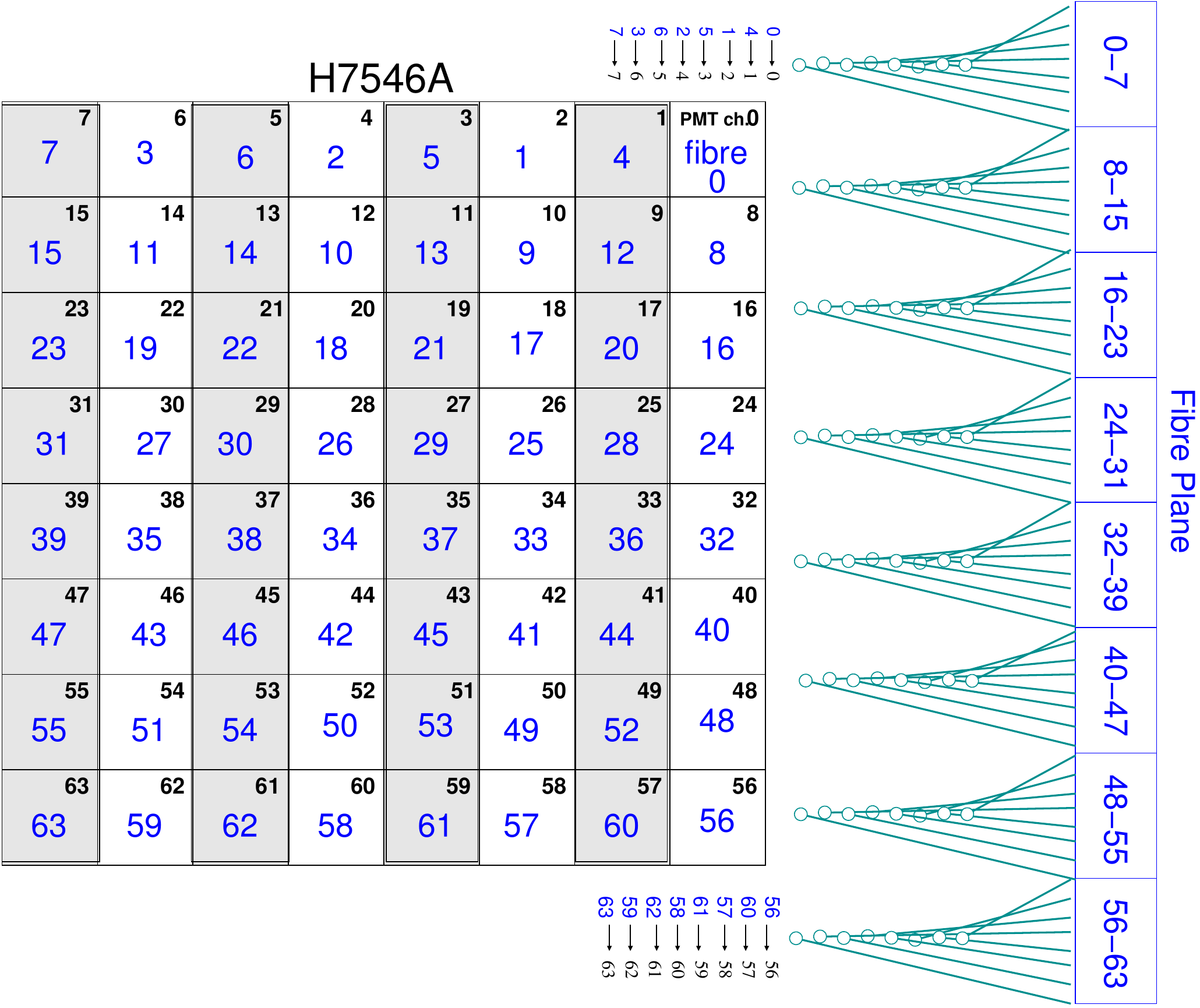} 
\caption{A proposed fibre channel to MA-PMT channel mapping such that adjacent MA-PMT channels will not crosstalk into adjacent fibre channels. }
\label{fig:newmapping}
\end{figure}

This work has illuminated several areas where a future fibre tracker of similar design should be improved over the current prototype design such that a position resolution of less than 0.200~mm can be achieved. A summary of new design resolutions and efficiencies is shown in Table~\ref{tab:summary4rnd}. The conclusions from this work are as follows:
 
\begin{itemize}
\item  A spectrum in the green wavelength (\textgreater{}500~nm) would be less affected radiation damage in the polystyrene and have a longer bulk attenuation length. The shorter blue wavelengths in the fibres also attenuate much faster than the longer wavelength and are affected more by radiation damage.
\item The single clad round fibres produced approximately 10 photoelectrons in 3.5~mm (18 pe/MeV) of fibre at 0~cm , as seen by the H8500 MA-PMT (22\% QE). Round, double clad fibres as readout by a super-bialkali H7456-100 (-200) MA-PMT with 34\% (40\%)  QE would produce 23 (28) pe at 0~cm for a similar fibre arrangement, or 44 (52) pe/MeV.
\item A smaller channel width improves the position resolution in both digital and analog readout modes. A 0.7~mm diameter fibre would be sufficient to reach the position resolution goal of less than 0.2~mm.
\item The variation in energy deposit should be minimized by close packing of the fibres. Squares fibres are better than round fibres for this. 
\item A channel overlap of 1/2 fibre width provides no more information in an analog readout mode than a digital mode, a result of the limited single photoelectron resolution of any proposed bi-alkali photodetector.
\item Adjacent fibre channels must not be placed near to each other on the photodetector. Crosstalk is the largest limiting factor in a digital readout. A better fibre channel to MA-PMT channel mapping is shown in Fig.~\ref{fig:newmapping}. Optical crosstalk at the track point may also present a problem. Only fibres with deposited energy should contribute to a position hit cluster.
\item The single photoelectron efficiency and resolution  must be better than that of the H8500/9500 ($\sigma_{1pe}/G_{1pe} = 0.6$) for an analog readout to be more effective than any digital system. SiPMs will typically meet this requirement  ($\sigma_{1pe}/G_{1pe} = 0.2$) but the noise may be a problem.
\item The H7456 family of MA-PMTs appears promising with its lower crosstalk rate and higher quantum efficiency for improving the tracker efficiency and resolution. An H7456-100 and -200 will be used in upcoming tests.  
\end{itemize}

\begin{table}[h!tbp]
\centering
\caption{Summary of all proposed designs with efficiency, $\epsilon_{2.7}$, and operating resolutions, $\sigma$, over four trigger positions. Double-clad fibre designs are indicated by (dc) or single-clad (sc).}
\begin{tabular}{|r|c c|c c|c c|c c|}
\hline & \multicolumn{2}{c|}{60~cm} & \multicolumn{2}{c|}{90~cm} & \multicolumn{2}{c|}{150~cm} & \multicolumn{2}{c|}{230~cm} \\ 
\hline design & $\sigma$/mm & $\epsilon_{2.7}$& $\sigma$/mm & $\epsilon_{2.7}$& $\sigma$/mm & $\epsilon_{2.7}$& $\sigma$/mm & $\epsilon_{2.7}$ \\ 
\hline 4rnd(sc)1mm(dig) &0.204&0.972&0.205 & 0.962 & 0.236 &0.936 & 0.236 &0.920 \\
\hline 6rnd(sc)1mm(dig)&0.189&0.986 & 0.189 & 0.983 & 0.202  & 0.974 & 0.203 & 0.963 \\
\hline 6rnd(dc)0.7mm(dig)& 0.122  &0.981 & 0.123 &0.977 &0.130  &0.966 &0.130 & 0.951\\ 
\hline 4sqr(dc)1mm(ana)& 0.170 &0.985 &0.174  &0.982 &0.182  &0.974 & 0.182 &0.960\\
\hline 4sqr(dc)0.7mm(dig)&0.118  &0.976 &0.118  &0.972 & 0.119 &0.956 &0.119  &0.941\\
\hline
\end{tabular} 
\label{tab:summary4rnd}
\end{table}

\section{Acknowledgements}

We would like to thank  Stefano Miscetti for his valuable time, effort and  suggestions when  preparing this technical report. Also, thanks to Alessandro Di Virgilio, who contributed to the
building of the fibre detector, and to Andrea Zossi, who helped with the mechanical setup of the experimental setup.

\pagebreak
\section*{Appendix A: Drift Tube Track Finding}
\addcontentsline{toc}{subsection}{Appendix A}
\subsection*{Track Reconstruction}

The measure time for the signal to reach the TDC after the trigger, $t_{meas.}$, is the sum of propagation time in the wires, the time for the trigger to be produced by the discriminator and the drift time for the first ions in the gas to reach the anode wire.
\begin{equation}
t_{meas.} = t_{delay} + t_{trig} + t_{drift}
\label{eqn:t1}
\end{equation}
Given that the delay timing and the trigger timing should be nearly constant for similar signals for a given tube, and only the drift time will vary, Eq.~\ref{eqn:t1} can be written as

\begin{equation}
t_{meas.} = t_{0} + t_{drift} \, .
\label{eqn:t2}
\end{equation}

\begin{figure}[htbp]
\centering
\includegraphics[angle=90,width=0.5\textwidth]{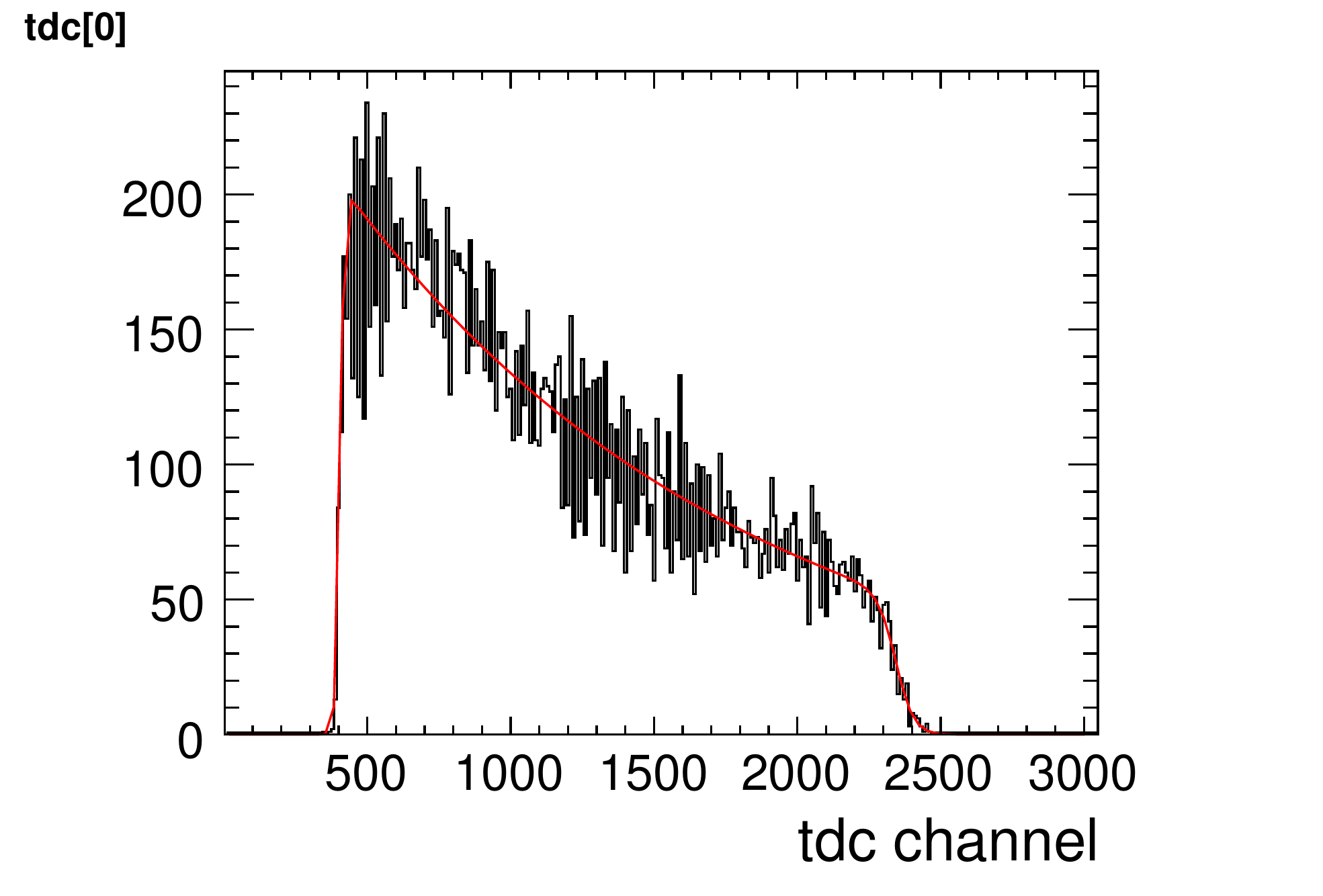}
\caption{A typical drift time distribution at nominal operating conditions. The data are fit to the function in Eq.~\ref{eqn:tdcfit}.}
\label{fig:tdc0}
\end{figure}

A typical timing signal distribution for a drift tube is shown in Fig.~\ref{fig:tdc0}. The distribution can be fitted with the emperical function~\cite{ref_bran}: 

\begin{equation}
\centering
f(t) = p0 + \frac{p1(1+p2e^{-\frac{t-p4}{p3}})}{(1+e^{\frac{-t+p4}{p6}})(1+e^{\frac{t-p5}{p7}})}
\label{eqn:tdcfit}
\end{equation}

The parameter $p0$ is the uncorrelated background, $p1$, $p2$ and $p3$ describe the shape of the central part of the distribution; $p4$ and $p5$ are the turning points of the rising and trailing edge, respectively; $p6$ and $p7$ describe the amplitudes of the rising and trailing edges. The parameter $p4$ is generally taken as the definition of $t_{0}$ in Eq.~\ref{eqn:t2}.

The difference between $p4$ and $p5$ for all eight tdc have an average value of 1927 channels, but this was seen to change over different data sets, most likely due to a change in gas temperature. A simple calibration of the tdc showed a slope of 5.36$\pm$0.15 ch/ns. This means that the width of the drift tube in time is 360~ns, assuming that there are no signal threshold effects near the edges of the tube.

\begin{figure}[htbp]
	\centering
 \includegraphics[angle=90,width=0.5\textwidth]{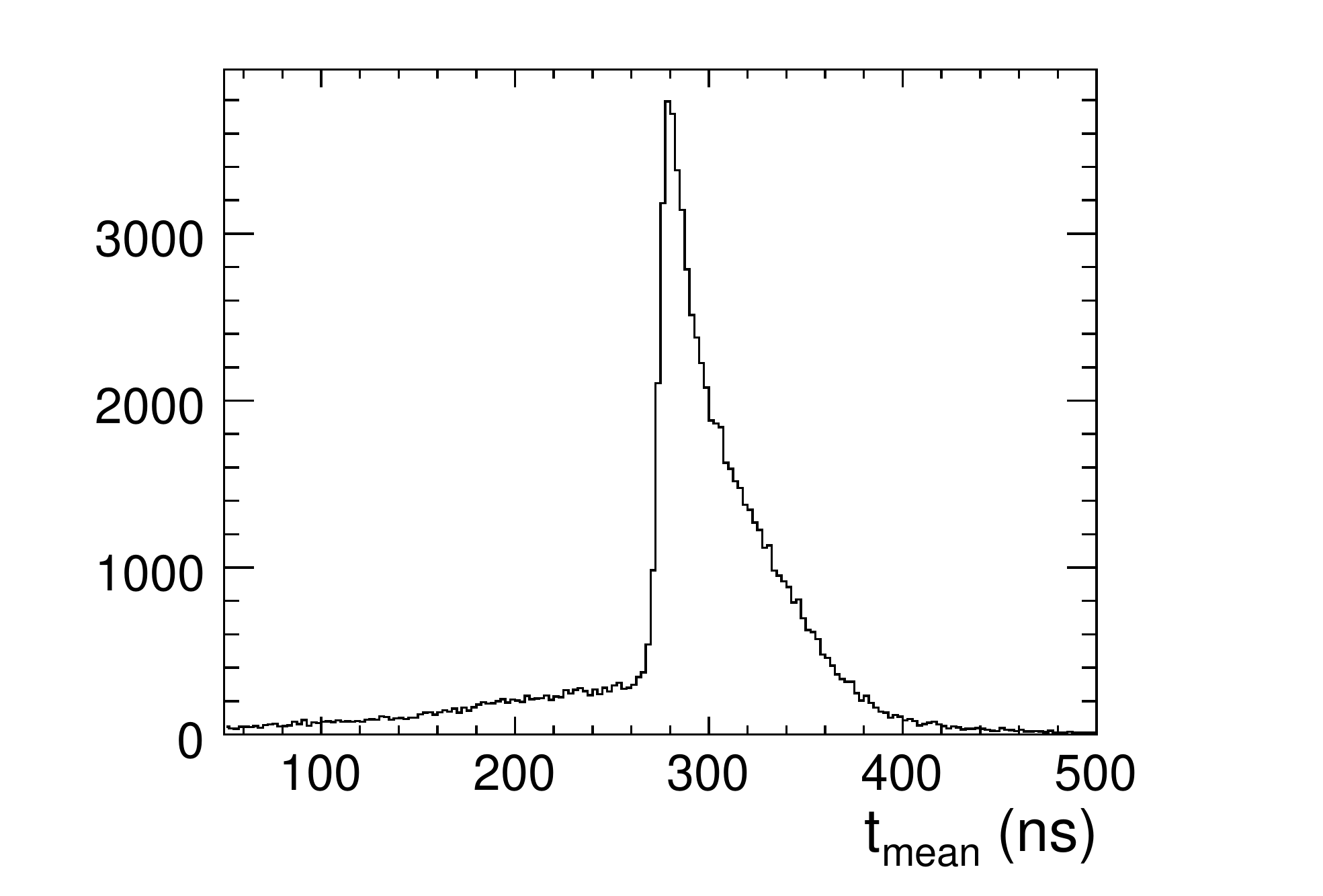}

\caption{Mean drift time of the three tube hits.}
 \label{fig:meandrift}
\end{figure}

\begin{figure}[htbp]
\centering
 \includegraphics[angle=90,width=0.5\textwidth]{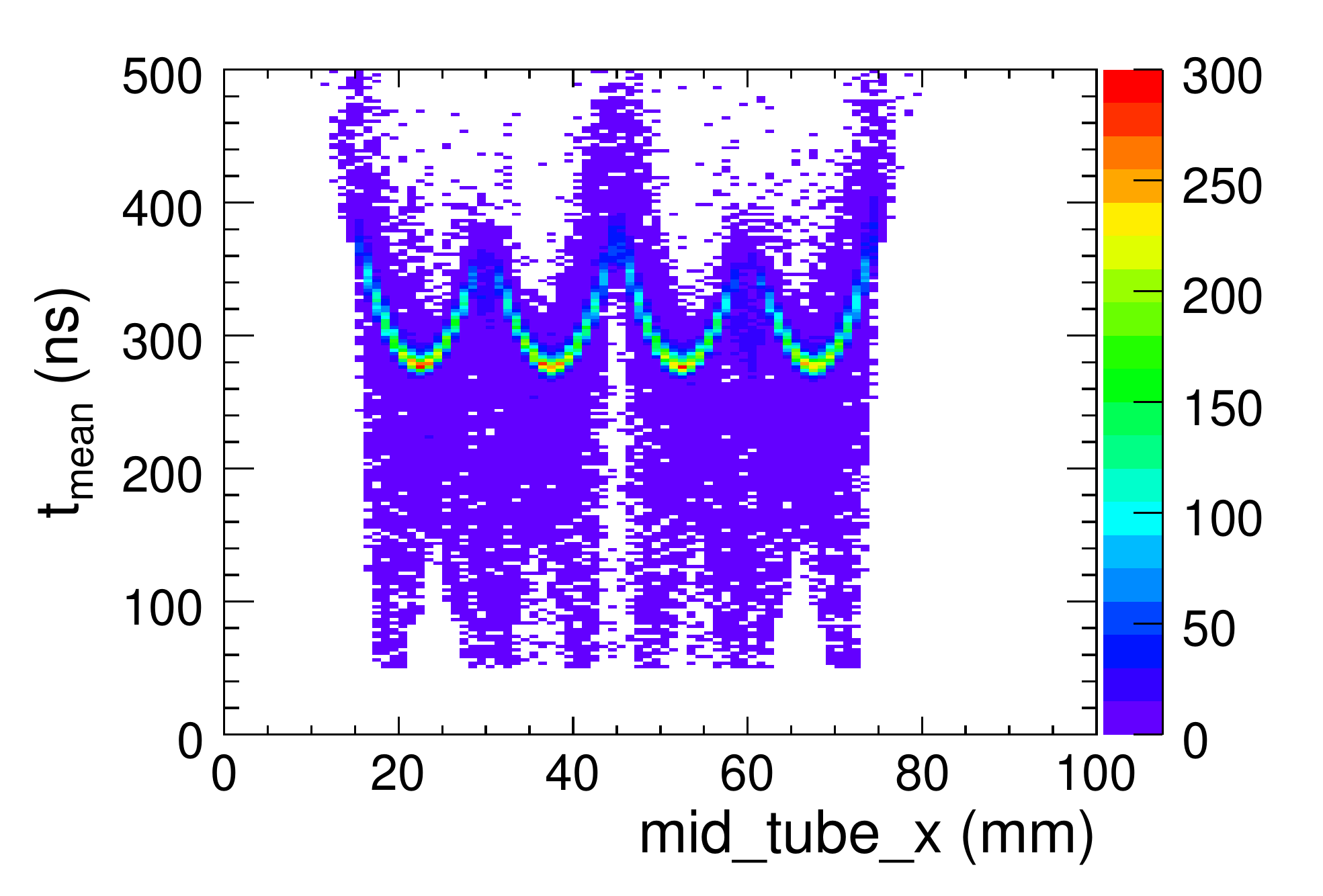}
 
\caption{Mean drift time of the three tube hits over the middle wire plane.}
 \label{fig:meandriftmid}
\end{figure}

For tubes arranged as in Fig.~\ref{fig:schem}, the mean drift time should be nearly constant or somewhat larger than the maximum drift time of a single tube, for a minimum ionizing particle traversing three tubes. Events where the mean drift time falls below the maximum drift time indicates a problem. Most likely, this is due to a secondary charged particle that has traveled close to the drift wire of one or more tubes, resulting in a reduced drift time. Figs.~\ref{fig:meandrift} and ~\ref{fig:meandriftmid} show the mean drift time distribution and mean drift time resolution for tracks traversing the middle tube plane, respectively. Cuts to the data are applied to remove the events with a mean drift time less than 250~ns. 

\begin{equation}
t_{mean} = \left(t_{row-0} + 2t_{row-1} + t_{row-2} \right)
\end{equation}

\subsection*{Minimization and Auto-calibration}

\subsubsection*{Minimization}
A track in the plane perpendicular to the wires can be written as 

\begin{equation}
\centering
 x = a\cdot{}y + b 
\label{eqn:line}
\end{equation}

The minimum distance between a line and the centre of a drift circle, or the impact parameter, is described by:

\begin{equation}
\centering
r_{i,impact} = \frac{|a\cdot{}h_{i} + b -k_{i} | }{\sqrt{1+a^{2}}}
\label{eqn:imp}
\end{equation}
where ($h_{i},k_{i}$) is the centre of the drift circle.

The parameters $a$ and $b$ are found by minimizing the $\chi{}^{2}$ function 

\begin{equation}
\centering
 \chi^2 = \sum_{i=1}^{N} \left( r_{i,impact}- r(t)_{i,drift} \right)^{2} / \sigma_{i}^{2} = \sum_{i=1}^{N} \frac{d_{i}^{2}}{\sigma_{i}^{2}}
\label{eqn:chi2}
\end{equation}

  The sum runs over the number of hits associated with the track. In this analysis N is required to be three, where there is one hit in each layer. The residual for each hit is $d_{i}$. The error associated with the residual is $\sigma_{i}$.

\subsubsection*{Auto-calibration}

The relation between the drift-time and the drift-radius, $r(t)$, of a tube is not known \textit{a priori}. It depends on several factors such as gas composition and EM-field distributions within the tube. The relationship can be derived, however, from timing information from the tubes themselves. The \textit{auto-calibration}, which will determine the $r(t)$ relation, uses data only from the tubes themselves.

The \textit{auto-calibration} method is based on the assumption that the muon tracks are straight lines that pass through the drift tubes. The assumed drift radius in the tube is then the point of closest approach to the wire of the tube. If $t_{0}$ is known for each tube, then the drift time is also known. The drift time and drift radius are related by the function $r(t)$. The auto-calibration is then an iterative process of minimizing the difference, the \textit{residual}, between the point of closest approach, or the impact parameter, and the drift radius calculated from $r(t)$.

The $r(t)$ function is a Chebyshev polynomial of the first kind as in Eq.~\ref{eqn:cheb}:

\begin{align}
r(t) &=c_{0} \times ( c_{1}t' \nonumber\\
     &+ c_{2}(2t'^{2}-1) \nonumber  \\
     &+ c_{3}(4t'^{3}-3t')  \nonumber\\	
     &+ c_{2}(8t'^{4}-8t'^{2}+1)\nonumber\\
     &+ c_{4}(16t'^{5} - 20t'^{3} +5t' ) \nonumber\\
     &+ c_{5}(32t'^{6} - 48t'^{4} +18t'^{2} -1 ) \nonumber\\
     &+ c_{6}(64t'^{7} -112t'^{5} +56t'^{3} -7t') \nonumber\\
     &+ c_{5}(128t'^{8} -256t'^{6} + 160t'^{4} -32t'^{2} +1) )\nonumber\\
     \label{eqn:cheb}
     \end{align}

where the function is constrained such that $r(t=0) = 0$. The normalized time variable is defined  $t'=t/t_{max}$. The value for $t_{max}$ was chosen to be 347~ns.  The coefficients   $c_{n}$ are the fit parameters with typical values listed below:     
     
\begin{align*}
c^{fit}_{0}&=1.47\times{}10^{+01}  \\ 
c^{fit}_{1}&=9.72\times{}10^{-01}\\
c^{fit}_{2}&=2.52\times{}10^{-01}\\
c^{fit}_{3}&=-4.27\times{}10^{-01}\\
c^{fit}_{4}&=-8.45\times{}10^{-02}\\
c^{fit}_{5}&=-1.56\times{}10^{-02}\\
c^{fit}_{6}&=2.71\times{}10^{-02}\,.\\
\end{align*}

The algorithm to minimize the residual begins with an initial $r(t)$ which is used to convert drift times to radii. My initial assumption was that the relationship was linear and that the width of the tdc distribution corresponded to the radius of the tube ($r = v\cdot{}t$). Since the test setup uses three MDT, the initial guess at the line tangent to the drift circles uses the points on a line tangent to the top and bottom drift circles which are hit ($t_{drift}>0$). This will generate four possible tangent lines. The tangent line with the smallest residual to the initial middle tube radius is chosen to be minimized with respect to the drift radii. This will sometimes cause left-right ambiguities, where the residual is nearly identical, for certain cases and must be dealt later with using other parameters.

The track line is then found by the tangent to the top and bottom tube with the minimal $\chi^{2}$ in Eq.~\ref{eqn:chi2}. The distance from the track to the middle wire is compared to the drift-time radius. The previous $r(t_{drift})$ relation is then corrected by fitting the difference, $\delta{}r$, with the Chebyshev polynomial function that is constrained so that $r(t=0) = 0$.  Repeated iterations with corrections should converge on the correct $r(t)$ such that $\delta{}r=0$ or within tolerance (10-20 microns).
The difference between drift radius and track distance is corrected to equal zero by fitting $\delta{}r$ to a second function like Eq.~\ref{eqn:cheb} such that mean value of $\delta{}r(t) \le 25~\mu{}m$. This is sometimes difficult to achieve for near the drift wire and drift tube edge, due to the poor resolution and efficiency in these regions.

\end{document}